\documentclass[aps,amsmath,amssymb,prd,showpacs,floatfix,preprint,superscriptaddress,nofootinbib,12pt]{JHEP3}
\usepackage[utf8]{inputenc}
\usepackage{amsmath}
\usepackage{graphicx}
\usepackage{color}
\usepackage{amsmath,epsfig}
\usepackage{amssymb,amsfonts}
\usepackage{latexsym}

\usepackage{tocvsec2}
\usepackage{subeqnarray}
\usepackage{xcolor}

\usepackage{graphicx}
\usepackage{longtable}
\def\nn{\nonumber}
\def\half{{1\over 2}}
\relax

\usepackage[vcentermath,enableskew]{youngtab}
\def\hri#1#2{\href{http://arxiv.org/abs/#1}{[ArXiv:#1]#2}}
\def\hre#1#2{\href{http://arxiv.org/abs/#1/#2}{[ArXiv:#1/#2]}}

\def\hree#1#2{\href{https://doi.org/#1}{#2}}
\def\hrj#1#2{\href{www.doi.org/#1}{#2}}
\relax
\renewcommand{\theequation}{\arabic{section}.\arabic{equation}}

\def\be{\begin{equation}}
\def\ee{\end{equation}}

\newcommand{\bear}{\begin{eqnarray}}
\newcommand{\bea}{\begin{eqnarray}}
\newcommand{\eear}{\end{eqnarray}}
\newcommand{\eea}{\end{eqnarray}}
\def\hre#1#2{\href{http://arxiv.org/abs/#1/#2}{[ArXiv:#1/#2]}}

\newbox\pippobox

\renewcommand{\b}[1]{\textbf{#1}}

\def\II{\relax{\rm I\kern-.18em I}}
\def\mt{{\tilde m}}

\def\e{\epsilon}
\def\m{\mu}
\def\n{\nu}
\def\r{\rho}
\def\s{\sigma}
\def\pa{\partial}
\def\t{\theta}
\def\sp{\;\;\;,\;\;\;}

\def\p{\partial}

\def\a{\alpha}
\def\b{\beta}

\def\tr{\ensuremath{\mathrm{Tr}}}

\def\l{\lambda}
\def\g{\gamma}
\def\d{\delta}

\def\II{{\cal I}}
\def\JJ{{\cal J}}

\def\LL{{\cal L}}

\def\NN{{\cal N}}
\def\OO{{\cal O}}
\def\PP{{\cal P}}

\def\WW{{\cal W}}

\def\cc{{\mathfrak c}}

\newcommand{\h}[1]{{\widehat{\rm #1}}}
\newcommand{\vis}[1]{{\rm{#1}}}
\newcommand{\viss}[1]{{\bf{#1}}}
\def\mt{{\mathcal T}}

\def\pmet{{\mathfrak{h}}}
\newcommand{\shift}[1]{{\mathfrak{#1}}}

\def\bg{{\bf g}}

\def\th{{\tilde h}}
\def\ar{\Rightarrow}

\title{Emergent gravity from hidden sectors and TT deformations}
\author{P. Betzios$^{\natural}$, E. Kiritsis$^{\flat,\natural}$, V. Niarchos$^{\flat,\dagger}$\\
~\\
$^\natural$ \href{http://www.apc.univ-paris7.fr}{APC, AstroParticule et Cosmologie}, Universit\'e Paris Diderot, CNRS/IN2P3, CEA/IRFU,
Observatoire de Paris, Sorbonne Paris Cit\'e,\\
 10, rue Alice Domon et L\'eonie Duquet, 75205 Paris
Cedex 13, France.

~\\
$^\flat$ \href{http://hep.physics.uoc.gr}{Crete Center for Theoretical Physics}, Institute for Theoretical and Computational Physics,
Department of Physics\\
University of Crete, Heraklion, Greece

~\\
$^\dagger$ \href{https://www.dur.ac.uk/cpt/}{Department of Mathematical Sciences and Centre for Particle Theory},\\
Durham University, Durham DH1 3LE, UK
}

\preprint{CCTP-2020-11\\
ITCP-2020/11\\
DCPT-20/15}
\abstract{We investigate emergent gravity extending the paradigm of the AdS/CFT correspondence. The emergent graviton is associated to the (dynamical) expectation value of the energy-momentum tensor. We derive the general effective description of such dynamics, and apply it to the case where a hidden theory generates gravity that is coupled to the Standard Model. In the linearized description, generically, such gravity is massive with the presence of an extra scalar degree of freedom. The propagators of both the spin-two and spin-zero modes are positive and well defined.
The associated emergent gravitational theory is a bi-gravity theory, as is (secretly) the case in holography. The background metric on which the QFTs are defined, plays the role of dark energy and the emergent theory has always as a solution the original background metric. In the case where the hidden theory is holographic, the overall description yields a higher-dimensional bulk theory coupled to a brane. The effective graviton on the brane has four-dimensional characteristics both in the UV and IR and is always massive.}

\begin{document}

\section{Introduction and results}
\label{intro}

There is a widespread belief among experts that gravity remains todate the weak link in our attempts to understand the universe.
On one hand, in many contexts (cosmology, black holes),  gravity and quantum mechanics seem to give results that violently clash against each other.
On the other hand, in all the cases in which only one, or the other of these frameworks seems relevant (as  in gravitational physics in the solar system, or particle physics at terrestrial experiments) both theories pass tests with flying colors.

Both gravity and the gauge theories that describe all the known interactions, are theories that rely on local symmetries. It is not understood why this is so. It has been  advocated  that this fact may  not be an accident, \cite{nielsen}.
However, the similarity stops at that point: gauge theories can be UV complete theories\footnote{Modulo complications with interactions mediated by scalars.},
while the theory of Einstein gravity is non-renormalizable.

In the case of gauge theories, we can say more. If we decree that experiment commands vector fields to mediate interactions, and that vector fields are good dynamical variables all the way to the UV (i.e.. weakly coupled), then we know that their interactions must be controlled by a gauge theory (possibly spontaneously broken). Sometimes, a similar  argument can also be made for composite low-energy vectors, as is the case near  the lower end of the conformal window in N=1 sQCD, \cite{Komar}.

Many different proposals have been put forward in the past in an effort to overcome the UV problems of gravity and which are reviewed in \cite{Sindoni,Carlip,Dreyer}.
One particular set of ideas involves the concept of ``emergent gravity" which is quite vague and has many different incarnations, \cite{ohanian}-\cite{Carone1}.
This is to be contrasted with the concept of induced gravity pioneered by Sakharov, \cite{Sakharov} and reviewed in \cite{Adler,Visser}.
In this second line of thought, the graviton is a priori a dynamical field that is coupled to ``matter". It is however the  matter that determines the gravitational action via Wilsonian renormalization and quantum effects.

It is reasonable to have a definition of emergent gravity as the theory where the graviton is a composite of the ``fundamental" fields of the theory. This definition can become confusing when dualities are at play. Indeed, dualities relate two theories written in terms of different ``fundamental" degrees of freedom. However, in most cases this is a good working definition.

Therefore from this point of view, and in contrast with induced gravity,  the emergent graviton is not a fundamental field at high energy, but only an effective low-energy field.
 In many cases, it emerges in the context of standard relativistic  quantum field theory (QFT). There are however non-relativistic approaches, typically inspired by condensed matter theories with or without an explicit cutoff, where a low-energy graviton is claimed to emerge, \cite{volovik}-\cite{SSLee}.

Perturbative string theory, is a case that shares some features of the emergence phenomenon. The standard NSR description of the theory has dynamical fields (the coordinates of the string) that do not involve directly the space-time metric.  The theory however, has explicit invariance under space-time diffeomorphisms\footnote{And world-sheet diffeomorphisms as well.}. In retrospect, a dynamical space-time metric (and many other fields) emerge in the theory, making it a quantum theory of gravity.

String theory, at the same time does something,  other approaches to quantizing gravity cannot do. It provides a perturbatively well-defined theory of quantum gravity that is semiclassical\footnote{In the limit of weak coupling, where we know how to handle the theory.}.

String theory\footnote{We consider string theory liberally,  as defined by conformal field theories in up to six dimensions, \cite{love,novel}.} provides a quantized theory of gravity at a cost: it introduces an infinite number of ``new" degrees of freedom, the excited states of the string and a scale, the string scale $M_s\sim \ell_s^{-1}$ at which these new states become effective, and resolve some of the problems of the  gravitational and other interactions.

The theory of strings is UV-finite but not UV-complete. The structure of closed string  perturbation theory suggests a theory with a smart cutoff at the string scale, containing therefore no UV divergences (but plenty of IR divergences), \cite{book}.
 Perturbative string theories are well-defined for energies well below the Planck scale, but
ill-defined at energy transfers at or above the Planck scale where string perturbation theory breaks down, \cite{gross,vene2}\footnote{This state of affairs changes when we consider non-perturbative definitions of the theory, using the AdS/CFT correspondence.}.

The presence of a non-renormalizable interaction in an effective QFT is not new in the field. We have two well-known, and historically important cases, whose resolution taught us interesting lessons.
The first is the Fermi interaction. Its non-renormalizability  was resolved by introducing new degrees of freedom (the W and Z bosons) and a related mass scale.
In a sense, for gravity, string theory comes somewhat close (but is not identical) to this paradigm.
It introduces new degrees of freedom and ``resolves" the gravitational interaction.
The new scale, $M_s$, is related to the Planck scale $M_P$ of the non-renormalizable gravitational interaction as $M_s^8=g_s^2 M_P^8$.
In the perturbative formulation therefore we always find $M_s\ll M_P$.

Another idea that is in a sense similar, is the approach of quantizing gravity starting with an $R^2$ Lagrangian, \cite{Stelle,Tomboulis,FKL}.
In that case, renormalizability is ensured by a new degree of freedom that is a ghost. Recently, modifications of this formalism have been proposed, \cite{strumia,anselmi}, but it is not clear yet, what is their interpretation.

The other example is QCD. Here, at low energy we have a non-renormalizable IR-free theory of pions. Its characteristic scale is the pion decay constant, $f_{\pi}$, and it is analogous to the Planck scale in the gravity example.
The effective theory that governs the pions is, much like gravity, mostly dictated by symmetries and their breaking. Here the symmetry is chiral symmetry. This effective theory is non-renormalizable, and quantizing it, introduces an a priori infinite set of counter terms that are cutoff dependent.
The quantum effects in chiral perturbation theory were studied over several decades with the result that as long as one asks questions for energies $\lesssim 1 $ GeV, the characteristic energy of strong interactions, the low-energy theory gives sensible results (that do however depend on the cutoff).

In this case, we know that the way to move up in energy, does not come via adding new fundamental fields to the low-energy fields (the pions)\footnote{There is a different general view in this case that was proposed in \cite{Dvali}.}.
There is a complete reorganization of the dynamical degrees of freedom of the theory: at high energy, the degrees of freedom are weakly interacting gluons and quarks, while at low-energy the quarks and gluons are tightly bound into hadrons including pions. The transition between the two descriptions is beyond analytical reach, although a sort of perturbation theory applies both in the UV and the IR, albeit to different degrees of freedom.
It is however only the UV formulation in terms of quarks and gluons that is amenable to first principles, numerical (lattice) computations.

Therefore,  the QCD paradigm of UV-resolutions suggests that a similar phenomenon might also happen in gravity.
The related  idea of composite  gravity is not new, and several approaches were attempted in the direction of making the graviton a composite of more elementary fields,  that would be part of a conventional QFT with good high-energy behaviour.

It has long been held as an analogy, to compare also with hydrodynamics.
This analogy is similar to the QCD paradigm mentioned above but may be closer in spirit to gravity. Hydrodynamics is a very successful, non-linear and dissipative effective theory, applicable to a host of quantum  theories, with dynamical variables that describe collective degrees of freedom. Viewed as a quantum theory, it is non-renormalizable (and also dissipative, i.e. non-unitary). It is clear that  the proper quantum theory that would extend the validity of the theory in the UV must involve different, more fundamental degrees of freedom, the nature of which depends on the UV details
of the system in question.\footnote{This is contrary to recent approaches in defining quantum versions of hydrodynamics, motivated by the AdS/CFT correspondence and its avatars, \cite{liu,rangamani}. However, these approaches attempt to define leading quantum effects for hydrodynamic theories and not a full UV resolution. }
The similarity with quantizing gravity and hydrodynamics does {not} stop there. Gravity indeed has many of the dissipative features of hydrodynamics, associated to the presence of horizons, a fact that has provided a deep connection between gravity and hydrodynamics in the context of holography, \cite{shiraz} and blackfolds, \cite{black}.

In composite approaches to gravity, the major difficulties lie in producing a diffeomorphism invariant theory, as well as providing a graviton that has dynamics at low energy in the quantum theory.
The past popularity of such attempts has also motivated the well-known Weinberg-Witten (WW) Theorem, \cite{WW} that provides strong constraints on composite gravitons (and composite gauge bosons).
Under a set of assumptions that include Lorentz invariance, well-defined particle states, a conserved covariant energy momentum tensor and a Lorentz covariant and gauge-invariant conserved global currents,  the theorem excludes

$\bullet$ massless particles of spin $s>{1\over 2}$ that can couple to a conserved current, and

$\bullet$ massless particles with spin $s>1$ that can couple to the energy momentum tensor.

Its assumptions however, allow for several loop-holes that help evading the theorem in known cases.
In particular, in does not exclude Yang-Mills theory, as in that case the global conserved currents are not gauge-invariant.
It also does not exclude a ``fundamental" graviton coupled to matter, as  the Lorentz-covariant energy-momentum tensor is not conserved (but only covariantly conserved).

Another counterexample is presented by the massless $\rho$-mesons at the lower-end of the conformal window of N=1 sQCD, \cite{Komar}.
These are the dual magnetic gauge bosons of Seiberg, \cite{seiberg}.
They evade the WW theorem because of the emergent gauge invariance, associated to them.
As we shall see in the sequel, there are other ways of bypassing the WW theorem. In particular Lorentz invariance is crucial. The notion of masslessness changes in the absence of Lorentz invariance and it is quite distinct even in spaces other than flat space but with high symmetry as de Sitter or Anti-de Sitter.

In any relativistic quantum field theory, there is at least one state with the quantum number of a graviton. It is the state generated out of the vacuum by the action of the (conserved) energy-momentum tensor. In weakly-coupled theories this state is unique (and it is a multiparticle state), while in strongly coupled theories it is a linear combination of one-particle states that are  generated by the action of the energy-momentum tensor. If a theory possesses a  large-N, strong coupling limit, then the width of such states vanishes and there is an infinite number of them.

In weakly-coupled theories this is a multi-particle state and therefore its effective interactions are expected to be non-local. In the opposite case, where the interactions are strong, we expect such a state to be tightly bound. If its ``size" is $L$, then we might hope that at distances $\gg L$ the effective interactions of such a state may generate gravity.

In particular, in a theory with an infinite coupling, we expect to have an emergent graviton with a point-like structure.
If the theory is not conformal, then we expect a discrete spectrum of such states, associated to the (generically complex) poles of the two-point function of the energy -momentum tensor. In a generic strongly-coupled theory, most such states will be unstable. In YM, to pick a concrete example, such states are in the trajectory of the $2^{++}$ glueball. The lightest such state is massive, and is unstable to decay to two $0^{++}$ scalar glueballs. The higher $2^{++}$ states are more {massive} and have larger decay widths.
If instead the strongly-coupled theory is conformal, then the spectrum of graviton bound-states forms a continuum.

The notion of a bound state graviton has been realized in a rather convincing way using the paradigm of the AdS/CFT correspondence, \cite{malda}.
The gauge-invariant operators associated to the bulk string/gravity description are composites of the N=4 sYM (super)gluons.
The major surprises that came together with Maldacena's conjecture, involved the emergence of extra dimensions in the string theory dual and the fact that the theory has a higher-dimensional diffeomorphism invariance\footnote{Another surprise for large-N practitioners was that the string equivalence  was expected for YM-like theories,  which are confining in the conventional sense but not for theories in a Coulomb phase as N=4 sYM.}.

The set of states generated by the sYM stress tensor out of the vacuum, can be organized into a ten-dimensional ``massless" graviton. Although the presence of emergent dimensions in holographic theories can be now  qualitatively understood\footnote{Emergent dimensions are associated with eigenvalue distributions of adjoint matrix fields in the large-N limit, \cite{stein}. Symmetry plays a role on both the number of emergent dimensions and the geometry they span, \cite{bere,kir}.} the emergence of a higher-dimensional diffeomorphism invariance remains a mystery.

The masslessness of the higher-dimensional graviton is explained by the conservation of the energy-momentum tensor of the dual QFT. Once this conservation is violated, the energy-momentum tensor obtains an anomalous dimension and the bulk graviton a mass\footnote{See also \cite{porrati} for an earlier realization in terms of modified AdS boundary conditions.} , \cite{e1,a1}. Therefore the ``masslessness" of the higher-dimensional graviton is the avatar of diffeomorphism invariance. Diffeomorphism invariance in turn is the avatar of the translational invariance of the dual QFT and the associated conservation of the energy-momentum tensor. This does not however imply that the four-dimensional graviton is massless.
Indeed, in N=4 sYM defined on Minkowski space we obtain a continuum of spin-two states starting at zero mass\footnote{This is the reason this theory evades the WW theorem which assumes among other things a isolated bound-state. The WW theorem involves a subtle limit to define the helicity amplitudes
that determine the couplings of massless states to the stress tensor or a
local current.  This limiting procedure is not valid in theories where the states form a continuum.}.
If instead we define the theory on $R\times S^3$, then  the graviton spectrum is discrete, but the theory has lost Lorentz invariance.

In non-conformal YM-like theories, like Witten's non-supersymmetric D$_4$ theory, \cite{d4} or IHQCD, \cite{ihqcd}, the graviton is massless in 5 dimensions but the spectrum of the four-dimensional gravitons (ie. the $2^{++}$ glueballs) is massive and discrete. This is analogous to the Higgs effect and is due to the non-trivial gravitational background.
In \cite{nitti2} it was attempted to make the lowest spin-two state  massless  in an asymptotically AdS setup with a negative result.

 The very strong interactions in a holographic theory are responsible for the gauge-theory bound-states being tightly bound and their effective theory being local in the emergent dimensions. At finite coupling, stringy effects become important and the interactions delocalize.
The large N limit is important in order for the bound-states to interact weakly. In a theory with finite N, $M_s\sim M_P$, and gravity and other interactions are strong.

We therefore learn from the holographic duality that,

$\bullet$ Strong coupling in QFT makes composite gravitons tightly-bound states.

$\bullet$  Large N makes gravitons weakly interacting.

Both properties are essential in obtaining a semiclassical and local theory of (composite) quantum gravity.

The AdS/CFT correspondence has also given us a non-perturbative definition of string theory, via the dual QFT. For standard dual  QFTs that are  holographic (strong coupling+large N), then the dual string theory is weakly-coupled and semiclassical. However in principle, this definition adresses questions in string theory that go beyond its perturbation theory. In a sense, it is a non-perturbative definition. This definition is generalizable to different dimensions and using different types of CFTs, \cite{novel}.

The AdS/CFT intuition therefore suggests that semiclassical effective gravity with composite gravitons is expected to emerge from {certain types of } holographic quantum field theories, \cite{Seiberg,Berenstein,Nitti,SMGRAV}.

If we are to describe the gravity we observe in terms of semiclassical composite graviton that is (nearly) massless, it is clear that we must seek its origin outside the Standard Model (SM) of particle interactions.
The simplest way\footnote{There may be more exotic variations on this theme. The SM could be part of the semiclassical holographic theory, and its elementary fields, to be composites of more elementary fields. Or it could be that parts of the SM are composite and others elementary. Crude holographic translations of these possibilities have been considered in the past in the context of the RS realizations of the SM, \cite{csaki}.} is to postulate that, \cite{SMGRAV},

$\bullet$ The whole of physics is described by a four-dimensional QFT.

$\bullet$ The total UV QFT contains a possibly holographic part, that is distinct from the (UV limit of the) SM. We shall call it the ``hidden" (holographic) theory.

$\bullet$ This hidden theory is coupled in the UV to the standard model via a messenger sector. It consists of fields transforming as bi-fundamentals under the gauge group of the ``hidden" theory, and the gauge group of the SM.
We call these fields the gravitational messengers. Their mass M  is assumed to be much larger than any of the SM scales.

$\bullet$ At energies $\ll M$ we may integrate out the gravitational messengers and we end up with an effective theory consisting of the SM coupled to the hidden holographic theory, via irrelevant interactions\footnote{There is one possible exception to this statement and it is connected to the gauge hierarchy problem, as it involves couplings to the Higgs mass term, the only relevant coupling of the SM.}.

$\bullet$ Although all operators of the hidden theory are coupled weakly at low energies to the SM, the SM quantum corrections generate ${\cal O}(M)$ masses for all of them with a few notable exceptions that are protected by symmetries: the graviton, the universal axion, \cite{axion}, and exactly conserved global currents, \cite{u12}.

Some of the relevant issues of this rather general setup have been discussed in \cite{SMGRAV}.
In this paper, we shall undertake a closer look at the emergence of gravity and its associated symmetries, in QFTs.

\subsection{Results}

We start by setting up  the effective description of an (emergent)  graviton associated to the energy-momentum tensor of a {single} QFT, in section \ref{simple}. The starting point is the renormalized and diffeomorphism-invariant Schwinger functional associated to a general background metric source.  We then introduce a suitably-defined effective action for the emergent graviton that is essentially capturing the 1PI vertices of the energy-momentum tensor of the theory. The dynamical emergent graviton is the classical field (expectation value) associated to the energy-momentum tensor.

We then describe the structure of this effective action using two approximations:

\begin{itemize}

\item In the first approximation, we consider the linearized Schwinger functional up to quadratic order in sources, capturing the one- and two-point functions of the energy-momentum tensor (to all orders in the derivative expansion).
    We then construct the effective graviton action that involves essentially the inverse of the two-point function.

From general properties of the spectral decomposition of the two-point function of the energy-momentum tensor, we can show that the emergent gravitational interaction is transmitted by a (generically massive) spin-two field as well as by a (generically massive) scalar. Both degrees of freedom have positive kinetic terms and non-negative masses$^2$ if the QFT is unitary.

\item In the second approximation, we consider a theory with a mass gap, and we parametrize the Schwinger functional in the IR, using a derivative expansion, keeping the full non-linear diffeomorphism invariance. We then construct the effective action for the emergent graviton, which gives rise to a bi-gravity theory. The dynamical metric is the emergent graviton field, while the fixed background metric is the fiducial metric on which the QFT is defined. The effective action has an overall (non-linear) diffeomorphism invariance that transforms both the dynamical and the background metric.

    The emergent graviton satisfies at the two-derivative level an Einstein-like equation, (\ref{f9}). The linearized analysis indicates that the interaction is mediated by a generically massive spin-two field as well as a scalar in (\ref{c42bbis}). The scalar is always ghostlike\footnote{This is always a feature of an Einstein description of a non-fine tuned bi-gravity theory, \cite{Hi,dR}.} while either the scalar or the spin-two graviton are tachyonic.

The discrepancy in this description, compared to the previous linearized approach (that is robust) is traced to the fact that truncating a derivative expansion in the Schwinger functional, and {then} constructing  the effective action mixes contact terms with pole terms and gives the wrong results for the proper residues and masses of the propagating fields.

\end{itemize}

We then proceed to consider a ``hidden" theory,  coupled to the Standard Model (SM) at some high scale $M$, via non-renormalizable interactions of the form
\be
\label{intr}
S_{int} =\lambda \int d^4 x\, \left( \vis{T}_{\mu\nu}(x)\, \h{T}^{\mu\nu} (x) +  \mathfrak{c} ~\vis{T}(x) \h{T}(x) \right)\sp \l\sim {1\over M^4}
~.
\ee
along the lines advocated in \cite{SMGRAV}. $\h{T}^{\mu\nu}$ is the energy-momentum tensor of the hidden theory while $\vis{T}_{\mu\nu}$ is the SM energy-momentum tensor. Both theories are defined on flat Minkowski space and $\vis{T}$, $\h{T}$ are the traces of the two energy-momentum tensors using the Minkowski metric.

We show that at the linearized level, the coupling of the two theories induces an extra interaction between energy-momentum tensors of the SM, mediated by an emergent metric. This metric  is associated to the expectation value of the hidden energy-momentum tensor.

We study this emergent gravitational interaction between SM sources using various (linearized) approximations. We assume for this that the hidden theory is either a perturbative theory, or even better,  a large N theory.

\begin{itemize}

\item The linearized effective action of the emergent metric has as kinetic term the inverse of the hidden theory energy-momentum tensor two-point function.

\item For the two-point function to be invertible, a nonzero vev for the energy momentum tensor is necessary. Otherwise the two-point function must be inverted on the space orthogonal to the zero modes and the description becomes non-local.  It is not clear if this is related to the issues discussed in \cite{Marolf}.

\item Using the general spectral representation of the two-point function we show that here as well, the new interaction given in (\ref{tt1}),  is mediated (up to contact terms) by a generically-massive spin-two field whose spectral density is the spin-2 spectral density of the hidden energy-momentum tensor.

    Similarly, there is also an extra scalar contribution arising from the trace of the energy-momentum tensor and is proportional to the spin-0 spectral density. Both contributions have positive norm and positive masses if the hidden theory is unitary.

Interestingly, if the $TT$ coupling in (\ref{intr}) is of the special type used in two-dimensions, \cite{TT,To}, and which in four-dimensions corresponds to, \cite{Tay},
\be
\mathfrak{c}=-{1\over 3}
\ee
 then the scalar contribution to the emergent gravitational force vanishes.

\item Expanding the energy-momentum tensor in momentum space to order $k^2$, we construct the linearized effective action for the emergent graviton. By a field redefinition (rotation and shift) of the emergent graviton field and by a constant shift of the SM energy momentum tensor, we can map the effective action, at the two-derivative level, to (linearized) Einstein gravity coupled to a non-zero cosmological constant, and to the (shifted) SM energy-momentum tensor.

    The shift in the SM energy-momentum tensor appears as a ``dark energy" whose origin is the hidden theory.
The parameters of the gravitational theory, the emergent Planck scale $M_{P}$, the cosmological constant $\Lambda$ and the ``dark energy" $\Lambda_{dark}$,  are given in terms of the input data, the high scale $M$, the number of colors  N of the hidden theory and the mass scale $m$ of the hidden theory, as in (\ref{eq1}), (\ref{eq11}) that we summarize below,
\be
{\Lambda\over M_P^2}\sim \left({m\over M}\right)^8\sp
{\Lambda_{dark}\over M_P^2}\sim max\Big\{N{m^4\over M^4}, 1\Big\}\left({m\over M}\right)^8
~,
\ee
\be
 {\Lambda_{dark}\over \Lambda}\sim max\Big\{N{m^4\over M^4}, 1\Big\}\sp {\Lambda\over M^2}\sim {m^2\over M^2}\sp  {M\over M_P}=\left({m\over M}\right)^3\;.
\ee
When $m\ll M$ both $\Lambda, \Lambda_{dark}\ll M_P$.

Here, the emergent Planck scale, $M_P$ in this approximation is determined by the ${\cal O}(k^2)$ contact term  $\h{b}_0$ in the energy-momentum tensor two-point function, (\ref{linearam}), whose sign is not (obviously) controlled by the unitarity of the spectral densities, as analyzed in \ref{momissues}.
$\h{b}_0$ should be negative for a positive Planck scale, and this is the result we find in the calculation of the relevant coefficient in theories of free bosons and fermions.

However, as in the case of the single theory, the effective action thus constructed in the momentum expansion,  differs from the one obtained
from the massive poles of the energy-momentum two-point function and which is controlled by the residues of the poles.

\item The linearized description of emergent gravity can be extended to a fully non-linear description using a generalization of the effective action in section  \ref{secNL}. In section \ref{expli} this effective action is analyzed in the derivative expansion and the associated Einstein-like  equations are derived. It is shown how the energy-momentum conservation of the combined hidden+SM theory in flat space,  transforms into the covariant conservation of the total energy-momentum tensor in the emergent metric.

    \item  In all formulations, there is always a solution to the emergent gravity equations that is equal to the background metric on which the original QFTs are defined. When both the hidden QFT and the SM are defined on the flat Minkowski metric, this implies that the flat metric is always a solution of the emergent gravity equations, if the combined theory is in its ground state. In particular in our linearized analysis of section \ref{linear}, despite the fact that the quadratic theory is a theory with an effective cosmological constant, the presence of the dark energy addition to the SM energy-momentum tensor makes the flat metric a solution.

          This fact indicates that the standard form of the cosmological constant problem does not exist. Here the flat metric being a solution to the emergent gravity equations, is correlated with its QFT progenitor being defined on a flat metric. Therefore, the existence of a flat metric solution in emergent gravity becomes a technically natural problem. However, this does not necessarily mean that the cosmological constant problem is innocuous in a cosmological time dependent setting.

\item It is a well known fact that in a theory of gravity there is no conserved energy in the bulk. This fact is evident in our formulation, as gravity emerges from a hidden theory interacting with the SM and therefore the SM energy is not conserved. However, in a gravitational theory, there is typically a conserved energy associated to asymptotic boundaries. This is well known to be the case for asymptotically flat as well as asymptotically AdS spacetimes. In our setup this can be understood as follows: The asymptotics of the metric are determined by the ``vacuum" of the combined theory and in such a case the asymptotic metric is that of the background QFT. In such a case there are no hidden theory sources asymptotically and therefore in this regime the SM energy is conserved. We have also shown that the conservation of the total energy in the fiducial QFT metric transforms to the ``Bianchi identity" of the emergent gravity equation which is a requirement for any consistent theory of gravity. In the case where the hidden theory is a holographic theory the situation is similar but even simpler to discern. If the hidden theory is in its ground state then the bulk solution is sliced by flat slices and the SM brane embedding gives it a flat metric. This remains true if the brane can slide along the holographic direction and this is the basis of the holographic self-tuning mechanism,  \cite{CKN}.
Interestingly, if the hidden theory is in the (translationally invariant) thermal state, the metric on the brane is still flat but the speed of light is renormalized, \cite{Betzios:2020zaj}.

\item We did not find a way to bridge the Einstein gravity plus cosmological constant description {with the linearized description containing the complete two-point function of the hidden theory}. {In particular, we cannot map the effective action to a linearisation of some of the ghost-free massive gravity actions }advocated in four dimensions, \cite{dRGT}. This leads us to the following important conclusion: We understand that even though the non-linear Einstein gravity plus cosmological constant description is a good description in the extreme IR as shown using a derivative expansion both in the linearised and the non-linear description, nevertheless it needs to be corrected with additional states coming from the poles of the two point function of the hidden theory energy-momentum  tensor (and non-linear effects coming from higher point correlation functions). Whilst we can show that the effective description capturing energy exchanges at the scale of such poles is that of ghost-free massive gravity, it is clear that these two descriptions have a different regime of validity. What we are currently missing, is a single non-linear effective gravitational action that can bridge both such descriptions with a wide regime of validity of energies ranging from the extreme IR up to the messenger scale $M$.

\end{itemize}

We finally study the special case where the large N, ``hidden" theory is a holographic theory in section \ref{holo}.

\begin{itemize}

\item We show that our setup of a holographic ``hidden theory"+SM is mapped to a holographic bulk (with Neumann boundary conditions)  plus a SM-brane in the gravitational description. Allowing this brane to move in the radial directions, this setup was recently studied in detail in \cite{CKN,GKNW,self-cosmo,HKNW}.

\item We have studied the quadratic action of the spin-two graviton on the SM-brane. The gravitational interaction is four-dimensional both at short and long distances. The short distance benchmark (DGP scale) is set by the size of the induced Einstein term on the brane, as in DGP, \cite{DGP}.
    The long distance benchmark is set by the typical curvature scale in the bulk as in AdS. This is qualitatively similar to brane-induced gravity in AdS, \cite{KTT}.
In the intermediate region, the gravitational interaction is five-dimensional. If the bulk curvature scale is smaller than the DGP scale, gravity is four-dimensional at all distances.

\item The graviton on the brane is always massive. Its mass is due to a resonance as in the DGP setup. The effective four-dimensional Planck scale and mass are given in (\ref{bb17}) and depend on bulk data (N, the scale of the bulk theory $m$ and the IR expansion of the bulk-to-bulk propagator at the position of the brane) as well as brane data (the induced Planck scale on the brane, $M$, and brane cosmological constant $\Lambda_4$). The graviton mass of the DGP pole, modulo self-tuning issues, is of natural order $1/N$ and therefore can be made arbitrarily small by making $N$ arbitrarily large.

\item There are also two generic scalar modes generating extra interactions on the brane, and their full analysis has been performed in \cite{CKN} with the following results.

There is a special scalar mode, that couples to the trace of the energy-momentum tensor on the brane. It is a linear combination of the bulk scalar mode that couples to the trace of the energy momentum tensor of the hidden holographic theory, and the brane-bending mode.
This mode has couplings that are (generically) parametrically of the same order as the effective gravitational coupling. The mass is also generically of the same order as the graviton mass. It is the spin-zero mode that was found in the general case before.

All other scalar modes (except axions) have effective masses that are parametrically much higher.

\end{itemize}

\subsection{Outlook}

Our analysis puts the intuition of emergent gravity as understood from the AdS/CFT paradigm, in a more general context.
It exemplifies how gravity is generated in general, and connects various earlier attempts, along similar lines.

Our setup exhibits many interesting features and these are discussed in section \ref{disc}.
However, it leaves several key questions unanswered.

\begin{itemize}

\item The mass spectrum of the spin-two part of the two-point function of the energy momentum tensor must be appropriate, in order to agree with observational constraints. The non-linearities must set in at the appropriate scale to diffuse the vDVZ discontinuity, \cite{vDVZ}. It is not clear if bootstrap (S-matrix) constraints in QFT allow such spectra.
    In any case, the hidden theory, in this context, must be most probably a large N theory with a very low characteristic scale.

\item The emergent gravitational interaction comes almost always  packaged with the scalar interaction associated with the trace of the energy-momentum tensor.  This generically provides for a gross violation of the equivalence principle and corresponds to the dilaton interaction in string theory.  Ways to avoid this are necessary and may include weaker interactions for the scalar, and/or heavier masses.
    This issue is not new and possible resolutions  have been discussed in the context of string theory, \cite{Damour}.
    Again S-matrix bootstrap constraints may be crucial to ascertain whether such a possibility is viable.
Interestingly, if the hidden theory has the special TT coupling studied in two-dimensions, the scalar interaction decouples.

\item  The dilaton interaction is associated with the trace part of the energy-momentum tensor. Experience from cosmology and holography suggests that  this is a special field that shares a non-trivial relationship to the conservation of the energy-momentum tensor and the conformal anomaly. In holography it can be shown that in QFTs with a single scale, its effective potential can be uniquely determined and is related to the UV conformal anomaly, \cite{Migdal,Li}. This is probably the case also in multi-scale theories.

\item Although there are reliable ways of studying the linearized gravitational interaction and also its non-linearities perturbatively, to all orders in the momentum expansion,  the quantitative study of the non-linear theory seems problematic. Attempts to study it in the derivative expansion are marred by mixing of pole terms with contact terms and lead to problematic non-linear effective bi-gravity theories. A better idea is needed in order to quantitatively study the non-linear theory.

\item This is even more important as the most puzzling features of standard gravity involve the non-linear effects and black holes, or more generally horizons.  In emergent gravity, we have a good reason to believe that some of the puzzles of {quantum} gravity will be resolved \`a la AdS/CFT, although  so far this resolution is difficult to be justified on the gravity side.

The emergent gravity picture {indicates that} the degrees of freedom hidden inside black holes, could be associated to the (large number) of degrees of freedom of the hidden theory.

In holographic theories, this translates to brane-world black holes,  a topic studied since 20 years ago, \cite{Emparan}, but where progress in four-dimensions is still numerical, \cite{Figueras}.

\item A related question is the interpretation of the black hole solution of GR in the context of emergent gravity. A collection of masses will collapse in emergent gravity, as we can follow in the linearized computation. The non-linear dynamics are difficult to ascertain, but we have a certain puzzle that emerges from massive graviton theories.
    In such theories static black holes do not seem possible due to the non-zero graviton mass\footnote{Unless they have a null apparent horizon.}, \cite{rosen}.
However, a graviton with a cosmological size mass, may make the decay time of such black holes extremely long.
The horizon in this context would appear as a caustic or a vanishing of the hidden energy-momentum tensor expectation value. However, this phenomenon must be explained by the non-linear theory.

There is another point of view however in holography that seems to tell a different story. Brane-bulk holographic systems like the ones we are discussing here seem to have stable black hole solutions, although in some of them the effective graviton on the brane is massive via a variant of the DGP mechanism, \cite{KR}. In such systems brane black hole solutions can be constructed that are static. In three (brane) dimensions such solutions are analytic, \cite{Emparan}, while in four dimensions they are only numerical, \cite{Figueras}.
It is interesting to understand this issue further and the differences if any between the holographic case and the non-holographic case.

Hawking evaporation, as is well known, would be the avatar of detailed balance of the many-body quantum state {of the combined system} associated to the black hole. {Recent works have also indicated how the Page curve for the fine-grained entropy during the black hole evaporation process can be obtained in systems comprised out of two sectors, that are specific examples of our general setup, } for more details see the review~\cite{Almheiri:2020cfm}.

 \item Another issue worth mentioning concerns the presence and fate of global symmetries in emergent gravity.

In the context of a single QFT generating gravity, in the generic case global symmetries are present. However the generic case also does not provide for a semiclassical and weakly-coupled gravity. We expect this to happen only when the QFT is holographic.

In that case, the gravitational theory is higher-dimensional and we know that global symmetries turn to bulk gauge symmetries in this description. In the context of a hidden theory coupled to the SM, we may make similar comments with global symmetries that arise in the hidden theory. In particular, in the hidden theory, when it is holographic,  the global symmetries become local and may couple to the SM, as discussed in detail in \cite{u1,u12}. The interesting question concerns global symmetries in the SM like B-L. If it is mixed with a hidden global symmetry during the coupling of the hidden and the visible theory, as discussed in detail in \cite{u1,u12} then it also becomes a local symmetry. However, we seem to have a priori the option to keep it intact and in that case it seems that it remains a global symmetry. On the other hand, in string theory realizations of this setup in which the SM is realized on a stack of branes, this option is not possible. We therefore remain with a non-conclusive issue concerning the SM global symmetries.

\item Cosmology in the emergent gravity context obtains a new face associated with instabilities that arise when probing the combined system hidden theory+SM through the lens of the SM alone. ``Dark" aspects of cosmology have now a natural place to reside: the hidden theory\footnote{This does not preclude part of dark matter to be due to matter in another small $N$ theory that is directly coupled to the hidden large N theory, but not the SM.}.

     It is also interesting that cosmological evolution in massive theories of gravity suggests that it is   linked to a dark energy of the order of what is measured today, \cite{review}.

Intriguingly, the framework of emergent gravity has the potential of combining two approaches of the dark matter problem that so far have been deemed different. The first and standard is the presence of new forms of matter from hidden sectors. The second is modifications of gravity, starting with the MOND idea, \cite{Milgrom} and ending with relativistic modified theories of gravitation, \cite{Skordis}.

\end{itemize}

All of the above and other {questions} are interesting to address in the near future.

%%%%%%%%%%%%%%%%%%%%%%%%%%%%%%

\section{A general setup for emergent gravity }
\label{setup}

Our starting point is the setup described in \cite{SMGRAV}, namely two QFTs interacting with each other, defining a UV-complete QFT. One of them is the ``visible" QFT (we shall be eventually interested in a variant of the SM) and the other we would like it to be eventually   a large-$N$ (hidden) QFT. The two are coupled with massive messenger fields of mass $M$. $M$  is assumed to be much larger than the characteristic scales of both QFTs.
At energy scales much smaller than the messenger mass, $M$, the messengers can be integrated-out leaving the hidden QFT interacting with the SM via a series of non-renormalizable interactions.

The setup we  describe here is a bit more general but similar in spirit to the one in \cite{SMGRAV}.

 We start with  a local relativistic quantum field theory on a fixed space-time background. For concreteness, we may take the fixed background to be four-dimensional and its metric $\eta_{\mu\nu}$ to be the flat Minkowski metric. We assume that this quantum field theory has the following features:
\begin{itemize}

\item[$(a)$] It possesses two widely separated, characteristic mass scales $m \ll M$.

\item[$(b)$] At energies $E\gg M$ the dynamics is described by a well-defined ultraviolet (UV) theory. The most obvious context is a UV fixed point described by a four-dimensional conformal field theory.\footnote{One could in principle imagine more exotic UV behavior involving higher dimensional QFTs or some form of string theories.}

\item[$(c)$] At energies $E\ll M$ there is a description of the low-energy dynamics in terms of two separate sets of degrees of freedom and two corresponding distinct quantum field theories interacting with each other via irrelevant interactions. This infrared (IR) splitting of the degrees of freedom is not unique, but once a split in a specific set of conventions is provided, we can use it as the starting point of our analysis. We  call the first quantum field theory, the {\it visible} QFT and denote all quantities associated with that theory with normal font notation. We  call the second quantum field theory, the {\it hidden} QFT and denote all its quantities with a hat notation. Schematically, the low-energy description is in terms of an effective action of the form
\be
\label{setupaa}
S_{IR} = S_{visible}(\Phi) + S_{hidden}(\h{\Phi}) + S_{int}(\Phi,\h{\Phi})
~,
\ee
where $\Phi$ are collectively the fields of the visible QFT and $\h{\Phi}$ the fields of the hidden QFT. The interactions in $S_{int}$ are formally a sum of irrelevant interactions of increasing scaling dimension of the form
\be
\label{setupab}
S_{int} = \sum_i \int d^4 x\, \lambda_i \, \vis{O}_i(x) \h{O}_i (x)
~,
\ee
where $\vis{O}_i$ are general  operators of the visible QFT and $\h{O}_i$ are general  operators of the hidden QFT. $S_{int}$ arises by integrating out massive degrees of freedom of the UV QFT with characteristic mass scale $M$. From the low-energy effective field theory point of view, $M$ is a scale that defines a natural UV cutoff. This is however a physical scale, and is determining the point in energy where the theory splits into two sectors, weakly interacting with each other at low energies.

It should be stressed that generically a UV completion of (\ref{setupaa}), (\ref{setupab}) involves new degrees of freedom, that we  call gravitational messenger fields, with masses of order $M$.
If the SM at large scales, is similar to the one we observe, there is one more possibility of coupling it to a hidden theory directly using a relevant coupling, if the hidden theory has a scalar dimension $\leq 2$ gauge invariant operator and this operator couples to the Higgs mass operator, \cite{quiros}.
We shall not explore further this possibility here.

\item[$(d)$] We assume an additional special feature that implicates the lower mass scale $m$. We require that the hidden QFT that appears in the low-energy effective description \eqref{setupaa} is a theory with mass gap $m$.
At energies $E$ in the range $m\ll E \ll M$ we can employ the description \eqref{setupaa} to describe a general scattering process involving both visible and hidden degrees of freedom. In this case all the degrees of freedom are relatively light and it is convenient to keep them both in the low-energy description. At energies $E\ll m \ll M$ it is more natural to integrate out the hidden degrees of freedom and obtain an effective field theory in terms of the visible degrees of freedom only.

\end{itemize}

In this paper we are interested in the low-energy $(E\ll M)$ behaviour of observables {\it defined exclusively in terms of elementary or composite fields in the visible QFT}. This restriction is ad hoc. It is not imposed by energetic reasons. Physically it would be relevant for an observer who can only access visible QFT fields. In that case, it is sensible to integrate out the hidden sector completely to obtain an effective theory of scattering amplitudes solely in the visible QFT at all energies $E\ll M$. It is that effective theory that we want to understand.

More explicitly, in this paper we are interested in the generating functional of correlation functions (Schwinger functional) in the visible QFT defined as
\be
\label{setupac}
e^{- W(\JJ)} = \int [D\Phi] [D\h{\Phi}] \, e^{-S_{visible}(\Phi,\JJ) - S_{hidden}(\h{\Phi}) - S_{int}(\vis{O}_i, \h{O}_i)}
~.
\ee
We have Wick-rotated the theory to Euclidean signature and $\JJ$ is collective notation that denotes the addition of arbitrary sources in the visible QFT. The path integral is understood as a Wilsonian effective action below the UV cutoff scale $M$. Performing the path integral over the hidden sector fields $\h{\Phi}$ we obtain
\be
\label{setupad}
e^{- W(\JJ)} = \int [D\Phi] \,  e^{-S_{visible}(\Phi,\JJ) - \WW (\vis{O}_i) }
~,
\ee
where the new quantity $\WW$ is the generating functional in the hidden QFT,
\be
 e^{-{\cal W}(\h{J})}\equiv \int  [D\h{\Phi}] \, e^{- S_{hidden}(\h{\Phi}) - \int \h{O}\h{J}}
~.
\ee

 From the point of view of the hidden QFT, the operators $\vis{O}_i$ appearing in the interaction $S_{int}$ in \eqref{setupaa}, \eqref{setupab} are dynamical sources.

In the context of \eqref{setupad}, a visible sector observer registers a formal series of increasingly irrelevant interactions. Is it possible to reformulate these interactions by integrating-in a set of classical fields? In the following sections we answer this question in the affirmative and show that one of the most natural by-products of this reformulation is dynamical gravity.

\section{The effective action for the emergent graviton: a simple example\label{simple}}

In this section we are going to first analyse a ``proof of concept": we show that we can summarise part of the dynamics of a {\em single} QFT in terms of a dynamical metric. This metric will reflect  the state (or states) that are generated by the conserved energy-momentum tensor out of the vacuum,  discussed already in the introduction.

We consider a single QFT in the presence of a (non-dynamical) background metric $\bg_{\m\n}$ and a {\em scalar source} $\phi$. We shall then determine the emergent gravity equations associated with the metric defined by the expectation value of the total energy-momentum tensor.

Before we embark on this calculation, we would like to discuss the definition, renormalization and symmetries of the Schwinger functional
\be
e^{-W(g_{\m\n},\phi)}=\int{\cal D}\chi~e^{-S(\chi,g_{\m\n})+\int \phi(x) ~O(\chi(x))}
\label{s1}\ee
where we used sketchy notation above. $\chi$ collectively denotes the quantum fields of the theory, $g_{\m\n}(x)$ is a fixed {but otherwise arbitrary} background metric and $\phi(x)$
is a scalar source that couples to the operator $O(\chi(x))$.

We now discuss how unique and well-defined is the  Schwinger functional (\ref{s1}).
\begin{enumerate}

\item Once we are given the action $S(\chi,\eta_{\m\n})$ of a QFT in a flat Minkowski metric $\eta_{\m\n}$, there are in principle many actions, $S(\chi,g)$ one can write for its extension to an arbitrary metric $g_{\m\n}$. In most cases, there is a minimal choice, where all quantum fields are minimally coupled to the metric.

    It is however well-known that there is a possibility of adding non-minimal couplings. Simple non-minimal couplings with two derivatives amount to alternative definitions of conserved stress tensors (in a flat background) that are known as ``improvements". For example, an addition of the diffeomorphism  invariant action
\be
\delta S=\int \sqrt{-g} ~V(\chi)~R
\label{f12}\ee
to the action, implements a redefinition of the energy momentum tensor that is
\be
\delta T_{\m\n}=V(\chi)\left[R_{\m\n}-{1\over 2}g_{\m\n}R\right]-(\nabla_{\m}\nabla_{\n}-g_{\m\n}\square)V(\chi)
\label{f13}\ee
This redefinition preserves  the property that the total stress-tensor is covariantly conserved. Moreover, in the flat limit,  $g_{\m\n}\to \eta_{\m\n}$,  the energy-momentum change is not zero.
Higher terms in curvature can also be added, they affect the energy momentum tensor in curved backgrounds but not in flat space.
We must choose therefore and fix the energy momentum tensor that is gauged in such an source functional, and the subsequent steps will be dependent on this choice. The number of choices is infinite. Any diff-invariant Lagrangian of the quantum fields and the metric that vanishes when the metric is flat would do.

Therefore, we must fix once and for all the action in a general metric so that it is classically diff-invariant and without gravitational anomalies.
This fixes the stress tensor in any background metric.
In this case, the only ambiguities that can appear in the Schwinger functional are associated to renormalization of the UV divergences. The terms affected are finite, and this issue is not unlike any other QFT renormalization.

\item It is simple to show that the linearized curved space action has a linearized (local) diffeomorphism invariance, \cite{lorentz}.
    Moreover, the action can be completed so that this can be upgraded to a complete non-linear diffeomorphism invariance, (\cite{lorentz} and references therein).

\item The path integral must be regulated in order for the Schwinger functional for the metric and other sources to be evaluated.
    This has always been tricky business, as diffeomorphism invariance clashes with various type of regulators. This has to be done carefully, if we want the renormalized Schwinger functional to be diffeomorphism invariant. In theories that have gravitational anomalies this cannot be done. However, all four-dimensional theories are free from gravitational anomalies, \cite{wa}.

For practical purposes, we can imagine using dimensional regularisation that has been shown to respect diffeomorphism invariance, if properly used.
It is however plausible that one could use a momentum cutoff. It is well known that in this case there is an explicit breaking of diffeomorphism invariance. It might be possible however that the offending terms can be controlled and subtracted as one renormalizes and then removes the cutoff.
This procedure was successful in using a momentum cutoff to regularise and renormalize chiral non-abelian four-dimensional gauge theories without violating gauge invariance in the continuum limit, \cite{warr}, although gauge invariance is violated at a finite cutoff. A similar procedure is used on the lattice.

Something similar may be possible for gravity as well, but that remains to be seen\footnote{A promising first step in that direction has been done recently in \cite{morris}.} . Holography  suggests that this should be possible. In a holographic setup, a hard cutoff (known as the displaced boundary) provides an appropriate way to regularize and renormalise while keeping the diffeomorphism invariance of the renormalized Schwinger functional.
In the meantime, our regularization of choice will be dimensional regularization.

\item Once we renormalise, keeping diffeomorphism invariance intact, we are in possession of a Schwinger functional that is finite and diff-invariant.

\end{enumerate}

We shall now proceed to start from this finite and diff-invariant Schwinger functional $S_{\rm Schwinger}(g,\phi)$ and define the effective action for composite operators along the lines described in \cite{tom}, albeit with some changes.

We first define the expectation value of the energy-momentum tensor from the
Schwinger functional as\footnote{We shall follow consistently this definition throughout the paper. One can go back to a more conventional definition by multiplying the one-point function by $-2$ and the two-point function by a factor of $4$.}
\be
h_{\m\n}\equiv {1\over \sqrt{g}}{\delta  S_{\rm Schwinger}(g,\phi)\over \delta g^{\m\n}}
\label{f14}\ee
where $\phi$ denotes collectively other sources that might have been turned-on in the Schwinger functional.
We then define a modified Legendre transform as follows
\be
\Gamma(h_{\m\n},\phi,{\bf g}_{\m\n})\equiv \int d^4x~\sqrt{g} ~h_{\m\n}(g^{\m\n}-{\bf g}^{\m\n})-S_{\rm Schwinger}(g,\phi)
\label{f15}\ee
where ${\bf g}_{\m\n}$ is the fiducial metric of the original QFT.
$\Gamma(h_{\m\n},\phi,{\bf g}_{\m\n})$ can be thought as a functional of $h_{\m\n}$, by expressing $g_{\m\n}$ in terms of $h_{\m\n}$ from (\ref{f14}).
We may now calculate
\be
{1\over \sqrt{g}}{\delta \Gamma\over \delta h_{\m\n}}=g^{\m\n}-{\bf g}^{\m\n}+{1\over \sqrt{g}}{\delta\sqrt{g}\over \delta h_{\m\n}}~h_{\r\s}(g^{\r\s}
-{\bf g}^{\r\s})+h_{\r\s}{\delta g^{\r\s}\over \delta h_{\m\n}}-{1\over \sqrt{g}}{\delta S_{\rm Schwinger}(g,\phi)\over \delta h_{\m\n}}
\label{f18}\ee
and use the chain rule to write
\be
{1\over \sqrt{g}}{\delta S_{\rm Schwinger}(g,\phi)\over \delta h_{\m\n}}=
{1\over \sqrt{g}}{\delta S_{\rm Schwinger}(g,\phi)\over \delta  g^{\r\s}}{\delta g^{\r\s}\over \delta h_{\m\n}}=h_{\r\s}{\delta g^{\r\s}\over \delta h_{\m\n}}
\label{f19}\ee
where in the last step we used the definition (\ref{f14}).
Substituting (\ref{f19}) into (\ref{f18}) we obtain
\be
{1\over \sqrt{g}}{\delta \Gamma\over \delta h_{\m\n}}=g^{\m\n}-{\bf g}^{\m\n}+{1\over \sqrt{g}}{\delta\sqrt{g}\over \delta h_{\m\n}}~h_{\r\s}(g^{\r\s}
-{\bf g}^{\r\s})
\label{f20}\ee
and evaluating $g_{\m\n}$ on the background metric we find that the effective action $\Gamma$ is extremal under variations of the expectation value $h_{\m\n}$
\be
{\delta \Gamma\over \delta h_{\m\n}}\Bigg|_{g_{\m\n}={\bf g}_{\m\n}}=0
\label{f21}\ee
Moreover, if $h^*_{\m\n}$ is the solution to (\ref{f21}) then
\be
\Gamma(h,\phi,{\bf g})\Bigg|_{h=h^*}=S_{\rm Schwinger}({\bf g},\phi)
\ee
We can therefore view $\Gamma(h,{\bf g})$ as an effective action for a dynamical field $h_{\m\n}$  that contains  the states generated by the energy-momentum tensor of the original theory out of the vacuum.

Indeed, if $\Gamma$ is computed perturbatively in $h_{\m\n}$ (or equivalently, perturbatively in $g_{\m\n}$) then its quadratic kernel is the inverse of the two-point function of the stress-tensor, viewed as an $\infty\times \infty$ matrix. In momentum space, it is the geometric inverse of the Fourier transform of the two-point function. The poles of the two-point function (when isolated), correspond to the particles generated from the vacuum by the energy momentum tensor.

The position of the poles give the masses of the associated states. In $\Gamma$, the inverse of the two-point function indicates that the quadratic term has zeros at the positions of the poles and this corresponds to the kinetic terms of the associated particles viewed as dynamical fields\footnote{This argument is less clear in a CFT, where the spectrum is continuous down to zero mass. But as we shall see later on, the {\em local} emergent gravity description we are going to develop, breaks down for theories with exact conformal invariance. If however the CFT is holographic, then AdS/CFT experience suggests the resolution of this problem by lifting $\Gamma$ to dynamics and gravity  in higher dimensions.}.

The expectation value of the stress tensor plays, essentially\footnote{Up to some importants details that we shall describe in the sequel.}, the role of a dynamical graviton in the theory, and whose dynamics is determined by $\Gamma(h)$.
It should be stressed that the construction of $\Gamma(h,\phi)$ above is well defined in the presence of sources of energy and momentum, and this is the reason we introduced also other sources like $\phi$. In their absence, the two-point function of the stress-tensor has zero modes and is therefore not invertible. In that case, we can only invert in the space orthogonal to the zero modes, and the procedure becomes quickly complicated\footnote{A similar phenomenon happens also in the case of emergent vector bosons. In that case the massless case leads to a non-local action that can be written down and analyzed, \cite{u1}.}.

\subsection{Gapped theories}

We shall now be a bit more explicit, by considering a context  where one can say something more about the Schwinger functional.
In particular, in a theory with a mass gap $m>0$,  we can expand the  Schwinger functional in a derivative expansion, valid up to distances of order $m^{-1}$ as follows\footnote{We keep only a single scalar source $\phi$ for simplicity.},
\begin{equation}
S_{\rm Schwinger}(g,\phi)\,=\,\int d^4x \sqrt{g}\,\left[-V(\phi)\,+ M^2(\phi) \,R\,-\,Z(\phi)\,(\partial \phi)^2\,+{\cal O}(\pa^4)\right]\label{f1}
\end{equation}
where the ellipsis indicates higher derivative terms.
Note that $V,M^2,Z$ are scalar source dependent functions that can be obtained by a direct  calculation of the path integral of the theory, in the presence of a non-trivial metric and other sources.  They depend on the initial QFT as well as the prescription of implementing diff-invariance (ie. on the precise form of the curved space action in (\ref{s1})).

We define for notational simplicity  the (energy-momentum) tensor of the scalar
\begin{equation}
 T^{\phi}_{\m\n}\,\equiv \,Z\,\partial_\m \phi \partial_\n \phi\,-\,\frac{1}{2}\,g_{\m\n}\,Z\,(\partial \phi)^2\sp {(T^{\phi})_{\m}}^{\m}=-Z(\pa\phi)^2
\label{f2}\end{equation}
as well as the ``improved"  tensor as
\be
{\cal T}^{\phi}_{\m\n}\equiv T^{\phi}_{\m\n}+(\nabla_{\m}\nabla_{\n}- g_{\m\n}\square)M^2\sp {({\cal T}^{\phi})_{\m}}^{\m}=-Z(\pa\phi)^2-\square M^2
\label{f2a}\ee

The total energy-momentum tensor expectation value is
\be
\langle T_{\m\n}\rangle \equiv {1\over \sqrt{g}}{\delta S_{\rm Schwinger}(g,\phi)\over \delta g^{\m\n}}={V\over 2}g_{\m\n}+M^2G_{\m\n}-{\cal T}^{\phi}_{\m\n}+{\cal O}(\pa^4)
\label{f6}\ee
where
\be
G_{\m\n}\equiv R_{\m\n}-{1\over 2}g_{\m\n}R
\ee
 is the standard Einstein tensor.
The energy-momentum tensor is covariantly  conserved
\be
\nabla_{g}^{\m}\langle T_{\m\n}\rangle =0
\label{f6a}\ee
provided that $\phi$ is constant or the following extremization equations hold on the source $\phi$ for arbitrary $g_{\m\n}$,
\be
V'-2Z\square \phi-Z'(\pa\phi)^2-(M^2)'~R+{\cal O}(\pa^4)=0
\label{f3}\ee
where primes stand for $\phi$ derivatives. This calculation  is detailed in appendix \ref{cc}.

A point that may cause confusion is the following.
The conservation of the energy-momentum tensor in flat space is associated with the translational invariance of the QFT. It is this conservation that is responsible for the diffeomorphism invariance for the Schwinger functional, \cite{lorentz} and this is also detailed in appendix \ref{cc}. In the presence of sources, translational invariance is broken, and we would naively expect that diffeomorphism invariance of the Schwinger  functional is gone. However, as shown in appendix \ref{cc}, although the conservation of the energy-momentum tensor is modified, the diffeomorphism invariance of the Schwinger  functional is still intact.

We now define $h_{\m\n}$ as  the expectation value of the stress tensor
\be
h_{\m\n}\equiv \langle T_{\m\n}\rangle={V\over 2}g_{\m\n}+M^2G_{\m\n}-{\cal T}^{\phi}_{\m\n}+{\cal O}(\pa^4)
\label{f4}\ee
and the (emergent) metric $\tilde h_{\m\n}$ as
\be
\tilde h_{\m\n}\equiv {2\over V}h_{\m\n}=g_{\m\n}+{2M^2\over V}G_{\m\n}(g)-{2\over V} {\cal T}^{\phi}_{\m\n}
+{\cal O}(\pa^4)\;.
\label{f5}\ee
Note that $\tilde h_{\m\n}$ is dimensionless and will eventually become the emergent metric of the theory.

Equation (\ref{f5}) can be solved for $g_{\m\n}$ as a function of $\tilde h_{\m\n}$, order by order in the derivative expansion, as
\be
g_{\m\n}=\tilde h_{\m\n}-\delta\tilde h_{\m\n}+{\cal O}(\pa^4)
\label{f7}\ee
with
\be
\delta\tilde h_{\m\n}={2\over V}\left({M^2}\tilde G_{\m\n}-{\tilde {\cal T}}^{\phi}_{\m\n}\right)
\label{f8}\ee
where we used a tilde over various tensors to indicate they are evaluated in the metric $\tilde h_{\m\n}$.

Equation (\ref{f5}) gives us the first dynamical equation for the metric $\tilde h$
\be
M^2\tilde G_{\m\n}={V\over 2}\left(\tilde h_{\m\n}-{\bold g}_{\m\n}\right)+{\tilde {\cal T}}^{\phi}_{\m\n}+{\cal O}(\pa^4)
\label{f9}\ee
where we substituted $g_{\m\n}={\bold g}_{\m\n}$ in (\ref{f5}).
Equation (\ref{f9})  is valid independent of the validity or not of (\ref{f3}). Note that it depends also on the original fixed background metric ${\bold g}_{\m\n}$.

{This Einstein equation is equivalent to the equation we shall obtain when we vary $\Gamma(\tilde h)$ once the other sources are extremal. This is shown in detail in appendix \ref{totalstresseffective} where the effective action $\Gamma$ is constructed up to two derivatives.

Equation (\ref{f9})  describes the dynamics of the emergent graviton, generated by the energy-momentum tensor.
Moreover, (\ref{f9}) is a form of a bi-gravity theory where $\tilde h_{\m\n}$ is the dynamical metric while ${\bold g}_{\m\n}$ is a fixed background metric. Both transform as tensors under the same diffeomorphisms.

At this stage, we  can confirm some general expectations described in the introduction. The first concerns the large $N_c$ limit: the action in (\ref{f1}) is of order ${\cal O}(N_c^2)$. It is simple to verify that the effective action $\Gamma(h)$ is of the same order, if $h_{\m\n}$ is appropriately rescaled. Therefore the composite graviton is weakly coupled as argued.
If on the other hand, the theory is gapped, then at energies well below the gap, the composite graviton is point-like, and this agrees with the local expansion of the action in (\ref{f1}) and the associated effective action.
If the theory at large $N_c$ is strongly coupled and gapless, then the local effective action approximation is invalid.
But at large $N_c$, the otherwise non-local Schwinger functional, can be written as a local functional in a higher dimension if the theory is holographic.
Another gapped case, where the low-energy functional description is incomplete is the case with a gap and a tower of higher massive poles that are not hierarchically separated from the the leading massive pole. This is the case of four-dimensional YM at large $N_c$. Such cases can be described better in a holographic setup that resums the tower of massive poles as KK states.

The conservation of the total stress tensor in (\ref{f6a}) now becomes
\be
\nabla^{\m}_g~h_{\m\n}=0
 \label{f10}\ee
which after converting the $g$-covariant derivative to the one in the $\tilde h$ metric, as in (\ref{3d3}) in appendix \ref{cem}, transforms to
$$
\nabla^{\m}_g~h_{\m\n}=\tilde\nabla^{\m}\left[{V\over 2}(\tilde h_{\m\n}+\delta\tilde h_{\m\n})+{\cal O}(\pa^4)\right]=
$$
\be
=\tilde\nabla^{\m}\left[{V\over 2}\tilde h_{\m\n}+{M^2}\tilde G_{\m\n}-{\cal T}^{\phi}_{\m\n}+{\cal O}(\pa^4)\right]=\tilde\nabla^{\m}\tilde T_{\m\n}
 \label{f11}
\ee
 where $\tilde T_{\m\n}$ is exactly the stress tensor expectation value of the theory in (\ref{f6}) (up to order ${\cal O}(\pa^2)$), but now evaluated,  instead of the background metric $g_{\m\n}$,  at the metric $\tilde h_{\m\n}$.

 Using (\ref{ccc7}) we have
 \be
\tilde \nabla^{\m}\left[ {M^2}\tilde G_{\m\n}-{\cal T}^{\phi}_{\m\n}\right]=-{1\over 2}(\tilde\nabla_{\n}M^2)~\tilde R-\tilde\nabla^{\m}T^{\phi}_{\m\n}=
\label{f17}\ee
$$-
\left[{1\over 2}(M^2)'\tilde R+Z\tilde \square\phi +{1\over 2}Z'(\pa\phi)^2\right]\pa_{\n}\phi+{\cal O}(\pa^4)
$$
and therefore (\ref{f11}) becomes
\be
\tilde\nabla^{\m}\tilde T_{\m\n}=
V'-2Z\tilde\square \phi-Z'(\tilde\nabla\phi)^2-(M^2)'~\tilde R+{\cal O}(\pa^4)=0
\label{fff3}
\ee
which is identical to (\ref{f3}) but in the emergent metric $\tilde h_{\m\n}$.
Therefore,  the conservation of energy-momentum of the original theory in the non-dynamical  background metric ${\bf g}_{\m\n}$ has transformed into the conservation of energy-momentum in the theory with dynamical metric $\tilde h_{\m\n}$.

It is important to stress that the covariant derivative in the emergent metric uses the standard Christoffel connection and this is imposed by the original theory. It is not something we have to decide.

Using (\ref{f9}) in (\ref{f11}) we obtain also an equivalent relation
\be
\tilde\nabla^{\m}\left[{V\over 2}(2\tilde h_{\m\n}-{\bf g}_{\m\n})+{\cal O}(\pa^4)\right]=0
\label{bb11}\ee
which is also equivalent to
\be
2\pa_{\n}V=\tilde\nabla_{\m}(V{\bf g}_{\m\n})
\label{b11}\ee

However, a study of the equation in (\ref{f9}) at the linearized level indicates that (not surprisingly) it contains ghost degrees of freedom.
This linearized analysis is performed in appendix \ref{aF1}. It is seconded by a study of symmetric solutions to the equations in appendix \ref{sol-ein} that is in agreement.
From the appendix we reproduce (\ref{c42c})
\be
S_{int}(T,T')={T^{\m\n}T'_{\m\n}-{1\over 3}TT'\over M^2(p^2-\Lambda)}-{1\over 6}{TT'\over M^2\left(p^2+{\Lambda\over 2}\right)}\sp \Lambda={V\over M^2}\;,
\label{c42bbis}\ee
which gives the interaction energy between two energy-momentum sources $T_{\m\n}$, $T'_{\m\n}$, and where $\Lambda$ is the effective cosmological constant of the Schwinger functional.
From this interaction we conclude that the spin-zero mode is always a ghost. Moreover, depending on the sign of the vev $\Lambda$, either the spin-2 or the spin-0 exchange behaves as a tachyon.
However, as we shall show in the next subsection,  if we compute the linearized interaction without performing a low-energy expansion of the Schwinger functional, we will find perfectly healthy interactions associated both to spin-0 and spin-2 components.

The reason that the (local) IR expansion of the Schwinger functional gives misleading results is that it mixes contact terms and pole terms. This can be indicated by the following toy example.
We consider a quadratic source functional
\be
W(J)=\int d^4 p~J(-p)G(p)J(p)\sp G(p)=G_0+{R\over p^2-m^2}
\label{k1}\ee
where we took the two-point correlator to have a pole and a constant contact term\footnote{We assume Lorentzian signature but drop $i\e$ terms}. It is clear that the interaction of the source $J$ contains an innocuous contact term contribution and the effect of the exchange of a particle of mass $m$ and residue $R$.
Consider now the following sequence of steps. Expand $W(J)$ up to ${\cal O}(p^2)$, construct the effective action $\Gamma$ to order ${\cal O}(p^2)$ and then recompute  the interaction of sources.
We have
\be
W(J)=\int d^4 p~J(-p)J(p)\left[\tilde G_0-{Rp^2\over m^4}+{\cal O}(p^4)\right]\sp \tilde G_0=G_0-{R\over m^2}
\label{k2}\ee
\be
h(p)={\delta W\over \delta J(-p)}=2J(p)\left[\tilde G_0-{Rp^2\over m^4}+{\cal O}(p^4)\right]
\label{k3}\ee
\be
\Gamma(h)=\int Jh-W={1\over 4}\int d^4 p~ h(-p)\left[\tilde G_0-{Rp^2\over m^4}+{\cal O}(p^4)\right]^{-1}h(p)=
\label{k4}\ee
$$
={1\over 4\tilde G_0}\int d^4p~h(-p)\left[1+{Rp^2\over m^4\tilde G_0}+{\cal O}(p^4)\right]h(p)
$$
Recomputing the original interaction in (\ref{k1}) from (\ref{k4}) we obtain instead
\be
W(J)= {m^4 \tilde G_0^2\over R}\int d^4 p~{J(-p)J(p)\over p^2+{m^4\over R}\tilde G_0}+{\cal O}(p^4)
\label{k5}\ee

Comparing (\ref{k5}) with (\ref{k1}) we observe that now both the residue and the position of the pole has changed.
The reason is that the position of the pole in (\ref{k5}) is now not reliable in the momentum expansion.
Moreover, depending on the sign and size of the initial contact term, $G_0$, the pole now may become a tachyon. The momentum expansions and subsequent inversions mix contact terms with pole data, and obscure the properties of the interaction. A more detailed discussion of these issues has appeared recently in \cite{Wet}.

\subsection{The linearized induced interaction}

Integrating out the emergent graviton at the quadratic order induces the quadratic interaction of sources, captured  by the Schwinger functional and the interaction is given by the two-point function of the stress-tensor.
The fluctuation of the background metric source $g_{\m\n}=\eta_{\m\n}+\delta g_{\m\n}$ is essentially an external energy source and we shall rename it $\delta g_{\m\n}=t_{\m\n}$. Using (\ref{c111})-(\ref{c114}) and (\ref{nn20}) we obtain the quadratic interaction of the external (conserved) energy-momentum source $t_{\m\n}$
\be
W_2(t)={1\over 2}\int {d^4 p\over (2\pi)^4}t^{\m\n}(p)~Q_{\m\n\r\s}(p)~ (p)~t^{\r\s}(-p)=
\label{bf1}\ee
$$
=\int {d^4 p\over (2\pi)^4}\left[{\Lambda\over 4}\left(t^{\m\n}(k)t_{\m\n}(-k)-t(k)t(-k)\right)
+2B_2(k)\left(t^{\m\n}(k)t_{\m\n}(-k)-{1\over 3}t(k)t(-k)\right)+\right.
$$
$$
+\left.{B_0(k)\over 3}t(k)t(-k)\right]
$$
where $\langle T_{\m\n}\rangle =\Lambda \eta_{\m\n}$ and we assumed that $k^{\m}t_{\m\n}=0$.

According to the discussion in appendix \ref{spectral}, the functions $B_{2,0}$ contain two more contact terms, one that is ${\cal O}(k^2)$ and another at ${\cal O}(k^4)$.
Therefore, we can split the interaction in (\ref{bf1}) into a contact part
\be
W_2^{contact}= \int {d^4 p\over (2\pi)^4}\left[{\Lambda\over 4}\left(t^{\m\n}(k)t_{\m\n}(-k)-t(k)t(-k)\right)  +\right.
\label{bf2}\ee
$$
+2\delta_2 k^2\left[t_{\m\n}(k)t^{\m\n}(-k)-t(k)t(-k)\right]+
$$
$$
+A_2k^4\left[ t_{\m\n}(k)t^{\m\n}(-k)\left.
-{1\over 3}t(k)t(-k)\right]+A_0k^4 t(k)t(-k) \right]
$$
and a part that depends on the nontrivial ({renormalised}) spectral densities that are both UV and IR finite,
\be
W_2^{non-local}={1\over 2}\int {d^4 p\over (2\pi)^4}\left[2B_2^{nl}(k)\left(t^{\m\n}(k)t_{\m\n}(-k)-{1\over 3}t(k)t(-k)\right)
+{B^{nl}_0(k)\over 3}t(k)t(-k)\right]
\label{bf3}\ee
The non-trivial interaction mediated by the energy momentum tensor is captured in (\ref{bf3}).
It depends crucially on the structure of $B^{nl}_{2,0}$.
If there is a mass gap and discrete states then near a pole we can approximate
\be
B_{2,0}\simeq {R_{2,0}\over k^2+m_{2,0}^2}
\label{bffs}\ee
where the residue $R_{2,0}$ has mass dimension six as $B$ has mass dimension four.
The resulting interaction in (\ref{bf3}) involves a massive spin-2 particle of mass $m_2$ and a massive spin-0 particle with mass $m_0$.
Note that for a unitary theory all residues are positive and the exchanges are never ghostlike.
Moreover by an appropriate rescaling of the interacting densities, we find the associated Planck scales to be given by
\be
M_{2,0}^2\sim N^2~{V^2\over R_{2,0}}
\label{planck}\ee
where we have also indicated a possible (large) N factor and $V$ is the vev.

In general, the static potential due to the spin-2 and spin-0 spectral densities in four dimensions is given by
\be
V_{2,0}(r)\sim - \frac{1}{r} \int_0^\infty d \mu^2  \mu^4 e^{- r \mu}  \,
 \rho_{2,0}(\mu^2)  \, .
\ee
We can explore the long distance structure of the  static interactions as a function of the structure of the low-energy behavior of the spectral densities.
Apart from the case of isolated low-lying poles analysed in appendix~\ref{staticpotential}, the spectral densities may have other behaviors in the IR, and this affects the long-distance asymptotics of the static emergent interactions.
For example in the gapless case we can parametrize the spectral densities as $\rho\sim \mu^a$, with $a\geq 0$\footnote{A four-dimensional CFT has $\rho_2=$ constant and $\rho_0=0$.}. In this case the long-distance asymptotics of the static potential are
\be
V\sim {1\over r^{6+a}}
\ee
Another case involves the existence of a gap, $\mu_0$ and a continuum above the gap,
\be
\rho\sim (\mu-\mu_0)^b~~~~\ar~~~~ V\sim {\mu_0^5\over r^{b+1}}e^{-\mu_0 r}
\ee
Finally the contribution of massive isolated poles like (\ref{bffs}) to the static potentials has the standard Yukawa form
\be
V\sim {1\over r}e^{-mr}\;\;.
\ee

We conclude this section with the following remarks.
\begin{itemize}

\item The effective action for the energy-momentum  tensor, summarizes the dynamics of states generated from the energy-momentum  tensor acting on the vacuum.

\item The associated physics  is summarized in terms of a dynamical  (emergent) metric.

\item The IR dynamics is local, if the theory has a mass gap and if the cosmological constant (ie the potential V in (\ref{f1})) is non-zero.
In (finely-tuned) cases where $V$ vanishes, then the effective gravitational theory becomes non-local. For example, in a supersymmetric QFT, if we turn on scalar supersymmetry breaking sources, then $V\not=0$ and the gravitational description is local.
If we only turn-on supersymmetry-preserving sources then the description becomes non-local.

\item $\Gamma(\tilde h_{\m\n})$ is a bigravity theory with a fixed background metric that is the fixed QFT metric, and a fluctuating metric that is the classical field associated with the stress-energy tensor.
    The full theory is diffeomorphism invariant, once we also transform the background metric.
As is typical of bigravity theories, the graviton is massive and this is compatible with the WW theorem.

\item  For a theory like YM, with small $N_c$, the propagator of the emergent graviton, being the inverse of the two-point function of the energy-momentum tensor, has a sequence of poles associated to the tower of $2^{++}$ glueballs.
All but the lowest one are unstable to decay to the lightest $0^{++}$ glueball. The lowest $2^{++}$ does not have enough mass to decay.
Therefore, we have a stable massive graviton and a tower of unstable cousins, having widths of the order of their masses. They do not qualify as particles in the effective field theory. Therefore the effective field theory here is composed of a scalar ($0^{++}$), a pseudoscalar ($0^{+-}$) and massive graviton ($2^{++}$).
In this theory all parameters, like $V$, $M^2$ etc are of the same order, given by $\Lambda_{YM}$. Therefore, the effective theory (and gravity) is strongly coupled.

At large $N_c$ however, there are a few differences. First and foremost, the decay widths are suppressed, and therefore the towers of the $2^{++}$ are nearly stable massive gravitons. They behave as the KK states of a 5-dimensional theory, and therefore, the natural formulation of this large-N theory is as a string theory in five dimensions.
Still, both in four and five dimensions, the effective action and therefore the relevant parameters scale as $N_c^2$, and therefore all interactions of glueballs are weak.

\item Although, all of the above are suggestive of emergent gravity, the gravity we observe in nature cannot be just that. $\Gamma$ summarizes the dynamics of a subset of states of the original QFT (those generated by the energy momentum tensor).  However, some of the techniques and ideas developed here will be used in subsequent sections where external sources of emergent gravity will be studied.

\end{itemize}

%%%%%%%%%%%%%%%%%%%%%%%%%%%%%%
\section{Linearized emergent gravity from a hidden sector}
\label{linear}

As argued in the previous section, states generated by the energy-momentum tensor of QFT have the properties of bona-fide gravitons, and this was analyzed in some concrete contexts. However, when we want to describe observable gravity , we need the graviton to be sourced from a sector outside the SM, as its SM siblings, are both loosely bound (and therefore have nonlocal interactions) and moreover are already accounted for in our description of particle interactions.
It is clear that observable gravity, if it arises as emergent gravity, it must arise from a QFT other than the SM. We shall call this theory the ``hidden theory".

In the sequel, we assume that the ``hidden theory" is a generic QFT, that is coupled at some high scale $M$ to the SM\footnote{We refer to the SM also as the ``visible sector".}. Both theories should be embedded in a UV complete four-dimensional quantum field theory. For this to happen, as argued in \cite{SMGRAV}, the two theories must interact via messengers with masses of order $M$. $M$ is assumed to be much larger that all the other scales of the SM or of the hidden theory. Eventually, we shall take the hidden theory to be a holographic (large N, strongly coupled) theory.

We start by considering an ad hoc irrelevant deformation of the IR effective theory that mediates interactions between the visible and hidden sectors via the action
\be
\label{linearaa}
S_{int} =\int d^4 x\, \left( \lambda \vis{T}_{\mu\nu}(x)\, \h{T}^{\mu\nu} (x) +\lambda'  \vis{T}(x) \h{T}(x) \right)
~.
\ee

In this expression, we have defined $\vis{T} \equiv  \eta^{\mu\nu} \vis{T}_{\mu\nu}$ and $\h{T} \equiv  \eta^{\mu\nu} \h{T}_{\mu\nu}$ to be the traces of the energy-momentum tensors $\vis{T}_{\mu\nu}$ and $\h{T}_{\mu\nu}$ in the visible and hidden QFTs respectively. The space-time indices are contracted with the flat (and fixed) background metric $\eta_{\mu\nu}$. Each of these energy-momentum tensors is assumed to be separately conserved before the irrelevant\footnote{{At the cutoff, both couplings, $\lambda$ and $\lambda'$, have negative mass dimension slightly above four}.} couplings $\lambda$, $\lambda'$ are turned on. In the presence of the interaction \eqref{linearaa},  only a single combination of the energy-momentum tensors $\vis{T}_{\mu\nu}, \h{T}_{\mu\nu}$ remains conserved.

Interactions like \eqref{linearaa} can appear naturally in the effective theory of the general setup of the previous section, but typically they will not appear alone. Other irrelevant interactions will also appear, including others of dimension 8, as well as interactions that deform separately the visible and hidden QFTs. The full set of irrelevant interactions that deform and couple the visible and hidden sectors in the IR is dictated by the details of the RG flow from the microscopic high-energy theory. Concrete examples with some level of technical control can be found in supersymmetric quantum field theories. An example with two decoupled CFTs in the extreme IR arises in the Coulomb branch of $\NN=4$ SYM theory (see, e.g. \cite{a2,a3} for some of the early discussions in this case, and \cite{e1,a1,e2} from the point of view of multi-gravity).

It is also important to mention that independent of the details of the coupling between the two theories, the effective interactions in (\ref{linearaa}) will always appear, and as such they are generic. To see this in a simple example, assume a coupling of two scalar operators, $O(x)$ and $\widehat O(x)$ of the form, $g\int d^4x O(x)~\widehat O(x)$. Then, the couplings in (\ref{linearaa}) will appear via the ${\cal O}(g^2)$ contribution
\be
\langle T_{\m\n}(x)\widehat T_{\r\s}(y)\rangle=g^2\int d^4w_1 d^4 w_2~
\langle T_{\m\n}(x)O(w_1)O(w_2)\rangle ~\langle \widehat T_{\m\n}(x)\widehat O(w_1) \widehat O(w_2)\rangle
\label{b1}
\ee
after expanding the correlator at long distances.

In this section, we treat the specific interactions \eqref{linearaa} as a warmup, toy example. Ignoring momentarily the specifics of the RG flow from the UV theory, we  proceed to analyze the effects of the irrelevant deformation \eqref{linearaa} at quadratic order in the couplings $\lambda, \lambda'$ in perturbation theory. Since the deformations are irrelevant, eventually our computation will exhibit a sensitivity on the details of the UV completion. With the setup of section \ref{setup} in mind, we assume that there is a natural UV cutoff scale $M$. We shall deal with any regularization issues by adopting a regularization scheme that respects the expected symmetries of the UV completion. This cutoff scale is the mass scale associated to the messenger sector that UV-completes the description as described in the previous section.

In this toy example, we show that it is possible to re-express the IR effective dynamics of the visible QFT by first integrating-out the hidden sector and then integrating-in a spin-2 field that emerges as a bona fide graviton. At quadratic order in the irrelevant couplings, the emergent theory of the spin-2 field is a linearised theory of gravity with a non-vanishing cosmological constant and a specific coupling to the visible QFT.

\subsection{Integrating-in a metric in perturbation theory\label{4.1}}

In what follows, it will be convenient to define the constant
\be
\label{linearaaa}
\mathfrak{c} \equiv \frac{\lambda'}{\lambda}
\ee
and the tensor
\be
\label{linearaab}
\viss{T}^{\mu\nu} \equiv \vis{T}^{\mu\nu} + {\mathfrak c}\, \vis{T}\, \eta^{\mu\nu}
~.
\ee
We assume $\lambda$ is non-vanishing so that $\mathfrak c$ is well-defined. In this notation the interaction \eqref{linearaa} takes the more compact form
\be
\label{linearaac}
S_{int} = \lambda \int d^4 x\,   \viss{T}^{\mu\nu}(x)\, \h{T}_{\mu\nu} (x)
~.
\ee

In the presence of \eqref{linearaac}, the generating functional of correlation functions in the visible QFT is
\begin{align}
\begin{split}
\label{linearab}
e^{- W(\JJ)} &= \int [D\Phi] [D\h{\Phi}] \, e^{-S_{vis}(\Phi,\JJ) - S_{hid}(\h{\Phi}) - \lambda \int d^4 x\, \viss{T}^{\mu\nu}(x)\, \h{T}_{\mu\nu} (x)}
\\
&=
\int [D\Phi] [D\h{\Phi}] \, e^{-S_{vis}(\Phi,\JJ) - S_{hid}(\h{\Phi})}
\bigg[ 1 - \lambda \int d^4 x\,  \viss{T}^{\mu\nu}(x)\, \h{T}_{\mu\nu} (x)
\\
&\hspace{1.5cm} + \frac{1}{2} \lambda^2 \int d^4 x_1 d^4 x_2 \, \viss{T}^{\mu\nu}(x_1) \viss{T}^{\rho\sigma}(x_2) \h{T}_{\mu\nu}(x_1) \h{T}_{\rho\sigma}(x_2) +\OO(\lambda^3) \bigg]
~,
\end{split}
\end{align}
where $\Phi$ and $\widehat \Phi$ denote collectively the quantum fields of the two theories and in the second equality we expanded the path integral perturbatively in $\lambda$ up to second order.
The second term on the second line involves the one-point function of $\h{T}_{\mu\nu}$ in the undeformed hidden theory. We have assumed that in the absence of the interaction \eqref{linearaac} $\h{T}_{\mu\nu}$ is the conserved energy-momentum tensor of a Lorentz-invariant QFT with a mass gap. In such a theory $\h{T}_{\mu\nu}$ has, in general, a one-point function of the form
\be
\label{linearai}
\langle \h{T}_{\mu\nu}(x) \rangle_{hid}^{(0)} = \h{\Lambda} \, \eta_{\mu\nu}
~,
\ee
where $\h{\Lambda}$ is a dimensionfull constant. The superindex $(0)$ and the subindex $hid$ denote that in this equation we evaluate the one-point function in the undeformed hidden theory. We henceforth assume that $\h{\Lambda}$ is non-vanishing. Similarly, the third line in \eqref{linearab} involves the two-point function of $\h{T}_{\mu\nu}$ denoted by
\be
\label{linearaia}
\h{G}_{\mu\nu\rho\sigma}(x_1-x_2) = \langle \h{T}_{\mu\nu} (x_1) \, \h{T}_{\rho\sigma} (x_2) \rangle_{hid}^{(0)}
~.
\ee
The dependence on $x_1-x_2$ is a consequence of translation invariance. We also use an upperscript to denote the connected part of this two-point function as $\h{G}^{(c)}_{\mu\nu\rho\sigma}(x_1-x_2)$. This connected part is the one obeying the Ward identity of Appendix \ref{Ward}, see \eqref{genpertudf} and \eqref{genpertudi}.

Then, denoting the partition function of the undeformed hidden theory as $e^{-W^{(0)}_{hid}}$ we can recast \eqref{linearab} as
\begin{align}
\begin{split}
\label{linearad}
e^{-W(\JJ)} & = e^{-W^{(0)}_{hid}} \int [ D \Phi ]\, e^{- S_{vis}(\Phi, \JJ)}
\bigg[  1 - \lambda \h{\Lambda} \int d^4 x \, \viss{T}(x)
\\
&\hspace{1.5cm}
+\frac{1}{2} \lambda^2 \h{\Lambda}^2 \int d^4 x_1 d^4 x_2 \, \viss{T}(x_1) \viss{T}(x_2)
\\
&\hspace{1.5cm}+
\frac{1}{2}\lambda^2
\int d^4 x_1 d^4 x_2 \, \viss{T}^{\mu\nu}(x_1)\, \viss{T}^{\rho\sigma} (x_2) \h{G}_{\mu\nu\rho\sigma}(x_1-x_2) +\OO(\lambda^3) \bigg]
~.
\end{split}
\end{align}
This expression reveals that from the point of view of the visible theory, the interaction \eqref{linearaac} with the hidden theory has induced effective interactions for the visible stress tensor. Working up to quadratic order in $\lambda$, we can exponentiate these interactions in an effective action of the form
\be
\label{linearaf1}
\delta S_{vis}  = \lambda \h{\Lambda} \int d^4 x \, \left(  \viss{T}(x) -  \viss{T}_{\m \n}(x)  \viss{T}^{\m \n}(x) \right)
-
\ee
$$
- \frac{1}{2} \lambda^2 \int d^4 x_1 d^4 x_2 \, \viss{T}^{\mu\nu}(x_1)\, \viss{T}^{\rho\sigma}(x_2)\, \h{G}^{(c)}_{\mu\nu\rho\sigma}(x_1-x_2)=
$$
\be
\label{linearaf2}
=   \lambda \h{\Lambda} \int d^4 x \,  \viss{T}(x)
- \frac{1}{2} \lambda^2 \int d^4 x_1 d^4 x_2 \, \viss{T}^{\mu\nu}(x_1)\, \viss{T}^{\rho\sigma}(x_2)\, \h{Q}_{\mu\nu\rho\sigma}(x_1-x_2)
~.
\ee
Notice that in the last line we have split the action into a separate linear and quadratic piece in $ \viss{T}_{\mu\nu}$ using an operator that is the connected correlator with the addition of a few contact terms
\be\label{linearae}
\h{Q}_{\m\n \r\s}(x_1-x_2) ={ \hat{\Lambda} \over 2}  \left(\eta_{\m\r}\eta_{\n\s}+\eta_{\n\r} \eta_{\m\s}+\eta_{\m\s}\eta_{\n\r}+\eta_{\n\s}\eta_{\m\r} \right) \delta^{(4)}(x_1-x_2) +  \h{G}^{(c)}_{\mu\nu\rho\sigma}  (x_1-x_2)
\ee
For more details on this operator and a derivation of this formula see Appendix \ref{sl} and in particular \eqref{c111}-\eqref{c141}.

We concentrate for a moment on the second term of this interaction, which is expressed more conveniently in momentum space in the form
\be
\label{linearafa}
\delta S^{TT}_{vis} \equiv - \frac{\lambda^2}{2}  \int \frac{d^4 k }{(2\pi)^4}  \, \viss{T}^{\mu\nu}(-k)\, \viss{T}^{\rho\sigma}(k)\, \h{Q}_{\mu\nu\rho\sigma}(k)
~.
\ee
Our Fourier transform conventions are summarised here:
\be
f(x) = \int \frac{d^4 k}{(2\pi)^4} e^{-ik\cdot x} f(k)\sp f(k) = \int d^4 x \, e^{ik\cdot x} f(x)\;.
\ee
This part can be reformulated as an interaction with a classical spin-2 field\footnote{There is no a priori relation with a similar $h_{\m\n}$ introduced in the previous section.} $h_{\mu\nu}$
\be
\label{linearag}
\delta S^{TT}_{eff} = \int  \frac{d^4 k }{(2\pi)^4} \left[ - h_{\mu\nu}(-k) \viss{T}^{\mu\nu}(k) + \frac{1}{2\lambda^2} \PP^{\mu\nu\rho\sigma}(k) h_{\mu \nu}(-k) h_{\rho \sigma}(k) \right]
~.
\ee
The inverse propagator $\PP^{\mu\nu\rho\sigma}(k)$ of the emergent spin-2 field is defined as the inverse of the hidden sector two-point function $\h{Q}_{\mu\nu\rho\sigma}(k)$ (which is the connected correlator with the addition of the contact terms given by \eqref{linearae}). It is straightforward to verify that by extremizing $\delta S^{TT}_{eff}$ with respect to $h_{\mu\nu}$ we obtain an equation of motion for $h_{\mu\nu}$, which we can insert back to $\delta S^{TT}_{eff}$ to recover $\delta S^{TT}_{vis}$ in \eqref{linearafa}. It remains to examine under what circumstances $\PP^{\mu\nu\rho\sigma}(k)$ is well-defined and what tensor structures it involves. This information follows from the properties of the hidden theory two-point function  $\h{Q}_{\mu\nu\rho\sigma}(k)$ and therefore from those of the connected correlator $\h{G}^{(c)}_{\mu\nu\rho\sigma}(k)$.

{Under the assumption that the hidden theory is a Lorentz-invariant QFT,} the Ward identities associated with translations imply that the general form of the connected two-point function $\h{G}^{(c)}_{\mu\nu\rho\sigma}(k)$ (in momentum space) is\footnote{Here we continue to work in Euclidean signature and in a slight abuse of notation we still denote the Euclidean metric $\delta^{\mu\nu}$ as $\eta^{\mu\nu}$. In Minkowski signature, the LHS of \eqref{linearaj} contains an overall factor of $i$ and $\eta^{\mu\nu}$ is the Minkowski metric ${\rm diag}(-1,1,1,1)$.}
\be
\label{linearaj}
%-i
\h{G}^{(c)}_{\mu\nu\rho\sigma}(k) = - {\h{\Lambda} \over 2} \big( \eta_{\mu\nu} \eta_{\rho\sigma} + \eta_{\mu\rho} \eta_{\nu\sigma} + \eta_{\mu\sigma} \eta_{\rho\nu} \big) + \h{b}(k^2) \Pi_{\mu\nu\rho\sigma}(k) +  \h{c}(k^2) \pi_{\mu\nu}(k) \pi_{\rho\sigma}(k)
~.
\ee
In \eqref{linearaj} there are two independent transverse tensor structures proportional to the arbitrary functions $\h{b}(k^2)$ and $\h{c}(k^2)$. They are  defined in the standard fashion as
\be
\label{linearak}
\pi_{\mu\nu} = \eta_{\mu\nu} - \frac{k_\mu k_\nu}{k^2}\sp {\pi_{\m}}^{\r}\pi_{\r\nu}=\pi_{\m\n}\sp {\pi^{\m}}_{\m}=3
~,
\ee
\be
\label{linearal}
\Pi_{\mu\nu\rho\sigma}(k) = \pi_{\mu\rho}(k) \pi_{\nu\sigma}(k) + \pi_{\mu\sigma}(k) \pi_{\nu\rho}(k)
~.
\ee

The Ward identity that leads to this result is summarized in Appendix \ref{Ward}. The first term on the RHS of \eqref{linearaj}, which is proportional to the constant $\h{\Lambda}$, arises from a contact term contribution to the Ward identity. $\h{\Lambda}$ is the same constant that appears in the expectation value of the energy momentum tensor of the hidden theory in eq.\ \eqref{linearai}.

We notice that the only combination of tensor structures which is analytic at  quadratic order in momentum, in the long-wavelength limit $k^2\to 0$, is the one that has
\be
\label{linearam}
\h{b}(k^2) = \h{b}_0 k^2 + \OO(k^4)~, ~~ \h{c}(k^2) = - 2 \h{b}_0 k^2 + \OO(k^4)
~.
\ee
This particular form is a consequence of diffeomorphism invariance and is proven in appendix \ref{free} and \ref{sl}.
$\h{b}_0$ is proportional to the $\delta_{2,0}$ contact terms discussed in detail in appendix \ref{momissues},
\be
\h{b}_0={3\pi^2\over 60}\delta_2\;.
\label{bb0}\ee
Interestingly, viewed as a part of the spectral densities this contact terms indicates a massless pole. However, it is multiplied by $k^4$ and is really a contact term.
The sign  of $\h{b}_0$ does not seem to be controlled by the spectral densities {and} the standard unitarity constraints.
In the free-field computations of appendix  \ref{free}, $\h{b}_0$ turns out to be negative which, as we shall soon see, implies a positive {Planck} scale for the spin-two mode.
On the other hand, as shown in appendix \ref{momissues}, in such a case the spin-zero mode is a ghost. We show that this is intimately linked to a similar statement in Einstein (non-tuned) bi-gravity.

The low-momentum expansion of $\h{Q}_{\mu\nu\rho\sigma}(k)$ up to quadratic order in momentum is
\begin{align}
\begin{split}
\label{linearama}
\h{Q}^{\mu\nu\rho\sigma}(k) &= { \h{\Lambda} \over 2} \big( - \eta^{\mu\nu} \eta^{\rho\sigma} + \eta^{\mu\rho} \eta^{\nu\sigma} + \eta^{\mu\sigma} \eta^{\rho\nu} \big)
\\
&+ \h{b}_0 \bigg[ k^2 \left(\eta^{\mu\rho} \eta^{\nu\sigma} + \eta^{\mu\sigma} \eta^{\nu\rho} - 2 \eta^{\mu\nu} \eta^{\rho\sigma} \right)
\\
&- \eta^{\mu\rho} k^\nu k^\sigma - \eta^{\nu\sigma} k^\mu k^\rho - \eta^{\mu\sigma} k^\nu k^\rho - \eta^{\nu\rho} k^\mu k^\sigma + 2 \eta^{\mu\nu} k^\rho k^\sigma + 2 \eta^{\rho\sigma} k^\mu k^\nu \bigg] +{\cal O}(k^4)
.
\end{split}
\end{align}
A term proportional to $\frac{k_\mu k_\nu k_\rho k_\sigma}{k^2}$ in ${\cal P}^{\m\n\r\s}$  cancels out when \eqref{linearam} is obeyed. This general formula is also proven in \eqref{c137}, by an expansion of the non-linear Schwinger functional. It is also \emph{universal} and holds for any theory as long as the
 Ward identities are obeyed.

As a simple check, consider the two-point function $\h{G}_{\mu\nu\rho\sigma}(k)$ in a theory of $N^2$ decoupled free massive bosons arranged as an $N\times N$ matrix $\h{\phi}$
\be
\label{linearao}
S_{hidden} = - \frac{1}{2}  \int d^4 x \, {\rm Tr} \left( \p_\mu \h{\phi}\, \p^\mu \h{\phi} + m^2 \h{\phi}^2 \right)
~.
\ee
In this normalisation the energy-momentum one- and two-point functions scale as $N^2$. Consequently, the natural large-$N$ scaling of the coupling of $\lambda$ in \eqref{linearaac} is $\OO(N^{-1})$.

In dimensional regularization $(\varepsilon = 4-d \to 0)$ the two-point function can be evaluated explicitly by performing the requisite Wick contractions. In the end, one recovers a result consistent with \eqref{linearam} with parameters
\be
\label{linearap}
\h{\Lambda} = - \frac{N^2}{64 \pi^2} \log \left( \frac{m^2 }{\mu_r^2}  \right)  m^4
~,
\ee
and
\be
\label{linearaq}
\h{b}(k^2) = - \frac{N^2}{16\pi^2} \log \left( \frac{m^2 }{\mu_r^2}  \right)   \left[ \frac{m^2}{12} k^2 + \frac{1}{120} k^4 \right]
~,\ee
\be
\label{linearaqa}
\h{c}(k^2) = - \frac{N^2}{16\pi^2} \log \left( \frac{m^2 }{\mu_r^2}  \right)  \left[ -\frac{m^2}{6} k^2 + \frac{1}{20} k^4 \right]
~\, ,
\ee
with $\mu_r$ a renormalisation mass scale. The details of this computation are summarized in appendix \ref{free}. This is the IR-limit $k^2/m^2 \rightarrow 0$ of the complete result provided in the appendix (evaluated using the MS-scheme). For $N^2$ free fermions there is a similar result with opposite sign in $\h{\Lambda}$, but the same sign in $\h{b}_0$. There also exist terms admitting a regular expansion in $k^2/m^2$ (no-logs), but all such terms are scheme dependent, for more details on this issue see appendix~\ref{momissues}. The dim-reg result is found to obey the Ward identities, in any subtraction scheme, since the divergent terms as well as the finite terms have the same tensor structure.

In a scheme with a hard UV cutoff $M$, we would find extra terms in the two-point function $\h{G}^{\mu\nu\rho\sigma}$ that violate the Ward identity for translation invariance.\footnote{Specifically, one finds quadratically and quartically divergent terms as well as finite, cutoff-independent, terms, which violate the Ward identity associated with translation invariance. The result of the computation for free bosons with a hard UV cutoff is also summarized in Appendix \ref{free}.} The logarithmically-divergent terms do not violate the Ward identities and the dictionary with dimensional regularization is $\Gamma\left( \frac{\varepsilon }{2} \right) \leftrightarrow \log\left( \frac{\mu_r^2}{m^2}\right)$. As a result, if we choose to work with a hard UV cutoff, we must do so in a renormalization scheme where appropriate counterterms subtract the offending terms that violate the Ward identities.

This is an important point that will be recurrent in this theme. Translation invariance, and its reincarnation as diffeomorphism invariance in the Schwinger functions and the associated effective action for the expectation value of the stress tensor (defined already in the previous section) prohibit power divergent contributions to the one-point function of the energy momentum tensor.
Therefore, this is {\em a direct link between emergent diffeomorphism invariance and the absence of power corrections to the one-point function of the stress tensor}.

A key feature of \eqref{linearaj}, is that the first term on the RHS is proportional to the constant $\hat \Lambda$, which has been assumed to be non-vanishing. In the above free-field example we observe that the value of $\h{\Lambda}$ is indeed a non-vanishing constant, which is proportional to the fourth power of the mass gap of the hidden QFT. This is expected more generally; in theories with a single scale  $m$, $\h{\Lambda}$ is non-vanishing and proportional to $m^4$, which is the main IR scale. In conformal field theories,  the energy-momentum one-point function vanishes when all symmetries are preserved.

The presence of $\h{\Lambda}$ facilitates a well-defined inversion\footnote{In the absence of  $\h{\Lambda}$, the two-point function has zero modes which are proportional to $k^{\mu}$ and is therefore  not invertible. In this case, one must invert in the space orthogonal to the zero modes. This gives rise to a non-local effective theory for the graviton. We shall discuss this issue later on in this paper. A similar phenomenon happens in the case of an emergent vector and has been analysed in \cite{u1}.} of $\h{Q}_{\mu\nu\rho\sigma}(k)$ in the long-wavelength limit $k^2 \to 0$, when we go from the expression \eqref{linearafa} to the expression \eqref{linearag}. In appendix \ref{inv2} we perform this inversion exactly to all orders in $k^2$ for a general two-point function. Here we are interested in the IR limit of this exact expression (up to quadratic order in the momentum expansion) which is
\begin{align}
\begin{split}
\label{linearar}
\PP^{\mu\nu\rho\sigma}(k) =&-\frac{1}{2} \h{\Lambda}^{-1} \left( \eta^{\mu\nu}\eta^{\rho\sigma} - \eta^{\mu\rho} \eta^{\nu\sigma} - \eta^{\mu\sigma} \eta^{\nu\rho} \right)
\\
&+  \h{b}_0 \h{\Lambda}^{-2} \bigg[ k^2 \left( \eta^{\mu\nu} \eta^{\rho\sigma} - \eta^{\mu\rho} \eta^{\nu\sigma} - \eta^{\mu \sigma} \eta^{\nu\rho} \right)
\\
&\hspace{1.4cm}+  ( \eta^{\nu \sigma} k^\mu k^\rho +\eta^{\nu \rho} k^\mu k^\sigma + \eta^{\mu \sigma} k^\nu k^\rho +\eta^{\mu\rho} k^\nu k^\sigma ) \bigg]+\ldots
~,
\end{split}
\end{align}
where $\h{b}_0$ is the coefficient of the quadratic term in the expansion \eqref{linearam}. It is straightforward to check that up to the given order
\be
\label{linearara}
\h{Q}^{\mu\nu\kappa \lambda} \PP_{\kappa \lambda \rho\sigma}
= \frac{1}{2} \left( \delta^\mu_\rho \delta^\nu_\sigma + \delta^\mu_\sigma \delta^\nu_\rho \right)
~.
\ee
Indices are raised and lowered with the background metric $\eta^{\mu\nu}$. An important point once again is that this IR expansion holds for any hidden theory obeying the Ward identities.

Recall now that the perturbative expansion of standard Einstein gravity (with a cosmological constant term) around flat space\footnote{In standard cases, in the presence of a cosmological constant this expansion will not make sense as the flat metric is not a solution. However, here, because of the dark energy component, the flat metric  a solution if the additional energy-momentum contribution is trivial.} $g_{\mu\nu} = \eta_{\mu\nu} + \pmet_{\mu\nu}$, is up to quadratic order
\begin{align}
\begin{split}
\label{linearas}
&S_{GR}= \frac{1}{16\pi G} \int d^4 x \, \sqrt g \, \left( R - \Lambda \right)
\\
&=\frac{1}{16\pi G} \int d^4 x \, \bigg \{ -\Lambda - \frac{1}{2} \Lambda \pmet_{\mu}^\mu
- \frac{1}{8} \Lambda \left( \eta^{\mu\nu}\eta^{\rho\sigma} - \eta^{\mu \rho} \eta^{\nu\sigma} - \eta^{\mu \sigma} \eta^{\nu\rho} \right) \pmet_{\mu\nu} \pmet_{\rho\sigma}
\\
&\hspace{1cm} - \bigg[
\frac{1}{4} \p_\rho \pmet_{\mu\nu} \p^\rho \pmet^{\mu\nu} - \frac{1}{4} \p_\rho \pmet^\mu_\mu \p^\sigma \pmet^\nu_\nu + \frac{1}{2} \p_\rho \pmet^\mu_\mu \p_\nu \pmet^{\rho \nu} - \frac{1}{2} \p_\rho \pmet_{\mu\nu} \p^\mu \pmet^{\rho\nu} \bigg]
~.
\end{split}
\end{align}
In momentum space the quadratic part of this action is
\begin{align}
\begin{split}
S_{GR}^{(2)}&= - \frac{1}{16\pi G}\int {d^4 k  \over (2\pi)^4} \, \bigg \{
\frac{1}{8} \Lambda \left( \eta^{\mu\nu}\eta^{\rho\sigma} - \eta^{\mu \rho} \eta^{\nu\sigma} - \eta^{\mu \sigma} \eta^{\nu\rho} \right) \pmet_{\mu\nu}(k) \pmet_{\rho\sigma}(-k)
\\
&\hspace{0.7cm} +  \bigg[
\frac{k^2}{8} \left( 2 \eta^{\mu\nu} \eta^{\rho\sigma} - \eta^{\mu\rho} \eta^{\nu\sigma} - \eta^{\mu \sigma} \eta^{\nu\rho} \right)
- \frac{1}{4} \eta^{\mu\nu} k^\rho k^\sigma - \frac{1}{4} \eta^{\rho\sigma} k^{\mu} k^\nu\label{b2}
\\
&\hspace{0.7cm}+ \frac{1}{8} ( \eta^{\nu \sigma} k^\mu k^\rho +\eta^{\nu \rho} k^\mu k^\sigma + \eta^{\mu \sigma} k^\nu k^\rho +\eta^{\mu\rho} k^\nu k^\sigma )
\bigg] \pmet_{\mu\nu}(k) \pmet_{\rho\sigma}(-k) \bigg\}
~.
\end{split}
\end{align}
Evidently, this expression does not coincide with \eqref{linearag}, \eqref{linearar} if we set $h_{\mu\nu}=\pmet_{\mu\nu}$. However, by setting in real space
\be
\label{linearat}
h_{\mu\nu} =  \pmet_{\mu\nu} - \frac{1}{2} \pmet\, \eta_{\mu\nu} +\lambda \h{\Lambda} \, \eta_{\mu\nu}
~, ~~ \pmet = \pmet^{\rho\sigma} \eta_{\rho\sigma}
\ee
and
\be
\label{linearau}
\shift{T}^{\mu\nu} \equiv \viss{T}^{\mu\nu} + \frac{ \lambda^{-1}}{2}  \left( 1 - \lambda^{-1} \h{\Lambda}^{-1} \right) \eta^{\mu\nu}~,~~
\shift{T}=\shift{T}^{\mu\nu}\eta_{\mu\nu}
~,
\ee
we find that the full effective action of the visible QFT at this order in the $\lambda$-expansion and at the two-derivative level is
\be
\label{linearav}
S_{eff} = S_{vis}
+\int d^4 x \, \left(  \pmet_{\mu\nu} \shift{T}^{\mu\nu} - \frac{1}{2} \pmet\, \shift{T} \right)
+ \frac{1}{16\pi G}\int d^4 x \, \bigg[ \sqrt g\left(   R -   \Lambda \right) \bigg]^{(2)}_{g_{\mu\nu} = \eta_{\mu\nu} + \pmet_{\mu\nu}} \, .
\ee
This transformation is performed in detail in the appendix in equations \eqref{h6b1} to \eqref{h6bbb}. The coupling of the graviton to the stress tensor seems non-standard. It can be written as a standard coupling to $ \shift{T}^{\mu\nu} - \frac{1}{2}\eta^{\m\n}~ \shift{T}$,
or   it can be seen to arise from an expansion around flat space of the covariant coupling $\int d^4 x \, \sqrt{g} \, g_{\m \n}  \shift{T}^{\mu \nu}  $.}
In this last expression, we can identify the effective gravitational parameters in terms of the original parameters, appearing in the two-point function of the stress tensor as
\be
\label{linearaw}
\Lambda = -  {\hat \Lambda\over 2  \h{b}_0}~, ~~
\frac{1}{16\pi G}\equiv M_P^2 = - { 4 \h{b}_0 \over   \lambda^{2} \h{\Lambda}^{2}}
~.
\ee
The superindex $(2)$ on the last term of \eqref{linearav} reminds us that in this expression we work up to quadratic order in the weak $\pmet_{\mu\nu}$ expansion.
Then,  this setup always predicts that the sign of the cosmological constant is the same to  that of the hidden vacuum energy.

For a more concrete example, consider a large-$N$ hidden theory with $N^2$ hidden degrees of freedom. In such a case, as shown in \cite{SMGRAV}, $\lambda\sim {\cal O}(1)$ when the hidden stress tensor is normalized so that its two-point function is $\sim {\cal O}(1)$. Here we use the normalization that all correlators are $\sim {\cal O}(N^2)$ and therefore we can parametrize  $\l$ as
\be
\l\sim {N^{-1}\over M^4}\;,
\ee
 where $M$ is a mass of the order of the messenger scale.

We therefore have the estimates\footnote{These estimates are not generic. We shall discuss this issue, in more detail in section \ref{fel}.}
\be
\h{\Lambda} \sim m^4 N^2 \sp \hat b_0\sim m^2 N^2
\label{b3}
~.
\ee
and
\be
\Lambda\sim m^2 \sp M_P^2\sim {M^8\over m^6}
\label{b4}
~.
\ee
with both last expressions scaling as ${\cal O}(N^0)$.

Interestingly, the effective four-dimensional Planck scale and the cosmological constant are enhanced with ${M\over m}\gg 1$.
We also have
\be
{\Lambda \over M_P^2}\sim \left({m\over M}\right)^8
\label{b5}
~.
\ee
It is interesting to observe that this qualitative analysis points to a parametrically small cosmological constant in Planck units.

The second term on the RHS of \eqref{linearav}, which describes the coupling of the visible QFT to the emergent graviton, can be expressed in terms of the original energy-momentum tensor of the visible QFT, $\vis{T}_{\mu\nu}$, as\footnote{To arrive at this expression we combined the definitions \eqref{linearaab} and \eqref{linearau}.}
\begin{align}
\begin{split}
\label{linearawa}
&\int d^4 x \, \left(  \pmet_{\mu\nu} \shift{T}^{\mu\nu} - \frac{1}{2} \pmet\, \shift{T} \right) =
\\
& \hspace{1cm}=  \int d^4 x \, \pmet_{\mu\nu} \left( \vis{T}^{\mu\nu} - \frac{1}{2} (1+ 2 {\mathfrak c}) \, \vis{T} \,\eta^{\mu\nu} - \frac{ \lambda^{-1} }{2}\left( 1 - \lambda^{-1} \h{\Lambda}^{-1} \right) \eta^{\mu\nu} \right)
~.
\end{split}
\end{align}
This coupling involves two extra terms besides the standard linear coupling of gravity to matter, $\pmet_{\mu\nu} T^{\mu\nu}$. The first is a coupling of the form $T \pmet$, which is explained  below  \eqref{linearav}. The second is an effective shift of the energy-momentum tensor of the visible QFT by a dark energy contribution proportional to the background metric $\eta^{\mu\nu}$ that we shall come back to in a moment. This particular type of coupling between matter and gravity is not in contradiction with the Bianchi identity in gravity. We  briefly elaborate on this point.

From the Euler-Lagrange variation of the action we obtain the following perturbative equations of motion for $\pmet_{\mu\nu}$
\begin{align}
\begin{split}
\label{bianchiaa}
&\frac{\Lambda}{2} \eta^{\mu\nu} + \frac{\Lambda}{4} \left(\eta^{\mu\nu} \eta^{\rho\sigma} - \eta^{\mu\rho} \eta^{\nu \sigma} - \eta^{\mu\sigma} \eta^{\nu\rho} \right) \pmet_{\rho\sigma}
\\
& - \frac{1}{2} \bigg( \partial^2 \pmet^{\mu\nu} - \eta^{\mu\nu} \p^2 \pmet
+ \eta^{\mu\nu} \p_\rho \p_\sigma \pmet^{\rho \sigma} +  \p^\mu \p^\nu \pmet
- \p^\mu \p_\rho \pmet^{\rho \nu} -\p^\nu \p_\rho \pmet^{\rho \mu} \bigg)
\\
&=16 \pi G   \left( \shift{T}^{\mu\nu} - \frac{1}{2} \shift{T}\eta^{\mu\nu} \right)
~.
\end{split}
\end{align}
Contracting both sides of this equation with a partial derivative eliminates the quadratic part on the second line of \eqref{bianchiaa} and gives the dynamical equation
\be
\label{bianchiab}
\p_\mu \left( \shift{T}^{\mu\nu} - \frac{1}{2} \shift{T}\eta^{\mu\nu} \right)
=   \frac{1}{16\pi G} \frac{\Lambda}{4} \left( \p^\nu \pmet - 2 \p_\rho \pmet^{\rho \nu} \right)
~,
\ee
which does not impose a constraint on the visible energy-momentum tensor $T^{\mu\nu}$. This equation is consistent with the perturbative expansion of the covariant conservation equation
\be
\label{bianchiac}
\nabla_\mu \left( \shift{T}^{\mu\nu} - \frac{1}{2} \shift{T}^{\rho\sigma} g_{\rho\sigma} \eta^{\mu\nu} \right) = 0
~.
\ee

To see this, notice that a perturbative expansion around flat space employs the expansion of the energy-momentum tensor
\be
\label{bianchiad}
\shift{T}^{\mu\nu}  - \frac{1}{2} \shift{T}^{\rho\sigma}g_{\rho\sigma} \eta^{\mu\nu} = t_{(0)}^{\mu\nu} +  t_{(1)}^{\mu\nu} + {\rm higher~order}
\ee
where we defined the contributions to the first two orders as
\be
\label{bianchiae}
t_{(0)}^{\mu\nu} =  \frac{1}{16\pi G} \frac{\Lambda}{2}\eta^{\mu\nu}
~,
\ee
which is the leading order term in \eqref{bianchiaa} when one switches off the perturbation ($\pmet_{\mu\nu}=0$) and
\be
\label{bianchiaf}
t_{(1)}^{\mu\nu} = \delta \shift{T}^{\mu\nu} -\frac{1}{2} \delta \shift{T}^{\rho\sigma} \eta_{\rho\sigma} \,  \eta^{\mu\nu}
-\frac{\pmet}{2} t_{(0)}^{\mu\nu}
~.
\ee
The combination
\be
\label{bianchiag}
\delta \shift{T}^{\mu\nu} -\frac{1}{2} \delta \shift{T}^{\rho\sigma} \eta_{\rho\sigma} \,  \eta^{\mu\nu}
\ee
is precisely the first order energy-momentum combination that appears on the RHS of eq.\ \eqref{bianchiaa}. Then, the perturbative expansion of eq.\ \eqref{bianchiac} gives
\begin{align}
\begin{split}
\label{bianchiai}
&\p_\mu t_{(1)}^{\mu\nu} + \left(\frac{1}{16\pi G} \frac{\Lambda}{2} \right)
\left[ \left( \Gamma^{(1)} \right)^\mu_{\rho \mu} \eta^{\rho \nu} +  \left( \Gamma^{(1)} \right)^\nu_{\rho \mu} \eta^{\rho \mu} \right] = 0
\\
\Rightarrow ~&  \p_\mu t_{(1)}^{\mu\nu}  + \left( \frac{1}{16\pi G} \frac{\Lambda}{2} \right)  \p_\mu \pmet^{\mu\nu} =0
\end{split}
\end{align}
which immediately translates to \eqref{bianchiab} after the implementation of \eqref{bianchiaf}.

 We summarise what we found in this subsection. When the interaction $S_{int}$ between the visible and hidden sectors is
\be
\label{linearawc}
S_{int} = \lambda \, \int d^4 x\, \left( \vis{T}_{\mu\nu}(x)\, \h{T}^{\mu\nu} (x) +\mathfrak c\,  \vis{T}(x) \h{T}(x) \right)
\ee
the effective gravitational description \eqref{linearav} up to quadratic order in the metric and up to two derivatives is
\begin{align}
\begin{split}
\label{linearawd}
S_{eff} =&\, S_{vis}
+ \int d^4 x \, \pmet_{\mu\nu} \left( \vis{T}^{\mu\nu} - \frac{1}{2} (1+ 2 {\mathfrak c}) \, \vis{T} \,\eta^{\mu\nu} -\frac{ \lambda^{-1}}{2} \left( 1 - \lambda^{-1} \h{\Lambda}^{-1} \right) \eta^{\mu\nu}
\right)
\\
&+ \frac{1}{16\pi G}\int d^4 x \, \bigg[ \sqrt g\left(  R - \Lambda  \right) \bigg]^{(2)}_{g_{\mu\nu} = \eta_{\mu\nu} + \pmet_{\mu\nu}}
~.
\end{split}
\end{align}
This computation also shows that Einstein gravity is a \emph{universal} effective IR description capturing the stress energy tensor exchanges between the two sectors, since its derivation was solely based on an infrared expansion and the use of the Ward identities for the general hidden theory stress energy correlator. However, there is an extra vacuum energy in this description which remembers the existence of the original flat metric.

To investigate a bit closer the parameters of the effective theory we set
\be
\lambda={1\over {N M^4}}\sp \hat \Lambda=\e m^4 N^2\sp
\hat b_0=-\kappa ~m^2~N^2
\label{ag1}
\ee
for hidden theories with $N^2$ degrees of freedom. $M$ is of the order of the messenger scale, and $m$ is the characteristic scale of the IR limit of the hidden theory (of the order of its mass gap).
$\e=\pm 1$ is a sign that determines the sign of the cosmological constant of the hidden theory.~~$\kappa$ is a dimensionless constant.

We observe that the emergent spin-2 theory \eqref{linearawd} has the standard quadratic interactions of a gravitational theory and exhibits both a cosmological constant
\be
\Lambda= {\e\over \kappa} m^2
\label{ag2} \ee
 in the gravitational sector and a dark energy contribution\footnote{$\Lambda_{dark}$ is defined so that gravity and matter couple through the linearised coupling $\pmet_{\mu\nu} \left( \widetilde T^{\mu\nu} - \frac{1}{2}(1+2\mathfrak c) \widetilde T \eta^{\mu\nu}\right)$ with $\widetilde T^{\mu\nu}\equiv T^{\mu\nu} + \Lambda_{dark} \eta^{\mu\nu}$. This shift is not possible in the special case $\mathfrak c =-\frac{1}{4}$, where the combination $T^{\mu\nu}-\frac{1}{2}(1+2\mathfrak c) T \eta^{\mu\nu}$ is traceless.} from (\ref{linearawa})
\be
\label{lineardark}
M_P^2\Lambda_{dark}= -\frac{(2\pi)^4}{1+4\mathfrak c} \lambda^{-1} \left( 1 +\frac{1}{2} \lambda^{-1} \h{\Lambda}^{-1} \right)
=-\frac{N}{1+4\mathfrak c}M^4\left(1+{\e x\over 2(2\pi)^4 N}\right)
\ee
\be
  x\equiv {M^4\over m^4}
\ee
to the energy-momentum tensor of the visible QFT.
The emergent Planck scale from (\ref{linearaw}) is
\be
M_P^2=-(2\pi)^8{{\hat b_0}\over \l^2\Hat \Lambda^2}= \kappa~{M^4\over m^2}x
\ee

We may now calculate the relevant ratios of scales
\be
{\Lambda\over M_P^2}=-{\e\over \kappa^2 x^2}\sp
{\Lambda_{dark}\over M_P^2}=-{{N\over x}+{\e \over 2(2\pi)^4}\over (1+4\mathfrak c)\kappa^2~x^2}
~,
\label{eq1}\ee
\be
 {\Lambda_{dark}\over \Lambda}={\left(\e ~{N\over x}+{1\over 2(2\pi)^4 } \right)\over (1+4\mathfrak c) }\sp {M^4\over M_P^4}={1\over \kappa^2 x^3}
\label{eq11}\ee

We usually assume that the messenger scale $M$ is much larger than the characteristic scale $m$ of the hidden QFT, ie. $x\gg 1$.

It should be also kept in mind that the quantum effects of the visible theory, including its contributions to the cosmological constant have not yet been taken into account.

\subsection{The induced gravitational interaction}\label{InducedTTinteraction}

In order to resolve the problems created by the mixing of contact terms with
pole data, we now directly analyse the induced gravitational interaction between two visible stress energy sources, using the general structure of the two-point function of the stress-tensor.

We start from (\ref{linearafa}) in momentum space
\be
L_{int}=-{\l^2\over 2}{\bf T}_{\m\n}(-k)\hat Q^{\m\n\r\s}(k){\bf T}_{\r\s}(k)=
\label{gg29}\ee
$$
=T_{\m\n}(k)Q^{\m\n\r\s}T_{\r\s}+\cc T(-k){Q_{\m}}^{\m\r\s}T_{\r\s}+\cc T_{\m\n}(-k){Q^{\m\n\r}}_{\r}T(k)+\cc^2 T(-k){{Q_{\m}}^{\m\r}}_{\r}T(k)
$$
Then from (\ref{linearae}) and (\ref{linearaj}) we have
\be
\label{h1n}
\hat Q^{\mu\nu\rho\sigma}(k) = -{a\over 2} \eta^{\mu\nu} \eta^{\rho\sigma} +{a\over 2}\big( \eta^{\mu\rho} \eta^{\nu\sigma} + \eta^{\mu\sigma} \eta^{\rho\nu} \big) + b(k^2) \Pi^{\mu\nu\rho\sigma}(k) +  c(k^2) \pi^{\mu\nu}(k) \pi^{\rho\sigma}(k)
~.
\ee
 with $a={\hat \Lambda\over 2}$, and substituting in (\ref{gg29})  we obtain,
\be
L_{int}=-{\l^2\over 2}\left[(a+2b)T^{\m\n}T_{\m\n}-4b{(kT)^{\m}(kT)_{\m}\over k^2}+(2b+c){(kkT)^2\over k^4}-2c{T(kkT)\over k^2}+\right.
\label{gg29a}\ee
$$\left.
+\left(c-{a\over 2}\right)T^2+\cc \left(2(2b+3c-a)T^2-2(2b+3c){T(kkT)\over k^2}\right)+\cc^2(6b+9c-4a)T^2\right]
$$
where
\be
(kT)^{\m}\equiv k_{\n}T^{\m\n}\sp (kkT)\equiv k_{\m}k_{\n}T^{\m\n}\;.
\ee
The functions $b$, $c$ above are $\widehat{b}$ and
$\widehat{c}$ in (\ref{linearaj}), and we dropped the hats for convenience.

Since the stress tensor is conserved to order ${\cal O}(\l^2)$ we have
\be
k^{\m}T_{\m\n}=0
\label{gg30}\ee
and using this, we obtain
\be
L_{int}=-{\l^2\over 2}\left[(2b+a)T_{\m\n}(-k)T^{\m\n}(k)+\left(c-{a\over 2}\right)T(-k)T(k)+\right.
\label{gg31}\ee
$$
\left.+2\cc(2b+3c-a)T(-k)T(k)+\cc^2(6b+9c-4a)T(-k)T(k)\right]
$$
$$
=-{\l^2\over 2}\left[(2 k^4 B_2+a)T_{\m\n}(-k)T^{\m\n}(k)+\phantom{2\over 3}\right.
$$
$$
\left.+\left({(1+3\cc)^2\over 3} k^4 B_0-{2\over 3} k^4 B_2-{a\over 2}(1+4\cc+8\cc^2)\right)T(-k)T(k)\right] \, ,
$$
where he have expressed everything in terms of the spin zero, $B_0$  and spin two $ B_2$ spectral densities of the two-point function of the energy-momentum tensor of the hidden theory,
\be
b\equiv k^4 B_2\sp  2b+3c\equiv k^4 B_0
\label{nh5}\ee
 as explained in appendix \ref{spectral}.

We may rewrite the important, non-contact part of (\ref{gg31}) as
\be
L_{int}= -{\l^2\over 2}\left[2 B_2\left(T_{\m\n}(-k)T^{\m\n}(k)-{1\over 3}T(-k)T(k)\right)+{(1+3\cc)^2\over 3}  B_0\right]+L_{\rm contact}
\label{tt1}\ee
The contact part contains contributions from the stress-tensor vev, as well as the contact contributions in $ B_{2,0}$ as explained in appendix  \ref{spectral}.
\be
L_{\rm contact}\equiv -{\l^2\over 2}\left[\left(a+\delta_2k^2+A_2^{ren}k^4\right)\left(T_{\m\n}(-k)T^{\m\n}(k)-{1\over 3}T(-k)T(k)\right)+\right.
\ee
$$
\left.+\left({(3\cc+1)^2\over 3}\left(k^4 A_0^{ren}-6\delta_2 k^2\right)-{24\cc^2+12\cc+1\over 6}a\right)T(-k)T(k)\right]
$$
Using the spectral representation in appendix~\ref{spectral}, we find that the coefficients can be expressed as integrals of the respective spectral densities
\be
\hat{B}_2(k)  \, = \, {3\pi^2k^4\over 80}\int^{\infty}_0d\mu^2 {\rho_{2}(\mu^2)\over k^2+\mu^2}
\label{gg27n}\ee
\be
{\hat{B}_0(k)\over 3} \, = \, {\pi^2k^4\over 40}\int^{\infty}_0d\mu^2 {\rho_{0}(\mu^2)\over k^2+\mu^2} \, .
\label{gg28n}\ee

The role of appropriate subtractions that need to be performed in order to make these expressions convergent, is discussed in appendix~\ref{momissues}.
For static sources of mass $m_s$, as in (\ref{c43}), we obtain in configuration space
\be
S_{int}=-{3\l^2 m^2_s \pi^2\over 320\pi }\square^2{1\over r}\left[{4\over 3}\int_0^{\infty}d\mu^2 \rho_{2}(\mu)e^{-\mu r}+{(1+3\cc)^2\over 3}\int_0^{\infty}d\mu^2 \r_{0}(\mu)e^{-\mu r}\right]
\label{gg32}\ee
where we have dropped a contact piece.

We now analyse a gapped theory with discrete spectrum, for which $c^{(0)} \sim \delta(\mu^2 - m_0^2)$ and $c^{(2)} \sim \delta(\mu^2 - m_2^2)$ are the lowest lying poles. Near such poles, the spin zero and spin two coefficients take the form
\be
\h{B}_2 \simeq \frac{R_2(m_2)}{k^2 + m_2^2} \sp \h{B}_0 \simeq \frac{R_0(m_0)}{k^2 + m_0^2} \, ,
\ee
with the residues $R_{2,0}$ being positive parameters (as long as the hidden theory is unitary) of mass dimension six.

A pole in the spin two part gives a contribution to (\ref{gg31}) which is
\be
L^{(2)}_{int}=-{\l^2 R_2}{\left[T_{\m\n}(-k)T^{\m\n}(k)-{1\over 3}T(-k)T(k)\right]\over k^2+m_2^2} \, .
\label{gg33}\ee
This is precisely the structure of the contribution of a propagating Fierz-Pauli (FP) graviton with mass $m_2$, as seen by contracting \eqref{FPpropagator} with two stress energy tensor sources and using the conservation equation as above.
Moreover, the associated Planck scale for the exchange is given from (\ref{gg33}) by
\be
{1\over M_2^2}\sim \l^2~R_2\sim {R_2\over M^8}
\ee
As $R_2$ is expected to be of order $R_2\sim m_2^6$ and therefore a mass much lower that the messenger scale, one obtains a ``large" effective Planck scale. In such a case the {messenger} scale {may} be allowed to be as low as $M\sim 10^{-3}$ eV, but such a low scale may be {problematic} due to the role of messenger mesons that are expected to be ultralight and carry SM quantum numbers.

On the other hand, poles in the spin zero part generate a dilaton like interaction of the form
\be
L^{(0)}_{int}=-{\l^2 R_0}{(1+3\cc)^2\over 6}{T(-k)T(k)\over k^2+m_0^2} \, .
\label{gg34}\ee
The associated Planck scale for the scalar exchange, is given from (\ref{gg34}) by
\be
{1\over M_0^2}\sim {\l^2~R_0~ (1+3\cc)^2}\sim {R_0~(1+3\cc)^2\over M^8}
\ee
and its size depends on the associated residue $R_0$ and $\cc$.

Equation (\ref{tt1})  is the cleanest formula that shows what is the nature of the emergent gravitational interaction.
As a function of the hidden theory, the structure of the spectral densities of the stress-tensor two-point function changes. In appendix \ref{staticpotential} we analyse the static gravitational interactions as a function of the structure of the spectral densities.

A few remarks are in order.
\begin{itemize}

\item First, (\ref{tt1}) shows that the vev $\hat \Lambda$ is only responsible for a contact interaction that can be removed by a redefinition.

\item
The spin-two density of the two-point function couples to a combination of stress tensors that are characteristic of massive gravity with a Fierz-Pauli mass term.

\item The positivity of this density in a unitary theory implies that this interaction is always attractive.

\item The spin-zero density associated with the trace of the (hidden)   energy-momentum tensor mediates an additional scalar interaction that couples the traces of the visible energy-momentum tensor. It is the ``dilaton" interaction.

 \item This interaction is typically associated with the conformal factor of the metric, and in our case this is precise, in terms of the emergent metric $h_{\m\n}$ we integrated in. However, unlike the case of non-Fierz-Pauli gravity where this mode is operational and ghostly, here the interaction is always attractive and stable (for a unitary theory).

 \item This spin-zero interaction decouples when $\cc=-{1\over 3}$. Interestingly, for this value of $\cc$, the $TT$ interaction is of the special form studied in two dimensions\footnote{The special combination in any dimension $d$ is $\cc=-{1\over d-1}$.}, \cite{TT}, and conjectured to also have special properties in higher dimensions, \cite{Tay}. We see here one such special property, the decoupling of the ``dilaton" interaction. The coupling of the dilaton is therefore proportional to $3\cc+1$.   The $TT$ coupling in two-dimensions was associated with massive two-dimensional gravity in \cite{To}.

     \item The effective action and parameters obtained in the previous {subsection} \ref{4.1} seem incompatible with the ones here. The reason was explained earlier: These are {affected by} contact terms that are not relevant for the interaction.
         The form of the effective action is however correct.

 \end{itemize}

\subsection{IR massive gravity actions}

So far we have observed the following

\begin{enumerate}

\item In a gapped theory,  the IR quadratic gravity (effective) action can be mapped to the Einstein action plus a cosmological constant, {as shown} in  subsection \ref{4.1}. There is also a ``dark energy" contribution to the visible stress tensor.

\item The induced gravitational interaction when the stress-tensor two-point function is dominated by poles, found in subsection \ref{InducedTTinteraction}, is given by the result of a  spin-2 massive graviton exchange as well as a scalar spin-0 massive exchange coupling to the trace of the stress tensor. Moreover, both degrees of freedom have positive contributions and both interactions are always attractive if the hidden theory is unitary.

\end{enumerate}

Our next step is to try to bridge these two observations by considering the effective theories of massive gravity at the linearized level.

We start with the Fierz-Pauli (FP) action describing a massive spin-2 particle in flat space, which is given by (we fix the overall normalisation to match \eqref{linearag}),
\be\label{Massive1}
S =  \frac{1}{ 2 \lambda^2} \int d^d x  \, h_{\m \n} \, \mathcal{P}^{\m \n \r \s}_{F.P.} \, h_{\r \s} \, ,
\ee
with the operator $\mathcal{P}_{F.P.}$ given by
\be\label{Massive2}
\mathcal{P}^{\m \n \r \s}_{F.P.} = \half \left(\eta^{\m \r} \eta^{\n \s} + \eta^{\m \s} \eta^{\n \r} - 2 \eta^{\m \n} \eta^{\r \s} \right) (\Box - m_2^2) + \partial^\m \partial^\n \eta^{\r \s} + \partial^\r \partial^\s \eta^{\m \n}  -
\ee
$$
- ( \eta^{\nu \sigma} \p^\mu \p^\rho +\eta^{\nu \rho} \p^\mu \p^\sigma + \eta^{\mu \sigma} \p^\nu \p^\rho +\eta^{\mu\rho} \p^\nu \p^\sigma )
$$
Comparing with \eqref{b2}, we find that the derivative terms match those of the quadratic Einstein action, without any redefinition of the metric $h_{\m \n}$.
The potential terms are however different
 from that of expanding the Einstein-Hilbert action with cosmological constant around flat space.
The equations of motion of FP massive gravity are
\be\label{Massive3}
\Box h_{\m \n} - \partial_\l \partial_\m h^\l_\n - \partial_\l \partial_\n h^\l_\m + \eta_{\m \n} \partial^\l \partial^\s h_{\l \s} + \partial_\m \partial_\n h - \eta_{\m \n} \Box h - m_2^2 (h_{\m \n} - \eta_{\m \n} h) = 0 \, ,
\ee
which are also equivalent to the set
\be\label{Massive4}
(\Box - m_2^2) h_{\m \n} = 0, \qquad \partial^\m h_{\m \n} = 0 , \qquad h =0 \, .
\ee
In total, we are left with 5 real space degrees of freedom of a four-dimensional spin 2 particle.
The specific tuning between the mass terms is adjusted so that the addition sixth scalar degree of freedom, that is ghost-like in Einstein gravity
 is absent.

If we invert the FP operator, we find in terms of projectors
\be\label{FPpropagator}
\hat{G}_{\m \n \r \s}^{F.P.}(k) = \frac{1}{k^2 + m_2^2} \left(\half \Pi^{(m)}_{\m \n  \r \s}  - \frac{1}{3} \pi^{(m)}_{\m \n} \pi^{(m)}_{\r \s} \right) \, ,
\ee
with
\be\label{Massive5}
\pi^{(m_2)}_{\mu\nu} = \eta_{\mu\nu} - \frac{k_\mu k_\nu}{m_2^2}\sp \Pi^{(m_2)}_{\mu\nu\rho\sigma}(k) = \pi^{(m_2)}_{\mu\rho}(k) \pi^{(m_2)}_{\nu\sigma}(k) + \pi^{(m_2)}_{\mu\sigma}(k) \pi^{(m_2)}_{\nu\rho}(k)
~,
\ee
Upon contracting \eqref{FPpropagator} with two stress energy tensor sources we find precisely the expression \eqref{gg33} for the effective interaction due to the exchange of the FP graviton.

The FP action does not capture the scalar interaction that couples to the trace of the visible stress-tensor.
For this, in principle, we should integrate-in an extra  scalar degree of freedom. For example \eqref{gg34} shows that one can describe such states via a scalar coupled to the trace of the stress energy tensor
\be
S_s = \frac{1}{2 \lambda^2 (1+3 \cc)^2} \int d^4 k \, \phi(k) (k^2+ m_0^2) \phi(-k) \, ,
\ee
where we chose to absorb constant parameters in the definition of $\phi$.
There is a scaling transformation that mixes this scalar with the trace of the emergent metric.
However, as our inversion of the interaction using a general metric shows, this extra interaction should be captured also by part of the induced metric.

 We summarize that a FP action plus a scalar dilaton action does not match the IR expansion of the gravitational action found in (\ref{linearawd}).
 It agrees, on the other hand, with the induced gravitational interaction near the poles of the stress-tensor two-point, function as seen in equations ~\eqref{gg33}, \eqref{gg34}.

The next step is to investigate the general massive gravity, without the FP tuning that is given by
\be
S =  \frac{1}{64 \pi G} \int d^d x  \, h_{\m \n} \, \mathcal{P}^{\m \n \r \s}_{F.P.cc} \, h_{\r \s} \, ,
\ee
with the operator given by\footnote{The limit, $m_0\to\infty$ gives back the FP action.}

\be\label{FPandCC}
\mathcal{P}^{\m \n \r \s}_{F.P.cc} = \half \left(\eta^{\m \r} \eta^{\n \s} + \eta^{\m \s} \eta^{\n \r} - 2 \eta^{\m \n} \eta^{\r \s} \right) \Box  + \partial^\m \partial^\n \eta^{\r \s} + \partial^\r \partial^\s \eta^{\m \n}  -
\ee
$$
 - ( \eta^{\nu \sigma} \p^\mu \p^\rho +\eta^{\nu \rho} \p^\mu \p^\sigma + \eta^{\mu \sigma} \p^\nu \p^\rho +\eta^{\mu\rho} \p^\nu \p^\sigma )-
$$
$$
- \frac{m_2^2}{2} \left(\eta^{\m \r} \eta^{\n \s} + \eta^{\m \s} \eta^{\n \r} \right) +m_2^2{m_2^2+2m_0^2\over 2(2m_2^2+m_0^2)} \eta^{\m \n} \eta^{\r \s}
$$
where we parametrized the mass terms in terms of the masses, $m_2$ of the spin-2 mode, and $m_0$ of the spin-0 mode (that in (\ref{FPandCC}) is actually a ghost).
Naively matching (\ref{linearawd}) with  (\ref{FPandCC})  would give
\be
m_2^2=m_0^2=-{\Lambda\over 2}
\ee
and a ghostly scalar mode.

However, this matching is ``illegal". The reason is the fluctuation in (\ref{linearawd}) is shifted from the vacuum fluctuation and is therefore not a fluctuation around a metric with a cosmological constant.
Moreover,  the regime of validity of our results in \ref{InducedTTinteraction} does not overlap with the regime of validity in (\ref{linearawd}).
Indeed, (\ref{linearawd}) is valid when all momenta are well below any massive poles in the two-point function, whereas in (\ref{tt1}) we expanded around a massive pole.

We can also discuss massive gravity (F.P.) in the presence of a non trivial cosmological constant or with an arbitrary FP tuning at the linearised level. Even though we know that generically such an action by itself does not describe a well-defined theory, since there is a propagating ghost mode, nevertheless from the present point of view, it could merely serve as an IR expansion of the complete theory, and needs to be supplemented with extra states that can in principle restore the unitarity of the total theory. The complete description is in terms of the hidden theory stress energy correlators as advocated in subsection~\ref{InducedTTinteraction} and is perfectly healthy and unitary.
On the other hand, the local effective gravitational actions only contain a part of this information at specific energy scales and therefore might appear to be non-unitary.

\subsection{Generalization to arbitrary interactions}

In the previous discussion we assumed a very special interaction between the two theories, namely (\ref{linearaa}) that involved the two energy-momentum tensors of the individual theories that were conserved in the decoupled theories.
In the most general case, the interactions at energies well below $M$, will contain an infinite sequence of multi-trace couplings of the form
\be
S_{int}=\int d^4x\left[\sum_i \left(\l_i ~O_i\widehat O_i+\tilde \l_i ~ W_{\m}^i\widehat W^{\mu,i}+\hat \l_i ~T^{i}_{\m\n}\widehat T^{\m\n,i}+\cdots\right)\right]
\label{w1}\ee
where $O_i$ denote scalar operators, $W_{\mu}^i$ vector operators, $T_{\m\n}^i$ symmetric tensor operators and so on. The higher the dimensions of these operators, the more suppressed the couplings $\l_i$ are, as they are proportional to appropriate inverse powers of the messenger scale, M.

The calculation of the effective action, emerging from integrating out the hidden theory proceeds along similar lines as we have done for the stress tensor interaction. For the scalar interactions this has been done in detail in \cite{axion}, where  composite (emergent) axions were discussed.
Integrating out the hidden theory will induce, to leading order,  linear and quadratic interactions among the relevant SM operators. These will be resolved by introducing emergent fields.
For each operator  appearing in (\ref{w1}) an emergent field will appear with the same tensor properties as the associated operators.

The induced interactions, to quadratic order, will have linear terms where the emergent fields will couple linearly to the appropriate standard model operators (shifted by hidden-sector vevs). They will also have quadratic terms proportional to the inverse of the appropriate hidden tensor connected two-point functions.

As was analyzed in detail in \cite{axion}, the associated mass terms that will appear in the quadratic part of the emergent fields will be of the order of the UV-cutoff (messenger) scale $M^2$, except if protected by symmetries.
In \cite{axion} one such type of symmetry was explored in detail: the topological symmetry of instanton density correlators, (known already from QCD) which prohibits powers of the cutoff to appear in the relevant correlators. Therefore, the associated mass is controlled by the mass scales of the low-energy theories.

In this paper we have seen a similar effect for the {gravitational} case. Indeed, the effective cosmological constant that corresponds to the ``mass" part of the emergent graviton propagator, is protected from translation Ward identities and is logarithmic in the cutoff rather than quartic. This we have seen by an explicit calculation above. The same applies to the emergent Planck scale that is also logarithmic in the cutoff rather than quadratic\footnote{A similar phenomenon appears for conserved global currents of the hidden theory. They should give rise to emergent gauge fields, \cite{u1,u12}.}.

\subsection{Features of the emergent linearised effective theory\label{fel}}

\begin{itemize}
\item The generic low-energy IR effective action dictated by Ward identities is that of standard model coupled to Einstein gravity via a non-standard gravitational coupling. The stress tensor suffers a shift due to the background metric.

\item The IR (bi-gravitational) theory by itself is not unitary, even though the starting point was a perfectly unitary QFT. The reason is that the IR expansion and computation of the effective action do not commute and introduce spurious values for the poles and residues.

  Therefore the effective action for the emergent graviton needs {supplementary information} at finite energies near the poles of the two-point function.

\item In particular, around the momentum space poles of the hidden stress tensor two-point function, the linearised effective gravitational description is that of a standard massive gravity theory (F.P.) together with {a positive} norm massive dilaton mode.

\item We might be able to say something also for the case of no-mass gap for the hidden sector theory (CFT) using the spectral representation. In such a case, the effective gravitational description is either non-local in four dimensions or local and higher dimensional, since we have a continuum of states at low momenta. There is a special case where the density of states is depleted extremely fast in the IR (exponentially). This  happens in random matrix theories near the edge of the support of the spectrum (Airy behaviour), see~\cite{Ginsparg:1993is,Betzios:2020nry}.

\end{itemize}

Our next problem is to understand, how and if the linearized emergent gravity that appeared because of the coupling of the two QFTs completes at the non-linear level and provides a diffeomorphism invariant theory for the emergent graviton, along the lines we have seen in the single theory case in section \ref{simple}. This is the goal  of the next section.

\vskip 0.5cm
%%%%%%%%%%%%%%%%%%%%%%%%%%%%%%%%%%%%%%%%%%%
\section{The non-linear theory and diffeomorphism invariance}\label{secNL}
\vskip 0.5cm

In this section we extend the analysis performed in section \ref{linear} to the nonlinear level. In particular, we aim at describing the emergence of gravity in its full nonlinear fashion using symmetries, most notably the translational invariance of the high-energy QFT.

Before we start, it is important to expose several inputs we learn from the realization of translation invariance, its relation to diffeomorphism invariance and holography.

It is well known that in holographic theories, translational invariance, a global invariance of the boundary theory, translates into a local invariance of the bulk theory, namely diffeomorphism invariance. This is one aspect of the general rule that global symmetries in QFT translate to local symmetries in the bulk holographic theory. This {rhymes} also well with the fact that global symmetries in QFT translate in local symmetries of the Schwinger source functional, if properly defined and renormalized.

 As long as translational invariance is exact, the bulk diffeomorphism invariance is exact. Despite this fact,  the graviton spectrum of the holographic theory is generically massive, as the diffeomorphism invariance is spontaneously broken by the saddle-point (vacuum) solution.
It is well known that in such cases none of the problems of massive gravity emerges\footnote{The same is true for massive spin two fields in string theory, \cite{e2}.}, \cite{porrati,DR,e2}. In the cases where the boundary translational invariance is broken there is an explicit mass generated for the bulk graviton. A breaking of translational invariance due to a domain wall was entertained in \cite{KR} and a massive graviton on the domain wall was found.
In the case of two interacting CFTs, the holographic picture is that of two AdS spaces ``interacting" via boundary conditions at their common boundary, \cite{e1,a1,e2}. In that case there are two initially massless gravitons corresponding to the two decoupled AdS spaces. Once they are coupled with interactions, only one massless graviton remains, and the other obtains a one-loop mass. This is in agreement with the anomalous dimension of a non-conserved boundary energy-momentum tensor, \cite{a1}.
A potentially interesting, alternative example is Euclidean wormholes. In such a case, if their dual is related to two interacting theories, as advocated in \cite{worm}, then a pair of such interacting theories generates a single bulk graviton. It is plausible that the wormhole case is the limit of the massive graviton  mass going to infinity\footnote{Such a limit completely reorganizes the 1/$N_c$ expansion.}.

We now proceed to extend our analysis of the previous section to the non-linear level. As in section \ref{linear}, the main object of investigation is the generating functional of the correlation functions in the visible QFT
\begin{equation}
e^{-W\left(\mathcal{J},\mathcal{\hat J},{\bf g}\right)}\,=\,\int\,\left[D \Phi\right] [D \h \Phi]\,e^{-S_{visible}\left(\Phi,\mathcal{J},{\bf g}\right)-S_{hidden}\left(\h \Phi,{\bf g},\mathcal{\hat J}\right)\,-\,S_{int}\left(\mathcal{O}^i,\h {\mathcal{O}}^i,{\bf g}\right)}\label{fun}
\end{equation}
where $\Phi^i$ and $\h \Phi^i$ are respectively the fields of the visible QFT and the hidden $\h{QFT}$.
$\mathcal{J}$ and ${\mathcal{\hat J}}$ are (scalar) sources in the visible and hidden theories respectively.
The interaction part is defined as:
\begin{equation}
S_{int}\,=\,\int\,d^4x\sqrt{{\bf g}}\,\sum_i\,\lambda_i\,\mathcal{O}_i(x)\,\h{\mathcal{O}}_i(x)
\label{12}\end{equation}
where $\mathcal{O}_i$ are operators of the visible QFT,  $\h{\mathcal{O}}_i$ operators of the hidden $\h{QFT}$ and the $\lambda_i$ are generic couplings.
Note that the theory as written in (\ref{fun}) has $M$ (the messenger mass scale) as a cutoff.
If we also include the sources of the hidden theory and add them to $S_{int}$ it becomes
\begin{equation}
S_{int}\,=\,\int\,d^4x\sqrt{{\bf g}}\,\sum_i\,\left[\lambda_i\,\mathcal{O}_i(x)+\mathcal{\hat J}^i\right]\,\h{\mathcal{O}}_i(x)
\label{121}\end{equation}

In (\ref{fun}) we made explicit the dependence of the full theory on a fixed, non-dynamical background metric ${\bf g}_{\m\n}$. Although our final interest is on this metric being the flat metric, we keep it explicit so that we track the dependence on the final result on the nature of this background metric.

It should be stressed that the functional integral in general is defined so that it produces the correlators in a given state of the total theory.
We have defined it here so that in the absence of sources, it reproduces correlators in the vacuum of the combined theory.
However, by turning on non-trivial sources we may access correlators in excited states.

For energies $E\ll M$,  we can integrate out the hidden theory and obtain
\begin{align}
e^{-W\left(\mathcal{J},\mathcal{\hat J},{\bf g}\right)}\,&=\,\int\,[D \Phi] [D\h \Phi]\,e^{-S_{visible}\left(\Phi,\mathcal{J},{\bf g}\right)-S_{hidden}\left(\h \Phi,{\bf g},\mathcal{\hat J}\right)\,-\,S_{int}}\,\nonumber\\&=\,\int\,[D \Phi] \,e^{-S_{visible}\left(\Phi,\mathcal{J},{\bf g}\right)\,-\,\mathcal{W}\left(\mathcal{O}^i+\mathcal{\hat J}^i,{\bf g},\right)}
\label{6}\end{align}
where the functional\footnote{According to (\ref{121}) the correct combination is $\mathcal{O}^i+{1\over \lambda_i}\mathcal{\hat J}^i$, but we have simplified the notation by rescaling the hidden theory sources with the appropriate couplings.}
 $\mathcal{W}\left(\mathcal{O}^i+\mathcal{\hat J}^i,{\bf g},\right)$ represents the generating functional for the hidden theory with the original fixed sources $\mathcal{\hat J}$ and $\bf g_{\m\n}$ and new dynamical sources $\mathcal{O}^i$ given by the operators of the visible theory as is implied by (\ref{12}).

The low-energy interactions of the visible theory are now  controlled by the following action
\be
S_{total}=S_{visible}\left(\Phi,\mathcal{J},{\bf g}\right)+\mathcal{W}\left(\mathcal{O}^i+\mathcal{\hat J}^i,{\bf g}\right)
\label{3}\ee

We now put the full theory (both the hidden and visible QFTs) on an arbitrary curved manifold with metric  $g_{\m\n}$ and define again the generating functional in the presence of the background metric as
\begin{equation}
e^{-W\left(\mathcal{J},g,\mathcal{\hat J}\right)}\,=\,\int\,[D \Phi] \,e^{-S_{visible}\left(\Phi,\mathcal{J},g\right)\,-\,\mathcal{W}\left(\mathcal{O}^i+\mathcal{\hat J}^i,g\right)}\label{func}
\end{equation}.

The general metric $g_{\m\n}$ appears as a source in the full theory. The coupling to a general metric is controlled by the diffeomorphism invariance of the full theory. It is however ambiguous as there are many different ways of defining the action with arbitrary $g_{\m\n}$. This has been discussed in detail in the previous section. We assume the minimal coupling, where we replace derivatives with covariant derivatives. This is well defined for the low-dimension fields but can also become ambiguous at higher derivative terms. We shall return to this later on.

The intuition from the previous section is that the expectation value of the energy-momentum tensor of the hidden theory can act as a metric for the visible theory. We define
\begin{equation}
h^*_{\mu\nu}\,\equiv\,\frac{1}{\sqrt{g}}\,\frac{\delta \mathcal{W}\left(\mathcal{O}^i,g,\mathcal{\hat J}\right)}{\delta g^{\mu\nu}}\,\Big|_{g_{\m\n}={\bold g}_{\m\n}}=\,\langle \h T_{\mu\nu}\rangle\label{defmetric}
\end{equation}
as the vacuum expectation value of the stress tensor of the hidden theory $T_{\mu\nu}^{hidden} \equiv \h T_{\mu\nu}$.
We should also note that the expectation value is calculated in an appropriate state that is part of the original definition of the path integral.
It depends also on the hidden theory sources $\mathcal{\hat J}^i$.
We shall be mostly interested in the case where this state is the vacuum state of the hidden theory but other cases can be described as well by taking for example $\mathcal{\hat J}^i$ to be non-trivial.
Note also that there is no expectation value taken in the visible theory and in particular that $h_{\m\n}$ depends explicitly on the fluctuating SM fields.

 The functional derivative must in the end be computed at $g_{\mu\nu}={\bf g}_{\mu\nu}$ where ${\bf g}_{\mu\nu}$ is a fixed background metric on which we define the original theory. We keep it arbitrary, although later we shall be interested to set ${\bf g}_{\mu\nu}=\eta_{\mu\nu}$.
We define and renormalize the Schwinger functional so that diffeomorphism invariance is manifest.

The diffeomorphism invariance of the functional $W(\mathcal{J},g,\mathcal{\hat J})$ is reflecting (as usual) the translational invariance of the underlying QFT, \cite{lorentz}.
Its consequence is the conservation of the total energy-momentum tensor
calculated from  \eqref{func}. Using  (\ref{defmetric}) we obtain\footnote{See the detailed discussing in appendix \ref{cc}.}
\begin{equation}
\nabla^{\m}_g\left(h_{\mu\nu}\,+\,T_{\mu\nu}\right)\,\sim\,E.M.\label{conserv}
\end{equation}
where $\nabla^{\m}_g$ is the covariant derivative with respect to the metric $g$ and we have also defined the stress tensor of the visible theory as
\begin{equation}
T_{\mu\nu}\,\equiv\,\frac{1}{\sqrt{g}}\,\frac{\delta S_{visible}\left(\Phi,g,\mathcal{J}\right)}{\delta g^{\mu\nu}}\,=\,T^{visible}_{\m\n}
\label{b6}\end{equation}
The right-hand side of equation (\ref{conserv}) is proportional to the equations of motion of the visible theory as well as those on the external sources $\mathcal{J}$ and $\mathcal{\hat J}$ . It vanishes as soon as an expectation value is taken and the external sources are put on shell (see appendix \ref{cc}).

We may now  invert equation \eqref{defmetric} to obtain:
\begin{equation}
g_{\m\n}\,=\,g_{\m\n}\left(\mathcal{O}^i+\mathcal{\hat J}^i,h_{\m\n}\right)
\label{b7}\end{equation}
As we show later on, to leading order in the long-distance expansion,  $h_{\m\n}$ and $g_{\m\n}$ are proportional to each other.

We also define the Legendre-transformed functional
\be
\Gamma\left(h,\mathcal{O}^i+\mathcal{\hat J}^i,\bf g\right)\,\equiv \,\,S_{visible}(h,\Phi)\,-\,S_{visible}({ g},\Phi)\,+
\label{b8}\ee
$$
\,+\,\int d^4x\,\sqrt{g(\mathcal{O}^i+\mathcal{\hat J}^i,h)}\,\,h_{\m\n}\left[g^{\m\n}(\mathcal{O}^i+\mathcal{\hat J}^i,h)\,-{\bf g}^{\m\n}\right]\,-
$$
\be
-\,\mathcal{W}\left(\mathcal{O}^i+\mathcal{\hat J}^i,g(\mathcal{O}^i+\mathcal{\hat J}^i,h)\right)\label{leg}
\ee
where ${\bf g}_{\m\n}$ is the  fixed reference metric on which the original theory is defined on.

We now define the ``effective action" $S_{eff}(h,\Phi)$, which is a functional of the SM dynamical fields and the ``induced metric"\footnote{$h_{\m\n}$ as defined has scaling dimension four. The appropriate dimensionless emergent metric will be defined in the next section.}  $h_{\m\n}$.
\begin{equation}
S_{eff}(h,\Phi,\mathcal{J},\mathcal{\hat J},{\bf g})\,=\,S_{visible}\left(\Phi,h,\mathcal{ J}\right)\,-\,\Gamma\left(h,\mathcal{O}^i+\mathcal{\hat J}^i,{\bf g}\right)=
\label{leg2}\end{equation}
$$
\,S_{visible}({ g(\mathcal{O}^i+\mathcal{\hat J}^i,h)},\Phi,\mathcal{ J})\,-\,\int d^4x\,\sqrt{g(\mathcal{O}^i+\mathcal{\hat J}^i,h)}\,h_{\m\n}\left[g^{\m\n}(\mathcal{O}^i+\mathcal{\hat J}^i,h)\,-{\bf g}^{\m\n}\right]\,+
$$
\be
+\,\mathcal{W}\left(\mathcal{O}^i+\mathcal{\hat J}^i,g(\mathcal{O}^i+\mathcal{\hat J}^i,h)\right)
\label{leg1}
\ee
When extremised with respect to the emergent metric $h_{\m\n}$, $S_{eff}$ satisfies the desired property, namely that on-shell\footnote{Here on-shell means solving the equations of motion for $h$ and substituting back in the action.} it is given by the original action.
To show this, we calculate the variation with respect to the metric $h_{\mu\nu}$ of the functional $\Gamma$ defined in \eqref{leg}
\begin{equation}
\frac{\delta \Gamma\left(h,\mathcal{O}^i+\mathcal{\hat J}^i,\bf g\right)}{\delta h_{\m\n}}\,=\,\sqrt{h}\,T^{\m\n}_{g}\,+\,\frac{1}{2}\sqrt{g}\,g_{\tau\omega}\,\frac{\delta g_{\tau \omega}}{\delta h_{\m\n}}\,h_{\rho\sigma}\,\left(g^{\rho\sigma}-{\bf g}^{\rho\sigma}\right)\,+\,\sqrt{g}\left(g^{\rho\sigma}-{\bf g}^{\rho\sigma}\right)
\label{1}\end{equation}
Two terms in this variation cancelled when we used \eqref{defmetric}.
The second term in the right-hand side of (\ref{leg}) did not contribute as it does not depend on $h_{\m\n}$.
We now have to set the source metric $g_{\m\n}$ equal to the original background metric ${\bf g}_{\m\n}$. Setting $g_{\m\n}={\bf g}_{\m\n}$ in (\ref{1}) we obtain
\begin{equation}
\frac{\delta \Gamma\left(h,\mathcal{O}^i+\mathcal{\hat J}^i,\bf g\right)}{\delta h_{\m\n}}\big|_{g_{\m\n}={\bf g}_{\m\n}}\,=\,\sqrt{h}\,T^{\m\n}_{\bf g}\label{EQdyn}
\end{equation}
which leads to
\begin{equation}
\frac{1}{\sqrt{h}}\frac{\delta \Gamma\left(h,\mathcal{O}^i+\mathcal{\hat J}^i,\bf g\right)}{\delta h_{\m\n}}=T^{\m\n}_{\bf g}\label{EQh}
\end{equation}
The variation of  $S_{eff}(h)$ with respect to $h_{\m\n}$ is now
\begin{equation}
\frac{\delta S_{eff}}{\delta h_{\m\n}}\big|_{g_{\m\n}={\bf g}_{\m\n}}\,=\sqrt{h}T^{\m\n}_{\bf g}-\frac{\delta \Gamma\left(h,\mathcal{O}^i+\mathcal{\hat J}^i,\bf g\right)}{\delta h_{\m\n}}=0\,.
\label{2}\end{equation}
where we used (\ref{EQh}). Therefore $S_{eff}$ is extremal with respect to  $h_{\m\n}$.

We also need to  show that when $S_{eff}$ is evaluated in the emergent metric $h_{\m\n}$, that solves (\ref{2}), then it reduces to the original induced action for the visible theory, namely  (\ref{3}).
We denote by $h^*_{\m\n}$ the solution of (\ref{EQh}). By construction, it is the vev of the hidden energy-momentum tensor from (\ref{defmetric})
\be
h^*_{\m\n}=\frac{1}{\sqrt{\bf g}}\,\frac{\delta \mathcal{W}\left(\mathcal{O}^i+\mathcal{\hat J}^i,{\bf g}\right)}{\delta {\bf g}^{\mu\nu}}\,\label{4}
\ee
We now evaluate $S_{eff}$ from (\ref{leg}) at $h=h^*$ (and $g={\bf g}$) to obtain the advertised result,
\be
S_{eff}(h^\star,\Phi)\,
=\,S_{visible}(\Phi,{\bf{g}})\,+\,\mathcal{W}\left(\mathcal{O}^i+\mathcal{\hat J}^i,{\bf g}\right)\equiv S_{total}
\label{b88}\ee

To summarize, we have shown that the effects of the hidden theory on the visible theory can be reformulated as the visible theory coupled to an emergent {\em dynamical} metric ($h_{\m\n}$ in our example). The visible theory has been now coupled to dynamical gravity. The gravitational dynamics encodes the influence of the hidden theory in the visible dynamics.
The combined theory is fully diffeomorphism invariant, apart from the coupling to the fixed external source ${\bf g}_{\m \n}$.

In the next section, we take a low-energy point of view to make the nature of the induced gravitational interactions more transparent.

\section{The low-energy emergent gravitational dynamics}\label{expli}

To make the general procedure described in the last section more explicit, we take a low-energy approach and parametrize the induced effective action ${\cal W}$ of the hidden theory, defined in \eqref{6}.

In order to write an expansion for the functional ${\mathcal{W}}(\mathcal{O}^i+\mathcal{\hat J}^i,g)$ we should classify the gauge-invariant operator content of the visible theory (that we would like to eventually identify with the SM).

\begin{itemize}
\item Scalar operators $\Phi^i$. In the SM, their dimension is $\Delta \geq 4$ apart from the operator $|H|^2$, where $H$ is the Higgs doublet,  having dimension $\Delta_{H^2}=2$;
\item Vector operators $J_\m^i$: they all have dimension $\Delta \geq 3$;
\item Tensor operators $\mathcal{T}^i_{\mu\nu}$: they have dimension $\Delta \geq 4$.

    \item There are also higher spin operators with $\Delta\geq 4$, but we shall neglect them here.
\end{itemize}
Taking in consideration all this information, the effective action $\mathcal{W}$ takes the form
\begin{equation}
{\mathcal{W}}(g,\Phi^i,\mathcal{\hat J}^i)\,=\,\int d^dx \sqrt{g}\,\left(-V(\Phi^i+\mathcal{\hat J}^i)\,+ \hat M_P^2(\Phi^i+\mathcal{\hat J}^i) \,R\,-\,\mathcal{L}\,\,+\,\dots\right)\label{lowW}
\end{equation}
where subsequently we ignore the currents of the theory and focus on the energy-momentum tensor.
In the potential $V(\Phi)$ above, non-derivative, gauge invariant  scalar operators appear\footnote{In the SM, these would include the Higgs scalar as well as the mass-generating scalar terms. On the other hand $\mathcal{ L}$ includes the leading operators with derivatives made out of the fields of the visible theory.}. $V(\Phi)$ controls essentially the {\it hidden vacuum energy}.
$R$ is the Ricci scalar with respect to the metric $g_{\mu\nu}$. The coefficient of the Einstein term can also be a function of the same scalar operators as $V(\Phi)$. We may remove the $\Phi$ dependence by an appropriate conformal transformation which will modify the potential as well as the other two-derivative terms. We shall not however perform this here.
$\mathcal{L}$ in (\ref{lowW}) is containing all gauge invariant  kinetic and interaction terms up to two derivatives of the visible theory.  Finally the ellipsis stands for corrections containing more than two derivatives.

In order to be able to write (\ref{lowW}) we have assumed that the hidden theory has a mass gap that controls the derivative expansion.
It should be stressed however that even in the absence of a gap, the leading non-local contributions in the gravitational sector start at four derivatives, with the conformal anomaly term. If the theory contains relevant scalar operators, then the non-local behavior can set in earlier.
On the other hand, ``dead-end" CFTs (theories without relevant scalar operators) have a non-local action starting in momentum space with $p^4\log p^2$ for small $p$.

We now set to zero the hidden theory sources $\mathcal{\hat J}^i=0$. They can always be reinstated in the subsequent formulae in a straightforward fashion.

Using the definition \eqref{defmetric} we obtain
$$
h_{\m\n}\,=\,-\,\frac{1}{2}\,g_{\m\n}\,\left[-V(\Phi^i)\,+\hat M_P^2(\Phi^i) \,R\,-\,\mathcal{L}\right]\,+\hat M_P^2(\Phi^i) \,R_{\m\n}\,-
$$
\be
-\,(\nabla_{\m}\nabla_{\n}-g_{\m\n}\square)~\hat M_P^2(\Phi^i)-\mathcal{L}_{\m\n}+\cdots
\label{hresult}
\ee
\be
={V\over 2}g_{\m\n}+\hat M_P^2 G_{\m\n}-T^{\mathcal{ L}}_{\m\n}-\,(\nabla_{\m}\nabla_{\n}-g_{\m\n}\square)~\hat M_P^2+\cdots
\label{hresult1}
\ee
where we defined:
\begin{equation}
\mathcal{L}_{\m\n}\equiv\frac{1}{\sqrt{g}}\frac{\delta \mathcal{L}}{\delta g^{\m\n}}\sp T^{\mathcal{ L}}_{\m\n}\equiv \mathcal{L}_{\m\n}-{1\over 2}\mathcal{L}g_{\m\n}\sp G_{\m\n}=R_{\m\n}-{1\over 2}g_{\m\n}R
\label{7}\end{equation}
and the covariant derivatives and Laplacian  are defined with respect to the metric $g_{\m\n}$.
We also define the ``improved" tensor
\be
{\cal T}^{\mathcal{ L}}_{\m\n}=  T^{\mathcal{ L}}_{\m\n}+\,(\nabla_{\m}\nabla_{\n}-g_{\m\n}\square)~\hat M_P^2
\label{e1}\ee
so that (\ref{hresult1}) can be rewritten as
\be
h_{\m\n}={V\over 2}g_{\m\n}+\hat M_P^2 G_{\m\n}-{\cal T}^{\mathcal{ L}}_{\m\n}+{\cal O}(\pa^4)
\label{e2}\ee

 As we have seen in the previous section, we must eventually construct the effective action for $h_{\m\n}$. For this, we would now like to invert  \eqref{hresult} order by order in the derivative expansion to obtain $g_{\m\n}$ as a function of $h_{\m\n}$ and the fields of the visible theory.
We obtain
\begin{equation}
g_{\m\n}\,=\,\tilde h_{\m\n}\,-\,\frac{2}{V}\, \hat M_P^2\tilde{G}_{\m\n}\,+\, {2\over V}\tilde {\cal T}^{\mathcal{L}}_{\m\n}+{\cal O}(\pa^4)\equiv \tilde h_{\m\n}-\delta\tilde h_{\m\n}+{\cal O}(\pa^4)
\label{res}\end{equation}
where we have defined the dimensionless tensor
\be
\tilde{h}_{\m\n}=\frac{2}{V}h_{\m\n}
\label{9} \ee
which now appears as a dimensionless metric. As we shall see below, it will play the role of the emergent metric in the visible theory and we shall call it the {\em emergent metric}.

In equation (\ref{res}) we also have that the curvatures and covariant derivatives are now with respect to the emergent metric $\tilde h_{\m\n}$,
\be
R(g)=R(\tilde h)+{\cal O}(\pa^4)\sp  R_{\m\n}(g)=R_{\m\n}(\tilde h)+{\cal O}(\pa^4)
\label{l3}\ee
where the ellipsis includes higher-derivative/higher dimension terms and we denoted $\tilde R\equiv R(\tilde h)$ and so on.
We also added tildes to all tensors to indicate they are evaluated in the metric $\tilde h$.

We can invert the metric $g_{\m\n}$ to obtain the contravariant metric as
\be
g^{\m\n}=\tilde h^{\m\n}+\delta\tilde h^{\m\n}+{\cal O}(\pa^4)\label{res1}
\ee
where $\tilde h^{\m\n}$ is the inverse of $\tilde h_{\m\n}$, $\delta\tilde h_{\m\n}$ is defined in (\ref{res})  and all indices on the right hand side are raised with $\tilde h^{\m\n}$.

We now evaluate equation (\ref{res}) at $g_{\m\n}={\bf g}_{\m\n}$, the original background metric, and rewrite it as
\begin{equation}
\hat M_P^2 \tilde{G}_{\m\n}\,=\,\frac{V}{2}\,\left(\tilde{h}_{\m\n}\,-\,{\bf g}_{\m\n}\right)\,+\tilde{\cal  T}^{\mathcal{L}}_{\m\n}+{\cal O}(\pa^4) \label{E1}
\end{equation}
This is the emergent Einstein equation and  is similar to (\ref{f9}) that we obtained in the  case of a single theory, with the difference that in place of $\tilde{\cal  T}^{\phi}_{\m\n}$ that was the total improved energy-momentum  tensor of all other sources, here we have the  partial improved energy-momentum tensor $\tilde{\cal T}^{\mathcal{L}}_{\m\n}$.
The reason is {that in the present example, we defined the emergent metric in (\ref{hresult}) to be related to the vev of a non-conserved energy-momentum tensor.}
We know from the linearized computation of the previous section, that $T^{\mathcal L}$ is essentially the visible stress tensor, up to some corrections due to the hidden theory and depends also on the original background metric on which we place the total system.

Equation  (\ref{E1}) is essentially the equation that emerges by varying the effective action $\Gamma (h)$ defined in (\ref{leg2}), (\ref{leg1})
\be
\Gamma(h,\mathcal{O}^i)=S_{visible}(h,\Phi)\,-\,S_{visible}({\bf g},\Phi)\,+\,
\nonumber\ee
\be
+\int d^4x\,\sqrt{g(\mathcal{O}^i,h)}\,h_{\m\n}\left[g^{\m\n}(\mathcal{O}^i,h)\,-{\bf g}^{\m\n}\right]\,-\,\mathcal{W}\left(\mathcal{O}^i,g(\mathcal{O}^i,h)\right)\label{leg3}
\ee
with $g$ related to $h$ by (\ref{hresult}).

There is again a Ward identity associated to the emergent Einstein equation in (\ref{E1})
\be
\tilde \nabla^{\m} \tilde G_{\m\n}= \tilde \nabla^{\m}\left(\frac{V}{2\hat M_P^2}\,\left(\tilde{h}_{\m\n}\,-\,{\bf g}_{\m\n}\right)\,+{1\over \hat M_P^2}\tilde{\cal  T}^{\mathcal{L}}_{\m\n}\right)+{\cal O}(\pa^5)=0
\label{e3}\ee
which is trivially satisfied because of the definition of $\tilde h_{\m\n}$.

We pause here to comment on an important point:
The Einstein equation (\ref{E1}) was obtained by solving (\ref{hresult}) that we rewrite here as
\be
\tilde h_{\m\n}
=g_{\m\n}+{2\over V}\left[\hat M_P^2 G_{\m\n}-T^{\mathcal{ L}}_{\m\n}-\,(\nabla_{\m}\nabla_{\n}-g_{\m\n}\square)~\hat M_P^2\right]+\cdots
\label{d17}
\ee
 using the derivative expansion.
 It gives the emergent metric $\tilde h_{\m\n}$ as a function of the background metric $g_{\m\n}\to \bg_{\m\n}$ and the various visible fields.
In particular for constant background fields and a flat background metric  $\bg_{\m\n}=\eta_{\m\n}$ we obtain a unique value for the emergent metric, $\tilde h_{\m\n}=\eta_{\m\n}$.

The inverse equation (\ref{E1}) however is, to the order we are working, a second order equation for the induced metric. This effect is well known as the effective action is {resumming} the Schwinger functional calculation.
It therefore allows for more solutions apart from the flat metric.

In section \ref{simple}, in the context of a single theory we have shown that the energy conservation law of the original theory, translates into a similar conservation law for the for the emergent $\tilde h_{\m\n}$ metric. This is also the case here.
 This is shown in detail in appendix \ref{cem} and is also visible in the simple example that we discuss in the next subsection.

\subsection{A simpler example}

To make things as explicit as possible, we work out here a simple example where the visible theory consists of a single scalar field, $\phi$.

 We assume an explicit expansion for the visible theory with an effective action
\begin{equation}
S_{visible}(g,\phi)\,=\,\int d^dx \sqrt{g}\,\left(-V_1(\phi)\,+ M_1^2(\phi) \,R\,-\,Z_1(\phi)\,(\partial \phi)^2\,+\,\dots\right)\label{lowv}
\end{equation}
along with the similar expansion for the hidden functional:
\begin{equation}
\mathcal{W}(g,\phi)\,=\,\int d^dx \sqrt{g}\,\left(-V_2(\phi)\,+ M_2^2(\phi) \,R\,-\,Z_2(\phi)\,(\partial \phi)^2\,+\,\dots\right)\label{lowv1}
\end{equation}
where in both expressions the ellipsis indicates higher derivative terms.
We define for simplicity the scalar tensors
\begin{equation}
 T^{\phi,(n)}_{\m\n}\,=\,Z_n\,\partial_\m \phi \partial_\n \phi\,-\,\frac{1}{2}\,g_{\m\n}\,Z_n\,(\partial \phi)^2\sp n=1,2
\label{l12}\end{equation}
as well as the ``improved" scalar tensors
\be
{\mathcal T}^{\phi,(n)}_{\m\n}= T^{\phi,(n)}_{\m\n}\,+(\nabla_{\m}\nabla_{\n}-g_{\m\n}\square)M_n^2\sp n=1,2
\label{l1212}\ee

We obtain for the visible energy-momentum tensor
\be
T_{\m\n}={V_1\over 2}g_{\m\n}+M_1^2G_{\m\n}-{\mathcal  T}^{\phi,(1)}_{\m\n}
\label{l17}\ee

The expectation value $h_{\m\n}$ of the hidden energy-momentum tensor defined in (\ref{hresult}) is given by
\be
h_{\m\n}={V_2\over 2}{g}_{\m\n}+M_2^2G_{\m\n}(g)-\mt^{\phi,(2)}_{\m\n}+\cdots
\label{10}\ee
and
\be
\tilde h_{\m\n}={2\over V_2}h_{\m\n}=g_{\m\n}+{2M_2^2\over V_2}G_{\m\n}(g)-{2\over V_2}  \mt^{\phi,(2)}_{\m\n}
+\cdots\;.
\label{11}\ee
It  can be solved for $g_{\m\n}$ as function of $\tilde h_{\m\n}$ as
\be
g_{\m\n}=\tilde h_{\m\n}-\delta\tilde h_{\m\n}+\cdots
\label{l19}\ee
with
\be
\delta\tilde h_{\m\n}={2\over V_2}\left({M_2^2}\tilde G_{\m\n}-\tilde\mt^{\phi,(2)}_{\m\n}\right)
\label{l18}\ee
where, from now on, we  use a tilde over various tensors to indicate they are evaluated in the metric $\tilde h_{\m\n}$.

According to our previous  definitions (\ref{lowW}) we have
\begin{equation}
\mathcal{L}\,=\,Z_2\,(\partial \phi)^2\,,\quad \mathcal{L}_{\m\n}\,=\,Z_2\,\partial_\m \phi \partial_\n \phi\sp T^{\mathcal{L}}_{\m\n}=T^{\phi,(2)}_{\m\n}\,.
\label{l10}\end{equation}

The emergent Einstein equation from sector $2$ is therefore
\begin{equation}
M_2^2 \tilde{G}_{\m\n}\,=\,\,\frac{V_2}{2}\left(\tilde h_{\m\n}-{\bf g}_{\m\n}\right)\,+\tilde\mt^{\phi,(2)}_{\m\n}\,\label{E21}
\end{equation}
At the two-derivative level, the conservation equation stemming from (\ref{E21}) is
\be
\tilde{\nabla}^\m\,\left(\frac{V_2}{2}\,(\tilde{h}_{\m\n}-{\bf g}_{\m\n})\,+\,\tilde\mt^{\phi,(2)}_{\m\n}\right)=\tilde\nabla^{\m}(M_2^2)\tilde G_{\m\n}\,
\label{l14a}
\ee
The conservation equation follows by acting with a covariant derivative  on both sides of (\ref{l14a}) and again is trivially satisfied due to the definition of $\tilde h_{\m\n}$.

We would like to analyze the fate of the conservation of energy in the original theory. For this we must study the form of the conservation equation for the total energy-momentum tensor
\be
\nabla^{\m}(h_{\m\n}+T_{\m\n})=(E.O.M)\nabla_{\nu}\phi
\label{l14b}\ee
as shown in appendix \ref{cc}.
The conservation (\ref{l14b}) is valid when $\phi$ is on-shell, ie. it satisfies
\be
2(Z_1+Z_2)\square \phi+(M_1^2+M_2^2)'R=(V_1+V_2)'+(Z_1+Z_2)'(\pa\phi)^2
\label{l15}\ee

As a first step we shall write the visible stress tensor $T_{\m\n}$ in terms of the emergent metric by expanding in derivatives
\begin{equation}
T_{\m\n}(g)\,=\,T^{(0)}g_{\m\n}+T^{(2)}_{\m\n}+\cdots
\label{l20}\ee
with
\be
T^{(0)}=
\frac{V_1}{2}\sp T^{(2)}_{\m\n}=M_1^2\,G_{\m\n}\,-\mt^{\phi,(1)}_{\m\n}+\cdots
\label{l9}\end{equation}

and can be expressed using the $\tilde h$ metric as
\begin{equation}
T_{\m\n}(g)\,=T^{(0)}\tilde h_{\m\n}+\tilde T^{(2)}(\tilde h)-T^{(0)}~\delta\tilde h_{\m\n}+\cdots
\label{l9a}
\end{equation}
$$=
\,\frac{V_1}{2}\tilde h_{\m\n}\,+\,\left(M_1^2-{V_1\over V_2}M_2^2\right)\,\tilde G_{\m\n}\,-\tilde\mt^{\phi,(1)}_{\m\n}+{V_1\over V_2}\tilde\mt^{\phi,(2)}_{\m\n}+\cdots
$$

Note that although the equation above should be valid in the $g$ metric, it is the same in the $\tilde h$ metric up to four-derivative corrections.

We can convert (\ref{l14b}) using the formulae of appendix \ref{cem} to obtain
\be
\tilde\nabla^{\m}\left[{V_1+V_2\over 2}\tilde h_{\m\n}+(M_1^2+M_2^2)\tilde G_{\m\n}-\tilde\mt^{\phi,(1)}_{\m\n}-\tilde\mt^{\phi, (2)}_{\m\n}\right]=0
\label{l16}\ee
It is crucial to note that the condition (\ref{l16}) that describes  the covariant energy-momentum conservation in the emergent metric is {\em equivalent} to (\ref{l15}) responsible for energy-momentum conservation in the background metric, after calculating the action of the covariant derivative in (\ref{l16}).

We conclude this section by stating that {we presented} a fully diff invariant formalism that captures the emergent graviton from the hidden sector. The original energy conservation survives in the emergent metric. {This is a conservation law that dictates how the total energy and momentum is transferred between the hidden and visible stress tensors a part of which is now described in terms of the emergent metric $\tilde{h}_{\m \n}$ and the associated Einstein tensor $\tilde{G}_{\m \n}$. We therefore notice that~\ref{l16} is symmetric in the labels denoting the visible and hidden functional as it should, while the emergent Einstein's equation~\ref{E21} treats them asymmetrically since only the hidden functional~\ref{lowv1} was used to define the emergent metric.}

\section{The holographic hidden QFT\label{holo}}

We  shall now investigate the special case where the hidden theory is a large $N$ holographic theory.

The general action can be written as
\be
S=S_{1}+S_{12}+S_2
\ee
where the interaction term has been defined in (\ref{linearaa}, \ref{linearaac}), $S_1$ is the action of the (hidden) holographic theory, and $S_2$ the action of the SM. This action has a UV cutoff at the scale $M$.
Applying the holographic correspondence, we can write
\be
\langle e^{iS_{12}}\rangle_{1}=\int_{\lim_{z\to z_0}G_{\m\n}(x,z)={\bf g}_{\m\n}} {\cal D}G~e^{iS_{\rm bulk}[G] + i \lambda \int d^4 x \, \sqrt{{\bf g}} \, \widehat T_{\m\n} {\bf T}^{\m \n}  }
\label{ee1}\ee
where on the left, the expectation value is taken in the holographic theory. $S_{\rm bulk}[G]$ is the bulk gravity action, $z$ is the holographic coordinate, $G_{MN}$ is the bulk graviton, dual  to the hidden theory stress-energy tensor operator $\widehat T_{\m\n}$ of dimension $\Delta=4$ and the gravitational path integral has boundary conditions for $G_{MN}$ to asymptote to the operator ${\bf g}_{\m\n}$ near the cutoff AdS boundary, where the metric becomes the background metric on which the system of coupled QFT's is originally defined. This boundary is denoted by $z_0 \sim 1/M$ and is thus related to the $UV$ cutoff of the original action. We have also neglected any other bulk fields.

Assuming that we have regulated the infinities of the bulk action by introducing appropriate counterterms at the boundary we find
\be
\delta S_{\rm bulk}[G]  = E.O.M +  \int_{z= z_0}  d^4 x \, \sqrt{{\bf g}} \, \widehat T_{\m\n} \delta {\bf g}^{\m \n} \, .
\label{ee11}\ee
We therefore observe that near the saddle point of the path integral, the role of the SM stress tensor ${\bf T}^{\m \n}$ is to shift the background metric for the hidden holographic theory. In other words we can write
\be
\langle e^{iS_{12}}\rangle_{1}=\int_{\lim_{z\to z_0}G_{\m\n}(x,z)={\bf g}_{\m\n} + \lambda {\bf T}_{\m \n}} {\cal D}G~e^{iS_{\rm bulk}[G]}
\label{ee111}\ee

By inserting a functional $\delta$-function  we may rewrite (\ref{ee111}) as
\be
\langle e^{iS_{12}}\rangle_{1}=\int {\cal D}\chi {\cal D} h \int\displaylimits_{\lim_{z\to z_0}G_{\m \n}(x,z)=  \chi_{\m\n} } {\cal D}G ~e^{iS_{\rm bulk}[G]-i \int d^4 x h^{\m\n}(x)(\chi_{\m\n}(x)- {\bf g}_{\m\n} - \lambda {\bf T}_{\m\n}(x))  }
\label{ee2}\ee
The total Schwinger functional is thus represented semi-holographically by substituting \eqref{ee2} into
\be
e^{i W({\bf g})} = \int {\cal D} \Phi_{SM} \, e^{i S_{SM}(\Phi_{SM}, {\bf g})} \, \langle e^{iS_{12}}\rangle
\label{ee21}
\ee
We now change perspective and integrate $\chi_{\m\n}(x)$ first in the path integral
\be
\langle e^{iS_{12}}\rangle_{1}= \int {\cal D}\chi \int {\cal D} h e^{ i \int d^4 x h^{\m\n}(x) ( {\bf g}_{\m\n} - \lambda {\bf T}_{\m\n}(x))  } \int_{\lim_{z\to z_0}G_{\m \n}(x,z)=  \chi_{\m\n} } {\cal D}G ~e^{iS_{\rm bulk}[G]-i \int d^4 x h^{\m\n}(x) \chi_{\m\n}(x)  } \, .
\label{ee211}\ee
This is equivalent to
\bea
\langle e^{iS_{12}}\rangle_{1} &=& \int {\cal D}h e^{ i \int d^4 x h^{\m\n}(x) ( {\bf g}_{\m\n} - \lambda {\bf T}_{\m\n}(x))  }  \int {\cal D} \chi \,  e^{ i W_{hid}(\chi) - i \int d^4 x h^{\m\n}(x) \chi_{\m\n}(x)  } \nn \\
  &=& \int {\cal D}h e^{ i \int d^4 x h^{\m\n}(x) ( {\bf g}_{\m\n} - \lambda {\bf T}_{\m\n}(x))  }  \,  e^{ i \Gamma^{eff}_{hid}(h)  } \, ,
\label{ee212}
\eea
that involves the effective action $\Gamma^{eff}_{hid}(h)$ of the (hidden) bulk theory. At the saddle point, this reduces to the  Legendre transform of the Schwinger functional of the bulk graviton. This corresponds in holography to switching boundary conditions at the AdS boundary from Dirichlet to Neumann for the graviton \cite{CM}.
We can then rewrite the effective action part, holographically, using Neumann boundary conditions (N.B.C)\footnote{For these boundary conditions $\lim_{z\to z_0}\partial_z G_{\m\n}(x,z)=(z-z_0)^3 h_{\m\n}(x)$.} and we therefore obtain
\be
\langle e^{iS_{12}}\rangle_{1}=\int_{G_{\m\n}(x,z_0) \, : \, N.B.C.} {\cal D}G_{MN}(x,z){\cal D}h_{\m\n}(x)~e^{iS_N[G]+i\int h^{\m\n}(x) ( {\bf g}_{\m\n} + \lambda {\bold T}_{\m\n}(x) )}
\label{e22}\ee
and hence
\be
e^{i W({\bf g})} =\int{\cal D}h_{\m\n}    \int_{G_{\m\n}(x,z_0) \, : \, N.B.C.} {\cal D}G_{MN} {\cal D} \Phi_{SM} ~e^{iS_N[G]+i\int h^{\m\n}(x) ( {\bf g}_{\m\n} + \lambda {\bold T}_{\m\n}(x) ) + i S_{SM}(\Phi_{SM}, {\bf g})} \, .
\label{e22a}\ee

This setup corresponds to our linearized computation in section \ref{linear} and describes a four-dimensional visible QFT, whose stress tensor (${\bf T}_{\m \n}$) is linearly coupled to a dynamical boundary graviton denoted by $h_{\m \n}(x)$. Its dynamics is governed by the holographic bulk dynamical  gravity in five dimensions. In addition, the original background metric ${\bf g}$, plays the role of a ``Dark" stress energy tensor that shifts the SM stress energy tensor ${\bf T}_{\m \n}$ . IR regularity of the bulk saddle point solution will give a condition that determines the leading metric asymptotics in terms of $h_{\mu \nu}(x)$ via  $\lim_{z\to z_0}\partial_z G_{\m\n}(x,z)=(z-z_0)^3 h_{\m\n}(x)$ (NBC's). This picture is a linearised approximation as long as we do not solve for ${\bf g}_{\m \n} = {\bf g}_{\m \n}(h) $ in the SM action. It turns out that the non-linear completion is quite simple and just involves setting ${\bf g}_{\m \n} = {\bf g}_{\m \n}(h) $ in the SM action so that the total system is self-consistently coupled to the dynamical boundary metric $h_{\mu \nu}(x)$.

To prove this, we  directly study the non-linear case and try to match the previous expression. In the non-linear case we can directly use \eqref{leg}, \eqref{leg1} for the Legendre transformed functional to represent it holographically in terms of Neumann boundary conditions for the metric and Dirichlet for any other operators that do not participate in the Legendre transform. Their dual bulk fields will be denoted collectively by $\Phi_{bulk}(x,z)$
\be
 e^{i S_{eff}(h, {\bf g})} = \int_{G_{\m\n}(x,z_0) \, : \, N. \, , \Phi_{bulk}  \, : \, D.   } {\cal D}G_{MN} {\cal D}\Phi_{bulk}~e^{i S_{bulk} [G, \Phi_{bulk}] \, + i\int h^{\m\n}(x){\bold g}_{\m\n}(x)   \, + i S_{SM}(\Phi_{SM}, \, {\bf g}(h))}
\label{e222}\ee
and therefore
\be
e^{i W({\bf g})} =    \int {\cal D}h_{\m\n}  {\cal D} \Phi_{SM} ~e^{i S^{eff}(h\, , \, {\bf g}) } \, .
\label{e222a}\ee
We therefore observe that  the non-linear completion of \eqref{e22a} is quite similar and it only involves the non-linear consistency of putting the SM action on the consistently determined solution ${\bf g}(h)$.

There are a few important comments now. The first issue  is how we determine the solution ${\bf g}(h)$. If we neglect the SM dynamics, this is solely determined by the regularity of the (hidden) holographic gravitational equations in $S_{bulk} [G, \Phi_{bulk}]$. If we include the SM dynamics, we need to self consistently solve for the total set of equations including those of the SM.

We may imagine the SM action as coupled at the radial scale $z_0\sim 1/M$ to the bulk action, where $M$ is the cutoff of the description and is identified with the messenger scale.
Following holographic renormalization \cite{solo,Bianchi:2001kw}, we may then rewrite the full bulk+brane action of the system as

\be
S_{total}=S_{bulk}+S_{brane}
\label{ee3}\ee
\be
S_{bulk}=M_P^3\int d^5x\sqrt{G}\left[-\Lambda_5+R_5(G)+\cdots\right]
\label{e4}\ee
\be
S_{brane}=M^2\int dz\delta(z-z_0) \left( \int d^4x\sqrt{ \gamma}\left[-\Lambda_{4}+R_4(\gamma)+\cdots\cdots\right] + S_{SM}(\gamma) \right)
\label{bb5}\ee
where $\gamma_{\m\n}(x)\equiv \hat G_{\m\n}(z_0,x)$ is the induced metric on the brane. In the bulk action we neglected other bulk fields and higher derivatives. On the brane action we used the SM action of fields coupled to the induced metric and included the localized cosmological constant and Einstein terms on the brane, as they are expected from loop effects of the SM. In both these expressions we neglect higher derivative terms. As we shall be interested at energies $E\ll M$, we can safely ignore these higher derivative terms on the brane. Finally we stress once more that the boundary conditions for the metric in the  bulk action are Neumann.

It should be noted that what we have here is a close analogue of the DGP mechanism, \cite{DGP}, with one difference: both the brane and bulk data are non-trivial. The case where the dynamics of bulk fields are included has been reconsidered recently in \cite{CKN,GKNW,self-cosmo} in the context of the self-tuning mechanism for the SM (brane) cosmological constant. We shall return to this later on.

A main difference in the physics of an emergent graviton, originating in a holographic theory, is that due to the strong coupling effects, there is an infinity of graviton-like resonances coupled to the SM energy and momentum densities. They correspond to the poles of the two-point function of the energy-momentum tensor of the ``hidden" holographic theory or to the continuum in cases of continuous spectra.

If the holographic theory is gapless, then there is a continuum of modes and, as mentioned earlier, in such a case the induced gravitational interaction is generically non-local\footnote{The RS paradigm, \cite{RS}, shows that the presence of bulk cutoffs can change this and produce local gravity in the IR, despite the absence of a mass-gap.}.
If the theory has a gap,  as large-N YM, then there is a tower of nearly stable states at large N that are essentially the 2$^{++}$ glueball trajectory and act as the KK modes of the bulk axion that couple with variable strengths to the SM instanton densities.

To investigate these interactions, we  analyze the propagator of the graviton on the SM brane.
To do this, we introduce a $\delta$-function source for the graviton on the brane and we  solve the bulk+brane equations in the linearized approximation, assuming a non-trivial profile for the bulk graviton  while the metric and other scalars have the holographic RG flow profile of a Lorentz-invariant QFT, in the conformal frame, namely
 \be
 ds^2=e^{2A(z)}(dz^2+dx_{\m}dx^{\m})
 \label{bb6}\ee
 while the bulk scalars depend on the holographic coordinate $z$ only.
 The detailed calculation in the context of a flat bulk was done in \cite{DGP}.
 The present detailed calculation including the tensor structures was carefully done in \cite{CKN} where the interested reader is referred for details. A similar analysis with RS boundary conditions and AdS bulk was done in \cite{KTT}.  We show only the simplified context here. We will also fix the boundary condition for the scale factor $A$ so that it vanishes at the position of the SM brane, $A_0=0$.

The equation for the spin-2 propagator in the presence of the SM brane, and a energy source on the SM brane is
\be
\left[\pa_z^2+\left(3A'\right)\pa_z +\square_4\right]G(x,z)+
\label{bb7}\ee
$$
+{M^2\over M_P^3}\delta(z-z_0)(\square_4 -\Lambda_4)G(x,z)=\delta(z-z_0)\delta^{(4)}(x)
$$
where we work in Euclidean 4d space and primes stand for derivatives with respect to $z$. We Fourier transform along the four space-time dimensions to obtain
\be
\left[\pa_z^2+\left(3A'\right)\pa_z -p^2\right]G(p,z)-{M^2\over M_P^3}\delta(z-z_0)(p^2 +\Lambda_4)G(p,z)=\delta(z-z_0)
\label{bb8}\ee
where $p^2=p^ip^i$ is the (Euclidean) momentum squared. Later on we  also use $p=\sqrt{p^2}$.

 To solve (\ref{bb8}), we must first solve this equation for $z>z_0$ and for $z<z_0$
 obtaining two branches of the bulk propagator, $G_{IR}(p,z)$ and $G_{UV}(p,z)$ respectively. The IR part, $G_{IR}(p,z)$ depends on a single multiplicative integration constant as the regularity constraints in the interior of the bulk holographic geometry fix the extra integration constant.
$G_{UV}(p,z)$ is defined with Neumann boundary conditions at the AdS boundary and depends on two integration constants. In the absence of sources and fluctuations on the SM brane, the propagator is continuous with a discontinuous $z$-derivative at the SM brane\footnote{For Randall-Sundrum branes this condition is replaced by $G_{UV}(p,z-z_0)=G_{IR}(p,z_0-z)$, which identifies the UV side with the IR side. This corresponds to a cutoff holographic QFT in the bulk.}
\be
G_{UV}(p,z_0;z_0)=G_{IR}(p,z_0;z_0)\sp \partial_zG_{IR}(p,z_0;z_0)-\partial_zG_{UV}(p,z_0;z_0)={1}
\label{bb9}\ee
where $M_P$ is the five-dimensional Planck scale in (\ref{e4}).
In this case, there is a single multiplicative integration constant left and the standard AdS/CFT procedure extracts from this solution the two-point function of the boundary stress tensor. We denote this bulk graviton propagator in the absence of the brane as $G_0(p,z;z_0)$ and it satisfies
\be
\left[\pa_z^2+\left(3A'\right)\pa_z -p^2\right]G_0(p,z;z_0)=\delta(z-z_0)
\label{bb10}\ee

In our case, the presence of an induced action on the SM brane changes the matching conditions to
\be
G_{UV}(p,z_0)=G_{IR}(p,z_0)\sp \partial_zG_{IR}(p,z_0)-\partial_zG_{UV}(p,z_0)={1+{M^2\over M_P^3}(p^2+\Lambda_4)G_{IR}(p,z_0)}
\label{bbb11}\ee
The general solution can be written in terms of the bulk propagator $G_0$ with Neumann boundary conditions at the boundary as follows\footnote{Recall that $G(p,z;z_0)$ and $G_0(p,z;z_0)$ are bulk scalar propagators in coordinate space in the radial/holographic direction $z$ and in Fourier space $p^\mu$ for the remaining directions $x^\mu$.}
\cite{CKN}
\be
G(p,z;z_0)={G_0(p,z;z_0)\over 1+{M^2\over M_P^3}(p^2+\Lambda_4)G_{0}(p,z_0;z_0)}
\label{bb12}\ee
 The propagator of the spin-2 graviton on the brane, is obtained by setting $z=z_0$ and becomes
\be
G(p,z_0;z_0)={G_0(p,z_0;z_0)\over 1+{M^2\over M_P^3}(p^2+\Lambda_4)G_{0}(p,z_0;z_0)}
\label{bb13}\ee
The general structure of the bulk spin-two graviton propagator $G_0$ is known, \cite{CKN} and is as follows. The position of the brane $z_0$ determines a bulk curvature energy scale, $R_0$. In the case of simple bulk RG flows\footnote{What is assumed is that the theory does not have multiple intermediate physics scales, but it is controlled by a single mass scale. In the presence of multiple scales a similar analysis is possible.} we obtain
\be
G_0(p,z_0;z_0)={1\over 2 M_P^3}
\left\{ \begin{array}{lll}
\displaystyle {1\over ~p}, &\phantom{aa} &p\gg R_0\\ \\
\displaystyle d_0-d_2p^2-d_4p^4+\cdots,&\phantom{aa}& p\ll R_0.
\end{array}\right.
 \label{bb14}
\ee
The coefficients $d_n$ above are finite and easily computable as multiple integrals of powers of the scale factor of the background solution, \cite{CKN}.
The IR expansion above is valid for all holographic RG flows. It starts having non-analytic terms starting at $p^4\log p^2$ as is the case with the bulk axion field. The expansion coefficients can be determined either analytically or numerically from the bulk holographic RG flow solution.
{The UV expansion in (\ref{bb14}) is given, expectantly,  by the flat space result.}

A careful expansion of the effective action shows that the interaction induced between two energy-momentum sources $T_{1,2}^{\m\n}$ localized on the brane, by the spin two excitation  is
given by, \cite{CKN}
\be
S_{int}=-{1\over 2M_P^3}\int d^4x\sqrt{\gamma}\int d^4x'\sqrt{\gamma'}~G(x,z_0;x',z_0)\left(\gamma_{\m\r}\gamma'_{\n\s}-{1\over 3}\gamma_{\m\n}\gamma'_{\r\s}\right)T_1^{\m\n}(x)T_2^{\r\s}(x')
\ee
where $G(x,z_0;x',z_0)$ is the Fourier transform of the propagator in (\ref{bb13}). Note that the tensor structure is appropriate for the exchange of a massive graviton.

 Using (\ref{bb14}) we now investigate the induced graviton interaction on the SM brane from (\ref{bb13}). It is known that $G_0(p,z_0;z_0)$ is monotonic as a function of $p$, vanishes at large $p$ and attains its maximum at $p=0$ compatible with (\ref{bb14}).
Therefore at short enough distances, $p\to\infty$, the graviton propagator becomes
\be
G_0(p,z_0;z_0)\simeq {1\over M^2}{1\over p^2}
 \label{b15}\ee
which is the propagator of a massless four-dimensional graviton.
For sufficiently small momenta, $p\ll m$,  we obtain
 \be
G^{-1}(p,z_0;z_0)\simeq  M^2\Lambda_4+2{M_P^3\over d_0}+\left(M^2-2M_P^3{d_2\over d_0^2}\right)p^2+{2M_P^3\over d_0}\left[\left({d_2\over d_0}\right)^2+{d_4\over d_0}\right]p^4+{\cal O}(p^6)
 \label{bb16}\ee

 These properties imply that there is single zero in the denominator of (\ref{bb13}) and therefore a single graviton pole.

 In a simple holographic theory  with a single mass scale $m$,  we have, \cite{CKN}
 \be
 d_0={\bar d_0 \over \ell^3m^4}\sp d_2={\bar d_2 \over \ell^3m^6}
 \sp d_4={\bar d_4 \over \ell^3m^8}
\label{bb19} \ee
 $\ell$ is the IR AdS length, $\bar d_n$ are dimensionless numbers, generically  of order ${\cal O}(1)$  and as usual $(M_P\ell)^3\sim N^2$.
{The expansion in (\ref{bb16}) is valid for $p\ll m$.
We may rewrite (\ref{bb16}) using (\ref{bb19}) as}
 \be
G^{-1}(p,z_0;z_0)\simeq  M^2\Lambda_4+2{(M_P\ell)^3\over \bar d_0}m^4+\left(M^2-2(M_P\ell)^3{\bar d_2\over \bar d_0^2}m^2\right)p^2+
 \label{bbb16}\ee
$$
+{2(M_P\ell)^3\over \bar d_0}\left[\left({\bar d_2\over \bar d_0}\right)^2+{\bar d_4\over \bar d_0}\right]p^4+{\cal O}(p^6)
$$

{We may then recast (\ref{bbb16}) as the propagator of a massive } four-dimensional graviton, with effective mass, $m_{eff}$,  and effective four-dimensional Planck scale, $M_{eff}$,
\be
M^2_{eff}=M^2+2(M_P\ell)^3{\bar d_2\over \bar d_0^2}m^2\sp m_{eff}^2=
{M^2\Lambda_4+2{(M_P\ell)^3\over \bar d_0}m^4\over M_{eff}^2}\;.
 \label{bb17}\ee
 Moreover, the coefficient of the $p^4$ term is dimensionless and of order ${\cal O}(N^2)$.

For $M\gg p\gg m$ we have instead
\be
 G^{-1}(p,z_0;z_0)\simeq M^2\Lambda_4+M^2 p^2+2M_P^3p+\cdots
 \ee

Depending on the hierarchy of the various scales of the problem at intermediate distances, it may be that $(M^2p^2+\Lambda^4)G_{0}(p,z_0;z_0)\ll 1$ and  the graviton propagator may behave as a 5-dimensional graviton
\be
G_0(p,z_0;z_0)\simeq {1\over 2M_P^3}{1\over p}
 \label{bb18}\ee
In such a regime, all resonances contribute equitably and the {resummed} result is as above.

In the scalar sector the story is a bit more complicated. The full analysis was performed in \cite{CKN} and the results are as follows.

There is a special scalar mode, that couples to the trace of the energy-momentum tensor of the brane. It is a linear combination of the bulk scalar mode that couples to the trace of the energy momentum tensor of the hidden holographic theory, and the brane-bending mode.
This mode has couplings that are (generically) parametrically of the same order as the effective gravitational coupling. The mass is also generically of the same order as the graviton mass. All other scalar modes (except axions) have effective masses that are parametrically much higher.

For such emergent gravitons to be candidates for describing the real life gravitons, their mass must be parametrically small, and in particular smaller that the inverse cosmological horizon of today, \cite{DN,DGZ}.
 This is is not impossible in the theories we are describing as the graviton mass is suppressed naturally at large $N$, \cite{CKN}. There are however many other finer problems that need to be addressed, before claiming victory.
The light scalar mode must have suppressed couplings compared to effective gravity, or better yet, much higher mass, or both.
The vDVZ discontinuity that exists for all such theories, \cite{vDVZ}, must be resolved via the analogue of the Vainshtein mechanism at the non-linear level, \cite{vain}.
For this a detailed analysis of the non-linear interactions is necessary as well as an analysis of the possible structures of the hidden theory.

\section{Discussion\label{disc}}

Our results paint a rather general picture of emergent gravity that contains however many dark areas.

\begin{enumerate}

\item On the foundational level, the structure of the two-point function of the energy-momentum tensor of a generic QFT is well understood.
    Of particular importance are the contact terms that can appear in momentum space, which we have carefully classified here.

At order ${\cal O}(k^0)$ the contact term is completely fixed in terms of the expectation value of the energy-momentum tensor via the translational Ward identities. At order ${\cal O}(k^2)$ the contact term has a unique arbitrary constant (scale) that controls both the spin-two and the spin-zero part.
The reason is connected to the existence of a single diffeomorphism-invariant term with two derivatives,  the Einstein term. The spin-zero part however,  has the opposite sign from the spin-two part. It is not clear {whether} this coefficient is constrained by unitarity. In some explicit examples we worked out,it is always of the same sign.

Finally, at ${\cal O}(k^4)$ there are two arbitrary constants, one multiplying the spin-two part and another the spin-zero part.

The rest of the energy-momentum tensor two-point function is controlled by the spin-two and spin-0 spectral densities that are positive definite in a unitary theory guaranteeing a healthy emergent gravitational interaction.

\item An immediate issue is what kind of hidden theory would provide an emergent graviton that seems to have the properties we measure for gravity in the real world.
    The first constraint is that the graviton mass should be {of the order of}, or smaller than  the inverse cosmological horizon today, $m_g\lesssim 10^{-33}$ eV.
This suggests that we should have a theory with a mass gap.
Then, the higher poles of the spin-two part of the two-point function must be sufficiently distant from the first pole. This would allow an effective description of the near-massless spin-two graviton.

It is not clear whether a small gap {in} the energy-momentum two point function can be accompanied by a much larger distance of the next heavier pole. In principle $S-matrix$ bootstrap constraints can shed light to this issue, but so far little is known as the four-point function of the energy-{momentum} tensor is complicated to handle.

Another possibility is that the higher poles have a substantial  imaginary part, as in the case of YM theory at small $N_c$. A quantitative analysis of this issue is important in order to address the viability of this case and a reliable model is needed.

\item Another important issue is the extra scalar that couples to the trace of the hidden theory energy-momentum tensor. This is to some extend the leading scalar candidate for violations of the equivalence principle.

    We have seen that the strength of this interaction is controlled by the details of the coupling of the hidden theory to the SM, and there is a fine tuning that can completely decouple this interaction. However in the spirit of our approach, this is not a desired resolution. Other possibilities include:

(a) That such a state is heavier that the graviton. In this case however, the mass must be $\gtrsim 10^{-3} $ eV in order not to interfere with equivalence principle experiments. Although these bounds may be relaxed by a few orders of magnitude, such a mass should  many orders of magnitude larger than that of the graviton. It is not at all clear if such a hierarchy is possible. This is again a question where bootstrap ideas may give bounds.
In theories like YM, the scalar mode is lighter {than} the $2^{++}$ state.

In holographic YM-like bottom up theories, that were analyzed in \cite{ihqcd} the scalar mode is also always lighter that the $2^{++}$ mode.
Although we do not have a proof that this is always true in holography, it seems difficult to engineer it otherwise.

(b) The coupling of the scalar mode to SM matter is much weaker than gravity. This is an intermediate case to decoupling completely the scalar mode choosing $\cc=-{1\over 3}$ in the setup of section \ref{linear}.
This may happen by a suppression of the associated spectral density in the IR.

Most probably the optimal case will be an appropriate combination of (a) and (b) but this remains to be seen.

\item A notable puzzle remains in the construction of the relevant effective actions for the graviton and its scalar partner and its connection to massive graviton effective gravity theories.

    The general arguments on the effective action for the stress tensor analyzed in section \ref{simple} indicate that both the massive graviton and its scalar partner can be described in terms of a two-index symmetric tensor, the classical vev of the energy-momentum stress tensor.
This is also the message we obtain from the linearized analysis of section \ref{linear}. Moreover, the theory is a bi-/gravity where the role of the fiducial metric is played by the background metric of the original QFT(s).

On the other hand, the construction of the effective action of the emergent graviton, in the derivative expansion (valid for gapped QFTs below the gap) in the case of a single theory gives a theory with ghosts and tachyons.
We have explained their presence by the fact that the IR expansion and the construction of the effective action messes up the poles and residues of the relevant operators. This is clear, as the quadratic interactions mediated by the emergent graviton are proportional to the spectral densities of the energy-momentum tensor that are healthy in a unitary stable theory.

The linearized theory, in the case of the two interacting QFTs, can be put in a standard gravitational form by field redefinitions. The resulting theory is  Einstein theory with a cosmological constant as well as a constant "dark" energy component. However, in this case only the quadratic and constant terms (in momenta) of the energy-momentum tensor two-point function are relevant and the non-trivial structure of the spectral densities including their poles seems invisible.

In none of the previous {cases} we obtain the healthy massive graviton theories of dRGT gravity, and it is fair to say  that the dRGT theory does not contain the extra scalar mode. One {would} imagine that it could be added by adding an extra scalar field, but the structure of its interactions with the graviton must be constrained.

On the other hand, the expansion of the effective emergent graviton theory in powers of the graviton is well-defined, controlled and has no obvious problems.

\item In the case of the hidden theory being a holographic QFT there is an effective description of the emergent graviton via a bulk plus brane gravitational action along the principles of D-branes and more general  brane-worlds. In this case, the analysis gives a generically massive graviton as well as a massive scalar, while in principle there is control over the non-linear action.

     The massive graviton in this case is due to the interplay between the hidden theory energy-momentum spectral density and the corrections to it due to the SM loops \`a la DGP, \cite{DGP}. This is an example where the interaction can deform substantially the mass of the emergent graviton from the poles of the hidden theory correlator.
          The analogue however of the Vainshtein phenomena in this case are under current investigation, \cite{vai}.

\item At this point we may return to the question: what is the difference between emergent gravity as described here and Zakharov's induced gravity? This can be shown by writing the kinetic term $K$ of the emergent graviton in momentum space,  in terms of the two-point functions, $G_{h}$ and $G_{SM}$ of the energy-momentum tensors
\be
K\simeq {1\over \lambda^2~G_{h}}+G_{SM}
\label{9.1}\ee
$G_h$ is the two-point function of the energy-momentum tensor of the hidden theory. $G_{SM}$ is the two-point function of the energy-momentum tensor of the SM and $\l$ is the coupling of the two theories.
We have suppressed indices in (\ref{9.1}).
The answer in (\ref{9.1}) is what we obtain in our case of emergent gravity. The induced gravity answer corresponds to the limit $\l\to\infty$ in (\ref{9.1}).

An interesting comment is relevant here, by writing the propagator of the emergent graviton that controls the interaction of sources, following (\ref{9.1}) as
\be
K^{-1}\simeq {\l^2 G_h\over 1+\l^2G_hG_{SM}}
\label{9.2}\ee
In (\ref{9.2}) we have two types of poles.

(a) Those emerging from the $G_h$, the correlator of the hidden theory.

(b) The zeros of the denominator $1+\l^2G_hG_{SM}$, that are the DGP-like poles. As we have shown in section \ref{holo} from general principles, only one such zero exists. It is also clear that the effective mass of this state is $m_g\sim {1\over ^N}$, modulo the effects of self-tuning.
Therefore, for sufficiently large $N$ this is the lowest mass of all the possible poles.

\item The emergent graviton theory is a bi-gravity theory. Despite this it has a standard conserved energy-momentum tensor, namely the one of the original combined total QFT\footnote{As we have shown, it has also a covariantly-conserved energy-momentum tensor related
to the stress energy tensor of the visible theory.}. Therefore, the Witten-Weinberg (WW) theorem should apply.
 However, we have seen that theories with a continuous spectrum, extending down to zero mass avoid the theorem. It also seems that our linearized effective action in section \ref{linear}  avoids the WW theorem, as the theory has a cosmological constant. Nevertheless, this theory does admit a flat metric solution, where Lorentz invariance is restored, due to the existence of dark energy.

 \item  We have found that there is always a solution to the emergent gravity equations that is equal to the background metric on which the original QFTs are defined. This happens if the combined theory is in its ground state.  When both the hidden QFT and the SM are defined in the flat Minkowski metric, this implies that the flat metric is always a solution of the emergent gravity equations in the ground state. In particular in our linearized analysis of section \ref{linear}, despite the fact that the quadratic theory is a theory with an effective cosmological constant, the presence of the dark energy addition to the SM energy-momentum tensor makes the flat metric a solution.

         Therefore the standard form of the cosmological constant problem does not exist. However, this does not necessarily mean that the cosmological constant problem is innocuous in a cosmological setting.

\item
It is clear that if the vacuum expectation value of the hidden energy-momentum tensor is zero, the emergent metric will start at the two-derivative level, and the related effective action will be non-local\footnote{This seems to be a different source of non-locality compared to the one discussed in \cite{Marolf}.}.
An analogous phenomenon was shown to happen also in the case of global symmetries and emergent gauge fields, \cite{u1}.
However, this can be avoided by a redefinition of the split between the hidden and the visible theory.

This is definitely the case when the hidden theory is a CFT.
However, in the case where the CFT is holographic, the non-localities can be resumed in the form of a higher-dimensional theory.

\item The separation of the interacting theory into a hidden sector and a visible sector, advocated and used in (\ref{6}) in section \ref{secNL} is not unique. As the emergent metric is given by the energy-momentum tensor of the hidden theory, it is also not unique.
    However, different choices are related by field redefinitions, in this case metric redefinitions.
In particular, this amounts to shifts of $\tilde h\to \tilde h+\delta\tilde h$ by terms that have always at least two derivatives. Note that the terms without derivatives always enter the definition of $\tilde h$ to leading order. Therefore the effective Einstein equations are modified by terms having at least four derivatives.
This is a standard field frame ambiguity that is present in all (gravitational) theories.

\item There are two (covariant) conservation equations that play a central role. One is (\ref{l15}) and controls the conservation of the total energy in the background metric ${\bf g}$. The other is (\ref{l16}) and is an identity related to the definition of the emergent metric, $\tilde h_{\m\n}$ and the conservation of the flat space energy-momentum tensor.

\item The analysis of the emergent dynamical gravity depends crucially on the state the hidden theory is in. This can be seen by turning-on sources also in the hidden theory. Then such sources will appear in the whole procedure and will affect the eventual observable gravitational equations. The presence of hidden theory sources, as the fact that the hidden theory may not be in its ground state, opens a host of possibilities for emergent gravity. We have already mentioned the extra ``dark energy" component that appears in the gravitational coupling to the SM energy-momentum tensor, but we can also have the analogue of dark matter played by ``hidden" matter.

 \item Our definition of the emergent metric prompts the question: what determines the signature of the ``classical" emergent metric?
If the hidden theory is in its vacuum state, then the emergent metric was found to be proportional to the flat background metric with Minkowski signature.
This is expected to still be true if we are close to the ground state of the theory.

Suppose however now that the hidden theory is filled with a thermal massless gas. In that case, the energy-momentum tensor is diagonal and of the form
$$T_{\m\n}\sim \left(\rho,{\rho\over 3},{\rho\over 3},{\rho\over 3}\right)
$$
This has Euclidean signature. Therefore we conclude that our setup allows signature change, although it is less clear what the implications are.

We may also envisage the opposite situation. Start with both theories defined on a Euclidean flat metric and consider states with negative pressure. This would generate a Lorentzian metric for the SM model.

Finally, we may consider mixed cases where the hidden and visible theories are defined on background metrics with different signatures.

\item  Integrating in a graviton is not the only option in the hidden theory. We can turn-on sources for all other operators and we can make their vevs dynamical via the partial Legendre transform, in the same way we have done with the hidden energy momentum tensor.
    Now, we have many more dynamical fields including scalars and vectors. Although this will give results equivalent to the previous case, the interpretation will be different. In particular, the gravitational description we have described above, will be obtained after ``integrating out" the other emergent fields. It is clear that fields that are effectively massless/light should be kept in order to simplify the effective description.

One is led therefore to the intuition described in \cite{SMGRAV} that only fields that will remain light after all quantum corrections are taken into account should be kept as dynamical sources in the effective visible theory.  Such field besides the metric, can include exact global symmetries~\cite{u1} (that can generate emergent gauge symmetries) as well as as instanton densities that have approximate shift symmetries and can generate axion couplings  to the visible sector~\cite{axion}.

    \item Note that from its very definition,  the emergent gravity is strictly {``classical"}. This is because we have used the effective action that resums automatically all quantum effects of the hidden theory, and our fluctuating field is an expectation value. As long as the hidden theory has a mass gap, this effective action is well defined, otherwise it will be non-local at all scales. In the visible theory, the only massless field is the photon but we understand how to deal with its quantum effects in the IR. Therefore the end result of our setup  is classical gravity coupled to a quantum visible theory.

    The interesting question in this context is: is emergent gravity a theory of quantum gravity? Although formulated in terms of an expectation value, the answer is yes, as the effective action has all the relevant information of the (quantum) correlations of the original theory.
In the case where the hidden theory is holographic, then the AdS-CFT correspondance makes clear that we have a quantum gravity theory (a string theory) in a higher dimension.

\item It is well known that in a translational-invariant QFT energy and momentum are conserved. This is not so when this theory is coupled to gravity. However, energy and momentum is still conserved when defined on the asymptotic boundary. In emergent gravity this phenomenon is understood as follows. Local energy and momentum conservation is violated due to the interaction with the hidden theory. However, the metric becomes asymptotically the fiducial background metric of the interacting QFTs. In such a regime the coupling is suppressed and the energy of the visible theory is conserved.

\item One can entertain the possibility of multiple hidden sectors, \cite{SMGRAV}. This setup makes sense when these different sectors do not interact directly, but only via their couplings to the SM.
    Each such sector will have its own messenger scale $M_i$. We can define individual metrics for each hidden sector separately, and they will all couple to the SM. This is a multi-gravity theory, coupled to the SM with a single diffeomorphism invariance.

The leading gravitational interaction will be due to the graviton which combines the smallest effective Planck scale (controlled by the relevant $N$ and the appropriate scale of the hidden QFT as in (\ref{planck})) together with the graviton mass. For masses smaller than about $10^{-3}$ eV the Planck scale of the second most important graviton must be well above the standard Planck scale.

\item An interaction mediated by spin-zero or spin-two particles is attractive once these particles are non-ghost-like.
In the SM  we are used to generically weak repulsive interactions as the mediators are weakly coupled gauge bosons. The exception is QCD, where due to strong coupling, the effective interactions are short range but mediated by scalars.
In the presence of graviphotons, \cite{u12}, emerging from the hidden theory, there are also repulsive interactions, competing with gravity and scalar interactions. Which wins depends on strengths but most importantly on the masses of the associated mediators.
It is an interesting question whether in this context we can address the weak gravity conjecture, \cite{weak}, advanced in string theory.
We should expect it to be valid for strongly-coupled versions of emergent gravity.
Presumably this  will emerge from bootstrap constraints on four -point-functions of currents and the energy momentum tensor.

\item It is interesting to speculate the type of dynamics that would provide some symmetric non-flat metrics in the effective visible theory coupled to emergent gravity.
    If both theories are defined on an arbitrary maximally symmetric background metric ${\bf g}_{\m\n}$, then both theories being in their ground state would amount to the emergent metric being equal to the background metric, $\tilde h_{\m\n}={\bf g}_{\m\n}$.

For the rest, a non-zero energy source in the SM, will induce a non-zero energy source in the hidden theory, due to the (weak) coupling of the two theories. For example, a non-zero vev of the energy-momentum tensor in the SM would interact with such a vev in the hidden theory. However it is difficult to imagine that this will generate a time-dependent solution, unless the state  of the combined theory is unstable that theory is induced to go to the true stable vacuum.

Although this process can be studied in the linearized limit, we do not have as yet a non-linear and reliable theory to follow it further.

\item A related question is the interpretation of the black hole solution of GR in the context of emergent gravity. A collection of masses will collapse in emergent gravity, as we can follow in the linearized computation. The non-linear dynamics are difficult to ascertain, but we have a certain puzzle that emerges from massive graviton theories.
    In such theories static black holes do not seem possible due to the non-zero graviton mass\footnote{Unless they have a null apparent horizon.}, \cite{rosen}.
However, a graviton with a cosmological size mass, may make the decay time of such black holes extremely long.
The horizon in this context would appear as a caustic or a vanishing of the hidden energy-momentum tensor expectation value. However, this phenomenon must be explained by the non-linear theory.

There is another point of view however in holography that seems to tell a different story. Brane-bulk holographic systems like the ones we are discussing here seems to have stable black holes although in some of them the effective graviton on the brane is massive via a variant of the DGP mechanism, \cite{KR}. In such systems brane black hole solutions can be constructed that are static. In three (brane) dimensions such solutions are analytic, \cite{Emparan}, while in four only numerical, \cite{Figueras}.
It is interesting to understand this issue further and the differences if any between the holographic case and the non-holographic case.

Hawking evaporation, as is well known, would be the avatar of detailed balance of the many-body quantum state {of the combined system} associated to the black hole. {Recent works have also indicated how the Page curve for the fine-grained entropy during the black hole evaporation process can be obtained in systems comprised out of two sectors, that are specific examples of our general setup, } for more details see the review~\cite{Almheiri:2020cfm}.

\item It is interesting that if the messenger scale is low (not far from the SM scales) and the gravitational interaction is as we observe it in nature, then there is no hierarchy problem in the combined theory\footnote{This however also implies that the SM must be made UV complete once embedded in the combined theory including the messengers.}.
An analogue of this statement was observed in large extra dimension scenarios, \cite{large}, as well as in the RS setup, \cite{RS}.

There are many other issues that we will not discuss further here. But we hope progress can be made on the above in the immediate future.

\end{enumerate}

%%%%%%%%%%%%%%%%%%%%%%%%%%%%%%%%%
\section*{Acknowledgements}\label{ACKNOWL}
\addcontentsline{toc}{section}{Acknowledgements}

We thank P. Anastasopoulos,  C. Bachas, V. Balasubramanian, T. Banks, M. Bianchi, G. Bossard, A. Bzowski, D. Consoli, B. Gout\'eraux, D. L\"ust, P. McFadden, F. Nitti, K. Papadodimas, O. Papadoulaki, J. Penedones, I. Saltas, C. Skordis, A. Tolley, N. Tsamis, M. Van Raamsdonk, G. Veneziano, L. Witkowski, R. Woodard for useful discussions. We would like to thank Adam Bzowski, Francesco Nitti and Kyriakos Papadodimas for a critical reading of the manuscript.
We would also like to thank Matteo Baggioli for participating in early stages of this work.

 This work was supported in part  by the Advanced ERC grant SM-GRAV, No 669288.

\newpage
\appendix
\renewcommand{\theequation}{\thesection.\arabic{equation}}
\addcontentsline{toc}{section}{Appendix\label{app}}
\section*{Appendices}

\section{Defining the stress energy correlators}\label{BridgingI}

In this Appendix we shall define the various stress-energy correlation functions used in the main text.

Consider first an action coupled to a metric $g_{\m\n}$\footnote{In this section we take the metric to be of Euclidean signature}. We have the following identity
\be
S(g+\delta g)=S(g)+\int d^4x ~{\delta S\over \delta g^{\m\n}(x)}\delta g^{\m\n}(x)+
\label{jj1}\ee
$$+{1\over 2}\int d^4x\int d^4 y~{\delta^2 S\over \delta g^{\m\n}(x)\delta g^{\r\s}(y)}\delta g^{\m\n}(x)\delta g^{\r\s}(y)+{\cal O}(\delta g^3)
$$
$$=S(g)+
\int d^4x \sqrt{g}~\hat T_{\m\n}(g)\delta g^{\m\n}+{1\over 2}\int d^4x \sqrt{g}\int d^4y~\left[{\delta \hat T_{\m\n}(x)\over \delta g^{\r\s}(y)}\delta g^{\m\n}(x)\delta g^{\r\s}(y)\right]-
$$
$$
-{1\over 4}\int d^4x \sqrt{g(x)}g_{\r\s}(x)\hat T_{\m\n}(x)\delta g^{\m\n}(x)\delta g^{\r\s}(x)+{\cal O}(\delta g^3)
$$
where we define
\be
\hat T_{\m\n}={1\over \sqrt{g}}{\delta S\over \delta g^{\m\n}} \, .
\label{b19}\ee
We would like now to pass from the action to the Schwinger functional that generates connected correlation functions. In particular, for an operator
$\mathcal{O}$ and a variational parameter $a$ we have the following relation to quadratic order in $a$
\be
\int ~e^{-S}=\langle e^{- a \mathcal{O}} \rangle = e^{- a  \langle \mathcal{O}  \rangle  + \half a^2 \langle \mathcal{O} \mathcal{O}  \rangle+{\cal O}(a^3)}  = e^{- W}
\ee
Applying this identity in the specific case we find up to quadratic order in $\delta g_{\m \n}$
\bea\label{quadr1}
W(g+\delta g) =W(g)+
\int d^4x \sqrt{g}~ \langle \hat T_{\m\n}(g) \rangle \delta g^{\m\n}+ \nn \\
+ {1\over 2}\int d^4x \sqrt{g}\int d^4y~\left[\langle {\delta  \hat T_{\m\n}(x) \over \delta g^{\r\s}(y)} \rangle  \delta g^{\m\n}(x)\delta g^{\r\s}(y)\right]- \nn \\
- {1\over 2}\int d^4x \sqrt{g}\int d^4y \sqrt{g} \langle  \hat T_{\m\n}(x) \hat T_{\r\s}(y) \rangle_{g} \delta g^{\m\n}(x)  \delta g^{\r \s}(y)- \nn \\
-{1\over 4}\int d^4x \sqrt{g(x)}g_{\r\s}(x) \langle \hat T_{\m\n}(x) \rangle \delta g^{\m\n}(x)\delta g^{\r\s}(x)  \nn \\
\eea
where we define\footnote{Our conventions are such that we minimise various factors of $2$ throughout. In order to pass to another common convention one has to multiply our one-point function with $-2$ and the two-point function with a factor of $4$ in all our formulae.}
\be
\sqrt{g(x)} \langle \hat T_{\m\n}(x) \rangle= {\delta W\over \delta g^{\m\n}(x)}\, \sp \sqrt{g(x)} \sqrt{g(y)} \langle \hat T_{\m\n}(x)  \hat T_{\r\s}(y) \rangle_g = {\delta W\over \delta g^{\m\n}(x) \delta g^{\r\s}(y)} \, .
\label{br2}\ee

On the other hand we can consider now directly the Schwinger functional depending on a metric $g_{\m\n}$ and take its variation to quadratic order.
We have the following identity
\be
W(g+\delta g)=W(g)+\int d^4x ~{\delta W\over \delta g^{\m\n}(x)}\delta g^{\m\n}(x)+
\label{br1}\ee
$$+{1\over 2}\int d^4x\int d^4 y~{\delta^2 W\over \delta g^{\m\n}(x)\delta g^{\r\s}(y)}\delta g^{\m\n}(x)\delta g^{\r\s}(y)+{\cal O}(\delta g^3)
$$
$$=W(g)+
\int d^4x \sqrt{g}~ \langle \hat T_{\m\n}(g) \rangle \delta g^{\m\n}+{1\over 2}\int d^4x \sqrt{g}\int d^4y~\left[{\delta \langle \hat T_{\m\n}(x) \rangle \over \delta g^{\r\s}(y)}\delta g^{\m\n}(x)\delta g^{\r\s}(y)\right]-
$$
$$
-{1\over 4}\int d^4x \sqrt{g(x)}g_{\r\s}(x) \langle \hat T_{\m\n}(x) \rangle \delta g^{\m\n}(x)\delta g^{\r\s}(x)+{\cal O}(\delta g^3)
$$
A matching between~ (\ref{br1}) and~ (\ref{quadr1}) gives
\be
\sqrt{g}(y) \langle  \hat T_{\m\n}(x) \hat T_{\r\s}(y) \rangle_{g} ={1\over 2}\left( \langle {\delta  \hat T_{\m\n}(x)  \over \delta g^{\r\s}(y)} \rangle_g
+\langle {\delta  \hat T_{\r\s}(y)  \over \delta g^{\m\n}(x)} \rangle_g  \right)- {1\over 2}\left(   {\delta \langle  \hat T_{\m\n}(x) \rangle_g \over \delta g^{\r\s}(y)}   + {\delta \langle  \hat T_{\r\s}(y) \rangle_g \over \delta g^{\m\n}(x)} \right) \, .
\label{match1}
\ee
In addition due to (\ref{br2}) we can also write the following expansion for (\ref{br1})
\be
W(g+\delta g)=W(g)+ \int d^4x \sqrt{g}~ \langle \hat T_{\m\n} \rangle_{g} \delta g^{\m\n}+{1\over 2}\int d^4x \sqrt{g}\int d^4y \sqrt{g} \langle  \hat T_{\m\n}(x) \hat T_{\r\s}(y) \rangle_{g} \delta g^{\m\n}  \delta g^{\r\s}  \,.
\label{br3}
\ee
Matching  (\ref{br1}) and (\ref{br3}) we obtain
\be
\sqrt{g}(y) \langle  \hat T_{\m\n}(x) \hat T_{\r\s}(y) \rangle_{g} = {1\over 2}\left({\delta \langle \hat T_{\m\n}(x) \rangle_g \over \delta g^{\r\s}(y)}+
 {\delta \langle \hat T_{\r\s}(y) \rangle_g \over \delta g^{\m\n}(x)}
 \right)-
\label{match2}
\ee
$$
-{1\over 4} \left(g_{\r\s}(x) \langle \hat T_{\m\n}(x) \rangle_g +g_{\m\n}(x) \langle \hat T_{\r\s}(x) \rangle_g\right)\delta^4 (x-y) \, .
$$
Matching (\ref{match1}) with (\ref{match2}) we find the following condition
\be
 {\delta \langle \hat T_{\m\n}(x) \rangle_g \over \delta g^{\r\s}(y)} + {\delta \langle \hat T_{\r\s}(y) \rangle_g \over \delta g^{\m\n}(x)} = \half\left(  \langle {\delta \hat T_{\m\n}(x) \over \delta g^{\r\s}(y)}  \rangle_g +
\langle {\delta \hat T_{\r\s}(y) \over \delta g^{\m\n}(x)}  \rangle_g \right)+ \label{match3}
\ee
$$+
\frac{1}{4} \left(g_{\r\s}(x) \langle \hat T_{\m\n}(x) \rangle_g+g_{\m\n}(y) \langle \hat T_{\r\s}(y) \rangle_g\right) \delta^4 (x-y) \, .
$$
In particular we observe that the various expressions differ through the same contact term proportional to the vev of the stress tensor. This will have as a consequence that the operator $\mathcal{P}_{\m \n \r \s}$ of the linearised computation \eqref{linearag}, \eqref{linearar}, and the analogous operator arising from a linearisation of the EOM's of the non linear computation \eqref{f9}, will also differ via this contact term due to equation (\ref{match2}). This is verified in appendix \ref{matchingfluctuations}.

\subsection{The $T \hat{T}$ interaction on an arbitrary background}\label{BridgingII}

We now set in (\ref{jj1})
\be
\delta g_{\m\n}=\l_1 T_{\m\n}(g)+\l_2 Tg_{\m\n}\sp T=g^{\m\n} T_{\m\n}
\label{jj2}\ee
and obtain
\be
S(g+\delta g)=S(g)+S_{int}(g)+\delta S_2(g)
\label{jj3}\ee
with
\be
S_{int}=\int d^4x \sqrt{g}\left[\l_1\hat T_{\m\n}(g)T^{\m\n} +\l_2 \hat T T\right]
\label{jj4}\ee
\be
\delta S_2=-{\l_1+4\l_2\over 4}\int d^4x \sqrt{g(x)}\hat T_{\m\n}(x)(\l_1 T^{\m\n}(x) +\l_2 T(x)g^{\m\n})T(x)+
\label{jj5}\ee
$$
+{1\over 2}\int d^4 x \sqrt{g(x)}\int d^4y{\delta \hat T_{\m\n}(x)\over \delta g^{\r\s}(y)}\left(\l_1T^{\m\n}(x)+\l_2 T(x)g^{\m\n}(x)\right)
\left(\l_1T^{\r\s}(y)+\l_2 T(x)g^{\r\s}(y)\right)
$$

We therefore have

\be
S(g)+\int d^4x \sqrt{g}\left[\l_1\hat T_{\m\n}(g)T^{\m\n} +\l_2 \hat T T\right] =S(g+\delta g)-\delta S_2
\label{jj6}\ee
Therefore
\be
\langle e^{-S-S_{int}}\rangle\equiv e^{-{\cal W}(g,T)}~=~\langle e^{-S(g+\delta g)+\delta S_2}\rangle ~=~\langle e^{\delta S_2}\rangle_{g+\delta g}~=~ e^{-W(g+\delta g)+W_1+{\cal O}(\l^3)}
\label{jj7}\ee
where $W(g)$ is the Schwinger functional of the unperturbed theory, and
\be
W_1(g,T)=-{\l_1+4\l_2\over 4}\int d^4x \sqrt{g(x)}\langle \hat T_{\m\n}(x)\rangle_{g+\delta g} (\l_1 T^{\m\n}(x) +\l_2 T(x)g^{\m\n}(x))T(x)+
\label{jj8}\ee
$$
+{1\over 2}\int d^4 x \sqrt{g(x)}\int d^4y~\langle {\delta \hat T_{\m\n}(x)\over \delta g^{\r\s}(y)}\rangle_{g+\delta g} \left(\l_1T^{\m\n}(x)+\l_2 T(x)g^{\m\n}(x)\right)
\left(\l_1T^{\r\s}(y)+\l_2 T(x)g^{\r\s}(y)\right)
$$
$$=
-{\l_1+4\l_2\over 4}\int d^4x \sqrt{g(x)}\langle \hat T_{\m\n}(x)\rangle_{g} (\l_1 T^{\m\n}(x) +\l_2 T(x)g^{\m\n}(x))T(x)+
$$
$$
+{1\over 2}\int d^4 x \sqrt{g(x)}\int d^4y~\langle {\delta \hat T_{\m\n}(x)\over \delta g^{\r\s}(y)}\rangle_{g} \left(\l_1T^{\m\n}(x)+\l_2 T(x)g^{\m\n}(x)\right)
\left(\l_1T^{\r\s}(y)+\l_2 T(x)g^{\r\s}(y)\right)+{\cal O}(\l^3)
$$
where the subscripts in the correlators indicate the metric that enters the action with which the expectation value is taken.

We now expand $W(g)$ in a derivative expansion
\be
W(g)=\int d^4x\sqrt{g}\left[-V+M^2 R+{\cal O}(\pa^4)\right]
\label{jj9}\ee
and we have
\be
\langle \hat T_{\m\n}(x)\rangle_{g}={1\over \sqrt{g}}{\delta W\over \delta g^{\m\n}}=M^2G_{\m\n}(x)+{V\over 2}g_{\m\n}(x)+\cdots
\label{jj11}\ee
with $M,V$ constants.
By manipulating the path integral we also obtain,
\be
{1\over \sqrt{g(y)}}{\delta \over \delta g^{\r\s}(y)}\langle \hat T_{\m\n}(x)\rangle_{g}={1\over \sqrt{g(y)}}\langle {\delta \hat T_{\m\n}(x) \over \delta g_{\r\s}(y)}\rangle_g-\langle \hat T_{\m\n}(x)\hat T_{\r\s}(y)\rangle_g
\label{jj12}\ee

so that
\be
\hat G_{\m\n;\r\s}(x,y)\equiv \langle \hat T_{\m\n}(x)\hat T_{\r\s}(y)\rangle_g=-{1\over \sqrt{g(y)}}{\delta \over \delta g^{\r\s}(y)}\langle \hat T_{\m\n}(x)\rangle_{g}+{1\over \sqrt{g(y)}}\langle {\delta \hat T_{\m\n}(x) \over \delta g_{\r\s}(y)}\rangle_g
=\label{jj13}
\ee
$$=
-{M^2\over \sqrt{g(y)}}{\delta G_{\m\n}(x)\over \delta g^{\r\s}(y)}-{V\over 4}\left(g_{\m\r}(x)g_{\n\s}(x)+g_{\m\s}(x)g_{\n\r}(x)\right)\delta ^{(4)}(x-y)+{1\over \sqrt{g(y)}}\langle {\delta \hat T_{\m\n}(x) \over \delta g_{\r\s}(y)}\rangle_g+\cdots
$$
From which we can calculate
\be
{1\over \sqrt{g(y)}}\langle {\delta \hat T_{\m\n}(x) \over \delta g_{\r\s}(y)}\rangle_g=\hat G_{\m\n;\r\s}(x,y)+{M^2\over \sqrt{g(y)}}{\delta G_{\m\n}(x)\over \delta g^{\r\s}(y)}+{V\over 4}\left(g_{\m\r}(x)g_{\n\s}(x)+g_{\m\s}(x)g_{\n\r}(x)\right)\delta ^{(4)}(x-y)
\label{jj14}\ee

We may now calculate
\be
\langle \hat T_{\m\n}(x)\rangle_{g+\delta g}=\langle \hat T_{\m\n}(x)\rangle_{g}-\int d^4y\sqrt{g(y)}\langle \hat T_{\m\n}(x)\hat T_{\r\s}(y)\rangle_{g}\delta g^{\r\s}(y)+{\cal O}(\delta g^2)
\label{jj15}\ee

We now rewrite (\ref{jj8}) as
\be
W_1(g,T)=-{\l_1+4\l_2\over 4}\int d^4x \sqrt{g(x)}(M^2G_{\m\n}+{V\over 2}g_{\m\n}) (\l_1 T^{\m\n}(x) +\l_2 T(x)g^{\m\n})T(x)+
\label{jj10}\ee
$$
+{1\over 2}\int d^4x\sqrt{g(x)}\int d^4y\sqrt{g(y)}\hat G_{\m\n ;\r\s}(x,y) \left(\l_1T^{\m\n}(x)+\l_2 T(x)g^{\m\n}(x)\right)
\left(\l_1T^{\r\s}(y)+\l_2 T(x)g^{\r\s}(y)\right)
$$

$$
-{1\over 2}\int d^4 x \sqrt{g(x)}\int d^4y~\left[M^2{\delta G_{\m\n}(x)\over \delta g^{\r\s}(y)}\right]
 \left(\l_1T^{\m\n}(x)+\l_2 T(x)g^{\m\n}(x)\right)
\left(\l_1T^{\r\s}(y)+\l_2 T(y)g^{\r\s}(y)\right)
$$
$$
+{V\over 4}\int d^4x\sqrt{g}\left(\l_1^2 T_{\m\n}T^{\m\n}+2\l_2(\l_1+2\l_2)T^2\right)
$$
$$
=-{\l_1+4\l_2\over 4}M^2\int d^4x \sqrt{g(x)}~G_{\m\n}~ (\l_1 T^{\m\n}(x) +\l_2 T(x)g^{\m\n})T(x)+
$$
$$
+{1\over 2}\int d^4x\sqrt{g(x)}\int d^4y\sqrt{g(y)}\hat G_{\m\n ;\r\s}(x,y) \left(\l_1T^{\m\n}(x)+\l_2 T(x)g^{\m\n}(x)\right)
\left(\l_1T^{\r\s}(y)+\l_2 T(x)g^{\r\s}(y)\right)
$$

$$
-{1\over 2}\int d^4 x \sqrt{g(x)}\int d^4y~\left[M^2{\delta G_{\m\n}(x)\over \delta g^{\r\s}(y)}\right]
 \left(\l_1T^{\m\n}(x)+\l_2 T(x)g^{\m\n}(x)\right)
\left(\l_1T^{\r\s}(y)+\l_2 T(y)g^{\r\s}(y)\right)
$$
$$
+{V\over 8}\int d^4x\sqrt{g}\left(2\l_1^2 T_{\m\n}T^{\m\n}+4\l_2(\l_1+2\l_2)T^2-(\l_1+4\l_2)^2 T^2\right)
$$

We may also calculate $W(g+\delta g)$ using
\be
\sqrt{\det(g+\delta g)}=\sqrt{g}\left[1+{1\over 2}{{\delta g}_{\m}}^{\m}-{1\over 4}{\delta g_{\m}}^{\nu}{\delta g_{\n}}^{\m}+{1\over 8}\left({{\delta g}_{\m}}^{\m}\right)^2+{\cal O}(\delta g^3)\right]=
\label{jj16}\ee
$$
=\sqrt{g}\left[1+{\l_1+4\l_2\over 2}T-{1\over 4}\left(\l_1^2T_{\m\n}T^{\m\n}+2\l_2(\l_1+2\l_2)T^2\right)+{(\l_1+4\l_2)^2\over 8}T^2+\cdots\right]
$$
\be
R(g+\delta g)=R(g)-(R^{\m\n}+g^{\m\n}\square-\nabla^{\m}\nabla^{\n})\delta g_{\m\n}+{1\over 2}\delta R^{\m\n ;\r\s}\delta g_{\m\n}\delta g_{\r\s}+\cdots
\label{jj17}\ee
where $\delta R^{\m\n ;\r\s}$ collects the second order contributions.
Putting  this together we have
\be
W(g+\delta g)=\int d^4x \sqrt{g}\left[-V+M^2R-\left(M^2G^{\m\n}+{V\over 2}g^{\m\n}\right)\delta g_{\m\n}+\right] + S_2
\label{jj18}\ee
where the quadratic part of the action is given by~%\cite{Deser:1974xq}
\be\label{p1}
S_2 = \frac{M^2}{2} \int d^4 x \sqrt{g}  \, \delta g_{\alpha \beta} \left( P^{\alpha \beta \rho \sigma} \nabla^2 + X^{\alpha \beta \rho \sigma } \right) \delta g_{\rho \sigma} + \left(\nabla^\nu \delta g_{\mu \nu} - \frac{1}{2} \nabla_\mu \delta g^\rho_\rho \right)^2 + \,
\ee
$$
+ \int d^4 x \sqrt{g} \, {V\over 8}\left(g^{\m\r}g^{\n\s}+g^{\m\s}g^{\n\r}-g^{\m\n}g_{\r\s}\right) \delta g_{\m\n}\delta g_{\r\s}\,
$$
\be
P^{\alpha \beta \rho \sigma} = \frac{1}{4} \left( g^{\alpha \rho} g^{\beta \sigma} + g^{\alpha \sigma} g^{\beta \rho} - g^{\alpha \beta} g^{\rho \sigma} \right)\,
\label{b21}\ee
\be
X^{\alpha \beta \rho \sigma } = P^{\alpha \beta \rho \sigma} R - g^{\alpha \rho} R^{\beta \sigma} + g^{\alpha \beta} R^{\rho \sigma} + R^{\alpha \rho \beta \sigma} \, .
\label{b20}\ee
This result holds for an arbitrary background metric $g$.

If we set the background to be flat in~\ref{p1}, one finds the simple result
\be\label{p2}
S_2 = \frac{M^2}{2} \int d^4 x  \frac{\lambda_1^2}{2} T_{\mu \nu} \Box T^{\mu \nu} - \lambda_1^2 T_{\mu \rho} \partial^\rho \partial^\sigma T_{\sigma \nu} - \lambda_1(\lambda_1 - 2 \lambda_2)T \partial^\mu \partial^\nu T_{\mu \nu} +
\ee
$$
\frac{1}{2} (\lambda_2^2 +\lambda_1 \lambda_2 - \lambda_1^2 ) T \Box T
+ {V\over 8}\left(2\l_1^2 T_{\m\n}T^{\m\n}+4\l_2(\l_1+2\l_2)T^2-(\l_1+4\l_2)^2T^2\right)
$$
We define
\be
e^{-{\cal W}(g,T)}\equiv \langle e^{-\int d^4x \sqrt{g}\left[\l_1\hat T_{\m\n}(g)T^{\m\n} +\l_2 \hat T T\right]}\rangle
\ee
We may now compute,
\be
{\cal W}(\eta,T) =  W(\eta)- {V\over 2}(\l_1+4\l_2)\int d^4 x ~T - S_2 +
\ee
$$
+ {V\over 8}\int d^4x \left(2\l_1^2 T_{\m\n}T^{\m\n}+4\l_2(\l_1+2\l_2)T^2-(\l_1+4\l_2)^2 T^2\right) -
$$
$$
- {1\over 2}\int d^4x \int d^4y  \, \hat G_{\m\n ;\r\s}(x,y) \left(\l_1T^{\m\n}(x)+\l_2 T(x) \eta^{\m\n}(x)\right)
\left(\l_1T^{\r\s}(y)+\l_2 T(x) \eta^{\r\s}(y)\right)+
$$
\be
+ {1\over 2}\int d^4 x \int d^4y~\left[M^2{\delta G_{\m\n}(x)\over \delta \eta^{\r\s}(y)}\right]_{g= \eta}
 \left(\l_1T^{\m\n}(x)+\l_2 T(x) \eta^{\m\n}(x)\right)
\left(\l_1T^{\r\s}(y)+\l_2 T(y) \eta^{\r\s}(y)\right)
\label{p3}
\ee
We now notice some pairwise cancelations between the last term of (\ref{p2}) and the second line of (\ref{p3}) as well as the first line of (\ref{p2}) with the last line of (\ref{p3}). Then the result simply boils down to
\be
{\cal W}(\eta,T) =  W(\eta)- {V\over 2}(\l_1+4\l_2)\int d^4 x ~T  -
\label{p4}\ee
$$
- {1\over 2}\int d^4x \int d^4y  \, \hat G_{\m\n ;\r\s}(x,y) \left(\l_1T^{\m\n}(x)+\l_2 T(x)\eta^{\m\n}(x)\right)
\left(\l_1T^{\r\s}(y)+\l_2 T(x) \eta^{\r\s}(y)\right) \nn \\
$$
In fact by a slightly more tedious computation a similar formula holds around any curved background with the simple replacement of computing the expectation values/correlators on that background geometry $g_{\mu \nu}$. One therefore finds
\be
{\cal W}(g,T) =  W(g)- {V\over 2}(\l_1+4\l_2)\int ~  d^4 x ~ \sqrt{g} ~T  -
\label{p5}\ee
$$
- {1\over 2}\int d^4x \sqrt{g} \int d^4y \sqrt{g} \, \hat G_{\m\n ;\r\s}^{(g)}(x,y) \left(\l_1T^{\m\n}(x)+\l_2 T(x)g^{\m\n}(x)\right)
\left(\l_1T^{\r\s}(y)+\l_2 T(x)g^{\r\s}(y)\right)
$$
which is the general expression to quadratic order in the deformation
\be
\delta g^{\m \n} = \left(\l_1T^{\m\n}(x)+\l_2 T(x)g^{\m\n}(x)\right)  \, .
\ee

\section{Ward identities and the energy-momentum two-point function}
\label{Ward}

In this appendix we consider a diffeomorphism-invariant QFT on a curved manifold and review the constraints that are imposed by Ward identities on the two-point function of their energy-momentum tensor. At the end, we specialise these constraints to Poincar\'e invariant QFTs in flat space
and use them to derive equation \ \eqref{linearaj} in the main text.

We  consider a $d$-dimensional theory with Lagrangian $\LL$
on an arbitrary space-time metric $g_{\mu\nu}$. Under an infinitesimal diffeomorphism generated by a vector $\xi_\mu$ the variation of the action $S$ is
\be
\label{genpertuda}
\delta_\xi {S} = \sqrt{-g}  \left( \nabla_\mu \xi_\nu + \nabla_\nu \xi_\mu \right) {T}^{\mu\nu}
~,\ee
where our definition for the energy-momentum tensor is $T^{\mu\nu} = \frac{1}{\sqrt{-g}} \frac{\delta S}{\delta g_{\mu\nu}}$.
The variation of the energy momentum tensor itself reads
\be
\label{genpertudb}
\delta_\xi {T}^{\mu\nu} = \xi^\sigma \nabla_\sigma {T}^{\mu\nu} + {T}^{\sigma \nu} \nabla^\mu \xi_\sigma + {T}^{\mu\sigma} \nabla^\nu \xi_\sigma
~.
\ee
$\nabla_\mu$ is the Levi-Civita covariant derivative of the background metric $g_{\mu\nu}$ and $g$ its determinant.

The invariance of the partition function\footnote{As in the main text, we denote collectively by $\Phi$ all the dynamical fields in the QFT that are integrated over in the path integral.}
\be
\label{genpertudbab}
Z= \int D\Phi \, e^{i S}
\ee
under the infinitesimal diffeomorphism \eqref{genpertuda} implies the conservation equation
\be
\label{genpartudbac}
\nabla_\mu \langle T^{\mu\nu} \rangle=0
~.
\ee
Similarly, the invariance of the one-point function of the energy-momentum tensor
\be
\label{genpertudbaa}
\langle T^{\rho\sigma}(y) \rangle = \frac{\int D \Phi \, e^{i S}\, T^{\rho\sigma}(y)}{\int D \Phi \, e^{i S}}
\ee
under the infinitesimal transformations \eqref{genpertuda}-\eqref{genpertudb} implies that
\bea
\label{genpertudca}
&&  i \int d^d x\, \sqrt{-g(x)}\,  (\nabla_\mu \xi_\nu +\nabla_\nu \xi_\mu)(x) \, \langle {T}^{\mu\nu}(x) {T}^{\rho\sigma}(y) \rangle_c +
\nonumber\\
&& + \left( \xi^\mu \nabla_\mu \langle {T}^{\rho\sigma}(y) \rangle + \langle {T}^{\mu\sigma}(y) \rangle \nabla^\rho \xi_\mu  + \langle {T}^{\rho \mu}(y) \rangle \nabla^\sigma \xi_\mu    \right) = 0
,
\eea
where $\langle  {T}^{\mu\nu}(x) {T}^{\rho\sigma}(y) \rangle_c$ is a connected two-point function since we also had to vary the denominator of (\ref{genpertudbaa}). We can massage this expression using two basic formulas
\be
\frac{\delta \xi^\mu(x)}{\delta \xi^\nu(y)} = \delta^\mu_\nu \delta^d (x-y) \, , \quad \int d^d x \sqrt{-g(x)} \frac{\delta^d (x-y) }{\sqrt{-g(y)}} = 1
\ee
to obtain
\bea
\label{genpertudcaa}
& - 2 i \sqrt{-g(x)}\,  \nabla_\mu  \, \langle {T}^{\mu\nu}(x) {T}^{\rho\sigma}(y) \rangle_c + \nn \\
& + \, \delta^d(x-y) \nabla^\nu \langle {T}^{\rho \sigma}(y) \rangle + \langle {T}^{\nu \sigma}(y) \rangle \nabla^\rho \delta^d(x-y)  + \langle {T}^{\rho \nu}(y) \rangle \nabla^\sigma  \delta^d(x-y)  = 0 \, ,  \nn \\
\eea
or by writing the last three terms of (\ref{genpertudca}) as
\be
+ \int d^d x  \sqrt{-g(x)}\, \frac{\delta^{(d)}(x-y)}{\sqrt{-g(y)}} \left( \xi^\mu \nabla_\mu \langle {T}^{\rho\sigma}(x) \rangle +  \langle {T}^{\mu\sigma}(x) \rangle  \nabla^\rho \xi_\mu +  \langle {T}^{\rho \mu}(x)  \rangle \nabla^\sigma \xi_\mu \right)
\ee
and integrating by parts, this then gives rise to the final form of the Ward identity
\begin{align}
\begin{split}
\label{genpertudc}
& 2  i \sqrt{-g(x)} \langle \nabla_\mu {T}^{\mu\nu}(x) {T}^{\rho\sigma}(y) \rangle_c
+\nabla^\nu \left( \delta^{(d)}(x-y)  \right) \langle {T}^{\rho\sigma}(x) \rangle
\\
&+ \nabla^\rho \left( \delta^{(d)}(x-y) \langle {T}^{\nu\sigma}(x) \rangle \right)
+ \nabla^\sigma \left( \delta^{(d)}(x-y) \langle {T}^{\rho\nu}(x) \rangle \right)
=0
~.
\end{split}
\end{align}
Besides the first term that contains the divergence of the energy-momentum tensor, the Ward identity \eqref{genpertudc} contains several contact terms.

We proceed to apply the Ward identity \eqref{genpertudc} to a theory on flat space. In this case, Lorentz invariance implies that the one-point function of the energy-momentum tensor is space-time-independent and proportional to the Minkowski metric
\be
\label{genpertudd}
\langle {T}^{\mu\nu}(x) \rangle = a \eta^{\mu\nu}
~,
\ee
where $a$ is a dimensionfull constant. The Ward identity $\p_\mu \langle {T}^{\mu\nu} \rangle =0$ agrees with this statement. In a conformal field theory there is no intrinsic scale and the conformal Ward identities set $a=0$. In a QFT with a mass scale, however, the constant $a$ can be non-vanishing. Consequently, we set
\be
\label{genpertude}
\langle \p^\nu {T}^{\rho\sigma}(x) \rangle=0
~.
\ee
We can then simplify \eqref{genpertudc} to obtain
\begin{align}
\begin{split}
\label{genpertudf}
&2 i \p_\mu \langle  {T}^{\mu\nu}(x) {T}^{\rho\sigma}(y) \rangle_c
+ \p^\nu \left( \delta(x-y) \right) \langle {T}^{\rho\sigma}(x) \rangle
\\
&+ \p^\rho (  \delta(x-y) \langle {T}^{\nu\sigma}(x) \rangle )
+ \p^\sigma ( \delta(x-y) \langle {T}^{\rho\nu}(x) \rangle )
=0
~.
\end{split}
\end{align}
Integrating over $\int d^d x e^{-ik x}$ we find the corresponding expression in momentum space
\be
\label{genpertudg}
k_\mu \langle {T}^{\mu\nu}(k) {T}^{\rho\sigma}(-k) \rangle_c
= i {a \over 2} \left(  k^\nu \eta^{\rho\sigma} + k^\rho \eta^{\nu\sigma} + k^\sigma \eta^{\rho\nu} \right)
~.
\ee
This allows us to deduce the two-point function advertised in section \ref{linear}\footnote{With an explicit example derived from an expansion of the Schwinger functional in Appendix \ref{sl}, equation \eqref{c116a}.}
\begin{align}
\begin{split}
\label{genpertudi}
 &i\langle {T}^{\mu\nu}(k) {T}^{\rho\sigma}(-k) \rangle_c
\\
&= -  {a \over 2} \left(  \eta^{\mu\nu}\eta^{\rho\sigma} + \eta^{\mu\rho }\eta^{\nu\sigma} + \eta^{\mu \sigma} \eta^{\rho\nu} \right)
+ b(k^2)\, \Pi^{\mu\nu\rho\sigma}(k) + c(k^2) \,\pi^{\mu\nu}(k) \pi^{\rho\sigma}(k)
~,
\end{split}
\end{align}
where
\be
\label{genpertudja}
\Pi^{\mu\nu\rho\sigma}(k) = \pi^{\mu\rho}(k) \pi^{\nu\sigma}(k) + \pi^{\mu\sigma}(k) \pi^{\nu\rho}(k)
~,\ee
\be
\label{genpertudjb}
\pi^{\mu\nu} (k) = \eta^{\mu\nu} - \frac{k^\mu k^\nu}{k^2}
\ee
are the two independent transverse tensor structures that are available. $b$ and $c$ are arbitrary functions of $k^2$. Notice that $k^2 \Pi^{\mu\nu\rho\sigma}$ includes the term $2\frac{k^\mu k^\nu k^\rho k^\sigma}{k^2}$. The combination that does not include this term is
\be
\label{genpertudk}
k^2 \left( \Pi^{\mu\nu\rho\sigma} - 2 \pi^{\mu\nu} \pi^{\rho\sigma} \right)
~.\ee
This particular combination at quadratic order in the momenta is a cotact term that is discussed in appendix \ref{spectral}.
It can appear in QFTs and we verify this with explicit computations in free field theories in Appendix \ref{free}.

We also state that the Ward identity as derived here is valid for the energy-momentum tensor normalized so that all is correlators are ${\cal O}(N^2)$. For other normalizations, the Ward identity should be appropriately modified.

\section{Translational invariance constraints on the Schwinger functional\label{cc}}

\vskip 1cm

In a theory that is translational invariant, the energy-momentum tensor is conserved,
 \be
 \pa^{\m}~T_{\m\n}=0\;.
\label{cc13} \ee
 If however, we couple the theory to an external metric that is space-time dependent, this breaks translational invariance, and the stress tensor is not generically conserved. However, it is well-known that a simple modification of the conservation law is still in play, namely  covariant conservation
\be
\nabla^{\m}~T_{\m\n}=0
\label{cc14}\ee
This is an avatar of the fact that the renormalized Schwinger functional for the energy-momentum tensor defined as
\be
e^{-W(g)}=\int {\cal D}{\chi}~e^{-S(\chi,g)}
\label{cc15}\ee
is diff invariant, where $\chi$ denotes collectively the quantum fields.
If the QFT is translationally invariant, then the renormalized  $W(g)$ is finite and diffeomorphism invariant, \cite{porrati,lorentz}.
There can be many subtleties in the definition and diffeomorphism invariance of $W(g)$ and they have been discussed in \cite{lorentz} as well as in section \ref{simple}. We shall assume that the Schwinger functional has been regularised and renormalized without violating the diffeomorphism invariance.

The energy momentum tensor expectation value $h_{\m\n}$ is defined  as follows
\be
h_{\m\n}={1\over\sqrt{g}}{\delta W\over \delta g^{\m\n}}
\label{cc16}\ee
Upon an infinitesimal diffeomorphism generated by a vector $\e^{\m}$, the metric varies as
\be
g'^{\m\n}=g^{\m\n}-\nabla^{\m}\e^{\n}-\nabla^{\n}\e^{\m}+{\cal O}(\e^2)
\label{cc17}\ee
and we must have
\be
W(g')=W(g)
\label{cc18}\ee
which translates to the identity
\be
\int d^4x\sqrt{g}~h_{\m\n}(\nabla^{\m}\e^{\n}+\nabla^{\n}\e^{\m})=0
\label{cc19}\ee
where we have used (\ref{cc16}).
Integrating by parts in (\ref{cc19}) we obtain
\be
\int d^4x\sqrt{g}~\e^{\n}\nabla^{\m}h_{\m\n}=0
\label{cc20}\ee
which in turn implies that
\be
\nabla^{\m}h_{\m\n}=0
\label{cc21}\ee
To recapitulate what we have shown is that in the absence of sources other than a metric:

$\bullet$ Translational invariance of a QFT implies linearized diffeomorphism invariance of its Schwinger functional.

$\bullet$ Upon extending the coupling of the metric properly at the non-linear level, we have full diffeomorphism invariance of the  Schwinger functional (if renormalization respects the symmetry).

$\bullet$ The expectation value of the energy-momentum tensor is covariantly conserved.

We may consider breaking the translational invariance by turning on other sources. The simplest case involves  an arbitrary translational invariant QFT, coupled to a general background metric $g_{\m\n}$ and a general scalar source $\Phi$ that is coupled to a selected scalar operator $O(x)$.
Naively, we could expect that the breaking of the translational invariance by $\Phi(x)$ would ruin the diffeomorphism invariance of the Schwinger functional. Not surprisingly, this will not be the case. We shall only find that the energy-momentum expectation value satisfies a modified conservation law.

Consider the (renormalized) Schwinger functional of this theory
\be
e^{-W(g,\Phi)}=\int {\cal D}{\chi}~e^{-S(\chi,g)+\int d^4 x\sqrt{g}~\Phi(x)~O(x)}
\label{cc22}\ee
where $\chi$ again denotes collectively the quantum fields, and $W$ is the renormalized functional. We have been careful to include the extra scalar source in a way that it does not break the diffeomorphism  invariance of the action.

TO be able to parametrize the Schwinger functional in a local fashion, we now assume that the QFT has a mass gap and write its IR expansion as
\be
W(\Phi,g)=\int d^4 x\sqrt{g}\left[-V(\Phi)+M^2(\Phi)R-Z(\Phi)(\pa\Phi)^2+\cdots\right]
\label{cc0}\ee
where the ellipsis stands for higher derivative terms in the sources, controlled by the mass gap of the theory.

We compute the vev of the energy-momentum tensor as
$$
h_{\m\n}\equiv {1\over \sqrt{g}}{\delta W\over \delta g^{\m\n}}=
$$
\be
=
{1\over 2}\left(V-M^2 ~R+Z(\pa\Phi)^2\right)g_{\m\n}+M^2~R_{\m\n}-(\nabla_{\m}\nabla_{\n}-g_{\m\n}\square)M^2-Z\pa_{\m}\Phi\pa_{\n}\Phi+\cdots
\label{cc1}\ee
As we now have the source $\Phi(x)$ breaking the translational invariance, we would expect that in general the energy-momentum tensor expectation value will not be covariantly conserved, ie $\nabla^{\m}~h_{\m\n}\not= 0$.
It would interesting however to see if there are external sources that still preserve energy-momentum conservation.

Taking the covariant derivative of (\ref{cc1}) we obtain
\be
\nabla^{\m}~h_{\m\n}={1\over 2}V'(\Phi)\pa_{\n}\Phi-{1\over 2}\pa_{\n}\left(M^2 ~R\right)+{1\over 2}\pa_{\n}\left(Z(\pa \Phi)^2\right)+(\nabla^{\m}M^2)R_{\m\n}+M^2\nabla^{\m}R_{\m\n}-
\label{cc3}\ee
$$
-\nabla^{\m}\nabla_{\m}(\pa_{\n}M^2)+\pa_{\n}(\square M^2)-\left(Z\square \Phi+Z'(\pa\Phi)^2\right)~\pa_{\n}\Phi-{1\over 2}Z\pa_{\n}(\pa \Phi)^2
$$
where primes in $V$ and $Z$ stand for derivatives with respect to $\Phi$.
We now use the identity $\nabla^{\m}R_{\m\n}={1\over 2}\pa_{\n}R$
as well as
\be
[\nabla_{\m},\nabla_{\n}]T^{\rho}={R^{\rho}}_{\sigma~\m\n}T^{\sigma}
\sp
[\nabla_{\m},\nabla_{\n}]T_{\rho}=-{R^{\sigma}}_{\rho~\m\n}T_{\sigma}
\label{cc4}\ee
which implies
\be
[\nabla_{\m},\nabla_{\n}]\nabla^{\m}\Phi=R_{\m\n}\nabla^{\m}\Phi
\label{cc5}\ee
to simplify
\be
\nabla^{\m}\nabla_{\m}\nabla_{\n}M^2=\nabla^{\m}\nabla_{\n}\nabla_{\m}M^2=\nabla_{\n}\nabla_{\m}\nabla^{\m}M^2+R_{\m\n}\nabla^{\m}M^2=
\label{cc6}\ee
$$
=\pa_{\n}\square ~M^2+R_{\m\n}\nabla^{\m}M^2
$$
Therefore
\be
-\nabla^{\m}(\nabla_{\m}\nabla_{\n}-g_{\m\n}\square)M^2=-\nabla^{\m}\nabla_{\m}(\pa_{\n}M^2)+\pa_{\n}(\square M^2)=-R_{\m\n}\nabla^{\m}M^2
\label{cc7}\ee
\be
\nabla^{\m}\left[M^2 G_{\m\n}-(\nabla_{\m}\nabla_{\n}-g_{\m\n}\square)M^2\right]=-{1\over 2}\nabla_{\n}M^2~R
\label{ccc7}\ee

We may now  rewrite (\ref{cc3}) as
\be
\nabla^{\m}~h_{\m\n}=0={1\over 2}\left[V'-2Z\square \Phi-Z'(\pa\Phi)^2-(M^2)'~R\right]\pa_{\n}\Phi
\label{cc7a}\ee
Therefore, to maintain the Ward identity (\ref{cc21}) we have two options.
 The trivial  one is to have a translationally invariant scalar source: $\pa_{\m}\Phi=0$.
The second is to impose the equations obtained by varying the Schwinger functional with respect to the scalar source $\Phi$.

\section{Conservation of energy momentum and the emergent metric\label{cem}}

In this appendix we perform the detailed calculations of converting the original   conservation of the energy and momentum (\ref{conserv})
that we reproduce here
\begin{equation}
\nabla^\m_{(g)}\,\left(h_{\m\n}+\,T_{\m\n}(g)\right)\,=\,0\,,
\label{8ap}\end{equation}
to an equivalent equation in the emergent metric $\tilde h_{\m\n}$

We rewrite (\ref{res1}) as
\be
g^{\m\n}=\tilde h^{\m\n}+\delta \tilde h^{\m\n}+{\cal O}(\pa^4)\sp g_{\m\n}=\tilde h_{\m\n}-\delta \tilde h_{\m\n}+{\cal O}(\pa^4)
\label{d1}\ee
with
\be
\delta \tilde h_{\m\n}=\frac{2}{V}\left(\hat M_P^2\, \tilde{G}_{\m\n}
-\tilde T^{\mathcal{L}}_{\m\n}
-(\tilde \nabla_{\m}\tilde \nabla_{\n}-\tilde h_{\m\n}\tilde \square)~\hat M_P^2\right)+{\cal O}(\pa^4)
\label{d2}\ee
where the indices on $\delta \tilde h_{\m\n}$ are raised and lowered by $\tilde h_{\m\n}$.
We then calculate the Christoffel symbols in terms of the new metric as
$$
{\Gamma_{\m\n}}^{\r}(g)={{\tilde \Gamma}_{\m\n}}^{\hskip 9pt\r}(\tilde h)+{{\tilde \Gamma}_{\m\n}}^{\hskip 9pt\s}(\tilde h)~\delta {{\tilde h}^{\r}}_{\hskip 4pt\s}-{1\over 2}\tilde h^{\r\s}(\pa_{\m}\delta \tilde h_{\n\s}+\pa_{\n}\delta\tilde h_{\m\s}-\pa_{\s}\delta\tilde h_{\m\n})+{\cal O}(\pa^5)=
$$
$$
={{\tilde \Gamma}_{\m\n}}^{\hskip 9pt\r}(\tilde h)-{1\over 2}\tilde h^{\r\s}(\tilde \nabla_{\m}\delta \tilde h_{\n\s}+\tilde\nabla_{\n}\delta\tilde h_{\m\s}-\tilde \nabla_{\s}\delta\tilde h_{\m\n})+{\cal O}(\pa^5)
$$
\be
={{\tilde \Gamma}_{\m\n}}^{\hskip 9pt\r}(\tilde h)-{1\over 2}(\tilde \nabla_{\m}\delta \tilde h_{\n}^{\hskip 4pt\r}+\tilde\nabla_{\n}\delta\tilde h_{\m}^{\hskip 4pt \r}-\tilde \nabla^{\r}\delta\tilde h_{\m\n})+{\cal O}(\pa^5)
\label{d3}\ee
Using the above we obtain
\be
\nabla^\m_{(g)}\,h_{\m\n}=\nabla^\m_{(g)}\,\left({V\over 2}~\tilde h_{\m\n}\right)=\tilde\nabla^{\m}\left[{V\over 2}\left(\tilde h_{\m\n}+\delta \tilde h_{\m\n}\right)\right]+{\cal O}(\pa^{5})
\label{3d3}\ee

We may now consider the conservation of the total stress tensor (\ref{8ap})
and rewrite  it as
\be
T_{\m\n}(g)=T^{(0)}g_{\m\n}+T^{(2)}_{\m\n}(g)+ {\cal O}(\partial^4)
\label{d4}\ee
where $T^{(0)}$ is the part that contains no derivatives and is therefore independent of $g$, while $T^{(2)}_{\m\n}(g)$ is the part that contains two derivatives etc.
Therefore
\be
T_{\m\n}(g)= T^{(0)}\tilde h_{\m\n}-T^{(0)}\delta \tilde h_{\m\n}+ T^{(2)}_{\m\n}(\tilde h)+ {\cal O}(\partial^4)
=T_{\m\n}(\tilde h)-T^{(0)}\delta \tilde h_{\m\n}+ {\cal O}(\partial^4)
\label{l5}\ee

 We now convert  the original energy-momentum  conservation equation (\ref{8ap}) to a conservation with respect to the $\tilde h_{\m\n}$ covariant derivative using (\ref{d1})-(\ref{d3}). To do this,  consider first a symmetric two-tensor $A_{\m\n}$
 with
 \be
 A_{\m\n}=A^{(0)}_{\m\n}+A^{(2)}_{\m\n}+ {\cal O}(\partial^4)
 \label{d7}\ee
 where $A^{(0)}_{\m\n}$ contains no derivatives while $A^{(2)}_{\m\n}$ contains two derivatives.
Then
\be
\nabla^g_{\rho}A_{\m\n}=\tilde \nabla_{\rho}\left[A^{(0)}_{\m\n}+A^{(2)}_{\m\n}\right]+
 \label{d8}\ee
$$+
{1\over 2}\left[A^{(0)}_{\s\n}(\tilde\nabla_{\rho}\delta\tilde h_{\m}^{\hskip 5pt \s} +\tilde\nabla_{\m}\delta\tilde h_{\r}^{\hskip 5pt \s} -\tilde\nabla^{\s}\delta\tilde h_{\m\r})+(\m\leftrightarrow \nu)\right]+{\cal O}(\pa^5)
$$
We now contract the indices in (\ref{d8}) to obtain
\be
\nabla_g^{\m}A_{\m\n}=\tilde\nabla^{\m}\left[A^{(0)}_{\m\n}+A^{(2)}_{\m\n}\right]+
 \label{d9}\ee
$$
+\delta\tilde h^{\m\r}\tilde \nabla_{\r}~A^{(0)}_{\m\n}+{1\over 2}A^{(0)}_{\s\n}\left(2\tilde\nabla^{\m}\delta\tilde h_{\m}^{\hskip 5pt \s}-\tilde\nabla^{\s}\delta\tilde h_{\m}^{\hskip 5pt \m}\right)+{1\over 2}
A^{(0)}_{\s\m}\tilde\nabla_{\n}\delta \tilde h^{\m\s}+ {\cal O}(\partial^5)
$$

We now specialize to (\ref{8ap}) with
\be
A^{(0)}_{\m\n}=\left({V\over 2}+T^{(0)}\right)\tilde h_{\m\n}\sp A^{(2)}_{\m\n}=T^{(2)}_{\m\n}-T^{(0)}~\delta\tilde h_{\m\n}
 \label{d10}\ee to obtain
\be
0=\nabla^\m_{(g)}\,\left(h_{\m\n}+\,T_{\m\n}(g)\right)=
\ee
\be
=\tilde\nabla^{\m}\left[\left({V\over 2}+T^{(0)}\right)\tilde h_{\m\n}+T^{(2)}_{\m\n}-T^{(0)}~\delta\tilde h_{\m\n}
\right]+\delta\tilde h_{\m\n}\tilde\nabla^{\m}\left({V\over 2}+T^{(0)}\right)+
 \label{d11}\ee
$$
+\left({V\over 2}+T^{(0)}\right)\tilde\nabla^{\m}\delta\tilde h_{\m\n}+{\cal O}(\pa^5)=
$$
\be
=\tilde\nabla^{\m}\left[\left({V\over 2}+T^{(0)}\right)\tilde h_{\m\n}+T^{(2)}_{\m\n}-T^{(0)}~\delta\tilde h_{\m\n}+\left({V\over 2}+T^{(0)}\right)\delta\tilde h_{\m\n}
\right]+{\cal O}(\pa^5)
 \label{dd11}\ee
Therefore (\ref{8ap}) is equivalent to
\be
\tilde\nabla^{\m}\left[{V\over 2}(\tilde h_{\m\n}+\delta\tilde h_{\m\n})+T_{\m\n}(\tilde h)
+ {\cal O}(\partial^4)\right]=0
 \label{d12aa}\ee
and using (\ref{d2}) we finally obtain
\be
\tilde\nabla^{\m}\left[{V\over 2}\tilde h_{\m\n}+
\hat M_P^2\, \tilde{G}_{\m\n}
-\tilde T^{\mathcal{L}}_{\m\n}
-(\tilde \nabla_{\m}\tilde \nabla_{\n}-\tilde h_{\m\n}\tilde \square)~\hat M_P^2
+T_{\m\n}(\tilde h)
+ {\cal O}(\partial^4)\right]=0
 \label{d13aa}\ee
This is precisely the covariant conservation of the original stress tensor in (\ref{hresult1}) but now in the emergent $\tilde h_{\m\n}$ metric.

\section{The calculation of the effective action}\label{totalstresseffective}

In this appendix we proceed to a direct evaluation of the effective action discussed in section~\ref{simple} of the main text for a single theory. For ease of reference, we rewrite here the main formula that we need from the analysis of the Schwinger functional \eqref{f1} which reads
\be
W(g,J)=\int \sqrt{g}\left[-V(J)+M^2(J)R-{Z(J)\over 2}(\pa J)^2+\cdots\right]
\label{c10}\ee
where $R$ is the Ricci scalar of the metric $g$ and the ellipsis stands for higher derivative terms.
Here we chose to denote collectively any scalar fields or external sources with the $J$. This is because the analysis below holds either for dynamical or non-dynamical fields/sources $J$.
We then calculate
\be
h_{\m\n}={V\over 2}g_{\m\n}+M^2G_{\m\n}-{1\over 2}T^J_{\m\n}+\cdots \, ,
\label{c11}\ee
with
\be
T^J_{\m\n}=Z(J)\left(\pa_{\m}J\pa_{\n}J-{1\over 2}g_{\m\n}(\pa J)^2\right)+2(\nabla_{\m}\nabla_{\n}-g_{\m\n}\square)M^2 \, .
\label{c12}\ee
The inverted equation is
\be
g_{\m\n}=\tilde h_{\m\n}-\delta \th_{\m\n}\sp \tilde h_{\m\n}={2\over V}h_{\m\n} \, ,
\label{c13}\ee
that gives rise to the equations of motion for the emergent metric tilde $h_{\m\n}$
\be
M^2~\tilde G_{\m\n}={V\over 2}\left(\th_{\m\n}-\bg_{\m\n}\right)+{1\over 2}\tilde T^J_{\m\n}+\cdots
\label{c15}\ee
where we have set $g_{\m\n}=\bg_{\m\n}$.

At first, using (\ref{c13}) one finds,
\be
\sqrt{g}=\sqrt{\tilde h}\left(1-{1\over 2}{\delta\th^{\m}}_{\m}+{\cal O}(\pa^4)  \right)=
\sqrt{\tilde h}\left[1+{M^2\over V}\tilde R+{1\over 2V}
{({{\tilde T}^J})_{\m}}^{\m}+{\cal O}(\pa^4)  \right]
\ee
and
\be
{\delta\th^{\m}}_{\m}=-{2M^2\over V}\tilde R-{{({{\tilde T}^J})_{\m}}^{\m}\over V}+\cdots=-{2M^2\over V}\tilde R+{6\over V}\square M^2+{1\over V}Z(\pa J)^2 \, .
\label{c18}\ee
We can also compute the original Schwinger functional \eqref{c10} as a function of $\th_{\m\n}$
\be
W(g,J)=\int d^4x\sqrt{\th}\left[-V+{\cal O}(\pa^4)\right]
\label{c16}\ee
where we dropped a total derivative.

We may also compute
\be
\sqrt{g}~ h_{\m\n}~(g^{\m\n}-\bg^{\m\n})=\sqrt{g}~ {V\over 2}\tilde h_{\m\n}~(\tilde h^{\m\n}+\delta\th^{\m\n}-\bg^{\m\n})=
\sqrt{g}\left(2V+{V\over 2}{\delta\th^{\m}}_{\m}-{V\over 2}\tilde h_{\m\n}\bg^{\m\n}+{\cal O}(\pa^4)\right)=
\label{c17}\ee
$$=
\sqrt{\th}V\left[1+\left(1-{{\delta\th^{\m}}_{\m}\over 2}\right)\left(1-{1\over 2}\bg^{\m\n}\th_{\m\n}\right)+{\cal O}(\pa^4)\right]
$$
We finally obtain
\be
\int d^4x \sqrt{g}~ h_{\m\n}~(g^{\m\n}-\bg^{\m\n})=\int d^4x\sqrt{\th}\left[V+\left(V+M^2\tilde R-3\tilde\square M^2-{Z\over 2}(\pa J)^2+\cdots\right)\left(1-{1\over 2}\bg^{\m\n}\th_{\m\n}\right)\right]
\label{c19}\ee
and using (\ref{f15}) and the results above we obtain
\be
\Gamma(\th,J)=\int d^4x\sqrt{\th}\left[2V+\left(V+M^2\tilde R-3\tilde\square M^2-{Z\over 2}(\pa J)^2+\cdots\right)\left(1-{1\over 2}\bg^{\m\n}\th_{\m\n}\right)\right]
\label{c20}\ee

We now verify that despite the complicated form of the effective action the equations of motion are equivalent to (\ref{c15}).
For this, we vary the action with respect to $\tilde h_{\m\n}$ to obtain
\be
-V\th_{\m\n}-{V\over 2}\th_{\m\n}\left(1-{1\over 2}\bg^{\a\b}\th_{\a\b}\right)
+\left(M^2\tilde G_{\m\n}-{\tilde {T^J}_{\m\n}\over 2}+\left(\tilde \nabla_{\m}\tilde\nabla_{\n}-\th_{\m\n}\tilde\square\right)M^2\right)\left(1-{1\over 2}\bg^{\a\b}\th_{\a\b}\right)-
\label{c21}\ee
$$
-\left(\tilde \nabla_{\m}\tilde\nabla_{\n}-\th_{\m\n}\tilde\square\right)\left[M^2\left(1-{1\over 2}\bg^{\a\b}\th_{\a\b}\right)\right]+
$$
$$
+{1\over 2}\bar\bg_{\m\n}\left[V+M^2\tilde R-{Z\over 2}(\pa J)^2\right]-{3\over 2}\th_{\m\n}\th^{\g\d}\pa_{\g}M^2\pa_{\d}\left(1-{1\over 2}\bg^{\a\b}\th_{\a\b}\right)+
$$

$$
+{3\over 2}\left[\pa_{\m}M^2\pa_{\n}\left(1-{1\over 2}\bg^{\a\b}\th_{\a\b}\right)+(\m\leftrightarrow\n)\right]-{3\over 2}\bar\bg_{\m\n}\tilde\square M^2+{\cal O}(\pa^4)=
$$
\be
-V\th_{\m\n}
+\left(-{V\over 2}\th_{\m\n}+M^2\tilde G_{\m\n}-{\tilde {T^J}_{\m\n}\over 2}\right)\left(1-{1\over 2}\bg^{\a\b}\th_{\a\b}\right)+
\label{c21c}\ee

$$
+{1\over 2}\bar\bg_{\m\n}\left[V+M^2\tilde R+{{({\tilde T}^J)_{\r}}^{\r}\over 2}\right]+{1\over 2}\th_{\m\n}\th^{\g\d}\pa_{\g}M^2\pa_{\d}\left(1-{1\over 2}\bg^{\a\b}\th_{\a\b}\right)-
$$
$$
-M^2\left(\tilde \nabla_{\m}\tilde\nabla_{\n}-\th_{\m\n}\tilde\square\right)\left(1-{1\over 2}\bg^{\a\b}\th_{\a\b}\right)
+
$$
$$
+{1\over 2}\left[\pa_{\m}M^2\pa_{\n}\left(1-{1\over 2}\bg^{\a\b}\th_{\a\b}\right)+(\m\leftrightarrow\n)\right]+{\cal O}(\pa^4)=0
$$

where above $\bar\bg_{\m\n}$ is the inverse background metric $\bg_{\m\n}$ with the two indices lowered with $\th_{\a\b}$.

We  now show that if we use (\ref{c15}) the equation above is an identity as expected.
Using (\ref{c11}), (\ref{c13}) and (\ref{c15}) we obtain
\be
\left(1-{1\over 2}\bg^{\a\b}\th_{\a\b}\right)=-1+{M^2\over V}\tilde R-{3\over V}\tilde\square M^2-{Z\over 2V}(\pa J)^2+{\cal O}(\pa^4)
\label{c22}\ee
and
\be
\bar \bg_{\m\n}=\th_{\m\n}+\delta\th_{\m\n}+{\cal O}(\pa^4)=\th_{\m\n}+{2M^2\over V}\tilde G_{\m\n}-{1\over V}\tilde {T^J}_{\m\n}+{\cal O}(\pa^4)
\label{c23}\ee
Also
\be
\bar\bg_{\m\n}\tilde\square M^2=\th_{\m\n}\tilde\square M^2
+{\cal O}(\pa^4)
\label{c24}\ee
\be
\left[\pa_{\m}M^2\pa_{\n}\left(1-{1\over 2}\bg^{\a\b}\th_{\a\b}\right)+(\m\leftrightarrow\n)\right]=0+{\cal O}(\pa^4)
\label{c25}\ee
\be
\th_{\m\n}\th^{\g\d}\pa_{\g}M^2\pa_{\d}\left(1-{1\over 2}\bg^{\a\b}\th_{\a\b}\right)=0+{\cal O}(\pa^4)
\label{c26}\ee
\be
 \left(\tilde\nabla_{\m}\tilde\nabla_{\n}-\th_{\m\n}\tilde\square\right)\left(1-{1\over 2}\bg^{\a\b}\th_{\a\b}\right)=0+{\cal O}(\pa^4)
\label{c27}\ee
Using the above the equation (\ref{c21c}) simplifies to
\be
  -V\th_{\m\n}-{V\over 2}\th_{\m\n}\left[-1+{M^2\over V}\tilde R-{3\over V}\tilde\square M^2-{Z\over 2V}(\pa J)^2\right]-\left(M^2\tilde G_{\m\n}-{{\tilde T}^J_{\m\n}\over 2}\right)+
\label{c28}\ee
$$
+{1\over 2}\th_{\m\n}\left[V+M^2\tilde R+{{{({\tilde T}^J)}^{\r}}_{\r}\over 2}\right]+{V\over 2}\left[{2M^2\over V}\tilde G_{\m\n}-{1\over V}{\tilde T}^J_{\m\n}\right]
+{\cal O}(\pa^4)=0
$$
which can be easily seen to be an identity.

We can also check the second identity that should be valid: the effective action (\ref{c20}) when evaluated at a solution of the equations of motion
(\ref{c15}) gives back the original Schwinger functional evaluated in the background metric. To verify this we insert (\ref{c22}) into (\ref{c20}) to obtain
\be
\Gamma(\th^*,J)=\int d^4x\sqrt{\th^*} V(J)+{\cal O}(\pa^4)=-W(\bg,J)+{\cal O}(\pa^4)
\label{c28a}\ee
where $\th^*$ is a solution of the equations (\ref{c15}) and in the last step in (\ref{c28a}) we used (\ref{c16}).

Note also that the effective action is a bimetric theory of gravity with dynamical metric $\th$ and background metric $\bg_{\m\n}$. Its form is different from the form of effective bimetric theories written in the literature.

\section{Linearised analysis of the effective action and equations}

\subsection{The linearized analysis in the case of a single theory\label{aF1}}

We shall now consider the effective equation \eqref{f9} or (\ref{c15}) arising from the non-linear treatment of a single theory.
\be
M^2\tilde G_{\m\n}={V\over 2}\left(\tilde h_{\m\n}-{\bold g}_{\m\n}\right)+{\tilde {\cal T}}^{\phi}_{\m\n}+{\cal O}(\pa^4)
\label{fff9}\ee

It is always having as a solution the metric of the QFT, $\th_{\m\n}=\bg_{\m\n}$.

We now set the background metric to the flat metric $\bg_{\m\n}=\eta_{\m\n}$ and look for linearized solutions around the background metric
\be
\th_{\m\n}=\eta_{\m\n}+{w_{\m\n}} \, , \quad \delta \th_{\m\n} = w_{\m \n}
\label{c29}\ee
We also assume that $V,M$ do not depend on $J$ and are therefore constants. The linearised expansion we are going to perform around flat space therefore takes the form
\be
M^2~ \frac{ \delta \tilde G_{\m\n}}{\delta \th_{\r \s}} \bigg|_{\th = \eta} w_{\r \s}  = {V\over 2} w_{\m \n} +{1\over 2}  T_{\m\n}+\cdots
\label{c29b}\ee
We need the linearized Riemann tensor,
\be
R_{\m\a\n\b}={1\over 2}\left(\pa_{\b}\pa_{\m}w_{\a\n}+\pa_{\a}\pa_{\n}w_{\b\m}-\pa_{\a}\pa_{\b}w_{\m\n}-\pa_{\m}\pa_{\n}w_{\a\b}\right)
\label{c31}\ee
and the Ricci and Einstein tensors
\be
R_{\m\n}=\eta^{\a\b}R_{\m\a\n\b}={1\over 2}\left(\pa_{\m}\pa^{\a}w_{\a\n}+\pa_{\n}\pa^{\a}w_{\a\m}-\square w_{\m\n}-\pa_{\m}\pa_{\n}w\right)
\label{c32}\ee
\be
R=\eta^{\m\n}R_{\m\n}=\left(\pa\pa w-\square w\right)
\label{c33}\ee
\be
G_{\m\n}=R_{\m\n}-{R\over 2}\th_{\m\n}={1\over 2}\left(\pa_{\m}\pa^{\a}w_{\a\n}+\pa_{\n}\pa^{\a}w_{\a\m}-\square w_{\m\n}-\pa_{\m}\pa_{\n}w-\left(\pa\pa w-\square w\right)\eta_{\m\n}\right)
\label{c34}\ee
where we defined
\be
\pa\pa w\equiv \pa^{\m}\pa^{\n}w_{\m\n}\sp \square\equiv \pa^{\m}\pa_{\m}\sp w=\eta^{\m\n}w_{\m\n}
\label{c35}\ee
The linearized form of (\ref{c15}) using \eqref{c29b} is then
\be
\left(\pa_{\m}\pa^{\a}w_{\a\n}+\pa_{\n}\pa^{\a}w_{\a\m}-\square w_{\m\n}-\pa_{\m}\pa_{\n}w-\left(\pa\pa w-\square w\right)\eta_{\m\n}\right)={\Lambda}w_{\m\n}+{T_{\m\n}\over M^2}
\label{c30}
\ee
where we dropped the tilde from the stress tensor and defined
\be
\Lambda={V\over M^2}\;.
\label{c3000}\ee
Contracting this equation with a $\pa^{\m}$ derivative, the left-hand side vanishes, and we obtain
\be
M^2\Lambda\pa^{\m}w_{\m\n}+\pa^{\m}T_{\m\n}=0~~~\to~~~\pa^{\m}w_{\m\n}=-{\pa^{\m}T_{\m\n}\over M^2\Lambda}~~~\to~~~\pa\pa w=-{\pa\pa T\over M^2\Lambda}
\label{c36}\ee
Substituting (\ref{c36}) into (\ref{c30}) we obtain
\be
\left(\square+\Lambda\right) w_{\m\n}+\pa_{\m}\pa_{\n}w-\eta_{\m\n}\square w=-{T_{\m\n}\over M^2}+{\pa \pa T\eta_{\m\n}-\pa_{\m}\pa^{\a}T_{\a\n}-\pa_{\n}\pa^{\a}T_{\a\m}\over M^2\Lambda}
\label{c37}\ee
Taking the trace of (\ref{c36}) we obtain
\be
\left(\square -{\Lambda\over 2}\right)w={T\over 2M^2}-{\pa\pa T\over M^2\Lambda}\sp T\equiv \eta_{\m\n}T_{\m\n}
\label{c38}\ee
This can be solved in momentum space
where we define Fourier transforms as
\be
w(x)=\int {d^4p\over (2\pi)^4}w(p)~e^{ip\cdot x}
\label{c39}\ee
to obtain
\be
w(p)=-{T(p)+{2p^{\m}p^{\n}\over \Lambda}T_{\m\n}(p)\over M^2(2p^2+\Lambda)}
\label{c40}\ee
equation (\ref{c37}) becomes in momentum space
\be
\left(p^2-\Lambda\right)w_{\m\n}={T_{\m\n}\over M^2}+{p^{\a}p^{\b}T_{\a\b}\eta_{\m\n}-p_{\m}p^{\a}T_{\a\n}-p_{\n}p^{\a}T_{\a\m}
\over M^2\Lambda}+
\label{c41}\ee
$$
+(p_{\m}p_{\n}-\eta_{\m\n}~p^2){T(p)+{2p^{\m}p^{\n}\over \Lambda}T_{\m\n}(p)\over M^2(2p^2+\Lambda)s}
$$
after substituting (\ref{c40}).
Solving (\ref{c41}) we finally obtain
\be
w_{\m\n}(p)={T_{\m\n}\over M^2\left(p^2-\Lambda\right)}+{p^{\a}p^{\b}T_{\a\b}\eta_{\m\n}-p_{\m}p^{\a}T_{\a\n}-p_{\n}p^{\a}T_{\a\m}
\over M^2\Lambda\left(p^2-\Lambda\right)}+
\label{c42}\ee
$$
+{(p_{\m}p_{\n}-\eta_{\m\n}~p^2)\over \left(p^2-\Lambda\right)}{T(p)+{2p^{\m}p^{\n}\over \Lambda}T_{\m\n}(p)\over M^2(2p^2+\Lambda)}
$$
By a shift of the stress tensor we cannot cancel the bad poles.
The gravitational interaction of two sources $T_{\m\n}$ and $T'_{\m\n}$ is
\be
w^{\m\n}T'_{\m\n}={T^{\m\n}T'_{\m\n}-{1\over 2}TT'-{2\over \Lambda}p_{\rho}T^{\r\m}p^{\n}T'_{\m\n}+{2\over \Lambda}{\left((ppT)+{\Lambda\over 2}T\right)
\left((ppT')+{\Lambda\over 2}T'\right)\over 2p^2+\Lambda}\over M^2(p^2-\Lambda)}
\label{c42a}\ee
If the stress tensors are conserved, $p^{\m}T_{\m\n}=p^{\m}T'_{\m\n}=0$ and the interaction simplifies to
\be
w^{\m\n}T'_{\m\n}={T^{\m\n}T'_{\m\n}-{1\over 2}TT'+{\Lambda\over 2}{TT'\over 2p^2+\Lambda}\over M^2( p^2-\Lambda)}
\label{c42b}\ee
{Note that this has a smooth limit as $\Lambda\to 0$ and asymptotes to the massless gravity result.}
The same equation (\ref{c42b}) can be written as
\be
w^{\m\n}T'_{\m\n}={T^{\m\n}T'_{\m\n}-{1\over 3}TT'\over M^2(p^2-\Lambda)}-{1\over 6}{TT'\over M^2\left(p^2+{\Lambda\over 2}\right)}
\label{c42c}\ee
In this form it is a combination of a massive spin-2 graviton (Fierz-Pauli) and a massive spin zero particle which is ghost-like.
Moreover one of the two is always a tachyon.

We now consider static sources with
\be
T^{\m\n}(p)=2\pi m{\delta_{\m}}_0{\delta_{\n}}_0\delta(p^0)
\label{c43}\ee
Note that this source is conserved, $p^{\m}T_{\m\n}=0$ and therefore the solution in (\ref{c42}) can be simplified to
\be
w_{\m\n}={T_{\m\n}\over M^2\left(p^2-\Lambda\right)}+{(p_{\m}p_{\n}-\eta_{\m\n}~p^2)\over M^2\left(p^2-\Lambda\right)}{T(p)\over 2p^2+\Lambda}=
\label{c44}\ee
$$
={T_{\m\n}\over M^2\left(p^2-\Lambda\right)}+{(p_{\m}p_{\n}-\eta_{\m\n}~p^2)\over 3M^2\Lambda}
\left[{1\over \left(p^2-\Lambda\right)}-{2\over 2p^2+\Lambda}\right]T
$$
We therefore have
\be
w_{00}(p)=\left[{1\over \vec p^2-\Lambda}-{\vec p^2\over 3\Lambda}\left({1\over \left(\vec p^2-\Lambda\right)}-{2\over 2\vec p^2+\Lambda}\right)\right]{2\pi m\over M^2}\delta(p^0)
\label{c45}\ee
\be
w_{0i}(p)=0\sp w_{ij}(p)=-\left[{p_ip_j-\delta_{ij}\vec p^2\over 3\Lambda}\left({1\over \left(\vec p^2-\Lambda\right)}-{2\over 2\vec p^2+\Lambda}\right)\right]{2\pi m\over M^2}\delta(p^0)
\label{c46}\ee
In configuration space we have
\be
w_{00}(x^i)={m\over M^2}\int{d^3 p\over (2\pi)^3}e^{i\vec p\cdot \vec x}\left[{1\over \vec p^2-\Lambda}-{\vec p^2\over 3\Lambda}\left({1\over \left(\vec p^2-\Lambda\right)}-{2\over 2\vec p^2+\Lambda}\right)\right]=
\label{c47}\ee
$$
={m\over 3M^2}\int{d^3 p\over (2\pi)^3}e^{i\vec p\cdot \vec x}\left[{2\over \vec p^2-\Lambda}-{1\over 2\vec p^2+\Lambda}\right]
$$
\be
w_{ij}(x^i)=-{m\over 3M^2\Lambda}\int{d^3 p\over (2\pi)^3}e^{i\vec p\cdot \vec x}\left[{(p_ip_j-\delta_{ij}\vec p^2)}\left({1\over \vec p^2-\Lambda}-{1\over \vec p^2+{\Lambda\over 2}}\right)\right]=
\label{c48}\ee
$$=
-{m\over 3M^2\Lambda}\left(\delta_{ij}\square-\pa_i\pa_j\right)\int{d^3 p\over (2\pi)^3}e^{i\vec p\cdot \vec x}\left({1\over \vec p^2-\Lambda}-{1\over \vec p^2+{\Lambda\over 2}}\right)
$$
where above, $\square={\pa\over {\pa x^k}}{\pa\over {\pa x^k}}$.

We have
\be
D_+(m)(r)\equiv \int{d^3 p\over (2\pi)^3}{e^{i\vec p\cdot \vec x}
\over \vec p^2+m^2}={e^{-m r}\over 4\pi r}\sp r=\sqrt{x^ix^i}
\label{c49}\ee
We can also compute
\be
D^{ij}_+(m,r)\equiv \int{d^3 p\over (2\pi)^3}{e^{i\vec p\cdot \vec x}
\over \vec p^2+m^2}p^ip^j=
-\pa_i\pa_j\int{d^3 p\over (2\pi)^3}{e^{i\vec p\cdot \vec x}
\over \vec p^2+m^2}=-\pa_i\pa_j{e^{-m r}\over 4\pi r}=
\label{c49a}\ee
$$
=\delta^{ij}{e^{-mr}\over 4\pi r^3}(1+mr)-{x^ix^j\over 4\pi r^5}e^{-mr}(3+3mr+m^2r^2)
$$
Taking the trace
\be
\delta_{ij}D^{ij}_{+}=-{m^2\over 4\pi r}e^{-mr}=-m^2D_{+}(m,r)
\label{c49b}\ee
Doing this in (\ref{c49a}) we obtain instead
\be
\delta_{ij}D^{ij}_{+}=\int{d^3 p\over (2\pi)^3}e^{i\vec p\cdot \vec x}{p^2
\over \vec p^2+m^2}=\int{d^3 p\over (2\pi)^3}e^{i\vec p\cdot \vec x}\left[1-{m^2\over \vec p^2+m^2}\right]=\delta^{(3)}(\vec x)-m^2 D_{+}(m,r)
\label{c49e}\ee
This indicates that me missed a $\delta$-function in (\ref{c49a}), (\ref{c49b}). The correct formula is
\be
D^{ij}_+(m,r)=\delta^{ij}{e^{-mr}\over 4\pi r^3}(1+mr)-{x^ix^j\over 4\pi r^5}e^{-mr}(3+3mr+m^2r^2)+{\delta^{ij}\over 3}\delta^{(3)}(\vec x)
\label{c49c}\ee
\be
\delta_{ij}D^{ij}_{+}=\delta^{(3)}(\vec x)-{m^2\over 4\pi r}e^{-mr}=\delta^{(3)}(\vec x)-m^2D_{+}(m,r)
\label{c49d}\ee

We also have

\be
D_-(m)\equiv \int{d^3 p\over (2\pi)^3}{e^{i\vec p\cdot \vec x}
\over \vec p^2-m^2}={1\over 4\pi^2r}\int_{-\infty}^{\infty}{udu\over u^2-m^2r^2}\sin(u)
\label{c50}\ee
$D_-$ can only be defined by shifting the contour and there are two definitions that give a real answer:
\be
D^c_-(m)={1\over 4\pi r}\cos(mr)\sp D^s_-(m)={1\over 4\pi r}\sin(mr)
\label{c51}\ee
Consider first $\Lambda=k^2>0$.
Then
\be
w_{00}={m\over 12\pi M^2 r}\left[\cos(kr)-e^{-{k\over \sqrt{2}}r}\right]
\label{c52}\ee
\be
w_{ij}=-{m\over 12\pi M^2\Lambda}\left(\delta_{ij}\square-\pa_i\pa_j\right){1\over 4\pi r}\left[\cos(kr)-e^{-{k\over \sqrt{2}}r}\right]=
\label{c53}\ee
$$
-{m\over 12\pi M^2\Lambda}\left[{\delta^{ij}\over r^3}\left[(1-k^2r^2)\cos(kr)+ kr\sin(kr)-\left({k^2r^2\over 2}+{kr\over \sqrt{2}} + 1\right)e^{-{kr\over \sqrt{2}}}\right]\right.-
$$
$$
-\left.{x^ix^j\over r^5}\left[(3-k^2r^2)\cos(kr)+3kr\sin(kr)-\left(3+{3kr\over \sqrt{2}}+{k^2r^2\over 2}\right)e^{-{kr\over \sqrt{2}}}\right]\right]
$$
The metric therefore becomes
\be
\delta ds^2=w_{00}dt^2+w_{ij}dx^idx^j={m\over 12\pi M^2 r}\left[\cos(kr)-e^{-{k\over \sqrt{2}}r}\right]dt^2-
\label{c54}\ee
$$
-{m\over 12\pi M^2\Lambda}\left[-2\cos(kr)+2kr\sin(kr)+2 \left(1+{ kr \over \sqrt{2}} \right)e^{-{kr\over \sqrt{2}}}\right]{dr^2\over r^3}-
$$
$$
-{m\over 12\pi M^2\Lambda}\left[(1-k^2r^2)\cos(kr)+kr\sin(kr)-\left({k^2r^2\over 2}+{kr\over \sqrt{2}}+ 1 \right)e^{-{kr\over \sqrt{2}}}\right]{d\Omega_2^2\over r}
$$
where we used
\be
dx^idx^i=dr^2+r^2d\Omega_2^2\sp x^ix^jdx^idx^j=r^2 dr^2
\label{c55}\ee
These formulae  agree with the gravitational solutions studied in appendix (\ref{pert1}). The matching is through the relations:
\be
C_+ = 0 , \quad    C_- = \frac{m}{ 24 \pi}, \quad  \bar{C}_-= 0, \quad  \bar{C}_+ = \frac{m}{12 \pi },
\ee
Note also that the solution is singular when $\Lambda\to 0$ as expected.
When $\Lambda=-k^2<0$ the situation is reversed: the massive asymptotics become sinusoidal and vice versa.

We should contrast the previous calculation with standard GR.
In standard GR we have (\ref{c30}) with $\Lambda=0$.
In that case, we obtain that $T_{\m\n}$ is conserved, $\pa^{\m}T_{\m\n}=0$.
Because now the equation is invariant under the transformation
\be
w_{\m\n}\to w_{\m\n}+\pa_{\m}\xi_{\n}+\pa_{\n}\xi_{\m}
\label{c56}\ee
we may use this to make the metric transverse.
\be
w_{\m\n}=w^T_{\m\n}+\pa_{\m}\xi_{\n}+\pa_{\n}\xi_{\m}\sp \pa^{\m}w^T_{\m\n}=0
\label{c57}\ee
Then the equation above gives
\be
\pa^{\m}w_{\m\n}=\square\xi_{\n}+\pa_{\n}(\pa\cdot \xi)\sp \pa^{\m}\pa^{\n}w_{\m\n}=2\square(\pa\cdot \xi)
\label{c58}\ee
We can then solve
\be
\pa\cdot \xi={1\over 2}\square^{-1}\pa^{\m}\pa^{\n}w_{\m\n}
\label{c59}\ee
and we also have
\be
\pa_{\n}(\pa\cdot \xi)=\pa_{\n}{1\over 2}\square^{-1}\pa^{\m}\pa^{\n}w_{\m\n}={1\over 2}\square^{-1}\square \pa^{\m}w_{\m\n}={1\over 2}\pa^{\m}w_{\m\n}
\label{c60}\ee
which we use in (\ref{c58}) to obtain
\be
\xi_{\n}={1\over 2}\square^{-1}\pa^{\m}w_{\m\n}
\label{c61}\ee
If we now assume that we arrived in the transverse gauge, an extra diffeomorphism that leaves the gauge invariant must satisfy
\be
\square\xi_{\n}+\pa_{\n}(\pa\cdot \xi)=0
\label{c62}\ee
This then implies
\be
\pa\cdot \xi=0\sp \square \xi_{\n}=0
\label{c63}\ee with unique solution vanishing at infinity $\xi_{\m}=0$.

We now reconsider (\ref{c30}) with $\Lambda=0$ and a transverse metric
\be
\square w_{\m\n}+(\pa_{\m}\pa_{\n}-\eta_{\m\n}\square)w=-{T_{\m\n}\over M^2}
\label{c64}\ee
Taking the trace we obtain
\be
\square w={1\over 2}{T\over M^2}~~~\to ~~~w={1\over 2}\square^{-1}{T\over M^2}
\label{c65}\ee
Then (\ref{c64}) becomes
\be
\square w_{\m\n}=-{T_{\m\n}\over M^2}+{1\over 2}\left(\eta_{\m\n}-\square^{-1}\pa_{\m}\pa_{\n}\right){T\over M^2}
\label{c66}\ee
which in momentum space can be solved as
\be
w_{\m\n}(p)={T_{\m\n}(p)\over M^2 p^2}+{1\over 2p^2}\left(-\eta_{\m\n}+{p_{\m}p_{\n}\over p^2}\right){T(p)\over M^2}
\label{c67}\ee
The linearized gravitational interaction is
\be
w^{\m\n}T'_{\m\n}={T^{\m\n}T'_{\m\n}-{1\over 2}TT'\over M^2 p^2}
\ee

Taking again a static source like in (\ref{c43}) we have
\be
T=-2\pi m\delta(p^0)
\label{c68}\ee
From which we obtain
\be
w_{00}={2\pi m\over \vec M^2 p^2}\delta(p^0)-{\pi m\over M^2\vec p^2}\delta(p^0)={\pi m\over M^2 \vec  p^2}\delta(p^0)
\label{c69}\ee
\be
w_{i0}=0\sp w_{ij}={\pi m\over M^2 \vec p^2}\left(\delta^{ij}-{p^ip^j\over \vec p^2}\right)\delta(p^0)
\label{c70}\ee
In configuration space we obtain
\be
w_{00}={m\over 2M^2}\int {d^3 p\over (2\pi)^3}{e^{i\vec p\cdot \vec x}\over \vec p^2}={m\over 8\pi M^2 r}
\label{c71}\ee
\be
w_{ij}={m\over 16\pi M^2}\left({\delta^{ij}\over r}+{x^ix^j\over r^3}\right)
\label{c72}\ee
The metric is then
\be
\delta ds^2={m\over 8\pi M^2 r}dt^2+{m\over 16\pi M^2 r}\left(2dr^2+r^2d\Omega_2^2\right)
\label{c73}\ee
and the total metric
\be
ds^2=\left(-1+{m\over 8\pi M^2 r}\right)dt^2+\left(1+{m\over 8\pi M^2 r}\right)dr^2+\left(r^2+{m\over 16\pi M^2}r\right)d\Omega_2^2
\label{c75}\ee

We now define
\be
\hat r=\sqrt{r^2+{m\over 16\pi M^2}r}=r+{m\over 32\pi M^2}+{\cal O}(m^2)
\label{c76}\ee
\be
d\hat r=dr+{\cal O}(m^2)
\label{c77}\ee
and the metric becomes
\be
ds^2=-\left(1-{m\over 8\pi M^2 \hat r}\right)dt^2+{d\hat r^2\over \left(1-{m\over 8\pi M^2 \hat r}\right)}+\hat r^2d\Omega_2^2+{\cal O}(m^2)
\label{c78}\ee
which is the correct Schwarzschild metric of a point source.

\subsection{The linearized analysis of the $T \hat T$ deformation}

From the action (\ref{linearag}) and the definition (\ref{linearar}) we obtain the following linearized equation in momentum space in the far IR
\be
P^{\m\n\r\s}h_{\r\s}=\l^2{\bf T}^{\m\n}
\label{c78a}\ee
where
\be
P^{\m\n\r\s}=-{\l^2 M_P^2\over 4}\left[(p^2+\Lambda)\left(\eta^{\m\n}\eta^{\r\s}-\eta^{\m\r}\eta^{\n\s}-\eta^{\m\s}\eta^{\n\r}\right)+\right.
\label{c78b}\ee
$$+\left.
 ( \eta^{\nu \sigma} p^\mu p^\rho +\eta^{\nu \rho} p^\mu p^\sigma + \eta^{\mu \sigma} p^\nu p^\rho +\eta^{\mu\rho} p^\nu p^\sigma )\right]
$$

(\ref{c78a}) translates to
\be
\left(p^2 + {\Lambda}\right)h_{\m\n}-\left(p_{\m}p^{\r}h_{\r\n}+p_{\n}p^{\r}h_{\r\m}\right)-{1\over 2}\left(p^2 + {\Lambda}\right)h~\eta_{\m\n}={2 \over M_P^2}{\bf T}_{\m\n}
\label{c79}\ee
where the energy-momentum tensor is defined in (\ref{linearaab}) and $\Lambda, M_P^2$ were defined in (\ref{linearaw}).

On the other hand, by taking a trace in (\ref{c79}) and sequentially contracting with momenta, we obtain the following equations
\be
{\Lambda}p^{\r}h_{\r\n}-(pph)p_{\n}-{1\over 2}\left(p^2+{\Lambda}\right)h~p_{\n}={2\over M_P^2}p^{\r}{\bf T}_{\r\n}
\label{c80}\ee
\be
\left(p^2-{\Lambda}\right)(pph)+{p^2\over 2}\left(p^2+{ \Lambda}\right)h=-{2\over M_P^2}(pp{\bf T})
\label{c81}\ee
\be
2(pph)+\left(p^2+{ \Lambda}\right)h=-{2\over M_P^2}{\bf T}
\label{c82}\ee
where
$$pph\equiv p^{\m}p^{\n}h_{\m\n}\sp (pp{\bf T})\equiv p^{\r}p^{\n}{\bf T}_{\r\n}\;.$$
Solving (\ref{c80}) and (\ref{c81}) we obtain
\be
pph={2\over M_P^2\Lambda}\left[(pp{\bf T})-{p^2\over 2}{\bf T}\right]
\label{c83}\ee
\be
h=-{2\over M_P^2\Lambda}\left[{2\over \left(p^2+{\Lambda}\right)}(pp{\bf T})-{\left(p^2-{\Lambda}\right)
\over \left(p^2+{\Lambda}\right)}{\bf T}\right]
\label{c84}\ee
Substituting (\ref{c83}) and (\ref{c84}) in (\ref{c80}) we obtain
\be
p^{\r}h_{\r\n}={2\over M_P^2\Lambda}\left[p^{\r}{\bf T}_{\r\n}-{1\over 2}{\bf T}p_{\n}\right]
\label{c85}\ee
We finally substitute (\ref{c84}), (\ref{c85}) in (\ref{c79}) to obtain
\be
\left(p^2+{\Lambda}\right)h_{\m\n}={2\over M_P^2\Lambda}\left[
\left(p_{\m}(p{\bf T})_{\n}+p_{\n}(p{\bf T})_{\m}-{\bf T}p_{\m}p_{\n}-\eta_{\m\n}(pp{\bf T})\right)+{p^2-\Lambda\over 2}{\bf T}\eta_{\m\n}+\Lambda {\bf T}_{\m\n}\right]
\label{c86}\ee
We now use (\ref{linearaab}) to rewrite the above equation as
\be
\left(p^2+{\Lambda}\right)h_{\m\n}={2\over M_P^2\Lambda}
\left[
\left(p_{\m}(p{ T})_{\n}+p_{\n}(p{ T})_{\m}-(1+2\cc) {T}p_{\m}p_{\n}-\eta_{\m\n}(pp{ T})\right)+
\right.
\label{c87}\ee
$$
\left.+{1+2\cc\over 2}(p^2-\Lambda){T}\eta_{\m\n}+\Lambda {T}_{\m\n}\right]
$$
while (\ref{c84}) and (\ref{c85}) become
\be
p^{\r}h_{\r\n}={2\over M_P^2\Lambda}\left[p^{\r}T_{\r\n}-\left(\cc+{1\over 2}\right)Tp_{\n}\right]
\label{c88}\ee
\be
(p^2+\Lambda)h=-{2\over M_P^2\Lambda}\left[2(ppT)+\left(-(2\cc+1)p^2+(4\cc+1)\Lambda\right)T\right]
\label{c89}\ee

Note now that $\Lambda\sim {\cal O}(1)$ while ${M_P^{-2}}\sim {\cal O}(\l^2)$ and $p^{\m}{\bf T}_{\m\n}\sim {\cal O}(\l)$.
Therefore (\ref{c87}) becomes
\be
\left(p^2+{\Lambda}\right)h_{\m\n}={2\over M_P^2}\left[
{ T}_{\m\n}-{2\over \Lambda }\left(\cc+{1\over 2}\right)\left( p_{\m}p_{\n}-{p^2-\Lambda\over 2}\eta_{\m\n}\right)T\right]+{\cal O}(\l^3)
\label{c90}\ee
so that
\be
h_{\m\n}={2\over M_P^2(p^2+\Lambda)}\left[
{ T}_{\m\n}-(2\cc+1)\left( {p_{\m}p_{\n}\over \Lambda}+\eta_{\m\n}\right)T\right]+{2\cc+1\over M_P^2 \Lambda}\eta_{\m\n}T+{\cal O}(\l^3)
\label{c90a}\ee
To calculate now the leading gravitational interaction,
we must remember that it comes from
\be
-h_{\m\n}{\bf T'}^{\m\n}=-{2\over M_P^2(p^2+\Lambda)}\left[ T'_{\m\n}T^{\m\n}-{1\over 2}TT'+\right.
\label{c91}\ee
$$+\left.
(2\cc+1)\left[-2\cc +{2\cc+1\over 2}{p^2\over \Lambda}\right]TT'\right]+{\cal O}(\l^3)=
$$
\be
=-{2\over M_P^2(p^2+\Lambda)}\left[ T'_{\m\n}T^{\m\n}-{1\over 2}TT'-{(2\cc+1)(6\cc+1)\over 2}TT'\right]-{(2\cc+1)^2\over M_P^2 \Lambda}TT'+{\cal O}(\l^3)
\label{c91a}\ee
where the last piece is a contact interaction.
From inspection of (\ref{c91a}) we learn that:

$\bullet$ there is a vDVz discontinuity when $\cc\not=-{1\over 2}$ or $-{1\over 6}$.

$\bullet$ the graviton is massive when $\Lambda>0$ and a tachyon when $\Lambda<0$.

$\bullet$ When $\cc=-{1\over 2}$ or $c=-{1\over 6}$ there is no vDvZ discontinuity although the graviton is massive.

We now determine the static potential by again using (\ref{c43})
\be
h_{00}={2(2\pi)m\over M_P^2}\int {d^3p\over (2\pi)^3}{e^{i\vec p\cdot \vec x}\over (\vec p^2+\Lambda)}\left[-2\cc+{\vec p^2+\Lambda\over \Lambda}\right]=
\label{c102}\ee
$$
=-{4\cc (2\pi)m\over M_P^2}D_+(\sqrt{\Lambda},r)+{2m\left(2\cc+1\right)\over M_P^2}\delta^{(3)}(\vec x)
$$
\be
h_{ij}={2(2\pi)(2\cc+1)m\over M_P^2}\left({\pa_i\pa_j\over \Lambda}-\delta_{ij}\right)\int {d^3p\over (2\pi)^3}{e^{i\vec p\cdot \vec x}\over (\vec p^2+\Lambda)}-{(2\cc+1)(2\pi)m\over M_P^2\Lambda}\delta_{ij}\delta^{(3)}(\vec x)=
\ee
$$
={2(2\pi)(2\cc+1)m\over M_P^2}\left({\pa_i\pa_j\over \Lambda}-\delta_{ij}\right)D_+(\sqrt{\Lambda},r)-{(2\cc+1)(2\pi)m\over M_P^2\Lambda}\delta_{ij}\delta^{(3)}(\vec x)
$$
{For $\Lambda<0$ we have a negative mass-squared graviton.}

Finally, we shall  write (\ref{c79}) so that on the right-hand side there is only $T_{\m\n}$.
To do this we take the trace of (\ref{c79}) to obtain
\be
-(p^2+\Lambda)h-2(pph)={2\over M_P^2}(4\cc+1)T
\label{c92c}\ee
We now solve (\ref{c92c}) for $T$ and substitute in (\ref{c79}) to obtain
\be
\left(p^2 + {\Lambda}\right)h_{\m\n}-\left(p_{\m}p^{\r}h_{\r\n}+p_{\n}p^{\r}h_{\r\m}\right)+{2\cc\over 4\cc+1}(pph)\eta_{\m\n}-{2\cc+1\over 2(4\cc+1)}(p^2+\Lambda)h\eta_{\m\n}={2 \over M_P^2}{ T}_{\m\n}
\label{c93}\ee
By redefining in (\ref{c93})
\be
h_{\m\n}\to w_{\m\n}-{w\over 2}\eta_{\m\n}
\ee
(\ref{c93}) becomes
\be
(p^2+\Lambda)w_{\m\n}-\left(p_{\m}p^{\r}w_{\r\n}+p_{\n}p^{\r}w_{\r\m}\right)+p_{\m}p_{\n}w+{2\cc\over 4\cc+1}\left((ppw)-p^2 w\right)\eta_{\m\n}-{\cc\Lambda\over 4\cc+1}w\eta_{\m\n}={2T_{\m\n}\over M_P^2}
\label{c30b}
\ee

\subsubsection{Matching the fluctuation equations}\label{matchingfluctuations}

In this subsection we will match the linearized solutions of the single theory with those of the  TT-coupled theories.

 By redefining the emergent graviton so that we match the expansion of the Einstein-Hilbert action with cosmological constant~\ref{b2}
\be
h_{\mu\nu} =  \pmet_{\mu\nu} - \frac{1}{2} \pmet\, \eta_{\mu\nu} +\lambda \h{\Lambda} \, \eta_{\mu\nu}
\ee
we find from equation \eqref{c78a}
\bea
\left(p^2 + {\Lambda}\right)\pmet_{\m\n}-\left(p_{\m}p^{\r}\pmet_{\r\n}+p_{\n}p^{\r}\pmet_{\r\m}\right) + p_\mu p_\nu \pmet + \left( (pp \pmet) - p^2 \pmet \right) \eta_{\m \n} - \frac{\Lambda}{2} \pmet \eta_{\m \n} \nn \\
 = {2 \over M_P^2}  \left( \vis{T}_{\mu\nu} - \frac{1}{2} (1+ 2 {\mathfrak c}) \, \vis{T} \,\eta_{\mu\nu} - \frac{ \lambda^{-1} }{2}\left( 1 - \lambda^{-1} \h{\Lambda}^{-1} \right) \eta_{\mu\nu} \right)  \nn \\
\label{c79b}\eea
with $ \vis{T}_{\mu\nu}$ the original SM stress energy tensor. This is consistent then with~\ref{bianchiaa} that gives the fluctuation equation for the emergent graviton $\pmet_{\m \n}$. This means that the emergent graviton $\pmet_{\m \n}$ is coupled to the specific combination of the stress energy tensor. We also notice that for $\cc = - \half$, $-{1\over 6}$, the graviton is coupled in the usual way to the original stress tensor of the SM, with no extra coupling to its trace.

This should now be compared with the equation for the linearised fluctuations stemming from an expansion of the complete non-linear equation \eqref{f9}. This is also given in equations \eqref{c29b} and (\ref{c30}) that we reproduce here in a suggestive form in momentum space
\be
 \frac{ \delta \tilde G_{\m\n}}{\delta \th_{\r \s}} \bigg|_{\th = \eta} w_{\r \s} -  {\Lambda^{(n.l.)} \over 2} w_{\m \n} \,  = \,  {1\over 2  M_P^2}  T^{(n.l.)}_{\m\n}+\cdots
\label{c29bb}\ee
\be
(p^2-\Lambda^{(n.l.)})w_{\m\n}-\left(p_{\m}p^{\r}w_{\r\n}+p_{\n}p^{\r}w_{\r\m}\right)+p_{\m}p_{\n}w+\left((ppw)-p^2 w\right)\eta_{\m\n}={T^{(n.l.)}_{\m\n}\over M_P^2} \, ,
\label{c30a}
\ee
where with $T^{(n.l.)}_{\m\n}, \Lambda^{(n.l.)}$ we denote the stress energy tensor and cosmological constant appearing in the non-linear formulation of the problem, which we would like to relate with the stress energy tensor and cosmological constant of the linearised $T \hat{T}$ computation of section~\ref{linear} in the main text. One should be careful though since these last two expressions are a linearisation of the non-linear equations of motion, while \eqref{c79b} is coming from the variation of the quadratic part of a linearised effective action i.e. equations \eqref{linearag} and \eqref{linearav}.

We should therefore follow the analysis of appendix \ref{BridgingI} describing the expansion of the non-linear action. In particular taking into account equation \eqref{match2} of appendix \ref{BridgingI} and the symmetry in exchanging the indices $\m \n \leftrightarrow \r \s$, we find the relation between the variation of the following one-point stress tensor
\be
\tilde T_{\m \n} = \tilde G_{\m\n} - {\Lambda^{(n.l.)} \over 2} ( \tilde{h}_{\m \n} - \eta_{\m \n})
\ee
and the operator appearing in the quadratic expansion of the Einstein Hilbert term in the action around flat space
\be
\frac{ \delta \tilde T_{\m\n} (x) }{\delta \th^{\r \s}(y)}  \bigg|_{\th = \eta} = P_{\m \n \r \s}(x,y) -  {\Lambda^{(n.l.)} \over 2} \eta_{\r\s} \eta_{\m\n} \, \delta^4 (x-y) \, .
\label{match2b}
\ee
In this formula $ P_{\m \n \r \s}$ is the operator acting on the fluctuation $w_{\m \n} \equiv \pmet_{\m \n}$ in the quadratic part of the Einstein Hilbert action with a cosmological constant. This then results into \eqref{c30a} being written as
\be\label{c30aa}
\left(p^2 - {\Lambda^{(n.l.)}}\right)\pmet_{\m\n}-\left(p_{\m}p^{\r}\pmet_{\r\n} + p_{\n}p^{\r}\pmet_{\r\m}\right) + p_\mu p_\nu \pmet + \left( (pp \pmet) - p^2 \pmet \right) \eta_{\m \n} + \frac{\Lambda^{(n.l.)}}{2} \pmet \eta_{\m \n} = \, \nn
\ee
$$
 = P_{\m \n \r \s} \pmet^{\r \s} = {T^{(n.l.)}_{\m\n}\over M_P^2}
$$

We notice that this matches the fluctuation equation of the $T \hat{T}$ computation \eqref{c79b}, if the stress energy tensors of the linearised and the non-linear computation are related as follows
\be
T^{(n.l.)}_{\m\n} =  \left( \vis{T}_{\mu\nu} - \frac{1}{2} (1+ 2 {\mathfrak c}) \, \vis{T} \,\eta_{\mu\nu} - \frac{ \lambda^{-1} }{2}\left( 1 - \lambda^{-1} \h{\Lambda}^{-1} \right) \eta_{\mu\nu} \right)  \, ,
\ee
with the cosmological constants $\Lambda = - \Lambda^{(n.l.)}$ being related with a relative minus sign. The reason for this relative minus sign is the passage from the equations of motion for the Schwinger functional \eqref{f4} to the equation of motion of the Legendre transformed functional \eqref{f9}, so that an original positive cc. becomes a negative cc. of the effective equations of the emergent metric after the Legendre transform.

\subsection{The Schwinger functional and the effective action at the linearized level\label{sl}}

In this section we will perform that Legendre transform in the linearized approximation.

We start again from the Schwinger functional in (\ref{f1}) and expand it around the flat space metric to quadratic order in order to read the appropriate IR correlators. We assume that $V$ and $M$ are constants for simplicity.
Then we obtain
\be
g_{\m\n}\equiv \eta_{\m\n}+w_{\m\n}\sp g^{\m\n}=\eta^{\m\n}-w^{\m\n}+{w^{\m}}_{\a}w^{\a\n}+{\cal O}(w^3)
\label{c94}\ee
\be
\sqrt{-g}=1+{w\over 2}+{w^2\over 8}-{w^{\m\n}w_{\m\n}\over 4}+{\cal O}(w^3)
\label{c95} \ee
\def\G{\Gamma}
\be
{\Gamma_{\m\n}}^{\r}=g^{\r\s}[\m\n;\s]={1\over 2}\left(\pa_{\m}{w_{\n}}^{\rho}  +\pa_{\n}{w_{\m}}^{\rho}-\pa^{\r}w_{\m\n} \right)-
\label{c96}\ee
$$
-{w^{\r\s}\over 2}
\left(\pa_{\m}{w_{\n\s}}  +\pa_{\n}{w_{\m\s}}-\pa_{\s}w_{\m\n} \right)+{\cal O}(w^3)
$$
For the Riemann tensor:
\be
{R^{\l}}_{\m\n\r}=\p_{\n}{\G_{\m\r}}^{\l}-\p_{\r}{\G_{\m\n}}^{\l}
-{\G_{\m\n}}^{\s}{\G_{\s\r}}^{\l}+{\G_{\m\r}}^{\s}{\G_{\s\n}}^{\l}=
\label{c98}\ee
$$
={1\over 2}\pa_{\n}\left(\pa_{\m}{w_{\r}}^{\l}  +\pa_{\r}{w_{\m}}^{\l}-\pa^{\l}w_{\m\r} \right)
-{\pa_{\n}w^{\l\s}\over 2}
\left(\pa_{\m}{w_{\r\s}}  +\pa_{\r}{w_{\m\s}}-\pa_{\s}w_{\m\r} \right)-
$$
$$
-{w^{\l\s}\over 2}
\pa_{\n}\left(\pa_{\m}{w_{\r\s}}  +\pa_{\r}{w_{\m\s}}-\pa_{\s}w_{\m\r} \right)
$$
$$
-{1\over 2}\pa_{\r}\left(\pa_{\m}{w_{\n}}^{\l}  +\pa_{\n}{w_{\m}}^{\l}-\pa^{\l}w_{\m\n} \right)
+{\pa_{\r}w^{\l\s}\over 2}
\left(\pa_{\m}{w_{\n\s}}  +\pa_{\n}{w_{\m\s}}-\pa_{\s}w_{\m\n} \right)+
$$
$$
+{w^{\l\s}\over 2}
\pa_{\r}\left(\pa_{\m}{w_{\n\s}}  +\pa_{\n}{w_{\m\s}}-\pa_{\s}w_{\m\n} \right)-
$$
$$
-{1\over 4}\left(\pa_{\m}{w_{\n}}^{\s}  +\pa_{\n}{w_{\m}}^{\s}-\pa^{\s}w_{\m\n} \right)\left(\pa_{\s}{w_{\r}}^{\l}  +\pa_{\r}{w_{\s}}^{\l}-\pa^{\l}w_{\r\s} \right)+
$$
$$
+{1\over 4}\left(\pa_{\m}{w_{\r}}^{\s}  +\pa_{\r}{w_{\m}}^{\s}-\pa^{\s}w_{\m\r} \right)\left(\pa_{\s}{w_{\n}}^{\l}  +\pa_{\n}{w_{\s}}^{\l}-\pa^{\l}w_{\n\s} \right)+{\cal O}(w^3)
$$

\be
R_{\m\n}={R^{\l}}_{\m\l\n}={1\over 2}\pa_{\l}\left(\pa_{\m}{w_{\n}}^{\l}  +\pa_{\n}{w_{\m}}^{\l}-\pa^{\l}w_{\m\n} \right)
-{\pa_{\l}w^{\l\s}\over 2}
\left(\pa_{\m}{w_{\n\s}}  +\pa_{\n}{w_{\m\s}}-\pa_{\s}w_{\m\n} \right)-
\label{c97}\ee
$$
-{w^{\l\s}\over 2}
\pa_{\l}\left(\pa_{\m}{w_{\n\s}}  +\pa_{\n}{w_{\m\s}}-\pa_{\s}w_{\m\n} \right)
$$
$$
-{1\over 2}\pa_{\n}\left(\pa_{\m}{w_{\l}}^{\l}  +\pa_{\l}{w_{\m}}^{\l}-\pa^{\l}w_{\m\l} \right)
+{\pa_{\n}w^{\l\s}\over 2}
\left(\pa_{\m}{w_{\l\s}}  +\pa_{\l}{w_{\m\s}}-\pa_{\s}w_{\m\l} \right)+
$$
$$
+{w^{\l\s}\over 2}
\pa_{\n}\left(\pa_{\m}{w_{\l\s}}  +\pa_{\l}{w_{\m\s}}-\pa_{\s}w_{\m\l} \right)-
$$
$$
-{1\over 4}\left(\pa_{\m}{w_{\l}}^{\s}  +\pa_{\l}{w_{\m}}^{\s}-\pa^{\s}w_{\m\l} \right)\left(\pa_{\s}{w_{\n}}^{\l}  +\pa_{\n}{w_{\s}}^{\l}-\pa^{\l}w_{\n\s} \right)+
$$
$$
+{1\over 4}\left(\pa_{\m}{w_{\n}}^{\s}  +\pa_{\n}{w_{\m}}^{\s}-\pa^{\s}w_{\m\n} \right)\left(\pa_{\s}{w_{\l}}^{\l}  +\pa_{\l}{w_{\s}}^{\l}-\pa^{\l}w_{\l\s} \right)+{\cal O}(w^3)
$$

\be
R=g^{\m\n}R_{\m\n}=\eta^{\m\n}R_{\m\n}-w^{\m\n}R_{\m\n}+{\cal O}(w^3)=
\label{c99}\ee
$$
={1\over 2}\pa_{\l}\left(2\pa^{\m}{w_{\m}}^{\l}  -\pa^{\l}{w^{\m}}_{\m} \right)
-{\pa_{\l}w^{\l\s}\over 2}
\left(2\pa^{\m}{w_{\m\s}}-\pa_{\s}{w^{\m}}_{\m} \right)-{w^{\l\s}\over 2}
\pa_{\l}\left(2\pa^{\m}{w_{\m\s}} -\pa_{\s}{w^{\m}}_{\m} \right)-
$$

$$
-{1\over 2}\pa^{\m}\left(\pa_{\m}{w_{\l}}^{\l}  +\pa_{\l}{w_{\m}}^{\l}-\pa^{\l}w_{\m\l} \right)
+{\pa^{\m}w^{\l\s}\over 2}
\left(\pa_{\m}{w_{\l\s}}  +\pa_{\l}{w_{\m\s}}-\pa_{\s}w_{\m\l} \right)+
$$
$$
+{w^{\l\s}\over 2}
\pa^{\m}\left(\pa_{\m}{w_{\l\s}}  +\pa_{\l}{w_{\m\s}}-\pa_{\s}w_{\m\l} \right)-
$$
$$
-{1\over 4}\left(\pa_{\m}{w_{\l}}^{\s}  +\pa_{\l}{w_{\m}}^{\s}-\pa^{\s}w_{\m\l} \right)\left(\pa_{\s}w^{\m\l}  +\pa^{\m}{w_{\s}}^{\l}-\pa^{\l}{w^{\m}}_{\s} \right)+
$$
$$
+{1\over 4}\left(2\pa^{\m}{w_{\m}}^{\s}  -\pa^{\s}{w^{\m}}_{\m} \right)\pa_{\s}{w_{\l}}^{\l}-
$$
$$
-{w^{\m\n}\over 2}\pa_{\l}\left(\pa_{\m}{w_{\n}}^{\l}  +\pa_{\n}{w_{\m}}^{\l}-\pa^{\l}w_{\m\n} \right)
+{w^{\m\n}\over 2}\pa_{\n}\left(\pa_{\m}{w_{\l}}^{\l}  +\pa_{\l}{w_{\m}}^{\l}-\pa^{\l}w_{\m\l} \right)+{\cal O}(w^3)
$$

$$
=(\pa\pa w)-{1\over 2}\square w -(\pa w)^{\m}(\pa w)_{\m}+{1\over 2}(\pa w)^{\m}\pa_{\m}w-w^{\m\n}\pa_{\m}(\pa w)_{\n}+{1\over 2}w^{\m\n}\p_{\m}\pa_{\n}w-
$$

$$
-{1\over 2}\square w
+{1\over 2}\pa^{\m}w^{\l\s}\pa_{\m}{w_{\l\s}} +{1\over 2}w^{\m\n}\square{w_{\m\n}} +{1\over 4}\pa_{\n}w_{\m\r}\pa^{\n}w^{\m\r}-{1\over 2}\pa_{\n}\omega_{\m\r}\pa^{\rho}w^{\m\n}+
$$
$$
+{1\over 2}(\pa w)^{\m}\pa_{\m}w-{1\over 4}\pa^{\m}w\pa_{\m}w-w^{\m\n}\pa_{\m}(\pa w)_{\n}+{1\over 2}w^{\m\n}\square w_{\m\n}+{1\over 2}w^{\m\n}\pa_{\m}\pa_{\n}w+{\cal O}(w^3)
$$

$$
=(\pa\pa w)-\square w -(\pa w)^{\m}(\pa w)_{\m}+(\pa w)^{\m}\pa_{\m}w-2w^{\m\n}\pa_{\m}(\pa w)_{\n}+w^{\m\n}\p_{\m}\pa_{\n}w+
$$

$$
+{3\over 4}\pa^{\m}w^{\l\s}\pa_{\m}{w_{\l\s}} +w^{\m\n}\square{w_{\m\n}} -{1\over 2}\pa_{\n}\omega_{\m\r}\pa^{\rho}w^{\m\n}-{1\over 4}\pa^{\m}w\pa_{\m}w+{\cal O}(w^3)
$$
where
\be
(\pa w)_{\m}\equiv \pa^{\r}w_{\r\m}\sp w\equiv \eta^{\m\n}w_{\m\n}\sp \pa\pa w\equiv \pa^{\m}\pa^{\n}w_{\m\n}
\label{c100}\ee
\be
\int d^4x \sqrt{-g}R=\pa\pa w-\square w+{1\over 2}w(\pa\pa w)+{1\over 2}(\pa w)^{\m}(\pa w)_{\m}+
\label{c101}\ee
$$
+{1\over 4}w^{\m\n}\square{w_{\m\n}} -{1\over 4}w\square w+{\cal O}(w^3)
$$
\be
\sqrt{-g}V=V\left[1+{w\over 2}+{w^2\over 8}-{w^{\m\n}w_{\m\n}\over 4}+{\cal O}(w^3)\right]
\label{c108}\ee
\be
\sqrt{-g}{Z\over 2}(\pa J)^2={Z\over 2}\left[1+{w\over 2}+{w^2\over 8}-{w^{\m\n}w_{\m\n}\over 4}+{\cal O}(w^3)\right]\left[(\eta^{\m\n}-w^{\m\n}+(w^2)^{\m\n}+{\cal O}(w^3)\right]\pa_{\m}J\pa_{\n}J
\label{c109}\ee
$$
=-{1\over 2}{T^{J_0}_{\m}}^{\m}-{1\over 2}w^{\m\n}T^{J_0}_{\m\n}+{1\over 2}\left((w^2)^{\m\n}-{w\over 2}w^{\m\n}\right)T^{J_0}_{\m\n}-{1\over 8}\left(w_{\m\n}w^{\m\n}-{w^2\over 2}\right){{(T^{J_0})}_{\r}}^{\r}+{\cal O}(w^3)
$$
with
\be
T^{J_0}_{\m\n}={Z}\left(\pa_{\m}J\pa_{\n}J-{\eta_{\m\n}\over 2}\eta^{\r\s}\pa_{\r}J\pa_{\s}J\right)
\label{c110}\ee
which is the ${\cal O}(w^0)$ term in the expansion of the $T^J_{\m\n}$ stress tensor in (\ref{c12}) for $M$ constant,
\be
T^J_{\m\n}=T^{J_0}_{\m\n}-{Z\over 2}w_{\m\n}(\pa J)^2+{Z\over 2}\eta_{\m\n}(w^{\r\s}\pa_{\r}J\pa_{\s}J)+{\cal O}(w^2)
\label{c110a}\ee
$$
=T^{J_0}_{\m\n}+{w_{\m\n}\over 2}{({ T}^{J_0})_{\r}}^{\r}+{\eta_{\m\n}\over 2}w^{\r\s} {({ T}^{J_0})_{\r\s}}-{\eta_{\m\n}\over 4}w~{({T}^{J_0})_{\r}}^{\r}
$$
We also have

with
\be
\sp (\pa J)^2\equiv \eta^{\m\n}\pa_{\m}J\pa_{\n}J
\label{c110b}\ee
Therefore we obtain
\be
W(w,J)=\int d^4x\sqrt{-g}\left[-V+M^2 R-{Z\over 2}(\pa J)^2+\cdots\right]
\label{c111}\ee
$$
=W_0+W_1+W_2+{\cal O}(w^3)
$$
with
\be
W_0=\int d^4x\left[-V+{1\over 2}{T^J_{\m}}^{\m}\right]
\label{c112}\ee
\be
W_1=\int d^4x\left[-{V\over 2}w+{1\over 2}w^{\m\n}T^J_{\m\n}
+w^{\m\n}\pa_{\m}\pa_{\n}M^2-w\square M^2\right]
\label{c113}\ee
\be
W_2={1\over 2}\int d^4x\left[{V\over 4}\left(2w^{\m\n}w_{\m\n}-w^2\right)-\left((w^2)^{\m\n}-{w\over 2}w^{\m\n}\right)T^J_{\m\n}+{1\over 4}\left(w_{\m\n}w^{\m\n}-{w^2\over 2}\right){{(T^{J_0})}_{\r}}^{\r}+\right]+
\label{c114}\ee
$$
+{1\over 2}\int d^4x~M^2\left[w(\pa\pa w)+(\pa w)^{\m}(\pa w)_{\m}
+{1\over 2}w^{\m\n}\square{w_{\m\n}} -{1\over 2}w\square w
\right]=
$$
$$
={1\over 2}\int d^4 x~w^{\m\n}(x)\hat Q_{\m\n ;\r\s}(x) w^{\r\s}(x)
$$
where we split the action into linear and quadratic terms in fluctuations. The operator appearing in the quadratic terms is
\be
\hat Q_{\m\n ;\r\s}(x,y)=\left({V\over 4}+{{({T}^{J_0})_{\r}}^{\r}\over 8}\right)\left(\eta_{\m\r}\eta_{\n\s}+\eta_{\m\s}\eta_{\n\r}-\eta_{\m\n}\eta_{\r\s}\right)-
\label{c137}\ee
$$
-{1\over 4}\left(\eta_{\m\r}({ T}^{J_0})_{\n\s}+\eta_{\m\s}({ T}^{J_0})_{\n\r}
+\eta_{\n\r}({ T}^{J_0})_{\m\s}+\eta_{\n\s}({ T}^{J_0})_{\m\r}\right)+{1\over 4}\left(\eta_{\m\n}({ T}^{J_0})_{\r\s}+\eta_{\r\s}({ T}^{J_0})_{\m\n}\right)+
$$
$$
+{M^2\over 2}\left(\eta_{\m\n}\pa_{\r}\pa_{\s}+\eta_{\r\s}\pa_{\m}\pa_{\n}\right)+
$$
$$
-{1\over 8}\left[\eta_{\m\r}\left(\pa_{\n}(M^2\pa_{\s})+\pa_{\s}(M^2\pa_{\n})\right)+\eta_{\m\s}\left(\pa_{\n}(M^2\pa_{\r})+\pa_{\r}(M^2\pa_{\n})\right)
\right.
$$
$$
+\left.\eta_{\n\r}\left(\pa_{\m}(M^2\pa_{\s})+\pa_{\s}(M^2\pa_{\m})\right)+\eta_{\n\s}\left(\pa_{\m}(M^2\pa_{\r})+\pa_{\r}(M^2\pa_{\m})\right)\right]+
$$
$$
+{M^2\over 4}\left(\eta_{\m\r}\eta_{\n\s}+\eta_{\m\s}\eta_{\n\r}-2\eta_{\m\n}\eta_{\r\s}\right)\square
$$
{This agrees with the computation of section \ref{linear} namely (\ref{linearama}) and proves (\ref{linearam}) in general.}

If we now wish to interpret $W(g)$ as the generating functional of the connected correlation functions of the stress tensor in the non-linear regime we can use the results of Appendix \eqref{BridgingI}. In particular we use (\ref{br3}) and then perform an expansion around flat space  as in (\ref{c94}) to obtain
\be
W=W_0+\int d^4x ~\left[ \left( -w^{\m\n}+(w^2)^{\m\n} \right)~\langle T_{\m\n}\rangle \right]+
\label{c115a}\ee
$$
+{1\over 2}\int d^4x d^4y~w^{\m\n}(x)w^{\r\s}(y)~\langle T_{\m\n}(x)T_{\r\s}(y)\rangle+{\cal O}(w^3)
$$
$$
=W_0 - \int d^4x ~w^{\m\n}\langle T_{\m\n}\rangle+{1\over 2}\int d^4 x d^4 y~Q_{\m\n ;\r\s}(x,y)w^{\m\n}(x)w^{\r\s}(y)+{\cal O}(w^3) \, .
$$
Comparing (\ref{c115a}) with (\ref{c111}) up to (\ref{c114})  we can directly identify the one-point function
\be
\langle T_{\m\n}\rangle =  {V\over 2}\eta_{\m\n} - {1\over 2}(T^{J_0})^{\m\n} - \pa_{\m}\pa_{\n}M^2 + \eta_{\m\n}\square M^2
\label{c91c}\ee
and from the quadratic part that
\be
\hat Q_{\m\n ;\r\s}=Q_{\m\n ;\r\s} \, .
\label{c140}\ee
This then translates to the following relation between the connected correlator and the operator $Q_{\m\n ;\r\s}$
\be
\langle T_{\m\n}T_{\r\s}\rangle =-{1\over 2}\left[\left(\eta_{\m\r}\langle T_{\n\s}\rangle+\eta_{\n\r}\langle T_{\m\s}\rangle+\eta_{\m\s}\langle T_{\n\r}\rangle+\eta_{\n\s}\langle T_{\m\r}\rangle\right)\right]\delta^{(4)}(x-y)+Q_{\m\n ;\r\s}
\label{c141}\ee
We thus find that the difference between them is only in their contact term structure.

Substituting in (\ref{c141}) the vev from (\ref{c91c}) we finally obtain the explicit expression for the connected correlator
\be
\langle T_{\m\n}(x)T_{\r\s}(y)\rangle/\delta^{(4)}(x-y) =-{V\over 4}\left(\eta_{\m\r}\eta_{\n\s}+\eta_{\m\s}\eta_{\n\r}+\eta_{\m\n}\eta_{\r\s}\right)+
\label{c116}\ee
$$
+{{({T}^{J_0})_{\r}}^{\r}\over 8}\left(\eta_{\m\r}\eta_{\n\s}+\eta_{\m\s}\eta_{\n\r}-\eta_{\m\n}\eta_{\r\s}\right)+{1\over 4}\left(\eta_{\m\n}({ T}^{J_0})_{\r\s}+\eta_{\r\s}({ T}^{J_0})_{\m\n}\right)+
$$
$$
+{1\over 2}\left(\eta_{\m\r}(\pa_{\n}\pa_{\s}M^2)+\eta_{\m\s}(\pa_{\n}\pa_{\r}M^2)
+\eta_{\n\r}(\pa_{\m}\pa_{\s}M^2)+\eta_{\n\s}(\pa_{\m}\pa_{\r}M^2)\right)-
$$
$$
-\left(\eta_{\m\r}\eta_{\n\s}+\eta_{\m\s}\eta_{\n\r}\right)(\square M^2)-
$$
$$
-{1\over 8}\left[\eta_{\m\r}\left(\pa_{\n}(M^2\pa_{\s})+\pa_{\s}(M^2\pa_{\n})\right)+\eta_{\m\s}\left(\pa_{\n}(M^2\pa_{\r})+\pa_{\r}(M^2\pa_{\n})\right)
\right.
$$
$$
+\left.\eta_{\n\r}\left(\pa_{\m}(M^2\pa_{\s})+\pa_{\s}(M^2\pa_{\m})\right)+\eta_{\n\s}\left(\pa_{\m}(M^2\pa_{\r})+\pa_{\r}(M^2\pa_{\m})\right)\right]+
$$
$$
+{M^2\over 2}\left(\eta_{\m\n}\pa_{\r}\pa_{\s}+\eta_{\r\s}\pa_{\m}\pa_{\n}\right)
+{M^2\over 4}\left(\eta_{\m\r}\eta_{\n\s}+\eta_{\m\s}\eta_{\n\r}-2\eta_{\m\n}\eta_{\r\s}\right)\square
$$
We observe that
\be
\pa^{\m}\langle T_{\m\n}\rangle= \pa^{\m} T^J_{\m\n} - \half V'\pa_{\n}J
\label{c142}\ee

Setting $J=constant$, ie.  $T^J_{\m\n}\to 0$
The two-point function becomes
\be
\langle T_{\m\n}(x)T_{\r\s}(y)\rangle/\delta^{(4)}(x-y)=-{V\over 4}\left(\eta_{\m\r}\eta_{\n\s}+\eta_{\m\s}\eta_{\n\r}+\eta_{\m\n}\eta_{\r\s}\right)+
\ee
$$
-{M^2\over 4}\left[\eta_{\m\r}\pa_{\n}\pa_{\s}+\eta_{\m\s}\pa_{\n}\pa_{\r}
+\eta_{\n\r}\pa_{\m}\pa_{\s}+\eta_{\n\s}\pa_{\m}\pa_{\r}\right]+
$$
$$
+{M^2\over 2}\left(\eta_{\m\n}\pa_{\r}\pa_{\s}+\eta_{\r\s}\pa_{\m}\pa_{\n}\right)
+{M^2\over 4}\left(\eta_{\m\r}\eta_{\n\s}+\eta_{\m\s}\eta_{\n\r}-2\eta_{\m\n}\eta_{\r\s}\right)\square
$$

Going to momentum space we finally obtain
\be
\langle T_{\m\n}T_{\r\s}\rangle^{V}(p)=
-{V\over 4}\left(\eta_{\m\r}\eta_{\n\s}+\eta_{\m\s}\eta_{\n\r}+\eta_{\m\n}\eta_{\r\s}\right)+
\label{c116a}\ee
$$
-{M^2\over 2}\left(\eta_{\m\n}p_{\r}p_{\s}+\eta_{\r\s}p_{\m}p_{\n}\right)+{M^2\over 4}\left(\eta_{\m\r}p_{\n}p_{\s}+\eta_{\n\r}p_{\m}p_{\s}+\eta_{\m\s}p_{\n}p_{\r}+\eta_{\n\s}p_{\m}p_{\r}\right)+
$$
$$
-{M^2\over 4}\left(\eta_{\m\r}\eta_{\n\s}+\eta_{\m\s}\eta_{\n\r}-2\eta_{\m\n}\eta_{\r\s}\right)p^2
$$
This connected correlator agrees with the Ward identity results of Appendix \ref{Ward} and in particular with \eqref{genpertudi}.

A quadratic piece can also be obtained from four-derivative terms, so we shall briefly discuss this computation as well.
In four dimensions there are two independent ones
\be
S_4=\int D^4x\sqrt{-g}\left[M_1^2 R_{\m\n}R^{\m\n}+M_2^2R^2\right]
\label{c117}\ee
We use
\be
R_{\m\n}={1\over 2}\left(\pa_{\m}(\pa w)_{\n}+\pa_{\n}(\pa w)_{\m}\right)-{1\over 2}\square w_{\m\n}-{1\over 2}\pa_{\m}\pa_{\n}w+{\cal O}(w^2)
\label{c97c}\ee
\be
R=(\pa\pa w)-\square w+{\cal O}(w^2)
\label{c118}\ee
We finally obtain
\be
\int d^4 x\sqrt{-g}R_{\m\n}R^{\m\n}={1\over 4}\int d^4x\left[2(\pa w)^{\m}\square(\pa w)_{\m}-2(\pa\pa w)\square w+2(\pa\pa w)(\pa\pa w)+\right.
\label{c119}\ee
$$
\left.
+w^{\m\n}\square^2 w_{\m\n}+w\square^2 w+{\cal O}(w^3)\right]
$$
\be
\int d^4 x\sqrt{-g}R^2=\int d^4x\left[(\pa\pa w)(\pa\pa w)+w\square^2 w-2(\pa\pa w)\square w+{\cal O}(w^3)\right]
\label{c120}\ee
\def\t{\tau}

\section{Solutions to the effective Einstein equation of a single theory\label{sol-ein}}

In this appendix we will find solutions to the emergent Einstein equation (\ref{f9}) without extra sources.

The generic equation that we need to solve in the case of no sources takes the form:
\begin{equation}
M^2\tilde{G}_{\m\n}\,=\,  \frac{V}{2} \tilde{h}_{\m\n}\, - \frac{V}{2}\,{\bf g}_{\m\n}\,,\qquad \tilde{\nabla}^\m \left( V {\bf g}_{\m\n} - V \tilde{h}_{\m\n} \right)\,=\,0\label{eq2}
\end{equation}
We also define
\be
\Lambda\equiv - {V\over M^2} \, ,
\ee
so that positive $\Lambda$ agrees with the usual convention for a positive cosmological constant.

The first equation is an Einstein equation with an explicit stress energy source due to the background metric in which the QFT is defined. The second is just a compatibility condition that follows from the Einstein equation. We mainly focus in the case $V = const.$ from now on. This  simplifies the compatibility conditions. Similarly to the canonical case of Einstein equations in vacuum, we expect the existence of cosmological and spherically symmetric vacuum solutions.

\subsection{Einstein manifolds}

We first assume that the background metric of the QFT has constant curvature
\be
R_{\m\n} ({\bf g}) ={\lambda} {\bf g}_{\m\n}\sp R({\bf g})=d{\lambda} \sp
G_{\m\n}({\bf g}) =\left(1-{d\over 2}\right){\lambda} {\bf g}_{\m\n}
\label{j2}\ee
Making the ansatz for a solution to (\ref{eq2})
\be
\tilde h_{\m\n}=\kappa {\bf g}_{\m\n}
\label{j3}\ee
we obtain
\be
\kappa=1+(d-2){\lambda\over \Lambda}
\label{j4}\ee
In particular, if the metric ${\bf g}_{\m\n}$ is flat or Schwarzschild (ie. $\lambda=0$) the metric $\tilde h_{\m\n}$ is flat or Schwarzschild.

Moreover, depending on the sign of $\l$, and the sign of the cosmological constant $\Lambda$, there is screening or anti-screening, as $\kappa$ in (\ref{j4}) is larger or smaller than $1$.

\subsection{General spherically symmetric ansatz}

We search for spherically symmetric solutions using the ansatze for the metric :
\begin{equation}
ds^2\,=\,-f(r)\,dt^2\,+\,{g(r)}{dr^2}\,+\,\gamma(r) r^2 \,d \Omega^2
\label{kj4}\end{equation}
and a flat background metric
\be
\eta_{\m\n}dx^{\m}dx^{\n}=a^2(-dt^2+dr^2+r^2d\Omega_2^2)
\label{j5}\ee

The non trivial components of the equations (\ref{eq2})  reduce to

\be
tt~~:~~~4r^2{\g''\over \g}-r^2{\g'^2\over \g^2}-2r^2{\g'\over \g}{g'\over g}+12 r{\g'\over \g}-4r{g'\over g}-4{g\over \g}+4=2r^2\Lambda g\left[{a^2\over f}-1\right]
\label{j6}\ee
\be
rr~~:~~~r^2{\g'^2\over \g^2}+2r^2{\g'\over \g}{f'\over f}+4r{\g'\over \g}+4r{f'\over f}-4{g\over \g}+4=2r^2\Lambda g\left({a^2\over g}-1\right)
\label{j7}\ee

\be
\Omega~~:~~~ 2r\left({\g''\over \g}+{f''\over f}\right)+r{\g'\over \g}\left({f'\over f}-{g'\over g}-{\g'\over \g}+4\right)-r{f'\over f}{g'\over g}-2{g'\over g}-r{f'^2\over f^2}+2{f'\over f}=2r\Lambda g\left({a^2\over \g}-1\right)
\label{j8}\ee

It is clear that by a redefinition of the radial coordinate
\be
r\to {r\over \sqrt{\Lambda}}
\label{j9}\ee
$\Lambda$ disappears from the  equations but its sign remains.
It is also clear that if we rescale
\be
f\to a^2 f\sp g\to a^2 g\sp \g\to a^2\g\sp \Lambda \to {\Lambda\over a^2}
\ee
the equations are invariant. We therefore set $a=1$.

We may solve for $g$ from (\ref{j7})
\be
g(r)={\g\over (4-2r^2\Lambda \g)}\left[r^2{\g'^2\over \g^2}+2r^2{\g'\over \g}{f'\over f}+4r{\g'\over \g}+4r{f'\over f}+4-2r^2\Lambda \right]
\ee
and substitute it in (\ref{j6}) and (\ref{j8}).
We obtain two second order equations that involve two or less derivatives of $f$ and $\g$.
They are of the form
\be
a_{11}f''+a_{12}\g''+b_1=0\sp a_{21}f''+a_{22}\g''+b_2=0
\ee
with
\be
a_{11}=8 r^2 f^2 \gamma  \left(\Lambda  r^2 \gamma -2\right) \left(r \gamma '+2 \gamma \right)^2
\ee
\be
a_{12}=8 r^3 f^2 \gamma  \left(\Lambda  r^2 \gamma -2\right) \left(-r f' \gamma '-2 \gamma
   f'+2 \Lambda  r f \gamma\right)
   \ee

   \be
   a_{21}=4 r^2 f \gamma ^3 \left(\Lambda  r^2 \gamma -2\right) \left(-r f' \gamma '-2 \gamma
   f'+2 \Lambda  r f \gamma \right)
   \ee
   \be
   a_{22}=4 r^3 f \gamma^3 \left(\Lambda  r^2 \gamma -2\right) \left(f'^2+2 \Lambda  f^2\right)
   \ee
\be
{b_1\over 2 r}= -4 \Lambda  r^3 \gamma ^2 f'^2 \left(r \gamma '+2 \gamma \right)^2+4 \Lambda  r^2
   f \gamma f' \left(r \gamma '+2 \gamma \right) \left[-r^2 \gamma '^2+2 \gamma ^2
   \left(\Lambda  r^2-2\right)-4 r \gamma  \gamma'\right]+
\ee
$$   +
   f^2 \Big[2 r^3 \gamma '^3
   \left(\Lambda  r \gamma  \left(r f'-4\right)-6 f'\right)+4 r^2 \gamma  \gamma '^2
   \left(\Lambda  r \gamma \left(-r f'+\Lambda  r^2-6\right)-10 f'\right)-
$$
$$-
   8 r \gamma ^2
   \gamma ' \left(\Lambda  r \gamma  \left(r \left(\Lambda  r^2+2\right) f'-2 \Lambda
   r^2+4\right)-2 \left(\Lambda  r^2-4\right) f'\right)+
$$
$$+\left.
   4 \gamma ^3 \left(8 \left(\Lambda
   r^2-2\right) f'-\Lambda  r \gamma  \left(4 \Lambda  r^3 f'+\left(\Lambda
   r^2-2\right)^2\right)\right)-\Lambda  r^5 \gamma '^4\right]+
$$
$$  +
   2 \Lambda  r f^3 \left[-r^4 \gamma
   '^4-8 r^3 \gamma  \gamma '^3+r^2 \gamma  \left(10-3 \gamma  \left(\Lambda
   r^2+8\right)\right) \gamma'^2+4 r \gamma ^2 \left(\gamma  \left(\Lambda
   r^2-8\right)+2\right) \gamma '+\right.
$$
  $$+ \left.
   2 \gamma ^3 \left(\gamma  \left(\Lambda  r^2-4\right)
   \left(\Lambda  r^2+2\right)-2 \Lambda  r^2+12\right)\right]
$$
\be
{b_2\over 2 rf}= -\Lambda  r f \gamma  f' \left(r \gamma '+2 \gamma \right) \left[-r^2
   (\gamma -4) \gamma '^2+2 \gamma ^2 \left(\gamma  \left(3 \Lambda  r^2-2\right)-4 \Lambda
   r^2+8\right)-4 r (\gamma -4) \gamma  \gamma '\right]+
\ee
$$   +\gamma ^2 f'^2 \left(4 r \gamma
    \left(\Lambda  r^2 \gamma -4 \Lambda  r^2+2\right) \gamma '-r^2 \left(\Lambda  r^2 \gamma
   +4 \Lambda  r^2-6\right) \gamma '^2-2 \gamma ^2 \left(\Lambda ^2 r^4 \gamma +6 \Lambda
   r^2-8\right)\right)+
$$
$$+
   \Lambda  r f^2 \left[r^3 \gamma '^4-8 \gamma ^3 \left(\Lambda  r^2
   \gamma-2 \Lambda  r^2+4\right) \gamma '-2 r \gamma ^2 \left(3 \Lambda  r^2 \gamma -2
   \Lambda  r^2+8\right) \gamma '^2-8 r^2 \gamma  \gamma '^3+\right.
$$
$$+   \left.
   4 \Lambda  r (\gamma -1) \gamma
   ^4 \left(\Lambda  r^2-4\right)\right]
   $$
   The determinant
   \be
   \det(a)=a_{11}a_{22}-a_{12}a_{21}=-64 \Lambda  r^5 f^4 \gamma ^4 \left(\Lambda  r^2 \gamma -2\right)^2 \times
\ee
$$\times
 \left(-2 r^2 \gamma
   f' \gamma '-4 r \gamma ^2 f'-r^2 f \gamma '^2+2 \Lambda  r^2 f \gamma ^2-4 r
   f \gamma  \gamma '-4 f \gamma ^2\right)
   $$
 controls the singular points of the equations.

We also compute the basic scalar  curvature invariant
 \be
R= \frac{r^2 g \gamma ^2 f'^2+r f \gamma  \left[r \gamma  f' g'-2 g \left(r
   \gamma  f''+r f' \gamma '+2 \gamma  f'\right)\right]+f^2 \left[2 r \gamma
   g' \left(r \gamma '+2 \gamma \right)+\right.} {2 r^2 f^2
   g^2 \gamma ^2}
   \ee
   $$+
   {\left.g \left(r^2 \gamma '^2-4 r \gamma  \left(r
   \gamma ''+3 \gamma '\right)-4 \gamma ^2\right)+4 g^2 \gamma \right]\over 2 r^2 f^2
   g^2 \gamma ^2}
   $$

\subsubsection{Perturbation theory\label{pert1}}

We now proceed to solve the spherically symmetric ansatz equations in perturbation theory.

The trivial solution discussed in the beginning of this section is the background metric
\be
f(r)=g(r)=\g(r)=1
\label{j10}\ee
We now search for solutions with
\be
f(r)=1+\e f_1(r)+\e^2f_2(r)+{\cal O}(\e^3)\sp g(r)=1+\e g_1(r)+\e^2 g_2(r)+{\cal O}(\e^3)
\label{j11}\ee
\be
 \g(r)=1+\e \g_1(r)+\e^2\g_2(r)+{\cal O}(\e^3)
\label{j12}\ee
The first order equations become
\be
-2 r g_1'-2 g_1+2 r^2 \gamma _1''+6 r \gamma _1'+2 \gamma _1=-r^2 \Lambda f_1
\label{j13}\ee
\be
2 r f_1'-2 g_1+2 r \gamma _1'+2 \gamma _1=-r^2 \Lambda  g_1
\label{j14}\ee
\be
r^2 f_1''+rf_1'-rg_1'+r^2 \gamma _1''+2 r\gamma _1'=-r^2 \Lambda  \gamma _1
\label{j15}\ee
or in matrix form
\be
\left(\begin{matrix}0& -2r\pa_r-2 &2r^2\pa_r^2+6r\pa_r+2\\
2r\pa_r& -2& 2r\pa_r+2\\
r^2\pa_r^2+r\pa_r& -r\pa_r & r^2\pa_r^2+2r\pa_r\end{matrix}\right)\left(\begin{matrix} f_1\\ g_1\\\g_1\end{matrix}\right)=-r^2\Lambda\left(\begin{matrix} f_1\\ g_1\\\g_1\end{matrix}\right)
\ee
Changing variables to $r=e^x$ we obtain
\be
\left(\begin{matrix}0& -2\pa_x-2 &2\pa_x^2+4\pa_x+2\\
2\pa_x& -2& 2\pa_x+2\\
\pa_x^2& -\pa_x & \pa_x^2+\pa_x\end{matrix}\right)\left(\begin{matrix} f_1\\ g_1\\\g_1\end{matrix}\right)=-e^{2x}\Lambda\left(\begin{matrix} f_1\\ g_1\\\g_1\end{matrix}\right)
\ee
Again by a shift in $x$, $\Lambda$ can be set to 1.
These equations will have four integration constants. To see this , we can solve (\ref{j14}) for $g_1$ without introducing quadratures
\be
g_1=2{\dot f_1+\dot\g_1+\g_1\over 2-e^{2x}}
\ee
where dots are derivatives with respect to $x$.

Substituting at the other two equations we obtain
\be
2(e^{2x}-2)\ddot\g_1+4(e^{2x}-4)\dot\g_1+2(e^{2x}-6)\g_1+4(1-2e^{-2x})\ddot f_1-4(1+e^{-2x})\dot f_1+(2-e^{2x})^2f_1=0
\ee
\be
(e^{2x}-2)\ddot\g_1+(e^{2x}-6)\dot \g_1+e^{2x}(e^{2x}-4)\g_1+(e^{2x}-2)\ddot f_1-4\dot f_1=0
\ee
We can solve for $f_1$ as follows
\be
f_1=-{8\over e^{2x}(e^{4x}-2e^{2x}+16)}\g_1^{(3)}-{2\over e^{2x}}\ddot \g_1-{4(e^{4x}-3e^{2x}+6)\over e^{2x}(e^{4x}-2e^{2x}+16)}\dot \g_1+{2 e^{2x}\over (e^{4x}-2e^{2x}+16)}\g_1
\ee
while $\g_1$ satisfies the following linear fourth order homogeneous equation.
\be
\g_1^{(4)}-4{(e^x-2)(e^x+2)(e^{2x}+4)\over (e^{4x}-2 e^{2x}+16)}\g_1^{(3)}-{(e^{2x}+10)\over 2}\ddot\g_1-
\ee
$$
-
{(e^{6x}-34 e^{4x}+96 e^{2x}+192)\over 2(e^{4x}-2 e^{2x}+16)}\dot\g_1-{e^{4x}(e^{4x}-4e^{2x}+48)\over 2(e^{4x}-2 e^{2x}+16)}\g_1=0
$$

The structure of the equations indicates that the solutions for large $r$ must be of the form
\be
r^b~e^{k r^a}
\label{j16}\ee
Checking with the equations, we find $b=-1$ and $a=1$  when $a>0$.
We then parametrize
\be
rf_1(r)=\left[c_1+{c_2\over r}+{c_3\over r^2}+\cdots\right]e^{kr}
\label{j17}\sp
rg_1(r)=\left[a_1+{a_2\over r}+{a_3\over r^2}+\cdots\right]e^{kr}
\ee
\be
r\g_1(r)=\left[b_1+{b_2\over r}+{b_3\over r^2}+\cdots\right]e^{kr}
\label{j19}\ee
Substituting in the linearized equations we obtain to  leading order
\be
c_1\Lambda +2b_1k^2=0\sp a_1=0\sp (b_1+c_1)k^2+b_1V=0
\label{j20}\ee
A non-trivial solution of these equations exists when
\be
2k^4-\Lambda  k^2-\Lambda^2=0~~~\to~~~k^2=  \left\{ \begin{array}{l}
\displaystyle \Lambda ,\\ \\
\displaystyle -{\Lambda \over 2}
\end{array}\right.
\label{j21}
\ee

We now choose the case $\Lambda >0$ so that $k^2=\Lambda $.
We then compute the coefficients of the expansion
\be
c_1=-2b_1\sp c_{n>1}=0\sp a_1=0\sp a_2={2 b_1\over k}\sp a_3=-{2 b_1\over k^2}\sp a_{n>3}=0
\label{j22}\ee
\be
b_2=-{b_1\over k}\sp b_3={b_1\over k^2}\sp b_{n>3}=0
\label{j23}\ee
Therefore we obtain the exact solution
\be
rf_1=-2b_1~e^{kr}\sp rg_1={2b_1\over kr}\left[1-{1\over k r}\right]~e^{kr}
\sp r\g_1=b_1\left[1-{1\over kr}+{1\over k^2 r^2}\right]~e^{kr}
\label{j24}\ee

The solution with $k\to -k$ is also a solution and we therefore obtain a two parameter family of solutions.

As mentioned above there is another linearly independent solution of the $k^2=\Lambda $ branch, that corresponds to $k\to -k$. Therefore one solution grows with distance $k>0$, and the other ($k<0)$ decreases with distance.

Finally there is another branch of solutions that correspond to $k^2=-{\Lambda \over 2}$. For $\Lambda>0$ that we assumed these correspond to sinusoidal solutions that we will now construct. We solve the equations (\ref{j13})-(\ref{j15}) and we find the following exact solutions
\be
f_1={C_1\over r}\cos(k'r)+{C_2\over r}\sin(k'r)\sp k'^2={\Lambda \over 2}>0
\ee
\be
g_1={2\over r}\left(-{C_2\over k'r}+{C_1\over (k'r)^2}\right)\cos(k'r)+{2\over r}\left({C_1\over k'r}+{C_2\over (k'r)^2}\right)\sin(k'r)
\ee
\be
\g_1={1\over r}\left(C_1+{C_2\over k'r}-{C_1\over (k'r)^2}\right)\cos(k'r)+
{1\over r}\left(C_2-{C_1\over k'r}-{C_2\over (k'r)^2}\right)\sin(k'r)
\ee

We expect that this also will have higher order corrections that are regular.

We now write the general solution to the first order equations that has four integration constants:
\be
rf_1=-2C_+~e^{kr}-2C_-~e^{-kr}+{\bar C_+}\cos(k'r)+{\bar C_-}\sin(k'r)
\ee
\be
 rg_1={2C_+\over kr}\left[1-{1\over k r}\right]~e^{kr}-{2C_-\over kr}\left[1+{1\over k r}\right]~e^{-kr}+
\ee
$$
+
2\left(-{\bar C_-\over k'r}+{\bar C_+\over (k'r)^2}\right)\cos(k'r)+{2}\left({\bar C_+\over k'r}+{\bar C_-\over (k'r)^2}\right)\sin(k'r)
$$
\be
 r\g_1=C_+\left[1-{1\over kr}+{1\over k^2 r^2}\right]~e^{kr}+C_-\left[1+{1\over kr}+{1\over k^2 r^2}\right]~e^{-kr}+
\ee
$$
+
\left(\bar C_++{\bar C_-\over k'r}-{\bar C_+\over (k'r)^2}\right)\cos(k'r)+
\left(\bar C_--{\bar C_+\over k'r}-{\bar C_-\over (k'r)^2}\right)\sin(k'r)
$$
with
\be
k=+\sqrt{\Lambda}\sp k'=+\sqrt{\Lambda\over 2}\sp \Lambda>0
\ee
or
\be
k=+\sqrt{|\Lambda|\over 2}\sp k'=+\sqrt{|\Lambda|}\sp \Lambda<0
\ee
All solutions above except the one proportional to $C_+$ vanish as $r\to\infty$ and are therefore good approximations to the solutions of the non-linear equations.
On the other hand the $C_+$ solution is a good solution only as $r\to -\infty$.

The solutions above agree with the solutions found in the linearized analysis in appendix \ref{aF1} as they should. The analysis there indicated the presence of tachyonic modes that are associated with the trigonometric asymptotics here.

\subsubsection{Second order perturbation theory}

To establish the that the first order solution we found is a regular starting point of a well defined perturbative expansion we shall also compute the next order solution here.

The second order equations become (once we insert the first order solutions in (\ref{j24}))
\be
2r^2\g_2''+6 r\g_2'+2\g_2-2rg_2'-2g_2+k^2r^2f_2=
\label{j25}\ee
$$
={k^2C_1^2e^{2kr}\over 2} \left(13+{4\over kr}-{6\over k^2r^2}-{18\over k^3 r^3}+{69\over k^4 r^4}-{90\over k^5 r^5}+{45\over k^6 r^6}\right)
$$
\be
2r\g_2'+2\g_2+2rf_2'-2g_2+k^2r^2g_2=
\label{j26}\ee
$$
={k^2C_1^2e^{2kr}\over 2}\left(3+{12\over kr}-{14\over k^2 r^2}+{2\over k^3 r^3}-{13\over k^4r^4}+{18\over k^5 r^5}-{9\over k^6 r^6}\right)
$$
\be
r^2(\g_2''+f_2'')+2r\g_2'+rf_2'-rg_2'+k^2 r^2\g_2=
\label{j27}\ee
$$
=
{k^2C_1^2e^{2kr}\over 2}\left(19-{40\over kr}+{62\over k^2 r^2}-{70\over k^3 r^3}+{71\over k^4 r^4}-{54\over k^5 r^5}+{27\over k^6 r^6}\right)
$$

The homogeneous system has all the four possible solutions the first order system had. Choosing the one we chose, the homogenous solution will correct the arbitrary constant of the first order system. Therefore the non-trivial solution is the solution of the inhomogeneous system.

By setting
\be
f_2=(kC_1)^2 e^{2kr}\tilde f_2\sp g_2=(kC_1)^2 e^{2kr}\tilde g_2
\sp \g_2=(kC_1)^2 e^{2kr}\tilde \g_2
\label{j28}\ee
the equations become
\be
2r^2\tilde\g_2''+(6+8kr)r\tilde\g_2'+2(1+6 k r+4 k^2 r^2)\tilde\g_2-2r\tilde g_2'-(2+4kr)\tilde g_2+k^2r^2\tilde f_2=
\label{j29}\ee
$$
=-{13\over 2}-{45\over 2k^6 r^6}+{45\over k^5 r^5}-{69\over 2k^4 r^4}+{9\over k^3 r^3}+{3\over k^2 r^2}-{2\over k r}
$$
\be
2r(\tilde\g_2'+\tilde f_2')+(2+4kr)\tilde\g_2+4kr \tilde f_2+(-2+k^2r^2)\tilde g_2=\phantom{AAAAAAAAAAAAAAAAA}
\label{j30}\ee
$$
\phantom{AAAAAAAAAAAAAAAA}={3\over 2}-{9\over 2k^6 r^6}+{9\over k^5 r^5}-{13\over 2k^4 r^4}+{1\over k^3 r^3}-{7\over k^2 r^2}+{6\over k r}
$$
\be
r^2(\tilde \g_2''+\tilde f_2'')+(2+4kr)r\tilde \g_2'-r\tilde g_2'+(1+4kr)r\tilde f_2'-2kr\tilde g_2+kr(4+5kr)\tilde \g_2+2kr(1+2kr)\tilde f_2=
\label{j31}\ee
$$
=
{19\over 2}+{27\over 2k^6 r^6}-{27\over k^5 r^5}+{71\over 2k^4 r^4}-{35\over k^3 r^3}+{31\over k^2 r^2}-{20\over k r}
$$
The solutions are (apparently) infinite power series in inverse powers of $kr$
\be
\tilde \g_2={11\over 18 (kr)^2}+{140\over 81(kr)^3}+{359\over 729(kr)^4}+{13471\over 2187(kr)^5}+{623989\over 39366(kr)^6}+{15158639\over 177147(kr)^7}+
\label{j32}\ee
$$
+{554687923\over 1062882 (kr)^8}+{17740912960\over 4782969 (kr)^9}+{\cal O}\left((kr)^{-10}\right)
$$
\be
\tilde g_2={3\over 2(kr)^2}-{26\over 9 (kr)^3}+{833\over 81(kr)^4}-{8911\over 729 (kr)^5}+{64841\over 4374(kr)^6}+{629843\over 19683 (kr)^7}+
\label{j33}\ee
$$
+{58660213\over 354294 (kr)^8}+{541314920\over 531441(kr)^9}+{\cal O}\left((kr)^{-10}\right)
$$
\be
\tilde f_2={29\over 18(kr)^2}-{274\over 81(kr)^3}-{1441\over 729(kr)^4}-{20915\over 2187(kr)^5}-{1547153\over 39366(kr)^6}-{35244115\over 177147(kr)^7}-
\label{j34}\ee
$$
-{1275610535\over 1062882(kr)^8}-{40267694600\over 4782969(kr)^9}+{\cal O}\left((kr)^{-10}\right)
$$
Obviously, these are asymptotic series as the coefficients grow with the order.

\subsection{Cosmological solutions}

 We assume that the background metric is flat and search for cosmological solutions of the type:
\bea
d\tilde{s}^2\,=\,\tilde{h}_{\m\n}dx^\m dx^\n\,=\,-b(t)^2\,dt^2\,+\,a(t)^2\,\left(\frac{dr^2}{F(r)^2}\,+\,r^2\,d\Omega^2\right)& \, , \nn \\
{\bf{g}}_{\m\n} = \eta_{\m \n} = \text{diag} \left( -1 , 1, r^2, r^2 \sin^2 \theta \right)& \,.
\eea
 We have used spherical coordinates for the spatial part.

The scalar curvature of the 3-dim spatial submanifold is:
\begin{equation}
R\,=\,-\frac{2 \,\left(2 \,r\, F\, F'\,+F^2-1\right)}{r^2}
\end{equation}
We will hence solve the following equations:
\begin{equation}
\tilde{G}_{\m\n}\, + \, \frac{\Lambda }{2}\left(\tilde{h}_{\m\n}\,-\,\eta_{\m\n}\right) \, = 0 \,,\qquad \tilde{\nabla}^\m \eta_{\m\n}\,=\,0\label{left2}
\end{equation}
where $\Lambda $ is a constant.

The Einstein equations ($tt,rr$ and spherical part) are:
\begin{align}
&6 r^2 \dot a(t)^2+r^2 {\Lambda } a(t)^2 \left(b(t)^2-1\right)-2 b(t)^2 F(r) \left(2 r
   F'(r)+F(r)\right)+2 b(t)^2=0\\
   &4 r^2 a(t) \dot a(t) \dot b(t)-2 r^2 b(t) \left(2 a(t) \ddot a(t)+\dot a(t)^2\right)+b(t)^3
   \left(-r^2 {\Lambda } a(t)^2+F(r)^2 \left(r^2 {\Lambda }+2\right)-2\right)=0\\
   &r \left(4 r a(t) \dot a(t) \dot b(t)-2 r b(t) \left(2 a(t) \ddot a(t)+\dot a(t)^2\right)+b(t)^3
   \left(-r {\Lambda } a(t)^2+2 F(r) F'(r)+r {\Lambda }\right)\right)=0
\end{align}
Combining the last two, we obtain a simple equation for $F(r)$:
\begin{equation}
2 r F(r) F'(r)-F(r)^2 \left(r^2 {\Lambda }+2\right)+r^2 \Lambda +2=0\label{comb2}
\end{equation}
which is solved by:
\begin{equation}
F(r)=\sqrt{1\,+\,k\,e^{\frac{{\Lambda }}{2}r^2}\,r^2}
\end{equation}
This implies that the scalar curvature of the spatial submanifold is:
\begin{equation}
R\,=\,-2 \,k\, e^{\frac{r^2 \,{\Lambda }}{2}} \left(r^2 \,{\Lambda }+3\right)
\end{equation}
and it is constant for ${\Lambda }=0$.\\
The other non trivial equations coming from the compatibility condition are:
\begin{align}
&-b(t)^3 \left(F(r)^2+2\right) \dot a(t)+3 a(t)^2 b(t) \dot a(t)-2 a(t)^3 \dot b(t)=0\\
&r F(r) F'(r)+F(r)^2-1\,=\,0 \label{comb1}
\end{align}
Combining \eqref{comb1} with \eqref{comb2} sets $k=0$ and therefore we first obtain:
\begin{equation}
F(r)\,=\,1
\end{equation}
which means the spatial submanifold is flat space.

Introducing the solution $F(r)=1$ in the other equations we obtain
\begin{align}
&\frac{3 \dot a(t)}{a(t) b(t)^2}-\frac{3 \dot a(t)}{a(t)^3}-\frac{2 \dot b(t)}{b(t)^3}=0\\
&\frac{3 \dot a(t)^2}{a(t)^2}+\frac{1}{2} {\Lambda } \left(b(t)^2-1\right)=0\\
&4 a(t) \dot a(t) \dot b(t)-2 b(t) \left(2 a(t) \dot a(t)+\dot a(t)^2\right)+b(t)^3 {\Lambda }\left(1-
   a(t)^2\right)=0
\end{align}
Solving the first we obtain:
\begin{equation}
b(t)\,=\,\pm\,\frac{a(t)^{3/2}}{\sqrt{3 a(t)+c}}
\end{equation}
and substituting  we obtain the final equation for $a(t)$:
\begin{equation}
\boxed{\left(\frac{a'(t)}{a(t)}\right)^2\,=\,\frac{{\Lambda }}{6}\left(1-\frac{a(t)^3}{3 a(t)+c}\right)}\label{cosmo}
\end{equation}

The simplest solution for $c=0$ takes the form:
\begin{align}
&F(r)=1\\
&b(t)\,=\,\frac{a(t)}{\sqrt{3}}\\
&a(t)\,=\,\pm\,i \sqrt{3}\, \text{csch}\left(\sqrt{{\Lambda\over 6 }}~t
\right)\label{solu}
\end{align}
There is no problem with $a(t)$ being imaginary since only $a^2, b^2$ appears in the physical metric. However the metric has the inverse signature compared with the background metric.

We  now simplify and analyse this solution.
We can shift and rescale $t$ so as to bring the solution into the form:
\begin{equation}
d\tilde{s}^2\,=\,\,a(\tilde{t})^2\,\left(d \tilde{t}^2 \, - \, ds^2_{flat}\right) \, , \quad \tilde{t} \in (- \infty, \infty)
\end{equation}
with:
\begin{equation}
a(\tilde{t})^2\,=\,\mathcal{C}\,\text{csch}^2\left(\tilde{t} \sqrt{\Lambda /2}\right)
\end{equation}
where $\mathcal{C} = 1/3$ is just an overall constant governing the size of the universe. Now we use proper time coordinates $a(\tilde{t})d\tilde{t}=dT$ and obtain:
\be
d\tilde{s}^2\,= \, \mathcal{C}\left( \,dT^2\, - \,\sinh^2 \left(T\,\sqrt{{\Lambda }/2}\right)\,ds^2_{flat} \right)\, , \quad  {\Lambda }>0 \, , \quad T \in (-\infty, 0] \cup [0, \infty)
\ee
We can  therefore have  for ${\Lambda }>0$ a non-compact big bang or crunch universe that is exponentially expanding/contracting driven asymptotically from the positive CC. The point where the universe has zero size is $T=0$.

For $\Lambda  < 0$ there exist two options. We can either analytically continue this solution with $\Lambda  \rightarrow - \Lambda $ so that we find a Euclidean solution (this now has a different signature than ${\bf g}_{\m \n}$)
\be
d\tilde{s}_{E}^2\,= \, \mathcal{C}\left( \,dT^2\, +  \,\sin^2 \left(T\,\sqrt{{|\Lambda |}/2}\right)\,ds^2_{flat} \right)\, , \quad  {\Lambda }<0 \, , \quad T \in (-\pi, 0]  \, .
\ee
This looks like a series of compact in Euclidean time analogues of a bang-crunch geometry.
Or we can instead search directly for a Lorentzian solution that is found to be
\be
d\tilde{s}^2\,= \, \mathcal{C}\left( \,-dT^2\,+\,\cos \left(T\,\sqrt{{|\Lambda |}/2}\right)^2\,ds^2_{flat} \right)\, , \quad  {\Lambda }<0 \, , \quad T \in (-\pi/2, \pi/2)
\ee
This corresponds to a bang-crunch universe. It is also related to a Euclidean solution of the $\Lambda  > 0$ branch upon $\Lambda  \rightarrow - \Lambda $.

Another generic lesson we can learn is that the topology of the background metric descends to the topology of the emergent metric (for example if one wants dS-like metrics with spherical topology slices the background metric should also have such a topology of $\mathbb{R}\times S^3$). For example if we now use for the background metric
\begin{equation}
ds^2 = - dt^2 + d \Omega_3^2 \, ,
\end{equation}
we find that there exist solutions similar to those found before related by a simple replacement $ds_{flat}^2 \rightarrow d \Omega_3^2$ and ${\Lambda } \rightarrow {\Lambda }+2$ (this is due to the fact that the positive curvature of the $S^3$ slices contributes positively to the effective cosmological constant). In particular the simplest solution now takes the form
\be
ds^2\,=\mathcal{C}\left( \,dT^2\, - \,\sinh^2\left(T\,\sqrt{({\Lambda }+2)/2}\right)^2\,d \Omega_3^2 \right)\, , \quad  {\Lambda }> 0
\ee
and this represents again an exponentially expanding/contracting universe that now has compact $S^3$-slices.

\subsection{Conformally flat Cosmology}

One can find time dependent conformally flat cosmological solutions using the ansatz
\begin{equation}
ds^2\,=\, A(t) \left( -\,dt^2\,+\,dr^2\,+\, r^2 \,d \Omega^2 \right)
\end{equation}
and
\be
\eta_{\m\n}dx^{\m}dx^{\n}=-dt^2+dr^2+r^2d\Omega_2^2
\ee
The most general solution depends on a single integration constant
\be
A(t) = \sec^2 \left(\sqrt{\frac{\Lambda}{6}} t + C_1 \right)
\ee
The case $C_1=0$ is the simplest case, which one can transform to proper time to obtain
\be
ds^2 = \, - d T^2 \, + \cosh^2 \left(\sqrt{\frac{\Lambda}{6}} T  \right) \left(dr^2 \, + r^2 d \Omega_2^2 \right)
\ee
This is a universe that bounces, having a finite minimum size at the bounce.

\section{Inverting a general stress energy two-point function\label{inv2}}

In this appendix we provide calculational details in inverting the two-point function of the stress-tensor.

The general stress energy two-point function  in momentum space is given by the following formula (see also appendix \ref{spectral}),
\be
\label{h1}
G^{\mu\nu\rho\sigma}(k) = {a_2\over 2} \eta^{\mu\nu} \eta^{\rho\sigma} +{a_1\over 2}\big( \eta^{\mu\rho} \eta^{\nu\sigma} + \eta^{\mu\sigma} \eta^{\rho\nu} \big) + b(k^2) \Pi^{\mu\nu\rho\sigma}(k) +  c(k^2) \pi^{\mu\nu}(k) \pi^{\rho\sigma}(k)
~.
\ee
where
\be
b(k)={\hat b(k)}\sp c(k)={\hat c(k)}
\label{hh1a}\ee
to agree with our notation in \eqref{linearaj}. The functions $b(k)$ and $c(k)$ have mass dimension four.
 There are two independent transverse tensor structures defined in a standard fashion as
\be
\label{hh2}
\pi^{\mu\nu} = \eta^{\mu\nu} - {k^\mu k^\nu\over k^2}\sp {\pi_{\m}}^{\r}\pi_{\r\nu}=\pi_{\m\n}\sp {\pi^{\m}}_{\m}=3
~,
\ee
\be
\label{hh3}
\Pi^{\mu\nu\rho\sigma}(k) = \pi^{\mu\rho}(k) \pi^{\nu\sigma}(k) + \pi^{\mu\sigma}(k) \pi^{\nu\rho}(k)
~.
\ee
Following the analysis of appendix \ref{spectral} we parametrize  the low momentum expansions of $b,c$ as
\be
b(k)=\mu_0+{\mu}_1 k^2 + \mu_2 k^4 + \mathcal{O}(k^4\log k^2) \sp c(k)=\nu_0+\nu_1k^2 + \nu_2 k^4 + \mathcal{O}(k^4\log k^2)
\ee
Compatibility of (\ref{h1}) with (\ref{c116a}) or \eqref{c137} that are obtained from the low-energy diff-invariant effective action implies that
if the low momentum expansions of $b,c$ are
\be\label{ex0}
b(k)=\mu_0+{\mu}_1 k^2 + \mu_2 k^4 + \mathcal{O}(k^6) \sp c(k)=\nu_0+\nu_1k^2 + \nu_2 k^4 + \mathcal{O}(k^6) \, ,
\ee
then we must have
\be
\mu_0=\nu_0=0 \sp 2\mu_1=-{M^2\over 2}= -\nu_1 \, .
\label{ex1}\ee
Therefore
\be
2b+c={\cal O}(k^4)
\label{ex2}\ee
proving the relation (\ref{linearam}) in the main text. These results are then consistent with the Ward identities of Appendix \ref{Ward} and  (\ref{c116a})  for the connected correlator if we also set
\be
a_1=a_2=a \sp a = -\frac{V}{2}
\ee
In case we wish to describe the tensor $Q^{\m \n \r \s}$ introduced in \eqref{c137} and used in the main text, we need to set instead
\be
a_2 = - a_1= a \sp a = -\frac{V}{2} \, .
\ee
We can also decompose (\ref{h1}) into trace and traceless parts
\be
\label{h4}
G^{\mu\nu\rho\sigma}(k) = {a_2\over 2} \eta^{\mu\nu} \eta^{\rho\sigma} +{a_1\over 2}\big( \eta^{\mu\rho} \eta^{\nu\sigma} + \eta^{\mu\sigma} \eta^{\rho\nu} \big)  +  B_2 \left[\Pi^{\mu\nu\rho\sigma}-{2\over 3}\pi^{\mu\nu} \pi^{\rho\sigma}\right] +  {  B_0\over 3} \pi^{\mu\nu}\pi^{\rho\sigma}~.
\ee
where we have defined the spin-2 and spin-0 parts and
\be
 B_2\equiv b\sp   B_0\equiv 2b+3c
\label{h5}\ee

By using the spectral representation in appendix \ref{spectral} and in particular equations \eqref{nn20} and \eqref{bcg}, we may write in $d=4$
\be
\hat{B}_2(k) \,= \, {3\pi^2k^4\over 80}\int^{\infty}_0d\mu^2 {\rho_{2}(\mu^2)\over k^2+\mu^2}
\label{gg27}\ee
\be
{\hat{B}_0(k)\over 3} \, = \, {\pi^2k^4\over 40}\int^{\infty}_0d\mu^2 {\rho_{0}(\mu^2)\over k^2+\mu^2} \, .
\label{gg28}\ee
In this expression $\rho_{2,0}$ are the positive definite spectral weights of the spin-0 and spin-2 part of the correlator. Unitarity would imply that both of them are positive as integrals of positive definite quantities (as long as $k^2$ is positive). If $k^2$ is negative, the integral obtains most of its contribution from the pole $\mu^2 = |k^2|$ which still has a positive residue. Nevertheless these integrals are typically divergent and in a renormalised theory one needs to perform certain subtractions, see appendix~\ref{momissues}.

From (\ref{ex1}) we obtain
\be
 B_2=-{M^2\over 4}k^2 +{\cal O}(k^4)\sp  B_0=M^2 k^2+{\cal O}(k^4)
\label{ex3}\ee
We now parameterize the inverse, $P^{\m\n\r\s}$, as
\be
P^{\m\n\r\s}=A_{1}\eta^{\m\n}\eta^{\r\s}+A_2(\eta^{\m\r}\eta^{\n\s}+\eta^{\m\s}\eta^{\n\r})+
B_1(\eta^{\m\n}k^{\r}k^{\s}+\eta^{\r\s}k^{\m}k^{\n})+
\label{h6}\ee
$$
+B_2(\eta^{\m\r}k^{\n}k^{\s}+\eta^{\m\s}k^{\n}k^{\r}+\eta^{\n\r}k^{\m}k^{\s}+\eta^{\n\s}k^{\m}k^{\r})+C_1k^{\m}k^{\n}k^{\r}k^{\s}
$$
which is the most general parametrization compatible with symmetry in ($\m\leftrightarrow \n$) and ($\r\leftrightarrow \s$)as well as ($\m\n\leftrightarrow \r\s$).
The inverse is defined by
\be
G^{\m\n\r\s}P_{\r\s\e\t}={1\over 2}({\delta^{\m}}_{\e}{\delta^{\n}}_{\t}+{\delta^{\m}}_{\t}{\delta^{\n}}_{\e})
\label{h7}\ee

From (\ref{h7}) we find
\be
A_1={2a_1+a_2\over 3\bar B_0}-{1\over 3\bar B_2}
\sp
A_2={1\over 2\bar B_2}
\ee
\be
k^2 B_1=-{2(a_1+2a_2)\over 3\bar B_0}+{1\over 3\bar B_2}
\sp
k^2 B_2={1\over 2a_1}-{1\over 2\bar B_2}
\ee
\be
k^4 C_1=-{2(a_1+a_2)\over a_1(2a_1+a_2)}+{2\over 3\bar B_2}+{4(a_1+2 a_2)^2\over 3(2a_1+a_2)\bar B_0}
\ee
with
\be
\bar B_0\equiv (2a_1+a_2)  B_0+2a_1(a_1+2a_2)
\sp
\bar B_2\equiv 2 B_2+a_1
\label{bbar}\ee

 We now take directly the low-momentum expansions of these expressions for generic parameters $a_1$ and $a_2$ describing an arbitrary contact term structure. The low momentum expansion of these expressions using \eqref{ex3} is (up to $k^2$)
\bea
A_1 &=& - \frac{a_2}{2 a_1^2 + 4 a_1 a_2} - \frac{M^2(2a_1^2 + 4 a_1 a_2 + 3 a_2^2)}{4 a^2_1 (a_1 + 2 a_2)^2} ~k^2
\sp A_2 =  \frac{1}{2a_1} + \frac{M^2 }{4 a^2_1}k^2   \, , \nn \\
B_1 &=& \frac{M^2 (a_1 + a_2) }{2 a_1^2(a_1+ 2 a_2)} \sp B_2= - \frac{M^2}{4a_1^2}  \, ,
\label{lowmom}
\eea
with the expansion for $C_1$ irrelevant since it affects a term $O(k^4)$. This then means that the IR expansion of the inverted operator is
\be
P^{\m\n\r\s}= - \left(\frac{a_2}{2 a_1^2 + 4 a_1 a_2} + \frac{M^2 k^2(2a_1^2 + 4 a_1 a_2 + 3 a_2^2)}{4 a^2_1 (a_1 + 2 a_2)^2}  \right) \eta^{\m\n}\eta^{\r\s} +
\label{h6b}\ee
$$
+ \left( \frac{1}{2a_1} + \frac{M^2 k^2}{4 a^2_1} \right) (\eta^{\m\r}\eta^{\n\s}+\eta^{\m\s}\eta^{\n\r})+
$$
$$
+\frac{M^2 (a_1 + a_2) }{2 a_1^2(a_1+ 2 a_2)} (\eta^{\m\n}k^{\r}k^{\s}+\eta^{\r\s}k^{\m}k^{\n}) - \frac{M^2}{4a_1^2} (\eta^{\m\r}k^{\n}k^{\s}+\eta^{\m\s}k^{\n}k^{\r}+\eta^{\n\r}k^{\m}k^{\s}+\eta^{\n\s}k^{\m}k^{\r})
$$
If we wish this operator to describe the quadratic part of linearised  Einstein gravity (with no redefinition of the emergent metric) the kinetic terms should be compared with \eqref{linearas}, \eqref{b2}. This then gives the following constraints on the coefficients:
\bea
\frac{M^2}{a_1^2 \lambda^2} = \frac{1}{32 \pi G} \, , \quad (a_1 + a_2) = a_1 + 2 a_2 \,
\eea
which has no solution. This means that we inevitably have to redefine what we call the emergent metric using the trace. We assume that the operator \eqref{h6}, \eqref{h6b} acts on a metric labelled by $h_{\m \n}$ with which we define the original effective action \eqref{linearag} $S_{eff.}(h)$. By redefining
\be
h_{\m \n} = w_{\m \n} - \half C w \eta_{\m \n} \sp h = w(1-2 C)
 \ee
 we find how the various terms change
\bea\label{h6b1}
h h = (1- 2 C)^2 w^2 \, , \quad (k h)_\n (k h)^\n = (k w)_\n (k w)^\n + \frac{C^2}{4} w^2 k^2 - C w (kkw) \nn \\
 h_{\m \n} h^{\m \n} = w_{\m \n} w^{\m \n} + (C^2 - C) w^2 \, , \quad h(kk h) = (1- 2 C) w (kkw) - \frac{C(1-2C)}{2} k^2 w^2 \, ,  \nn \\
\eea
This then means that the new operator acting on the $w_{\m \n}$ metric reads
\be
P^{\m\n\r\s}_{(w)}= - \left(\frac{a_2 (1-2C)^2}{2 a_1^2 + 4 a_1 a_2} -  \frac{C^2 - C}{a_1}  \right)  \eta^{\m\n}\eta^{\r\s} -
\label{h6bb}\ee
$$
 - \frac{M^2 k^2 }{4 a_1^2} \left( \frac{(1-2C)^2 (2a_1^2 + 4 a_1 a_2 + 3 a_2^2)}{ (a_1 + 2 a_2)^2} + C(1-C) + \frac{2 C(1-2C)  (a_1 + a_2) }{(a_1+ 2 a_2)}  \right) \eta^{\m\n}\eta^{\r\s}   +
$$
$$
+ \left( \frac{1}{2a_1} + \frac{M^2 k^2}{4 a^2_1} \right) (\eta^{\m\r}\eta^{\n\s}+\eta^{\m\s}\eta^{\n\r})  - \frac{M^2}{4a_1^2} (\eta^{\m\r}k^{\n}k^{\s}+\eta^{\m\s}k^{\n}k^{\r}+\eta^{\n\r}k^{\m}k^{\s}+\eta^{\n\s}k^{\m}k^{\r}) +
$$
$$
\frac{M^2 }{2 a_1^2} \left( \frac{(1-2C) (a_1 + a_2) }{(a_1+ 2 a_2)} + C \right) (\eta^{\m\n}k^{\r}k^{\s}+\eta^{\r\s}k^{\m}k^{\n}) \, ,
$$
which can be further massaged into
\be
P^{\m\n\r\s}_{(w)}= - \left(\frac{a_2 + 2 a_1 C(1-C)}{2 a_1^2 + 4 a_1 a_2}  \right)  \eta^{\m\n}\eta^{\r\s} -
\label{h6bbb}\ee
$$
 - \frac{M^2 k^2 }{4 a_1^2} \left( \frac{a_1^2(2 - 4 C + 3 C^2) + 2 a_1 a_2 (2-C) + 3 a_2^2}{ (a_1 + 2 a_2)^2} \right) \eta^{\m\n}\eta^{\r\s}   +
$$
$$
+ \left( \frac{1}{2a_1} + \frac{M^2 k^2}{4 a^2_1} \right) (\eta^{\m\r}\eta^{\n\s}+\eta^{\m\s}\eta^{\n\r})  - \frac{M^2}{4a_1^2} (\eta^{\m\r}k^{\n}k^{\s}+\eta^{\m\s}k^{\n}k^{\r}+\eta^{\n\r}k^{\m}k^{\s}+\eta^{\n\s}k^{\m}k^{\r}) +
$$
$$
\frac{M^2 }{2 a_1^2} \left( \frac{(1-C) a_1 + a_2 }{(a_1+ 2 a_2)}  \right) (\eta^{\m\n}k^{\r}k^{\s}+\eta^{\r\s}k^{\m}k^{\n})
$$
We will now first fix the kinetic terms in order to obtain Einstein gravity and then see how much freedom is left for the contact terms. The conditions stemming from the kinetic terms are:
\bea
\frac{M^2}{a_1^2} = \frac{\lambda^2}{32 \pi G} \, , \quad  \frac{a_1^2(2 - 4 C + 3 C^2) + 2 a_1 a_2 (2-C) + 3 a_2^2}{ (a_1 + 2 a_2)^2} = 2 \, , \quad \frac{(1-C) a_1 + a_2 }{(a_1+ 2 a_2)} =1 \, \nn \\
\eea
The last two give a remarkably simple relation
\be
C = - {a_2 \over  a_1}
 \ee
 that fixes the constant $C$. Now,  all the kinetic terms agree with the expansion of the Einstein-Hilbert action. The structure of the constant contact terms becomes
\be
P^{\m\n\r\s}_{c.t's}(w)= + \frac{a_2}{2 a_1^2} \eta^{\m\n}\eta^{\r\s} + \frac{1}{2 a_1} (\eta^{\m\r}\eta^{\n\s}+\eta^{\m\s}\eta^{\n\r})
\ee

All this was achieved without the use of the relation between $a_2$ and $a_1$ from the Ward-identities.
These give a specific relation between them depending the operator one wishes to invert. In particular for the operator $Q_{\m \n \r \s}$ that we invert to obtain the IR expansion of the effective action, $a_1 = - a_2$ so that $C=1$. In this case the effective action of section \ref{linear} describes linearised Einstein gravity with a cosmological constant and is expected to be non-linearly completed according to the discussion of section \ref{secNL}. It also shows that Einstein gravity is \emph{universal} in the sense that its derivation in the IR was solely based on the use of the Ward identities and an infrared expansion for the general correlator.

Instead of the previous IR expansion, we could also expand \eqref{h6} near the poles of \eqref{h1}. Such states would correspond to massive gravitons. Since this is not an IR expansion for small $k^2$, it is most conveniently and clearly described in terms of the induced stress energy interaction coming from  (\ref{linearafa}) using the specific expression for the propagator \eqref{h1}. This is what we analyse in the main text in subsection \ref{InducedTTinteraction}.

\section{The spectral representation of the stress-tensor two-point function\label{spectral}}

In this Appendix we provide the general formula giving the spectral representation of the connected stress energy tensor two-point function. We will need two spectral densities, $\rho_{0}(\mu^2)$ and $\rho_{2}(\mu^2)$ for the spin-$0$ and spin-$2$ part of the correlator. In case the one-point function of the stress tensor is non-zero, we need three extra terms as shown in equation (\ref{genpertudi}). The total connected correlator in the presence of such a vev
\be
\langle T_{\mu \nu}(x) \rangle = a \delta_{\mu \nu}
 \ee
 acquires the contact terms
\be\label{K1}
\langle T_{\m \n}(x) T_{\rho \sigma}(0) \rangle_{\text{total}} = - { a \over 2} \left( \delta_{\m \n} \delta_{\rho \sigma} + \delta_{\m \rho} \delta_{\nu \sigma} + \delta_{\m \sigma} \delta_{\nu \rho} \right) \delta^d(x) +  \langle T_{\m \n}(x) T_{\rho \sigma}(0) \rangle \, .
\ee
The last piece admits a non trivial spectral decomposition. One finds (for a Euclidean metric)
\be\label{K2}
\langle T_{\m \n}(x) T_{\rho \sigma}(0) \rangle = {\cal A}_d \int_0^\infty d \mu^2 \rho_{0}(\mu^2) \Pi^{(0)}_{\mu \nu , \rho \sigma}(\partial) G(x,\mu^2)	 + {\cal A}_d \int_0^\infty d \mu^2 \rho_{2}(\mu^2) \Pi^{(2)}_{\mu \nu , \rho \sigma}(\partial) G(x,\mu^2)	
\ee
where the tensors are
\be\label{K3}
 \Pi^{(0)}_{\mu \nu , \rho \sigma}(\partial) = \frac{1}{\Gamma(d)} S_{\mu \nu} S_{\rho \sigma} \, , \quad  \Pi^{(2)}_{\mu \nu , \rho \sigma}(\partial) = \frac{1}{\Gamma(d-1)} \left(\frac{d-1}{2} S_{\mu (\rho} S_{\nu \sigma)} - S_{\mu \nu} S_{\rho \sigma} \right)
\ee
with $S_{\m \n} = \partial_\mu \partial_\nu - \delta_{\mu \nu} \partial^2$, and the scalar propagator is
\be\label{K4}
G(x,\mu^2) \, = \, \int \frac{d^d p}{(2 \pi)^d} \frac{e^{i p x}}{p^2 + \mu^2} \, = \, \frac{1}{2\pi} \left( \frac{\mu}{ 2 \pi |x|} \right)^{(d-2)/2} K_{(d-2)/2} \left(\mu |x| \right)
\ee
We also have
\be\label{K5}
\int_0^{\infty}d\mu^2 ~\mu^{d-4}~G(x,\mu^2)={2^{d-4}\Gamma\left({d\over 2}-1\right)\Gamma(d-2)\over \pi^{d\over 2}|x|^{2d-4}}
\ee
and
\be\label{K6}
S_{\m\n}{1\over |x|^{2d-4}}=4(d-1)(d-2){\left(x^{\m}x^{\n}-{1\over 2}\delta^{\m\n}|x|^2\right)\over |x|^{2d-2}}
\ee
This implies that in a CFT $\rho_{0}=0$ and $\rho_{2}\sim \mu^{d-4}$.
The multiplicative constant is
\be\label{K7}
{\cal A}_d = \frac{V}{(d+1) 2^{d-1}} \, , \qquad V = \frac{2 \pi^{d/2}}{\Gamma(d/2)}
\ee

By taking two possible traces on the energy momentum tensor correlator we can find the following two relations
\be\label{tr1}
\langle T_\mu^\mu(x) T_\nu^\nu(0) \rangle \, = \, \frac{{\cal A}_d}{\Gamma(d)} \int_0^\infty d \mu^2 \rho_{0}(\mu^2) \mu^4 G(x,\mu^2)
\ee
and
\be\label{tr2}
\langle T_{\mu \nu}(x) T^{\mu \nu}(0) \rangle \, - \, \frac{1}{d-1} \langle T_\mu^\mu(x) T_\nu^\nu(0) \rangle  \, = \, \frac{(d-2)V}{2^d \Gamma(d)} \int_0^\infty d \mu^2 \rho_{2}(\mu^2) \mu^4 G(x,\mu^2)
\ee
This spectral representation applies to renormalised field theory, where the spectral densities are matrix elements in the physical Hilbert space. In particular, the integrals over $\mu^2$ should be well defined.

Using equations \eqref{K2} and \eqref{K4}, we can also describe the spectral representation in momentum space. We have
\be\label{KK2}
\langle T_{\m \n}(p) T_{\rho \sigma}(-p) \rangle =  {\cal A}_d\, \Pi^{(0)}_{\mu \nu , \rho \sigma}(p) \,\bar {G}_{0}(p) \,	+ {\cal A}_d\, \Pi^{(2)}_{\mu \nu , \rho \sigma}(p) \, \bar {G}_{2}(p)  \, ,	
\ee
where the momentum dependent coefficients are expressed in terms of the spectral weights as
\be
\bar G_i(k) = \int_0^\infty d \mu^2 \frac{\rho_i(\mu^2 )}{k^2 + \mu^2}\sp i=0,2
\label{nn99}\ee
For more details on the momentum space decomposition, see also appendix~\ref{inv2} and in particular equations \eqref{h4} and \eqref{gg27}, \eqref{gg28}. Since these expressions are not always convergent, one has to perform appropriate subtractions as described in the next sub-appendix~\ref{momissues}.

We can also invert such relations via dispersion relations, so that we can express the spectral measures in terms of the momentum space correlators via
\be\label{disp1}
\rho_{0}(\mu^2) \, = \, \frac{\Gamma(d)}{\pi {\cal A}_d} \frac{1}{\mu^4} \Im \langle T_\mu^\mu(p) T_\nu^\nu(-p) \rangle_{p^2 = - \mu^2}
\ee
and
\be\label{disp2}
\rho_{2}(\mu^2) \, = \, \frac{2^d \Gamma(d)}{\pi V (d-2)} \frac{1}{\mu^4} \Im \left(\langle T_{\mu \nu}(p) T^{\mu \nu}(-p) \rangle   - \frac{1}{d-1}\langle T_\mu^\mu(p) T_\nu^\nu(-p) \rangle \right)_{p^2 = -\mu^2} \, .
\ee
We notice that only the imaginary parts of the momentum space correlators appear in (\ref{disp1}) and (\ref{disp2}), which are automatically physical and do not contain ambiguous contact-term divergences. In other words, by the knowledge of the spectral measures, we cannot reconstruct the full momentum space correlators, since we cannot interchange the $p$ and $\mu^2$ integrals. Expressing a momentum space correlator via the spectral representation, the integral over the spectral factor will generically exhibit divergences. These divergences will then have to be subtracted appropriately as described in the next subsection \ref{momissues}.

\subsection{Renormalization  in momentum space}\label{momissues}

As we discussed, there can exist various issues when trying to formulate a spectral representation of correlators in momentum space and the integral over the spectral factor will generically exhibit divergences. Since we can reduce the tensor structure using various operators, we need to study only the scalar parts of this representation.
In momentum space we have
\be
\langle T_{\m\n} T_{\r\s}\rangle (k)={(d-1)^2 {\cal A}_d\over 2\Gamma(d)}~k^4~\left[\pi_{\m\r}\pi_{\n\s}+\pi_{\m\s}\pi_{\m\r}-{2\over d-1}\pi_{\m\n}\pi_{\r\s}\right]\bar G_2+{{\cal A}_d\over \Gamma(d)}k^4~\pi_{\m\n}\pi_{\r\s}~\bar G_0
\label{nn20}\ee
$$
\equiv   B_2(k) \left[\pi_{\m\r}\pi_{\n\s}+\pi_{\m\s}\pi_{\m\r}-{2\over d-1}\pi_{\m\n}\pi_{\r\s}\right]+{ B_0(k)\over 3}\pi_{\m\n}\pi_{\r\s}
$$
with $\bar G_i(k)$ given in (\ref{nn99}) and
\be
\pi_{\mu\nu} = \eta_{\mu\nu} - \frac{k_\mu k_\nu}{k^2}\;,
\ee
as in (\ref{linearak}).

For $d=4$ we have also used the parametrization in (\ref{h4}) in terms of the spectral functions $ B_{2,0}$. They are related to the rest as
{\be
 B_0={\pi^2\over 40}k^4~\bar G_0(k)\sp  B_2={3\pi^2\over 80}k^4\bar G_2
\label{bcg}\ee}

Typically,  the integral over $\mu^2$ in (\ref{nn99}) does not converge either at zero or infinity.
We can rearrange the integral so that we can separate the UV and IR divergences by using the identity
\be
{\rho_i(\m^2)\over k^2+\mu^2}={\rho_i(\mu^2)\over \mu^2+m_{IR}^2}-(k^2-m_{IR}^2){\rho_i(\mu^2)\over (\mu^2+m_{IR}^2)(k^2+\mu^2)}
\label{nn5}\ee
and rewrite
\be
\bar G_i(k)=A_i-(k^2-m_{IR}^2)\int_0^{\infty}{d\mu^2\over (\mu^2+m_{IR}^2)}{\rho_i(\mu^2)\over (k^2+\mu^2)}
\label{nn6}\ee
with
\be
A_i\equiv \int_0^{\infty}d\mu^2{\rho_i(\mu^2)\over \mu^2+m_{IR}^2}
\label{nn7}\ee
$m_{IR}$ acts as an IR cutoff and is needed if the theory in question is massless\footnote{Convergence in the IR assumes that $\lim_{\mu\to 0} \mu^2\rho_i(\mu^2)=0$.}. This happens if the IR CFT is non-empty.
On the other hand, all UV divergences are now hidden in $A_i$.
We may introduce a UV cutoff $\Lambda$ and define
\be
A_i^c(\Lambda,m_{IR})\equiv  \int_0^{\Lambda^2}d\mu^2{\rho_i(\mu^2)\over \mu^2+m_{IR}^2}
\label{nn8}\ee
so that the cutoff spectral functions are
\be
\bar G^c_i(k)=A^c_i-(k^2-m_{IR}^2)\int_0^{\infty}{d\mu^2\over (\mu^2+m_{IR}^2)}{\rho_i(\mu^2)\over (k^2+\mu^2)}
\label{nn6a}\ee

As $\Lambda\to\infty$, we have a finite number of divergent terms, starting with a single logarithm in $d=4$,
\be
A_i^c\simeq c_i^{UV}\Lambda^{d-4}+d_i^{UV}\Lambda^{d-6}+\cdots +e_i^{UV}\log \Lambda^2+\cdots\sp d\geq 4\sp d=even
\label{nn13}\ee
or
\be
A_i^c\simeq c_i^{UV}\Lambda^{d-4}+d_i^{UV}\Lambda^{d-6}+\cdots +e_i^{UV}\Lambda+\cdots\sp d>4\sp d=odd
\label{nn144}\ee
 We then define the renormalized $A_i$ by subtracting the divergences and eventually a finite piece, and then taking the UV cutoff to infinity.
\be
A_i^{ren}(m_{IR})=\lim_{\Lambda\to\infty}(A_i^c-{\rm UV ~~divergences})
\label{nn14}\ee
$A_i^{ren}(m_{IR})$ is now a finite contact term that still depends in general on $m_{IR}$, if the IR theory is a non-trivial CFT. It is important to mention that the UV divergences do not depend on $m_{IR}$, and therefore the  subtracted piece does not depend on $m_{IR}$. This will guarantee that the final renormalized density is $m_{IR}$ -independent.

Finally the renormalized $\bar G_i$ is given by
\be
\bar G_i^{ren}\equiv A^{ren}_i(m_{IR})-(k^2-m_{IR}^2)\int_0^{\infty}{d\mu^2\over (\mu^2+m_{IR}^2)}{\rho_i(\mu^2)\over (k^2+\mu^2)}
\label{nn66}\ee
and is independent of $m_{IR}$.

For a CFT$_4$ we have $\rho_i(\mu^2)=c_i$ and we obtain
\be
A^c_i=c_i\log{\Lambda^2+m_{IR}^2\over m^2_{IR}}
\label{nn9}\ee
This can be renormalized by subtracting the leading UV divergence
\be
A_i^{ren}\equiv \lim_{\Lambda\to\infty} \left(A_i^c-c_i\log{\Lambda^2\over M^2}\right)=c_i\log{M^2\over m_{IR}^2}
\label{nn10}\ee
The scheme dependence is associated with the value of $M$.
The renormalized $\bar G$ for a CFT$_4$ is then
\be
\bar G_i^{ren}=A_i^{ren}-(k^2-m_{IR}^2)\int_0^{\infty}{d\mu^2\over (\mu^2+m_{IR}^2)}{\rho_i(\mu^2)\over (k^2+\mu^2)}=
\label{nn11}\ee
$$
=c_i\left[\log{M^2\over m_{IR}^2}-(k^2-m_{IR}^2)\int_0^{\infty}{d\mu^2\over (\mu^2+m_{IR}^2) (k^2+\mu^2)}\right]=-c_i\log{k^2\over M^2}
$$
where $M$ is a renormalization group scale and keeping in mind that $c_0=0$.
The appearance of the arbitrary scale $M$ in the momentum space correlator is another avatar of the conformal anomaly.

For a theory with a mass gap, we can set the scale $m_{IR}=0$ and we can rewrite (\ref{nn66}) as
\be
\bar G_i^{ren}\equiv A^{ren}_i-k^2\int_0^{\infty}{d\mu^2\over \mu^2}{\rho_i(\mu^2)\over (k^2+\mu^2)}
\label{nn666}\ee
In $d=4$ the $A_i^{ren}$ are dimensionless contact terms whose value depends on the renormalization scheme.
The low momentum expansion of (\ref{nn666}) becomes
\be
\bar G_i^{ren}\equiv A^{ren}_i-B_i~k^2+{\cal O}(k^4)\sp B_i\equiv \int_{m_0^2}^{\infty}{d\mu^2\over \mu^4}\rho_i(\mu^2)
\label{nn15}\ee
where $m_0$ is the mass gap of the correlator.

For a general four-dimensional theory without a mass gap we have that
$\rho_{i}(\mu^2)\simeq c_i^{UV}$ for $\mu\to\infty$ while
$\rho_{i}(\mu^2)\simeq c_i^{IR}$ for $\mu\to 0$.
We pick two scales, $m_1\to 0$ so that it is much smaller that all the scale of the theory, while $m_2\to\infty$ is much larger than all scales of the theory (except the UV cutoff) and write (\ref{nn66}) as
\be
\bar G_i^{ren}\equiv A^{ren}_i(m_{IR})-I^i_{IR}-I^i_{UV}-I^i_{inter}
\label{nn16}\ee
with
\be
I^i_{IR}\equiv (k^2-m_{IR}^2)\int_0^{m_1^2}{d\mu^2\over (\mu^2+m_{IR}^2)}{\rho_i(\mu^2)\over (k^2+\mu^2)}\simeq c_{i}^{IR} (k^2-m_{IR}^2)\int_0^{m_1^2}{d\mu^2\over (\mu^2+m_{IR}^2) (k^2+\mu^2)}=
\label{nn17}\ee
$$
=c_i^{IR}\left[\log{(m_1^2+k^2)\over k^2}+\log{m_{IR}^2\over (m_1^2+m_{IR}^2)}\right]
$$
\be
I^i_{UV}\equiv (k^2-m_{IR}^2)\int_{m_2^2}^{\infty}{d\mu^2\over (\mu^2+m_{IR}^2)}{\rho_i(\mu^2)\over (k^2+\mu^2)}\simeq c_{i}^{UV} (k^2-m_{IR}^2)\int_{m_2^2}^{\infty}{d\mu^2\over (\mu^2+m_{IR}^2) (k^2+\mu^2)}=
\label{nn19}\ee
$$
=c_i^{UV}\log{m_2^2+k^2\over m_2^2+m_{IR}^2}
$$
and
\be
I^i_{inter}\equiv (k^2-m_{IR}^2)\int_{m_1^2}^{m_2^2}{d\mu^2\over (\mu^2+m_{IR}^2)}{\rho_i(\mu^2)\over (k^2+\mu^2)}
\label{nn21}\ee
From these expressions, we deduce that $I_{inter}$ is a regular power series in $k^2$ for $k^2$ small. Therefore in $\bar G_i$  there is only a log $k^2$ divergence that is appearing due to the IR CFT. For a gapless theory, we can write a small $k^2$ expansion that is of the form
\be
\bar G_{i}=c_i^{IR}\log{M^2\over k^2}+{\rm regular~~ expansion~~ in~~ }k^2
\label{nn22}\ee
and where $M^2$ is some scale of the theory. It is ambiguous in the formula above as it can be changed by changing the regular part.

On the other hand, as $k^2\to\infty$ we obtain
\be
I^i_{IR}\simeq {\rm regular~~series~~in~~}{1\over k^2}\sp
I^i_{UV}\simeq c_{i}^{UV}\log k^2+{\rm regular~~series~~in~~}{1\over k^2}
\label{nn23}\ee
\be
I^i_{inter}={\rm regular~~series~~in~~}{1\over k^2}
\label{nn24}\ee
so that
\be
\bar G_{i}=c_i^{UV}\log{ k^2\over M'^2}+{\rm regular~~ expansion~~ in~~ }{1\over k^2}
\label{nn25}\ee
as $k^2\to \infty$.

So far we have seen that as $k^2\to 0$, $\bar G^{ren}_i$ are regular functions of $k^2$ with an exception of a $\log k^2$ appearance, if the theory is gapless.
There is, however, a set of contact terms, compatible with stress tensor conservation and IR regularity that are not included in (\ref{nn66}).
Indeed consider
\be
\bar G_i^{ren}(k)\to G_i^{ren}+{\delta_i\over k^2}
\label{nn26}\ee
Then
\be
\delta\langle T_{\m\n} T_{\r\s}\rangle (k)={3{\cal A}_4\over 4}~k^2~\left[\pi_{\m\r}\pi_{\n\s}+\pi_{\m\s}\pi_{\m\r}-{2\over 3}\pi_{\m\n}\pi_{\r\s}\right]{\delta_2}+{{\cal A}_4\over 6}k^2~\pi_{\m\n}\pi_{\r\s}~{\delta_0}
\label{nn20a}\ee
and the absence of the ${k_{\m}k_{\n}k_{\r}k_{\s}\over k^2}$ term implies that
\be
6\delta_2+\delta_0=0
\label{nn27}\ee
and inserting in (\ref{nn20a}) we obtain
\be
\delta\langle T_{\m\n} T_{\r\s}\rangle (k)={3{\cal A}_4\delta_2\over 4}\left[k^2(\delta_{\m\r}\delta_{\n\s}+\delta_{\m\s}\delta_{\m\r}-2\delta_{\m\n}\delta_{\r\s})-\right.
\label{nn28}\ee
$$
\left.-(\delta_{\m\r}k_{\n}k_{\s}+
\delta_{\n\s}k_{\m}k_{\r}+\delta_{\m\s}k_{\n}k_{\r}+\delta_{\n\r}k_{\m}k_{\s})+2\delta_{\m\n}k_{\r}k_{\s}+2\delta_{\r\s}k_{\m}k_{\n}\right]
$$
This is the same as the quadratic part of the Einstein-Hilbert term around flat space in (\ref{b2}). It is clear, that because of the relation (\ref{nn27}), if $\delta_2>0$, then $\delta_0<0$ and the spin zero piece of this particular term is ghost-like).

Summarizing, the explicit contact contributions in the renormalized stress tensor functions $\bar G_i^{ren}$ in four-dimensions are
\be
\bar G^{ren, contact}_2(k)=A_2^{ren}+{\delta_2\over k^2}\sp \bar G^{ren, contact}_0(k)=A_0^{ren}-{6\delta_2\over k^2}
\ee

Note that the $\delta$ contact term appears as a massless pole in the spectral functions $\bar G_2$ and $\bar G_2$.

We conclude this part as follows.
\begin{itemize}

\item The two-point function of conserved stress tensor is determined by the spin-2 and spin-0 spectral densities. For small momenta, they start at ${\cal O}(k^4)$

\item      The renormalized spin-2 and spin-0 spectral densities  are well-defined from (\ref{nn20}) and (\ref{nn66}) modulo two sets of contact terms. The spectral densities  are positive in positive theory.

\item  The first set of such contact terms are associated with the scheme-dependent {\em dimensionless} coefficients, $A^{ren}_{2}$ and $A_0^{ren}$ and they are ${\cal O}(k^4)$. In the Schwinger functional they are correlated with the curvature-squared terms in the metric. $A^{ren}_{2}$ and $A_0^{ren}$ are a priori unrelated and can be chosen to be both positive.

\item The second set of such contact terms are ${\cal O}(k^2)$ and are given  in (\ref{nn28}). They always contain a ghost-like contribution, that can be put  in the spin-0 part if $\delta_2>0$.
    As $\delta_2$ has dimension mass$^2$, such terms are typically quadratically divergent, in the presence of the cutoff. Upon renormalization, they depend on the masses of the theory. They are clearly absent in a CFT. Moreover, they look like massless poles in the spectral functions, $\bar G_2$ and $\bar G_2$..

\end{itemize}

\subsection{The static potential}\label{staticpotential}

Another quantity we can compute with the knowledge of the spectral densities, is the static potential for sources due to  stress energy exchange.  We can relate the static potential $\Phi(r)$ with the spin-$0$ and spin-$2$ spectral factors using equation \eqref{K2} as follows
\be
\Phi(r) = \int_{-\infty}^\infty dt  \langle T_{00}(t, \vec{r}) T_{00}(0,0) \rangle = \int_{-\infty}^\infty  \frac{ d^{d-1} \vec{p}}{(2 \pi)^{d-1}} \, e^{i \vec{p} \, \vec{r}} \, \langle T_{00} (p_0 = 0, \vec{p}) T_{00} (p_0 = 0, -\vec{p}) \rangle   =
\ee
$$
=    \left( \frac{{\cal A}_d}{\Gamma(d)} \int_0^\infty d \mu^2 \r_{0}(\mu^2)  + \frac{{\cal A}_d}{\Gamma(d-2)} \int_0^\infty d \mu^2 \r_{2}(\mu^2)  \right) \int_{-\infty}^\infty \frac{ d^{d-1} \vec{p}}{(2 \pi)^{d-1}} \frac{|\, \vec{p} \, |^4 e^{i \vec{p} \, \vec{r}}}{|\, \vec{p} \, |^2+ \mu^2}
$$
An equivalent result can be obtained if we use equation \eqref{nn20}, the advantage of the position space result is that all the integrals are manifestly convergent in the order they appear (if we first perform the momentum space integral and then the integral over the spectral weight).

In particular for $d=4$ we obtain from the spin zero spectral weight ($u=  |\, \vec{p} \,|$)
\be
\Phi_0(r) =  - \frac{{\cal A}_4}{\Gamma(4)} \int_0^\infty d \mu^2 \r_{0}(\mu^2) \int_{0}^\infty \frac{ d u }{2 \pi^2 r} \frac{ u^5 \, \sin(u r) }{u^2+ \mu^2}  = \label{j1}\ee
$$
=- \frac{{\cal A}_4}{\Gamma(4) 4 \pi r} \int_0^\infty d y  y^2 e^{- r \sqrt{y}}  \,
\left( R_i\sum_i \delta(y - m_i^2) + \theta(y> M^2) \rho_{0}(y) \right) \, ,
$$
 In the calculation above, we first extended the even in $u$ integral and performed it by picking the poles for $u = i \mu$. We finally set $y = \mu^2$ to simplify the expression in the integral. In the last line, we assumed the presence of poles and a continuum for the spectral measure.
We immediately find that poles give the expected massive boson behaviour
\be
\Phi_0^{pole}(r) = - \frac{\pi^2~R}{480 } m^2 {e^{- m r}\over r}
\label{nn1}\ee
where we used ${\cal A}_4={\pi^2\over 20}$.
The spin two part can be analysed in precisely the same fashion, the static potential leading again to an attractive force as for the spin zero part since they both carry the same sign.

We should also emphasize that such simple poles in the spectral measure cannot appear in free or weakly interacting quantum field theories. To obtain this behavior we need strong coupling in the hidden theory. A weakly coupled theory typically exhibits spectral branch cuts. Such a case is analysed in appendix~\ref{free}, and the resulting static potential is found in appendix~\ref{freestaticpotential}.

\subsection{Improvements}

In flat space, there is a unique improvement term that can be added to the stress tensor, namely
\be
T^I_{\m\n}=T_{\m\n}+(\partial_{\m}\partial_{\n}-\delta_{\m\n}\square)V_2
\ee
where $V_2$ is a scalar operator. It order, not to disturb scale covariance, it must have dimension two in the UV. Otherwise its coefficient has non-trivial dimensions. As the extra contribution to energy and momentum is a boundary term at spatial infinity, it leaves energy and momentum unaffected.
In a weakly coupled theory, that includes bosons, there is an obvious candidate for $V_2$ that is quadratic in scalars.

Using the spectral representation of the two-point function of
\be
\langle V_2 V_2\rangle (k)={{\cal A}_4\over \Gamma_{d}}\bar G_V(k)\sp \bar G_V(k)=\int_0^{\infty}d\mu^2 ~{\rho_V(\mu^2)\over k^2+\mu^2}
\ee
we may write the two-point function of $T_I$ as in (\ref{nn20}) with
\be
G^I_{2}=G_2\sp G^I_{0}=G_0+G_V
\ee

Interestingly, there is a non-local improvement that renders the stress-tensor traceless,
\be
T_{\m\n}'=T_{\m\n}-{1\over d-1}\left(\delta_{\m\n}-{\pa_{\m}\pa_{\n}\over \square}\right){T_{\r}}^{\r}\sp {T'_{\r}}^{\r}=0
\ee

\section{The Energy-momentum two-point function of free fields}
\label{free}

\subsection{Free bosons}

In this subsection we consider\footnote{In this appendix we drop the hat notation of the hidden QFT and compute the energy-momentum two-point function in Minkowski signature.} the energy-momentum two-point function for $N^2$ decoupled free massive real scalars arranged in an $N\times N$ matrix $\phi$
\be
\label{freeaa}
S = - \frac{1}{2} \int d^4 x \, {\rm Tr}\left( \p_\mu \phi \p^\mu \phi + m^2 \phi^2 \right)
~.
\ee
The corresponding energy-momentum tensor is defined as in the main text as $$T^{\mu\nu} =  \frac{2}{\sqrt{-g}} \frac{\delta S}{\delta g_{\mu\nu}}\;,$$ which gives
\be
\label{freeab}
T^{\mu\nu} = - {\rm Tr} \left[ \p^\mu \phi \p^\nu \phi - \frac{1}{2} \eta^{\mu\nu} \left( \p_\rho \phi \p^\rho \phi + m^2 \phi^2\right) \right]
~.
\ee
Notice that with  this normalisation, the one- and two-point functions of the energy-momentum tensor scales like $N^2$.

We compute the connected two-point function
\be
\label{spinmassak}
\langle {T}^{\rho_1\sigma_1}(k) {T}^{\rho_2 \sigma_2} (-k) \rangle_c
= \int d^d x \, e^{-i k x} \langle {T}^{\rho_1\sigma_1}(x) {T}^{\rho_2 \sigma_2} (0) \rangle_c
\ee
in momentum space in the free theory by performing all the pertinent Wick contractions with the use of the two-point function
\be
\label{spinmassal}
\langle {\phi}_{a_1 b_1} (x) {\phi}_{a_2 b_2} (0) \rangle = \delta_{a_1 b_2}\delta_{b_1 a_2} \int \frac{d^d p}{(2\pi)^d}  \frac{-i e^{-i p x}}{p^2 + m^2}
~.
\ee
$a,b=1,\ldots,N$ are indices labelling the free fields. For the moment we leave the space-time dimension $d$ as a free parameter. We will evaluate the ensuing momentum integrals by Wick rotating to the Euclidean signature, so the $i\epsilon$ prescription will be left implicit.

The implementation of the Wick contractions leads to the $d$-dimensional momentum integral
\begin{align}
\begin{split}
\label{spinmassam}
&\langle {T}^{\mu\nu}(k) {T}^{\rho \sigma} (-k) \rangle_c
=- \int \frac{d^d p }{(2\pi)^d} \frac{1}{(p^2+m^2)((p+k)^2+m^2)}
\\
&\hspace{1cm} \bigg[ p^\mu (p+k)^\nu \left( p^\rho (p+k)^\sigma + p^\sigma (p+k)^\rho\right)
\\
&\hspace{1cm} - \eta^{\rho\sigma} p^\mu (p+k)^\nu (p^2+m^2 + pk) - \eta^{\mu\nu} p^\sigma (p+k)^\rho (p^2+m^2 + pk)
\\
&\hspace{1cm} +\frac{1}{2}\eta^{\mu\nu} \eta^{\rho\sigma} \left( p^2+ m^2 + pk \right)^2 \bigg]
~.
\end{split}
\end{align}
In what follows, we evaluate this integral in two ways: $i)$ by using dimensional regularization, and $ii)$ by using a hard UV cutoff. Since both are standard textbook computations, we proceed to quote the result.

\paragraph{Dimensional regularization.} Applying dimensional regularization, $\varepsilon=4-d \to 0$, to the momentum integral \eqref{spinmassam} after using the master formula
\be
I(a,b;d)\, = \, \int \frac{d^d p}{(2 \pi)^d} \frac{p^b}{(p^2+ M^2)^a} \, = \, \frac{\pi^{d/2}}{(2 \pi)^d \, \Gamma(d/2)} \frac{\Gamma\left( \half (b+d) \right) \Gamma\left( a - \half(b+d) \right)}{  M^{a - (b+d)/2}\Gamma\left( a \right)}
\ee
we obtain (notice that we can add and subtract terms proportional to odd powers of $\sim 2x - 1$ since they integrate to zero)
\begin{align}
\begin{split}
\label{genperturesult}
&\langle {T}^{\mu\nu}(k) {T}^{\rho\sigma} (-k) \rangle_c
=  \frac{i}{16\pi^2} \int_0^1 d x \left[-\frac{2}{\epsilon} + \log \frac{c_1   \Delta}{\mu^2}   \right] \times
\bigg\{ \frac{\Delta^2}{4} \left( \eta^{\mu\rho}\eta^{\nu\sigma} + \eta^{\mu\sigma} \eta^{\nu\rho} \right) -
\\
&\hspace{1cm} - \frac{1}{4} \eta^{\mu\nu}\eta^{\rho\sigma} \left( m^4 + m^2 k^2 (1- 2 x(1-x) )\right) + 2 x^2(1-x)^2 k^\mu k^\nu k^\rho k^\sigma  -
\\
&\hspace{1cm} -  \left( \eta^{\mu\nu} k^\rho k^\sigma + \eta^{\rho\sigma} k^\mu k^\nu \right) x \left( (1-2x) m^2 + \frac{k^2}{2}(1-x)(1 - 8 (1-x)x) \right)
\\
&\hspace{1cm}+ \frac{\Delta}{2} x(1-2x)  \bigg(  \eta^{\mu\rho} k^\nu k^\sigma + \eta^{\mu\sigma} k^\nu k^\rho
+ \frac{1}{4} \eta^{\nu \rho} k^\mu k^\sigma + \frac{1}{4} \eta^{\nu \sigma} k^\mu k^\rho \bigg)
\bigg\}
~.
\end{split}
\end{align}
In these formulae $c_1 = e^\gamma /\pi$ and $\Delta = m^2 + k^2 x(1-x)$. This integral exhibits a branch-cut structure since we considered a free theory, that starts at $k^2 = - 4 m^2$. The general structure is as follows: There is a divergent $\OO(1/\varepsilon)$ part and a finite $\OO(\varepsilon^0)$ part. Both parts are momentum-dependent. The divergent part contributes a finite set of contact terms to the two-point function in real space.  We can then subtract the divergent terms and absorb $c_1$ (MS-scheme) into a renormalised scale $\mu_r$, leaving only a renormalised logarithmic term. The renormalised result can then be expanded in an infinite series of powers of momenta (a low-momentum expansion).  The IR-limit is then given by sending $k/m\rightarrow 0$ that results in the following expansion
\begin{align}
\begin{split}
\label{genpertueaaA}
&\langle {T}^{\mu\nu}(k) {T}^{\rho\sigma} (-k) \rangle_c
= \frac{i}{16\pi^2}
\bigg\{ \frac{m^4}{4} \left( -\eta^{\mu\nu}\eta^{\rho\sigma} + \eta^{\mu\rho}\eta^{\nu\sigma} + \eta^{\mu\sigma} \eta^{\nu\rho} \right)
\\
&\hspace{1cm} +\frac{m^2}{12} \bigg[ k^2 \left(-2 \eta^{\mu\nu} \eta^{\rho\sigma} + \eta^{\mu\rho} \eta^{\nu\sigma} + \eta^{\mu\sigma} \eta^{\nu\rho} \right)
+ 2 \eta^{\mu\nu} k^\rho k^\sigma + 2 \eta^{\rho\sigma} k^\mu k^\nu
\\
&\hspace{1cm}~~~~~~~~ - \eta^{\mu\rho} k^\nu k^\sigma - \eta^{\mu\sigma} k^\nu k^\rho - \eta^{\nu\rho} k^\mu k^\sigma - \eta^{\nu \sigma} k^\mu k^\rho \bigg]
\\
&\hspace{1cm}+\frac{1}{30} \bigg[ k^4 \left( \frac{3}{2} \eta^{\mu\nu} \eta^{\rho\sigma} + \frac{1}{4} \eta^{\mu\rho} \eta^{\nu\sigma} +\frac{1}{4} \eta^{\mu\sigma} \eta^{\nu \rho} \right) + k^2 \bigg( -\frac{3}{2} \eta^{\mu\nu} k^\rho k^\sigma -\frac{3}{2} \eta^{\rho\sigma} k^\mu k^\nu -
\\
&\hspace{1cm}   - \frac{1}{4} \eta^{\mu\rho} k^\nu k^\sigma -\frac{1}{4} \eta^{\mu\sigma} k^\nu k^\rho - \frac{1}{4} \eta^{\nu \rho} k^\mu k^\sigma - \frac{1}{4} \eta^{\nu \sigma} k^\mu k^\rho \bigg)
+ 2 k^\mu k^\nu k^\rho k^\sigma \bigg]
 +  \OO\left(\frac{k^{6}}{m^2} \right) \bigg\}
\\
&\hspace{1cm}~~~~~~~~~  \times  \left(  \log \frac{m^2}{\mu_r^2} + \OO\left(\frac{k^{2}}{m^2} \right) \right) \,
~.
\end{split}
\end{align}
We observe that both the divergent piece \eqref{genperturesult} and the renormalised result \eqref{genpertueaaA} obey the Ward identity \eqref{genpertudg} and can be organized in the form \eqref{genpertudi} with coefficients $a(\equiv \Lambda)$, $b$, $c$ as quoted in equations \eqref{linearap}-\eqref{linearaqa}. There exist also terms without logarithmic running denoted by $\OO\left(\frac{k^{2}}{m^2} \right)$ in the last parenthesis. These terms are part of a regular expansion in powers of momentum and correspond to scheme dependent contact terms as shown in appendix~\ref{momissues}. As such they will not be of further interest to us.

As a check of the Ward identity \eqref{genpertudi}, we can also verify by independent computation of the one-point function $\langle T_{\mu\nu}(x)\rangle$ that the coefficient
\be
\label{genpertueaaaa}
a =  \frac{m^4}{64\pi^2}  \log \frac{  m^2}{\mu_r^2}
\ee
of the momentum independent terms of equation \eqref{genpertueaaA}, is the same coefficient that appears in the one-point function. Indeed, for the one-point function we obtain in momentum space
\begin{align}
\begin{split}
\label{genpertueaaab}
\langle T_{\mu\nu}(k) \rangle
&= i \delta(k) \int \frac{d^d p}{(2\pi)^d} \left[ \frac{p_\mu p_\nu}{p^2 + m^2}
- \frac{1}{2} \eta_{\mu\nu} \frac{p^2+m^2}{p^2+m^2} \right]
\\
&=\delta (k) \eta_{\mu\nu} \left[\left( \frac{1}{d}-\frac{1}{2}\right) \int \frac{d^d p_E}{(2\pi)^d}\frac{p_E^2}{p_E^2+m^2} - \frac{m^2}{2} \int \frac{d^dp_E}{(2\pi)^d} \frac{1}{p_E^2+m^2} \right]
~,
\end{split}
\end{align}
where the integrals on the second line are performed in Euclidean space. Using the formulae
\be
\label{genpertueaaac}
\int \frac{d^d p_E}{(2\pi)^d}\frac{p_E^2}{p_E^2+m^2} \underset{\varepsilon\to 0}{\to} \frac{m^4}{(4\pi)^2} \left(  \frac{2}{\epsilon} - \log \frac{c_1   m^2}{\mu^2}  \right)~, ~~
\int \frac{d^dp_E}{(2\pi)^d} \frac{1}{p_E^2+m^2} \underset{\varepsilon\to 0}{\to}  \frac{m^2}{(4\pi)^2} \left(  \frac{2}{\epsilon} - \log \frac{c_1   m^2}{\mu^2}  \right)
\ee
and renormalising in MS scheme, we recover $\langle T_{\mu\nu}(k)\rangle = a \delta(k) \eta_{\mu\nu}$ with the value of $a$ quoted in \eqref{genpertueaaaa}. We stress that the finite part of the contact term of the stress energy tensor one-point function is scheme dependent, but once a choice of scheme is chosen for all correlation functions (here MS-scheme), the Ward identities are found to hold.

\paragraph{Hard cutoff.} Evaluating the momentum integral \eqref{spinmassam} with a hard UV cutoff regulator $M$ we obtain many contributions that violate the Ward identity \eqref{genpertudg}. Here we quote the result of this computation up to order $k^2$ omitting terms that vanish in the limit $M\to \infty$
\begin{align}
\begin{split}
\label{spinmassau}
& \langle {T}^{\mu\nu}(k) {T}^{\rho \sigma} (-k) \rangle_c \big |_{k=0}
= \frac{i}{32\pi^2} \bigg[ \frac{1}{12} \left( - \eta^{\mu\nu} \eta^{\rho \sigma} - \eta^{\mu\rho} \eta^{\nu\sigma} - \eta^{\mu\sigma} \eta^{\nu\rho} \right) M^4
\\
& -\frac{1}{3} \left( 2 \eta^{\mu\nu} \eta^{\rho \sigma} + \eta^{\mu\rho} \eta^{\nu\sigma} + \eta^{\mu\sigma} \eta^{\nu\rho} \right) m^2 M^2
\\
&+\frac{1}{2} \left(  \eta^{\mu\nu} \eta^{\rho \sigma} - \eta^{\mu\rho} \eta^{\nu\sigma} - \eta^{\mu\sigma} \eta^{\nu\rho} \right) m^4 \log\left( \frac{M^2}{m^2} \right)
\\
& +\frac{1}{6} \left( \eta^{\mu\nu} \eta^{\rho \sigma} + \eta^{\mu\rho} \eta^{\nu\sigma} + \eta^{\mu\sigma} \eta^{\nu\rho} \right) m^4
 \bigg]
\\
&+\frac{iN^2}{8\pi^2} \bigg[
\frac{M^2}{48} \bigg(
k^2(2 \eta^{\mu\nu} \eta^{\rho\sigma} + \eta^{\mu\rho}\eta^{\nu\sigma} + \eta^{\mu \sigma} \eta^{\nu\rho} )
- 2 ( \eta^{\mu\sigma} k^\nu k^\rho + \eta^{\mu\rho} k^\nu k^\sigma )
\bigg)
\\
&-\frac{m^2}{144}\bigg(
(-2 \eta^{\mu\sigma} \eta^{\nu \rho} -2 \eta^{\mu\rho} \eta^{\nu\sigma} +13 \eta^{\mu\nu} \eta^{\rho \sigma} ) k^2
- 16 (\eta^{\rho\sigma} k^\mu k^\nu  + \eta^{\mu\nu} k^\rho k^\sigma)
\\
&~~~~~+11 (\eta^{\nu\sigma} k^\mu k^\rho +\eta^{\nu\rho} k^\mu k^\sigma) - \eta^{\mu\sigma} k^\nu k^\rho - \eta^{\mu\rho} k^\nu k^\sigma )
\bigg)
\\
&+\frac{1}{24}m^2 \log\left( \frac{M^2}{m^2} \right) \bigg(
( \eta^{\mu\sigma} \eta^{\nu \rho} + \eta^{\mu\rho} \eta^{\nu\sigma} -2 \eta^{\mu\nu} \eta^{\rho \sigma} ) k^2
\\
&~~~~~+ 2(\eta^{\rho\sigma} k^\mu k^\nu + \eta^{\mu\nu} k^\rho k^\sigma) - \eta^{\nu\sigma} k^\mu k^\rho - \eta^{\mu\sigma} k^\nu k^\rho - \eta^{\nu\rho} k^\mu k^\sigma -\eta^{\mu\rho} k^\nu k^\sigma  )
\bigg)
\bigg]
~.
\end{split}
\end{align}
The first four lines capture the $\OO(k^0)$ part of the two-point function. The five last lines capture the $\OO(k^2)$ part. We notice that the logarithmically divergent terms reproduce the result found in dimensional regularization.
There are also terms that are power-divergent in $M$ with a tensor structure that does not obey the Ward identities  \eqref{genpertudf}. Such terms violate the diffeomorphism invariance of the generating function of energy-momentum correlation functions and should be subtracted by appropriate counterterms to recover the Ward identities.

We also quote for reference the result for the one-point function $\langle T_{\mu\nu}(k)\rangle$ in the hard cutoff regularization
\be
\label{hardonepoint}
\langle T_{\mu\nu}(k)\rangle
=  \frac{1}{64\pi^2} \left( \frac{M^4}{2} + m^2 M^2 - m^4 \log\frac{M^2+m^2}{m^2} \right) \delta^{(4)}(k) \, \eta_{\mu\nu}
~.
\ee
As expected, the logarithmically divergent term reproduces the result \eqref{genpertueaaaa}.

\subsection{Free fermions}

For quick reference we also list here the energy-momentum two-point function for $N^2$ decoupled free massive Dirac fermions arranged in an $N \times N$ matrix $\psi$
\be
\label{freefermaa}
S=\int d^4 x\, \tr \left [ \bar \psi \left( i \gamma^\mu \partial_\mu - m \right) \psi \right]
~.
\ee
In this subsection we only report the result obtained in dimensional regularisation.

Using the energy-momentum tensor
\be
\label{freefermab}
T^{\mu\nu} = \frac{i}{2} \tr \left[  \bar\psi \gamma^{(\mu} \partial^{\nu)} \psi -  \partial^{(\mu} \bar\psi \gamma^{\nu)} \psi \right]
-\frac{1}{2} \eta^{\mu\nu} \tr \left[ \bar \psi \left( i \gamma^\mu \partial_\mu - m \right) \psi \right]
\ee
and the free fermion two-point functions\footnote{$\alpha,\beta$ are four-dimensional Dirac spinors in this equation.}
\be
\label{freefermac}
\langle (\psi_\alpha)_{a_1a_2} (x) (\bar \psi_{\beta})_{b_1 b_2}(0) \rangle
=\delta_{a_1b_2} \delta_{b_1a_2} \int \frac{d^4p}{(2\pi)^4} \frac{-i (\not p_{\alpha\beta} + m \delta_{\alpha\beta} )}{p^2+m^2} e^{-i p x}
\ee
one finds the following renormalised result up to $\OO(k^4)$
\begin{align}
\begin{split}
\label{genpertueaa}
&\langle {T}^{\mu\nu}(k) {T}^{\rho\sigma} (-k) \rangle
= \frac{i}{16\pi^2} \int_0^1 d x
\bigg\{ - \frac{m^4}{2} \left( - \eta^{\mu\nu}\eta^{\rho\sigma} + \eta^{\mu\rho}\eta^{\nu\sigma} + \eta^{\mu\sigma} \eta^{\nu\rho} \right)
\\
&\hspace{1cm} +\frac{m^2}{6} \bigg[ k^2 \left(-2 \eta^{\mu\nu} \eta^{\rho\sigma} + \eta^{\mu\rho} \eta^{\nu\sigma} + \eta^{\mu\sigma} \eta^{\nu\rho} \right)
+ 2 \eta^{\mu\nu} k^\rho k^\sigma + 2 \eta^{\rho\sigma} k^\mu k^\nu
\\
&\hspace{1cm}~~~~~~~~ - \eta^{\mu\rho} k^\nu k^\sigma - \eta^{\mu\sigma} k^\nu k^\rho - \eta^{\nu\rho} k^\mu k^\sigma - \eta^{\nu \sigma} k^\mu k^\rho \bigg]
\\
&\hspace{1cm}+\frac{1}{30} \bigg[ k^4 \left( - \eta^{\mu\nu} \eta^{\rho\sigma} + \frac{3}{2} \eta^{\mu\rho} \eta^{\nu\sigma} +\frac{3}{2} \eta^{\mu\sigma} \eta^{\nu \rho} \right)
\\
&\hspace{1cm}+ k^2 \bigg(  \eta^{\mu\nu} k^\rho k^\sigma + \eta^{\rho\sigma} k^\mu k^\nu  - \frac{3}{2} \eta^{\mu\rho} k^\nu k^\sigma -\frac{3}{2} \eta^{\mu\sigma} k^\nu k^\rho
\\
&\hspace{1cm}~~~~~~~~~ - \frac{3}{2} \eta^{\nu \rho} k^\mu k^\sigma - \frac{3}{2} \eta^{\nu \sigma} k^\mu k^\rho \bigg)
+ 2 k^\mu k^\nu k^\rho k^\sigma \bigg]
\bigg\} \times  \log \frac{m^2}{\mu^2_r}
~.
\end{split}
\end{align}
The steps of the computation are exactly the same as in the case of the free bosons and hence we do not repeat them here.

\subsection{The correlator in real space}

We start again from (\ref{freeab}) and
\be
\langle \phi^i(x)\phi^j(y)\rangle=\delta_{ij}G(|x-y|^2)
\label{f1f}\ee
rotate to Euclidean space and calculate
\be
\langle T_{\m\n}(x)T_{\r\s}(y)\rangle_{conn}=\pa_{\m}\pa_{\r}G\pa_{\n}\pa_{\s}G+\pa_{\n}\pa_{\r}G\pa_{\m}\pa_{\s}G-
\label{f2f}\ee
$$
-\delta_{\r\s}\left[\pa_{\m}\pa_{\eta}G\pa_{\n}\pa_{\eta}G+m^2\pa_{\m}G\pa_{\n}G\right]
-\delta_{\m\n}\left[\pa_{\r}\pa_{\eta}G\pa_{\s}\pa_{\eta}G+m^2\pa_{\r}G\pa_{\s}G\right]+
$$
$$
+{1\over 2}\delta_{\m\n}\delta_{\r\s}\left[\pa_{\eta}\pa_{\l}G\pa_{\eta}\pa_{\l}G+2m^2\pa_{\eta}G\pa_{\eta}{G}+m^4G^2\right]
$$
where all functions and all derivatives are with respect to $z^{\m}=x^{\m}-y^{\m}$.
Using
\be
\pa_{\m}G(z^2)=2z^{\m} G'(z^2)\sp \pa_{\m}\pa_{\n}G(z^2)=2\delta^{\m\n}G'+4z^{\m}z^{\n} G''(z^2)\sp \pa_{\eta}\pa_{\eta}G=8G'+4z^2G''
\label{f3f}\ee
We obtain
\be
\langle T^{\m\n}(z)T^{\r\s}(0)\rangle_{conn}=4\left(\delta^{\m\r}\delta^{\n\s}+\delta^{\n\r}\delta^{\m\s}\right)(G')^2+
\label{f4f}\ee
$$
+{1\over 2}\delta^{\m\n}\delta^{\r\s}\left(m^4 G^2+8m^2z^2(G')^2+16 z^2 G'G''+16 z^4(G'')^2\right)+
$$
$$
+8\left(\delta^{\m\r}z^{\n}z^{\s}+\delta^{\n\r}z^{\m}z^{\s}+\delta^{\m\s}z^{\n}z^{\r}+\delta^{\n\s}z^{\m}z^{\r}\right)G'G''-
$$
$$
-16(\delta^{\r\s}z^{\m}z^{\n}+\delta^{\m\n}z^{\r}z^{\s})\left[G'G''+z^2(G'')^2\right]+32z^{\m}z^{\n}z^{\r}z^{\s}(G'')^2
$$
$$=
4\left(\delta^{\m\r}\delta^{\n\s}+\delta^{\n\r}\delta^{\m\s}+\delta^{\m\n}\delta^{\r\s}m^2z^2\right)(G')^2+{m^4\over 2}\delta^{\m\n}\delta^{\r\s}G^2+
$$
$$+8\left[\delta^{\m\r}z^{\n}z^{\s}+\delta^{\n\r}z^{\m}z^{\s}+\delta^{\m\s}z^{\n}z^{\r}+\delta^{\n\s}z^{\m}z^{\r}+z^2\delta^{\m\n}\delta^{\r\s}
-2(\delta^{\r\s}z^{\m}z^{\n}+\delta^{\m\n}z^{\r}z^{\s})\right]G'G''+
$$
$$
+8\left[(z^2)^2\delta^{\m\n}\delta^{\r\s}-2z^2(\delta^{\r\s}z^{\m}z^{\n}+\delta^{\m\n}z^{\r}z^{\s})+4z^{\m}z^{\n}z^{\r}z^{\s}\right](G'')^2
$$

We can separate the trace and traceless parts:
We define
\be
\Theta\equiv {T_{\m}}^{\m}\sp \bar T_{\m\n}=T_{\m\n}-{1\over 4}\eta_{\m\n}\Theta
\ee
We then obtain
\be
\langle \Theta(z) T^{\r\s}(0)\rangle=\left[2m^4 G^2+8(1+2m^2z^2)(G')^2\right]\delta^{\r\s}-32(G'G''+z^2(G'')^2)z^{\r}z^{\s}
\ee

\be
\langle \Theta(z)\Theta(0)\rangle=4\left[2m^4 G^2+8(1+2m^2z^2)(G')^2\right]-32z^2(G'G''+z^2(G'')^2)
\ee
\be
\langle \Theta(z)\bar T^{\r\s}(0)\rangle=-32(G'G''+z^2(G'')^2)\left[z^{\r}z^{\s}-{1\over 4}\delta^{\r\s}z^2\right]
\ee
\be
\langle \bar T^{\m\n}(z)\bar T^{\r\s}(0)\rangle=
4\left(\delta^{\m\r}\delta^{\n\s}+\delta^{\n\r}\delta^{\m\s}\right)(G')^2+
\label{f4ff}\ee
$$
+8\left(\delta^{\m\r}z^{\n}z^{\s}+\delta^{\n\r}z^{\m}z^{\s}+\delta^{\m\s}z^{\n}z^{\r}+\delta^{\n\s}z^{\m}z^{\r}-\delta^{\r\s}z^{\m}z^{\n}-\delta^{\m\n}z^{\r}z^{\s}+{1\over 4}\delta^{\m\n}\delta^{\r\s}z^2\right){G'G''}-
$$
$$
+32z^{\m}z^{\n}z^{\r}z^{\s}(G'')^2-8(\delta^{\r\s}z^{\m}z^{\n}+\delta^{\m\n}z^{\r}z^{\s})\left[z^2(G'')^2\right]+
$$
$$
+2\delta^{\m\n}\delta^{\r\s}\left(-(G')^2+ z^4(G'')^2\right)
$$
$$
=4(G')^2\left[\left(\delta^{\m\r}\delta^{\n\s}+\delta^{\n\r}\delta^{\m\s}-{1\over 2}\delta^{\m\n}\delta^{\r\s}\right)+\right.
$$
$$
+8\left(\delta^{\m\r}z^{\n}z^{\s}+\delta^{\n\r}z^{\m}z^{\s}+\delta^{\m\s}z^{\n}z^{\r}+\delta^{\n\s}z^{\m}z^{\r}-\delta^{\r\s}z^{\m}z^{\n}-\delta^{\m\n}z^{\r}z^{\s}+{1\over 4}\delta^{\m\n}\delta^{\r\s}z^2\right){G''\over G'}
$$
$$+
\left.32\left(z^{\m}z^{\n}-{1\over 4}z^2\delta^{\m\n}\right)\left(z^{\r}z^{\s}-{1\over 4}z^2\delta^{\r\s}\right){G''^2\over G'^2}\right]
$$

We will now evaluate the propagator $G$.
It satisfies
\be
(\pa_{\m}\pa_{\m}-m^2)G(x)=\delta^{(4)}(x)
\ee
The static propagator satisfies instead
\be
(\sum_{i=1}^3\pa_i\pa_i-m^2)G_{stat}(\vec r)=\delta^{(3)}(\vec r)
\ee

\be
G_{stat}(\vec r)\equiv -\int {d^3\vec p\over (2\pi)^3}{e^{i\vec p\cdot\vec r}\over \vec p^2+m^2}=-{1\over 4\pi^2}\int{p^2 dp\over p^2+m^2}\int_0^{\pi}\sin \theta d\theta ~e^{ipr\cos\theta}=
\label{f5f}\ee
$$
=-{1\over 4\pi^2}\int{p^2 dp\over p^2+m^2}\int_{-1}^{1} du ~e^{ipru}=-{1\over 2\pi^2 ~r}\int_0^{\infty} {pdp\over p^2+m^2}\sin(pr)
$$

The last integral is ill defined and we will define it as
\be
I(\e)\equiv \int_0^{\infty} dp\left[{p\over p^2+m^2}-{1\over p}\right]\sin(pr)+\int_0^{\infty} {dp\over p}\sin(pr)~e^{-\e^2 p^2}
\label{f6f}\ee
$$
={\pi\over 2}\left(e^{-mr}-1\right)+{\pi\over 2}Erf\left[{r\over 2\e}\right]
$$
Clearly
\be
\lim_{\e\to 0^+}I(\e)=\int_0^{\infty} {pdp\over p^2+m^2}\sin(pr)
\label{f7f}\ee
Using $Erf[\infty]=1$ we finally obtain that for $r>0$,
\be
\int_0^{\infty} {pdp\over p^2+m^2}\sin(pr)={\pi\over 2}~e^{-mr}
\label{f8f}\ee
and
\be
G_{stat}(\vec r)=-{e^{-mr}\over 4\pi ~r}\sp r>0
\label{f9f}\ee
We also have from the definition
\be
\int d^3\vec r ~G_{stat}(\vec r)=-\int {d^3\vec p\over \vec p^2+m^2}\int {d^3\vec r\over (2\pi)^3} ~e^{i\vec p\cdot\vec r}=-\int {d^3\vec p\over \vec p^2+m^2}\delta^{(3)}(\vec p)=-{1\over m^2}
\label{f10f}\ee

On the other hand if we integrate (\ref{f9f}) we obtain
\be
\int d^3\vec r~G_{stat}(\vec r)=-\int d\Omega_2\int_0^{\infty}r^2dr~{e^{-mr}\over 4\pi ~r}=-\int_0^{\infty}rdr~e^{-mr}=-{1\over m^2}
\label{f11f}\ee
This agrees with (\ref{f10f}) and we conclude that there are no contact terms in (\ref{f9f}).

We now compute the full Euclidean propagator
\be
G( x)\equiv -\int {d^4 p\over (2\pi)^4}{e^{i\vec p\cdot x}\over  p^2+m^2}=-{1\over 4\pi^3}\int_0^{\infty}{p^3dp\over p^2+m^2}\int_0^{\pi}\sin^2 \theta d\theta ~e^{ip|x|\cos\theta}=
\label{f12f}\ee
$$
=-{1\over 4\pi^2}{1\over |x|}\int_0^{\infty}{p^2 dp\over p^2+m^2}J_1(p|x|)=-{1\over 4\pi^2|x|^2}\int_0^{\infty}{u^2 du\over u^2+m^2|x|^2}J_1(u)
$$

We now define the regulated integral
\be
I_1(\e)=\int_0^{\infty}du\left[{u^2 \over u^2+m^2|x|^2}-1\right]J_1(u)+\int_0^{\infty}du~e^{-\e^2 u^2}J_1(u)
\label{f13f}\ee
$$
=-m^2|x|^2\int_0^{\infty}{du\over u^2+m^2|x|^2}J_1(u)+\int_0^{\infty}du~e^{-\e^2 u^2}J_1(u)=
$$
$$
=\left(-1+m|x|K_1(m|x|)\right)+\left(1-e^{-{1\over 4\e^2}}\right)
$$

Therefore
\be
G(x)=-{1\over 4\pi^2|x|^2}\lim_{\e\to 0^+}I_1(\e)=-{m\over 4\pi^2|x|}K_1(m|x|)
\label{f14f}\ee
At large distances, $|x|\to\infty$
\be
G(x)\to -{\sqrt{m}\over 4\sqrt{2}(\pi|x|)^{3\over 2}}e^{-m|x|}
\label{f15f}\ee
while at short distances $|x|\to 0$
\be
G(x)\to -{1\over |x|^2}-{m^2\over 4}\left(2\gamma-1+2\log{|x|\over 2}\right)+{\cal O}(|x|^2)
\label{f16f}\ee

From (\ref{f12f}) we can calculate
\be
\int d^4x ~G(x)=-\int {d^4 p\over (2\pi)^4(  p^2+m^2)}
\int d^4x~e^{i\vec p\cdot x}=-\int {d^4 p\over   (p^2+m^2)}\delta^{(4)}(p)=-{1\over m^2}
\ee
We can also calculate it using (\ref{f14f}).
\be
\int d^4x ~G(x)=-\int d\Omega_3\int_0^{\infty} r^3 dr{m\over 4\pi^2 r}K_1(mr)=-{\Omega_3\over 4\pi^2 m^2}\int_0^{\infty} u^2 du K_1(u)=-{1\over m^2}
\ee
Therefore we do not expect any contact terms in (\ref{f14f})

We may now compute the static potential from the full propagator as
\be
G_{stat}(r)=\int_{-\infty}^{\infty} dt ~G(|x|)=\int_{-\infty}^{\infty} dt ~G(\sqrt{t^2+\vec r^2})=-{m\over 4\pi^2}\int_{-\infty}^{\infty} dt ~{K_1(m\sqrt{t^2+\vec r^2})\over \sqrt{t^2+\vec r^2}}
\ee
Changing variables to $u={t\over r}$ we obtain
\be
G_{stat}(r)=-{m\over 2\pi^2}\int_{0}^{\infty} du ~{K_1(mr\sqrt{u^2+1})\over \sqrt{u^2+1}}=-{m\over 2\pi^2}\int_{1}^{\infty} dv ~{K_1(mr v)\over \sqrt{v^2-1}}=-{e^{-mr}\over 4\pi r}
\ee
where in the last step we changed variables to $v=\sqrt{u^2+1}$.

We may now compute the derivatives that enter the two-point correlator
\be
G'= {m^4\over 8\pi^2 u^3}\left[K_1(u)-uK'_1(u)\right]\sp u\equiv m|x|
\ee
\be
G''={m^6\over 16\pi^2u^5}\left[-3K_1(u)+3uK'_1(u)-u^2K''_1(u)\right]
\ee

The long distance expansions are
\be
G^2\simeq {m^4\over (4\pi^2)^2}{\pi\over 2}{e^{-2u}\over u^3}+\cdots\sp (G')^2\simeq {m^8\over (8\pi^2)^2}{\pi\over 2}{e^{-2u}\over u^5}\left[1+{15\over 4u}+\cdots\right]
\ee
\be
{G'\over G}=-{m^2\over 2u}-{3m^2\over 4u^2}+\cdots
\ee
\be
{G''\over G'}=-{m^2\over 2u}-{5m^2\over 4u^2}+\cdots
\ee
\be
{(G'')^2\over (G')^2}={m^4\over 4u^2}+{5m^4\over 4u^3}+\cdots
\ee

The long-distance behavior of the traceless part of the correlator is
\be
\langle\bar T^{\m\n}(x)\bar T^{\r\s}(0)\rangle={m^3\over 32\pi^3}{e^{-2m|x|}\over |x|^5}\left[\left(\delta^{\m\r}\delta^{\n\s}+\delta^{\n\r}\delta^{\m\s}-{1\over 2}\delta^{\m\n}\delta^{\r\s}\right)+\right.
\ee
$$
-{4\over |x|^2}\left(\delta^{\m\r}x^{\n}x^{\s}+\delta^{\n\r}x^{\m}x^{\s}+\delta^{\m\s}x^{\n}x^{\r}+\delta^{\n\s}x^{\m}x^{\r}-\delta^{\r\s}x^{\m}x^{\n}-\delta^{\m\n}x^{\r}x^{\s}+{|x|^2\over 4}\delta^{\m\n}\delta^{\r\s}\right)m|x|
$$
$$+
\left.8\left({x^{\m}x^{\n}\over |x|^2}-{1\over 4}\delta^{\m\n}\right)\left({x^{\r}x^{\s}\over |x|^2}-{1\over 4}\delta^{\r\s}\right)m^2|x|^2\right]
$$
\be
\langle \Theta(x)\bar T^{\r\s}(0)\rangle=-{m^5\over 16\pi^3}{e^{-2m|x|}\over |x|^3}\left({x^{\r}x^{\s}\over |x|^2}-{1\over 4}\delta^{\r\s}\right)
\ee

\be
\langle \Theta(x)\Theta(0)\rangle={19 m^5\over 16\pi^3}{e^{-2m|x|}\over |x|^3}+\cdots
\ee

Therefore the leading behaviour of all correlators is captured by ${m^5\over |x|^3}e^{-2m|x|}$.

\subsection{The static potential}\label{freestaticpotential}

We start from
\be
\langle T_{00}(x)T_{00}(0)\rangle_{conn}=8(G')^2+{1\over 2}\left[m^4 G^2+8m^2(t^2+r^2)(G')^2+16 (t^2+r^2) G'G''+\right.
\label{f4fff}\ee
$$
\left.+16 (t^2+r^2)^2(G'')^2\right]+32t^2 r^2(G'')^2
$$
The static potential is proportional to
\be
V_{stat}(r)=\int_{-\infty}^{+\infty} dt\langle T_{00}(t,\vec r)T_{00}(0)\rangle_{conn}
\ee
We first change variable to $v=t^2+r^2$
\be
V_{stat}(r)=2\int_{r^2}^{+\infty} {dv\over \sqrt{v-r^2}}\langle T_{00}(v-r^2,\vec r)T_{00}(0)\rangle_{conn}=
\ee
$$
=2\int_{r^2}^{+\infty} {dv\over \sqrt{v-r^2}}\left[{m^4\over 2}G^2(v)+4(2+m^2 v)G'^2+8v G'G''+32r^2(v-r^2)(G'')^2\right]
$$
$$
={2m^2\over (2\pi)^4}\int_{r^2}^{+\infty} {dv\over \sqrt{v-r^2}}\left[-{m^2\over 4v^4}(128r^4-128r^2v+13v^2)K_{0}(m\sqrt{v})^2-\right.
$$
$$
-{m\over v^{9\over 2}}(128 r^4-128r^2 v+16m^2r^4 v+13 v^2-16 m^2r^2v^2+m^2v^3)K_0(m\sqrt{v})K_1(m\sqrt{v})-
$$
$$
-{1\over 2v^5}(256r^4-256 r^2v+64 m^2r^4v+26v^2-64 m^2 r^2 v^2+4m^4 r^2 v^2+4m^2v^3-4m^4r^2v^3-m^4v^4)K_{1}(m\sqrt{v})^2+
$$
$$
\left.+{m^2\over 4v^2}(5+m^2v)K_{2}(m\sqrt{v})^2\right]
$$
We now change variables to $v=u^2r^2$ to obtain
\be
V_{stat}(r)
={4m^2r\over (2\pi)^4}\int_{1}^{+\infty} {udu\over \sqrt{u^2-1}}\left[
-{m^2\over 4r^4 u^8}(128-128 u^2+13 u^4)K_0(mru)^2-\right.
\ee
$$
-{m\over r^5u^9}(128 - 128 u^2 + 16 m^2 r^2 u^2 + 13 u^4 - 16 m^2 r^2 u^4 + m^2 r^2 u^6)K_0(mru)K_1(mru)+
$$
$$+
{ (256 (u^2-1) - 26 u^4 +
   64 m^2 r^2 (u^4-u^2) - 4 m^2 r^2 u^6 + 4 m^4 r^4 (u^6-u^4) +
   m^4 r^4 u^8)\over 2 r^6 u^{10}}K_1(mru)^2+
   $$
$$
\left.+{m^2\over r^4 u^4}(5+4m^2r^2u^2)K_2(mru)^2\right] \, .
$$
We can perform the integrals in terms of the Meijer-$G$ function. At short distances $r\to 0$  we obtain
\be
V_{stat}(r)\simeq {1\over 8\pi^3 r^7}
\ee
while at long distances $r\to\infty$
\be
V_{stat}(r)\simeq {3 m^{9\over 2}\over (\pi r)^{5\over 2}}~e^{-2mr} \, .
\ee
This is the expected behaviour for the exchange of a massive particle.

\vskip 1cm

%

%%%%%%%%%%%%%%%%%%%%%%%%%%%%%%%%%%%%%%


\addcontentsline{toc}{section}{References}

\begin{thebibliography}{199}

\bibitem{nielsen}
H. B. Nielsen, {\em Do we need fundamental laws of nature?}, Gamma 36 and 37 (1979);\\
  D.~Forster, H.~B.~Nielsen and M.~Ninomiya,
  {\em ``Dynamical Stability of Local Gauge Symmetry: Creation of Light from Chaos,''}
  \href{http://www.doi.org/10.1016/0370-2693(80)90842-4}{Phys.\ Lett.\  {\bf 94B} (1980) 135};\\
  D.~L.~Bennett, N.~Brene and H.~B.~Nielsen,
  {\em ``Random Dynamics,''}
  \href{http://www.doi.org/10.1088/0031-8949/1987/T15/022}{Phys.\ Scripta T {\bf 15} (1987) 158}.



\bibitem{Sindoni}
  L.~Sindoni,
  {\em ``Emergent Models for Gravity: an Overview of Microscopic Models,''}
  SIGMA {\bf 8} (2012) 027;
  \hri{1110.0686}{[gr-qc]}.
  %%CITATION = doi:10.3842/SIGMA.2012.027;%%
  %44 citations counted in INSPIRE as of 25 Apr 2018

\bibitem{Carlip}
  S.~Carlip,
  {\em ``Challenges for Emergent Gravity,''}
  Stud.\ Hist.\ Phil.\ Sci.\ B {\bf 46} (2014) 200;
  \hri{1207.2504}{[gr-qc]}.
  %%CITATION = doi:10.1016/j.shpsb.2012.11.002;%%
  %31 citations counted in INSPIRE as of 26 Apr 2018

  \bibitem{Dreyer}
  O.~Dreyer,
  {\em ``Emergent general relativity,''}
  In *Oriti, D. (ed.): Approaches to quantum gravity* 99-110
  \hre{gr-qc}{0604075}.
  %%CITATION = GR-QC/0604075;%%
  %36 citations counted in INSPIRE as of 18 Feb 2019

  \bibitem{ohanian} H. C. Ohanian, {\em ``Gravitons as goldstone bosons,"} Phys. Rev. 184, 1305 (1969).


\bibitem{Isham}
  C.~J.~Isham, A.~Salam and J.~A.~Strathdee,
  {\em ``F-dominance of gravity,''}
 \href{http://www.doi.org/10.1103/PhysRevD.3.867}{Phys.\ Rev.\ D {\bf 3} (1971) 867}.

  %%CITATION = doi:10.1103/PhysRevD.3.867;%%
  %292 citations counted in INSPIRE as of 18 Feb 2019

\bibitem{akama}  H. Terazawa, K. Akama and Y. Chikashige, {\em ``Unified Model of the Nambu-Jona-Lasinio Type
for All Elementary Particle Forces,"} Phys. Rev. D 15, 480 (1977);\\
K. Akama, Y. Chikashige,
T. Matsuki and H. Terazawa, {\em ``Gravity and Electromagnetism as Collective Phenomena: A
Derivation of Einstein's General Relativity,"},  Prog. Theor. Phys. 60, 868 (1978);\\
K. Akama, {\em ``An Attempt at Pregeometry: Gravity With Composite Metric," } Prog. Theor.
Phys. 60, 1900 (1978);\\
K.~Akama and T.~Hattori,
  {\em ``Dynamical Foundations of the Brane Induced Gravity,''}
  Class.\ Quant.\ Grav.\  {\bf 30} (2013) 205002;
  \hri{1309.3090}{[gr-qc]}.
  %%CITATION = doi:10.1088/0264-9381/30/20/205002;%%
  %8 citations counted in INSPIRE as of 25 Apr 2018


 \bibitem{Vene}
 D. Amati and G. Veneziano, {\em ``Metric From Matter,"} Phys. Lett. B
105, 358 (1981);\\
D. Amati and G. Veneziano, {\em ``A Unified Gauge and Gravity Theory With Only Matter Fields,"},
Nucl. Phys. B 204, 451 (1982).


  \bibitem{Zee}
  A.~Zee,
  {\em ``Einstein Gravity Emerging From Quantum Weyl Gravity,''}
  Annals Phys.\  {\bf 151} (1983) 431.
  %%CITATION = doi:10.1016/0003-4916(83)90286-5;%%
  %48 citations counted in INSPIRE as of 26 Apr 2018

      \bibitem{Visser2}
  C.~Barcelo, M.~Visser and S.~Liberati,
  {\em ``Einstein gravity as an emergent phenomenon?,''}
  Int.\ J.\ Mod.\ Phys.\ D {\bf 10} (2001) 799;
  \hre{gr-qc}{0106002}.
  %%CITATION = doi:10.1142/S0218271801001591;%%
  %77 citations counted in INSPIRE as of 26 Apr 2018

  \bibitem{Hebecker}
  A.~Hebecker and C.~Wetterich,
  {\em ``Spinor gravity,''}
  Phys.\ Lett.\ B {\bf 574} (2003) 269
  doi:10.1016/j.physletb.2003.09.010
  \hre{hep-th}{0307109}.
  %%CITATION = doi:10.1016/j.physletb.2003.09.010;%%
  %39 citations counted in INSPIRE as of 18 Feb 2019

    \bibitem{Jenkins}
  A.~Jenkins,
  {\em ``Topics in theoretical particle physics and cosmology beyond the standard model,''}
  \hre{hep-th}{0607239};\\
  %%CITATION = HEP-TH/0607239;%%
  %10 citations counted in INSPIRE as of 18 Feb 2019
  {\em ``Constraints on emergent gravity,''}
  Int.\ J.\ Mod.\ Phys.\ D {\bf 18} (2009) 2249;
  \hri{0904.0453}{[gr-qc]}.
  %%CITATION = doi:10.1142/S0218271809015941;%%
  %37 citations counted in INSPIRE as of 18 Feb 2019
  
  \bibitem{Lee}
S.~S.~Lee,
{\em ``Holographic description of quantum field theory,''}
\hrj{10.1016/j.nuclphysb.2010.02.022}{Nucl. Phys. B \textbf{832} (2010), 567-585};
\hri{0912.5223}{ [hep-th]};
%49 citations counted in INSPIRE as of 24 Feb 2023
{\em ``A model of quantum gravity with emergent spacetime,''}
\hrj{10.1007/JHEP06(2020)070}{JHEP \textbf{06} (2020), 070}
\hri{1912.12291 }{[hep-th]};
%2 citations counted in INSPIRE as of 24 Feb 2023
%38 citations counted in INSPIRE as of 24 Feb 2023
{\em ``Background independent holographic description : From matrix field theory to quantum gravity,''}
\hrj{10.1007/JHEP10(2012)160}{JHEP \textbf{10} (2012), 160};
\hri{1204.1780}{ [hep-th]}.
%50 citations counted in INSPIRE as of 24 Feb 2023


\bibitem{Verlinde}
  E.~P.~Verlinde,
  {\em ``On the Origin of Gravity and the Laws of Newton,''}
  JHEP {\bf 1104} (2011) 029;
  \hri{1001.0785}{[hep-th]};\\
  %%CITATION = doi:10.1007/JHEP04(2011)029;%%
  %685 citations counted in INSPIRE as of 25 Apr 2018
  E.~P.~Verlinde,
  {\em ``Emergent Gravity and the Dark Universe,''}
  SciPost Phys.\  {\bf 2} (2017) no.3,  016;
  \hri{1611.02269}{[hep-th]}.
  %%CITATION = doi:10.21468/SciPostPhys.2.3.016;%%
  %101 citations counted in INSPIRE as of 25 Apr 2018

    \bibitem{Mukohyama}
  S.~Mukohyama,
  {\em ``Emergence of time in power-counting renormalizable Riemannian theory of gravity,''}
  Phys.\ Rev.\ D {\bf 87} (2013) no.8,  085030
  \hri{1303.1409}{[hep-th]}.
  %%CITATION = doi:10.1103/PhysRevD.87.085030;%%
  %4 citations counted in INSPIRE as of 26 Apr 2018


   \bibitem{SMGRAV}
  E.~Kiritsis,
  {\em ``Gravity and axions from a random UV QFT,''}s
  EPJ Web Conf.\  {\bf 71} (2014) 00068;
  \hri{1408.3541}{[hep-ph]}.
  %%CITATION = doi:10.1051/epjconf/20147100068;%%
  %2 citations counted in INSPIRE as of 25 Apr 2018


\bibitem{CKN}
  C.~Charmousis, E.~Kiritsis and F.~Nitti,
  {\em ``Holographic self-tuning of the cosmological constant,''}
  JHEP {\bf 1709} (2017) 031;
  \hri{1704.05075}{[hep-th]}.
  %%CITATION = doi:10.1007/JHEP09(2017)031;%%
  %4 citations counted in INSPIRE as of 25 Apr 2018


\bibitem{Carone1}
  C.~D.~Carone, T.~V.~B.~Claringbold and D.~Vaman,
  {\em ``Composite graviton self-interactions in a model of emergent gravity,''}
  \hri{1710.09367}{[hep-th]};\\
  %%CITATION = ARXIV:1710.09367;%%
  %1 citations counted in INSPIRE as of 25 Apr 2018
  C.~D.~Carone, J.~Erlich and D.~Vaman,
  {\em ``Emergent Gravity from Vanishing Energy-Momentum Tensor,''}
  JHEP {\bf 1703} (2017) 134;
  \hri{1610.08521}{[hep-th]}.
  %%CITATION = doi:10.1007/JHEP03(2017)134;%%
  %3 citations counted in INSPIRE as of 25 Apr 2018


  \bibitem{Sakharov}  A. D. Sakharov, {\em `` Vacuum quantum
uctuations in curved space and the theory of gravitation},"
Sov. Phys. Dokl. 12, 1040 (1968) [Dokl. Akad. Nauk Ser. Fiz. 177, 70 (1967)] [Sov. Phys. Usp.
34, 394 (1991)] [Gen. Rel. Grav. 32, 365 (2000)].



  \bibitem{Adler}
  S.~L.~Adler,
  {\em ``Einstein Gravity as a Symmetry Breaking Effect in Quantum Field Theory,''}
  Rev.\ Mod.\ Phys.\  {\bf 54} (1982) 729
   Erratum: [Rev.\ Mod.\ Phys.\  {\bf 55} (1983) 837].
  %%CITATION = doi:10.1103/RevModPhys.54.729;%%
  %574 citations counted in INSPIRE as of 26 Apr 2018

\bibitem{Visser}
  M.~Visser,
  {\em ``Sakharov's induced gravity: A Modern perspective,''}
  Mod.\ Phys.\ Lett.\ A {\bf 17} (2002) 977;
  \hre{gr-qc}{0204062}.
  %%CITATION = doi:10.1142/S0217732302006886;%%
  %125 citations counted in INSPIRE as of 25 Apr 2018


\bibitem{volovik}  Volovik, G. (2001). {\em Super
uid analogies of cosmological phenomena},
\href{https://www.sciencedirect.com/science/article/pii/S0370157300001393?}
{Physics Reports, 351 (4), 195-348}.

\bibitem{wen}
  M.~A.~Levin and X.~G.~Wen,
  {\em ``Colloquium: Photons and electrons as emergent phenomena,''}
  Rev.\ Mod.\ Phys.\  {\bf 77} (2005) 871
  doi:10.1103/RevModPhys.77.871
  \hre{cond-mat}{0407140}];\\
  %%CITATION = doi:10.1103/RevModPhys.77.871;%%
  %63 citations counted in INSPIRE as of 18 Feb 2019
  Z.~C.~Gu and X.~G.~Wen,
  {\em ``A Lattice bosonic model as a quantum theory of gravity,''}
  \hre{gr-qc}{0606100}.
  %%CITATION = GR-QC/0606100;%%
  %28 citations counted in INSPIRE as of 18 Feb 2019

 \bibitem{Horava}
  C.~Xu and P.~Horava,
  {\em ``Emergent Gravity at a Lifshitz Point from a Bose Liquid on the Lattice,''}
  Phys.\ Rev.\ D {\bf 81} (2010) 104033;
  \hri{1003.0009}{ [hep-th]}.
  %%CITATION = doi:10.1103/PhysRevD.81.104033;%%
  %27 citations counted in INSPIRE as of 26 Apr 2018

    \bibitem{SSLee}
  S.~S.~Lee,
  {\em ``Emergent gravity from relatively local Hamiltonians and a possible resolution of the black hole information puzzle,''}
  \hri{1803.00556}{[hep-th]}.
  %%CITATION = ARXIV:1803.00556;%%



%\cite{Kiritsis:2007zza}
\bibitem{book}
  E.~Kiritsis,
  {\em ``String theory in a nutshell,''}
\href{https://press.princeton.edu/titles/133767.html}{second edition,   Princeton University Press, 2019}.
  %65 citations counted in INSPIRE as of 18 Feb 2019

  \bibitem{gross}
  D.~J.~Gross and P.~F.~Mende,
  {\em ``String Theory Beyond the Planck Scale,''}
 \href{http://www.doi.org/10.1016/0550-3213(88)90390-2}{Nucl.\ Phys.\ B {\bf 303} (1988) 407}.

\bibitem{vene2}
  D.~Amati, M.~Ciafaloni and G.~Veneziano,
  {\em ``Can Space-Time Be Probed Below the String Size?,''}
  \href{http://www.doi.org/10.1016/0370-2693(89)91366-X}{Phys.\ Lett.\ B {\bf 216} (1989) 41}.

\bibitem{Komar}
  Z.~Komargodski,
  {\em ``Vector Mesons and an Interpretation of Seiberg Duality,''}
  JHEP {\bf 1102} (2011) 019
  doi:10.1007/JHEP02(2011)019
  \hri{1010.4105}{[hep-th]}.
  %%CITATION = doi:10.1007/JHEP02(2011)019;%%
  %36 citations counted in INSPIRE as of 18 Feb 2019

\bibitem{Stelle}
  K.~S.~Stelle,
  {\em ``Renormalization of Higher-Derivative Quantum Gravity,''}
 \href{http://www. doi.org/10.1103/PhysRevD.16.953}{Phys.\ Rev.\ D {\bf 16} (1977) 953}.
  %%CITATION = doi:10.1103/PhysRevD.16.953;%%
  %1637 citations counted in INSPIRE as of 18 Feb 2019



\bibitem{Tomboulis}
  E.~T.~Tomboulis,
  {\em ``Unitarity in Higher Derivative Quantum Gravity,''}
  \href{http://www.doi.org/10.1103/PhysRevLett.52.1173}{Phys.\ Rev.\ Lett.\  {\bf 52} (1984) 1173}.
  %%CITATION = doi:10.1103/PhysRevLett.52.1173;%%
  %107 citations counted in INSPIRE as of 18 Feb 2019

  \bibitem{FKL}
  S.~Ferrara, A.~Kehagias and D.~Lust,
  {\em ``Bimetric, Conformal Supergravity and its Superstring Embedding,''}
  JHEP {\bf 1905} (2019) 100
  doi:10.1007/JHEP05(2019)100
\hri{1810.08147}{[hep-th]}.
  %%CITATION = doi:10.1007/JHEP05(2019)100;%%
  %5 citations counted in INSPIRE as of 27 Jun 2019

 \bibitem{strumia}
  A.~Salvio, A.~Strumia and H.~Veermae,
  {\em ``New infra-red enhancements in 4-derivative gravity,''}
  Eur.\ Phys.\ J.\ C {\bf 78} (2018) no.10,  842
  doi:10.1140/epjc/s10052-018-6311-1
  \hri{1808.07883}{[hep-th]}.
  %%CITATION = doi:10.1140/epjc/s10052-018-6311-1;%%



  \bibitem{anselmi}
  D.~Anselmi and M.~Piva,
  {\em ``Quantum Gravity, Fakeons And Microcausality,''}
  JHEP {\bf 1811} (2018) 021
  doi:10.1007/JHEP11(2018)021
  \hri{1806.03605}{[hep-th]}.
  %%CITATION = doi:10.1007/JHEP11(2018)021;%%
  %6 citations counted in INSPIRE as of 18 Feb 2019



  \bibitem{love}
  C.~Lovelace,
  {\em ``Stability of String Vacua. 1. A New Picture of the Renormalization Group,''}
  \href{http://www.doi.org/10.1016/0550-3213(86)90253-1}{Nucl.\ Phys.\ B {\bf 273} (1986) 413}.
  %%CITATION = doi:10.1016/0550-3213(86)90253-1;%%
  %275 citations counted in INSPIRE as of 20 Feb 2019

  \bibitem{novel}
  E.~Kiritsis,
  {\em ``On novel string theories from 4d gauge theories,''}
  EPJ Web Conf.\  {\bf 70} (2014) 00040
  doi:10.1051/epjconf/20147000040
 \hri{1301.6810}{[hep-th]}.
  %%CITATION = doi:10.1051/epjconf/20147000040;%%
  %5 citations counted in INSPIRE as of 20 Feb 2019


  \bibitem{Seiberg}
  N.~Seiberg,
  {\em ``Emergent space-time,''} in the proceedings of the 23rd Solvay Conference on Physics. Brussels, Belgium. 1 - 3 December 2005
  \hre{hep-th}{0601234}.
  %%CITATION = doi:10.1142/9789812706768_0005;%%

    \bibitem{Berenstein}
  D.~Berenstein,
  {\em``Sketches of emergent geometry in the gauge/gravity duality,''}
  Fortsch.\ Phys.\  {\bf 62} (2014) 776;
  \hri{1404.7052}{[hep-th]}.
  %%CITATION = doi:10.1002/prop.201400026;%%
  %15 citations counted in INSPIRE as of 26 Apr 2018

    \bibitem{Nitti}
  F.~Nitti,
  {\em ``Holography and Emergent 4D Gravity,''}
  Mod.\ Phys.\ Lett.\ A {\bf 23} (2008) 289;
  \hri{0801.4793}{[hep-th]}.
  %%CITATION = doi:10.1142/S021773230802642X;%%
  %1 citations counted in INSPIRE as of 26 Apr 2018


  \bibitem{Dvali}
  G.~Dvali, G.~F.~Giudice, C.~Gomez and A.~Kehagias,
  {\em ``UV-Completion by Classicalization,''}
  JHEP {\bf 1108} (2011) 108
  doi:10.1007/JHEP08(2011)108
\hri{1010.1415}{[hep-ph]}.
  %%CITATION = doi:10.1007/JHEP08(2011)108;%%
  %173 citations counted in INSPIRE as of 27 Jun 2019


   \bibitem{shiraz}
  S.~Bhattacharyya, V.~E.~Hubeny, S.~Minwalla and M.~Rangamani,
  {\em ``Nonlinear Fluid Dynamics from Gravity,''}
  JHEP {\bf 0802} (2008) 045
  doi:10.1088/1126-6708/2008/02/045
\hri{0712.2456}{[hep-th]}.
  %%CITATION = doi:10.1088/1126-6708/2008/02/045;%%
  %801 citations counted in INSPIRE as of 19 Feb 2019

  \bibitem{black}
R.~Emparan, T.~Harmark, V.~Niarchos and N.~A.~Obers,
{\em ``World-Volume Effective Theory for Higher-Dimensional Black Holes,''}
\hrj{10.1103/PhysRevLett.102.191301}{Phys. Rev. Lett. \textbf{102} (2009), 191301};
\hri{0902.0427}{[hep-th]}.

 \bibitem{WW} S. Weinberg and E. Witten, {\em
   ``Limits on Massless Particles}," Phys. Lett. B 96, (1980) 59.


   \bibitem{liu}
  H.~Liu and P.~Glorioso,
  {\em ``Lectures on non-equilibrium effective field theories and fluctuating hydrodynamics,''}
  PoS TASI {\bf 2017} (2018) 008
  doi:10.22323/1.305.0008
  \hri{1805.09331}{[hep-th]}.
  %%CITATION = doi:10.22323/1.305.0008;%%
  %7 citations counted in INSPIRE as of 19 Feb 2019


  \bibitem{rangamani}
  F.~M.~Haehl, R.~Loganayagam and M.~Rangamani,
  {\em ``Topological sigma models and dissipative hydrodynamics,''}
  JHEP {\bf 1604} (2016) 039
  doi:10.1007/JHEP04(2016)039
  \hri{1511.07809}{[hep-th]}.
  %%CITATION = doi:10.1007/JHEP04(2016)039;%%
  %42 citations counted in INSPIRE as of 19 Feb 2019



   \bibitem{seiberg}
  N.~Seiberg,
  {\em ``Exact results on the space of vacua of four-dimensional SUSY gauge theories,''}
  Phys.\ Rev.\ D {\bf 49} (1994) 6857
  doi:10.1103/PhysRevD.49.6857
  \hre{hep-th}{9402044}.
  %%CITATION = doi:10.1103/PhysRevD.49.6857;%%
  %732 citations counted in INSPIRE as of 19 Feb 2019





\bibitem{malda}
  J.~M.~Maldacena,
  {\em ``The Large N limit of superconformal field theories and supergravity,''}
  Int.\ J.\ Theor.\ Phys.\  {\bf 38} (1999) 1113
   [Adv.\ Theor.\ Math.\ Phys.\  {\bf 2} (1998) 231];
  \hre{hep-th}{9711200}.
  %%CITATION = doi:10.1023/A:1026654312961, 10.4310/ATMP.1998.v2.n2.a1;%%
  %13692 citations counted in INSPIRE as of 21 May 2018


 \bibitem{e1}
  E.~Kiritsis,
  {\em ``Product CFTs, gravitational cloning, massive gravitons and the space of gravitational duals,''}
  JHEP {\bf 0611} (2006) 049
  doi:10.1088/1126-6708/2006/11/049;
  \hre{hep-th}{0608088}.
  %%CITATION = doi:10.1088/1126-6708/2006/11/049;%%
  %52 citations counted in INSPIRE as of 18 Jun 2018

  \bibitem{a1}
  O.~Aharony, A.~B.~Clark and A.~Karch,
  {\em ``The CFT/AdS correspondence, massive gravitons and a connectivity index conjecture,''}
  Phys.\ Rev.\ D {\bf 74} (2006) 086006
  doi:10.1103/PhysRevD.74.086006
  \hre{hep-th}{0608089}.
  %%CITATION = doi:10.1103/PhysRevD.74.086006;%%
  %49 citations counted in INSPIRE as of 18 Jun 2018


\bibitem{nitti2}
E.~Kiritsis and F.~Nitti,
{\em ``On massless 4D gravitons from asymptotically AdS$_5$ space-times,''}
\hrj{10.1016/j.nuclphysb.2007.02.024}{Nucl. Phys. B \textbf{772} (2007), 67-102};
\hre{hep-th}{0611344}.
%46 citations counted in INSPIRE as of 06 Oct 2020

  \bibitem{d4}
  E.~Witten,
  {\em ``Anti-de Sitter space, thermal phase transition, and confinement in gauge theories,''}
  Adv.\ Theor.\ Math.\ Phys.\  {\bf 2} (1998) 505
  doi:10.4310/ATMP.1998.v2.n3.a3
  \hre{hep-th}{9803131}.
  %%CITATION = doi:10.4310/ATMP.1998.v2.n3.a3;%%
  %2837 citations counted in INSPIRE as of 20 Feb 2019

  \bibitem{ihqcd}
  U.~Gursoy and E.~Kiritsis,
  {\em ``Exploring improved holographic theories for QCD: Part I,''}
  JHEP {\bf 0802} (2008) 032
  doi:10.1088/1126-6708/2008/02/032
  \hri{0707.1324}{[hep-th]};\\
  %%CITATION = doi:10.1088/1126-6708/2008/02/032;%%
  %330 citations counted in INSPIRE as of 20 Feb 2019
  U.~Gursoy, E.~Kiritsis and F.~Nitti,
  {\em ``Exploring improved holographic theories for QCD: Part II,''}
  JHEP {\bf 0802} (2008) 019
  doi:10.1088/1126-6708/2008/02/019
  \hri{0707.1349}{[hep-th]}.
  %%CITATION = doi:10.1088/1126-6708/2008/02/019;%%
  %339 citations counted in INSPIRE as of 20 Feb 2019


  \bibitem{stein}
  H.~Steinacker,
  {\em ``Emergent Geometry and Gravity from Matrix Models: an Introduction,''}
  Class.\ Quant.\ Grav.\  {\bf 27} (2010) 133001;
  \hri{1003.4134}{[hep-th]}.
  %%CITATION = doi:10.1088/0264-9381/27/13/133001;%%
  %136 citations counted in INSPIRE as of 18 Feb 2019



  \bibitem{bere}
  D.~E.~Berenstein, M.~Hanada and S.~A.~Hartnoll,
  {\em ``Multi-matrix models and emergent geometry,''}
  JHEP {\bf 0902} (2009) 010
  doi:10.1088/1126-6708/2009/02/010
  \hri{0805.4658}{[hep-th]}.
  %%CITATION = doi:10.1088/1126-6708/2009/02/010;%%
  %51 citations counted in INSPIRE as of 20 Feb 2019

  \bibitem{kir}
  E.~Kiritsis,
  {\em ``Dissecting the string theory dual of QCD,''}
  Fortsch.\ Phys.\  {\bf 57} (2009) 396
  doi:10.1002/prop.200900011
  \hri{0901.1772}{[hep-th]}.
  %%CITATION = doi:10.1002/prop.200900011;%%
  %53 citations counted in INSPIRE as of 20 Feb 2019


  \bibitem{porrati}
  M.~Porrati,
  {\em ``Higgs phenomenon for 4-D gravity in anti-de Sitter space,''}
  JHEP {\bf 0204} (2002) 058
  doi:10.1088/1126-6708/2002/04/058
  \hre{hep-th}{0112166}.
  %%CITATION = doi:10.1088/1126-6708/2002/04/058;%%
  %88 citations counted in INSPIRE as of 20 Jun 2018



  \bibitem{axion}
  P.~Anastasopoulos, P.~Betzios, M.~Bianchi, D.~Consoli and E.~Kiritsis,
  {\em ``Emergent/Composite axions,''}
  \hri{1811.05940}{[hep-ph]}.
  %%CITATION = ARXIV:1811.05940;%%
  %1 citations counted in INSPIRE as of 21 Feb 2019


   %\cite{u1}
\bibitem{u1}
P.~Betzios, E.~Kiritsis, V.~Niarchos and O.~Papadoulaki,
{\em ``Global symmetries, hidden sectors and emergent (dark) vector interactions,''}
\hri{2006.01840}{ [hep-ph]}.
%1 citations counted in INSPIRE as of 30 Sep 2020

\bibitem{u12}
P.~Anastasopoulos, M.~Bianchi, D.~Consoli and E.~Kiritsis,
{\em ``String (gravi)photons, ''dark brane photons'', holography and the
hypercharge portal,''}
\hri{2010.07320}{[hep-ph]}.
%0 citations counted in INSPIRE as of 27 Oct 2020


  \bibitem{csaki}
  C.~Csaki, C.~Grojean, L.~Pilo and J.~Terning,
  {\em ``Towards a realistic model of Higgsless electroweak symmetry breaking,''}
  Phys.\ Rev.\ Lett.\  {\bf 92} (2004) 101802
  doi:10.1103/PhysRevLett.92.101802
  \hre{hep-ph}{0308038}.
  %%CITATION = doi:10.1103/PhysRevLett.92.101802;%%
  %500 citations counted in INSPIRE as of 21 Feb 2019



  \bibitem{Marolf}
  D.~Marolf,
  {\em ``Emergent Gravity Requires Kinematic Nonlocality,''}
  Phys.\ Rev.\ Lett.\  {\bf 114} (2015) no.3,  031104;
  \hri{1409.2509}{[hep-th]}.
  %%CITATION = doi:10.1103/PhysRevLett.114.031104;%%
  %18 citations counted in INSPIRE as of 26 Apr 2018

  %\cite{Hinterbichler:2011tt}
\bibitem{Hi}
K.~Hinterbichler,
{\em ``Theoretical Aspects of Massive Gravity,''}
\hrj{10.1103/RevModPhys.84.671}{Rev. Mod. Phys. \textbf{84} (2012), 671-710}\\
\hri{1105.3735}{[hep-th]}.
%769 citations counted in INSPIRE as of 05 Oct 2020

\bibitem{dR}
C.~de Rham,
{\em ``Massive Gravity,''}
\hrj{10.12942/lrr-2014-7}{Living Rev. Rel. \textbf{17} (2014), 7};

\hri{1401.4173}{ [hep-th]}.
%663 citations counted in INSPIRE as of 05 Oct 2020


\bibitem{TT}
S.~Dubovsky, R.~Flauger and V.~Gorbenko,
{\em ``Solving the Simplest Theory of Quantum Gravity,''}
JHEP \textbf{09} (2012), 133
doi:10.1007/JHEP09(2012)133
\hri{1205.6805}{[hep-th]};\\
%101 citations counted in INSPIRE as of 27 May 2020
F.~Smirnov and A.~Zamolodchikov,
{\em ``On space of integrable quantum field theories,''}
Nucl. Phys. B \textbf{915} (2017), 363-383
doi:10.1016/j.nuclphysb.2016.12.014
\hri{1608.05499}{[hep-th]};\\
%157 citations counted in INSPIRE as of 27 May 2020
A.~Cavagli\`a, S.~Negro, I.~M.~Sz\'ecs\'enyi and R.~Tateo,
{\em ``$T \bar{T}$-deformed 2D Quantum Field Theories,''}
\hrj{10.1007/JHEP10(2016)112}{JHEP \textbf{10} (2016), 112};
\hri{1608.05534}{[hep-th]};\\
%162 citations counted in INSPIRE as of 15 Oct 2020
L.~McGough, M.~Mezei and H.~Verlinde,
``Moving the CFT into the bulk with $ T\overline{T} $,''
JHEP \textbf{04} (2018), 010
doi:10.1007/JHEP04(2018)010
\hri{1611.03470}{[hep-th]}.
%136 citations counted in INSPIRE as of 27 May 2020




%\cite{Tolley:2019nmm}
\bibitem{To}
A.~J.~Tolley,
{\em ``$T \bar T$ Deformations, Massive Gravity and Non-Critical Strings,''}
\hrj{10.1007/JHEP06(2020)050}{JHEP \textbf{06} (2020), 050};
\hri{1911.06142}{[hep-th]}.
%10 citations counted in INSPIRE as of 27 May 2020


\bibitem{Tay}
M.~Taylor,
{\em ``TT deformations in general dimensions,''}
\hri{1805.10287}{[hep-th]}.
%64 citations counted in INSPIRE as of 27 May 2020

%\cite{Betzios:2020zaj}
\bibitem{Betzios:2020zaj}
P.~Betzios and O.~Papadoulaki,
{\em ``Brane Cosmology and the self-tuning of the cosmological constant in the presence of bulk black holes,''}
\hrj{10.1140/epjc/s10052-020-8185-2}{Eur. Phys. J. C \textbf{80} (2020) no.7, 660}
\hri{2003.05767}{ [hep-th]}.
%0 citations counted in INSPIRE as of 27 Oct 2020


%\cite{deRham:2010kj}
\bibitem{dRGT}
C.~de Rham, G.~Gabadadze and A.~J.~Tolley,
{\em ``Resummation of Massive Gravity,''}
\hrj{10.1103/PhysRevLett.106.231101}{Phys. Rev. Lett. \textbf{106} (2011), 231101};
\hri{1011.1232}{ [hep-th]};\\
%1327 citations counted in INSPIRE as of 14 Sep 2020
S.~F.~Hassan and R.~A.~Rosen,
{\em ``Resolving the Ghost Problem in non-Linear Massive Gravity,''}
\hrj{10.1103/PhysRevLett.108.041101}{Phys. Rev. Lett. \textbf{108} (2012), 041101};
\hri{1106.3344}{[hep-th]}.
%648 citations counted in INSPIRE as of 14 Sep 2020



\bibitem{GKNW}
  J.~K.~Ghosh, E.~Kiritsis, F.~Nitti and L.~T.~Witkowski,
  {\em ``De Sitter and Anti-de Sitter branes in self-tuning models,''}
  \hree{10.1007/JHEP11(2018)128}{JHEP {\bf 1811} (2018) 128},
  \hri{1807.09794}{[hep-th]}.
  %%CITATION = doi:10.1007/JHEP11(2018)128;%%
  %12 citations counted in INSPIRE as of 19 Aug 2019

\bibitem{self-cosmo}
  A.~Amariti, C.~Charmousis, D.~Forcella, E.~Kiritsis and F.~Nitti,
  {\em ``Brane cosmology and the self-tuning of the cosmological constant,''}
  \hri{1904.02727}{[hep-th]}.
  %%CITATION = ARXIV:1904.02727;%%
  %1 citations counted in INSPIRE as of 19 Aug 2019

  \bibitem{HKNW}
Y.~Hamada, E.~Kiritsis, F.~Nitti and L.~T.~Witkowski,
{\em ``The self-tuning of the cosmological constant and the holographic relaxion,''}
\hri{2001.05510}{ [hep-th]}.
%2 citations counted in INSPIRE as of 15 Sep 2020


\bibitem{DGP}
  G.~R.~Dvali, G.~Gabadadze and M.~Porrati,
  {\em ``4-D gravity on a brane in 5-D Minkowski space,''}
  \hree{10.1016/S0370-2693(00)00669-9}{Phys.\ Lett.\ B {\bf 485} (2000) 208},
  \hre{hep-th}{0005016}.
  %%CITATION = doi:10.1016/S0370-2693(00)00669-9;%%
  %2716 citations counted in INSPIRE as of 19 Aug 2019


\bibitem{KTT}
  E.~Kiritsis, N.~Tetradis and T.~N.~Tomaras,
  {\em ``Induced gravity on RS branes,''}
  \hree{10.1088/1126-6708/2002/03/019J}{HEP {\bf 0203} (2002) 019},
  \hre{hep-th}{0202037}.
  %%CITATION = doi:10.1088/1126-6708/2002/03/019;%%
  %100 citations counted in INSPIRE as of 19 Aug 2019



  \bibitem{vDVZ} H.~van Dam and M.~J.~G.~Veltman,
{\em ``Massive and massless Yang-Mills and gravitational fields,''}
\hrj{10.1016/0550-3213(70)90416-5}{Nucl. Phys. B \textbf{22} (1970), 397-411};\\
%1046 citations counted in INSPIRE as of 29 Aug 2020
V.~I.~Zakharov,
{\em ``Linearized gravitation theory and the graviton mass,''}
\href{http://www.jetpletters.ac.ru/ps/index-v-12_en.shtml}{JETP Lett. \textbf{12} (1970), 312};
%765 citations counted in INSPIRE as of 29 Aug 2020
 [Pisma Zh. Eksp. Teor. Fiz. 12 (1970) 447].

\bibitem{Damour}
T.~Damour and A.~M.~Polyakov,
{\em ``The String dilaton and a least coupling principle,''}
\hrj{10.1016/0550-3213(94)90143-0}{Nucl. Phys. B \textbf{423} (1994), 532-558};
\hre{hep-th}{9401069}.
%874 citations counted in INSPIRE as of 07 Oct 2020



\bibitem{Migdal}
A.~A.~Migdal and M.~A.~Shifman,
{\em ``Dilaton Effective Lagrangian in Gluodynamics,''}
\hrj{10.1016/0370-2693(82)90089-2}{Phys. Lett. B \textbf{114} (1982), 445-449}.
%207 citations counted in INSPIRE as of 08 Oct 2020


\bibitem{Li}
E.~Kiritsis, W.~Li and F.~Nitti,
{\em ``Holographic RG flow and the Quantum Effective Action,''}
\hrj{10.1002/prop.201400007}{Fortsch. Phys. \textbf{62} (2014), 389-454};
\hri{1401.0888}{ [hep-th]};\\
%47 citations counted in INSPIRE as of 08 Oct 2020
{\em ``On the gluonic operator effective potential in holographic Yang-Mills theory,''}
\hrj{10.1007/JHEP04(2015)125}{JHEP \textbf{04} (2015), 125};
\hri{1410.1091}{ [hep-th]}.
%6 citations counted in INSPIRE as of 08 Oct 2020


\bibitem{Emparan}
R.~Emparan, G.~T.~Horowitz and R.~C.~Myers,
{\em ``Exact description of black holes on branes,''}
\hrj{10.1088/1126-6708/2000/01/007}{JHEP \textbf{01} (2000), 007};
\hre{hep-th}{9911043}\\;
%325 citations counted in INSPIRE as of 08 Oct 2020
R.~Emparan, A.~Fabbri and N.~Kaloper,
{\em ``Quantum black holes as holograms in AdS brane worlds,''}
\hrj{10.1088/1126-6708/2002/08/043}{JHEP \textbf{08} (2002), 043};
\hre{hep-th}{0206155}\\;
%211 citations counted in INSPIRE as of 08 Oct 2020
R.~Emparan, A.~M.~Frassino and B.~Way,
{\em ``Quantum BTZ black hole,''}
\hri{2007.15999}{[hep-th]}.

\bibitem{Figueras}
P.~Figueras,
{\em ``Braneworld Black Holes,''}
\hrj{10.1007/978-3-642-40157-2\_3}{Springer Proc. Math. Stat. \textbf{60} (2014), 37-53};\\
%3 citations counted in INSPIRE as of 08 Oct 2020
P.~Figueras and T.~Wiseman,
{\em ``Gravity and large black holes in Randall-Sundrum II braneworlds,''}
\hrj{10.1103/PhysRevLett.107.081101}{Phys. Rev. Lett. \textbf{107} (2011), 081101};
[\hri{1105.2558}{ [hep-th]}.
%83 citations counted in INSPIRE as of 08 Oct 2020





%\cite{Almheiri:2020cfm}
\bibitem{Almheiri:2020cfm}
 A.~Almheiri, T.~Hartman, J.~Maldacena, E.~Shaghoulian and A.~Tajdini,
{\em ``The entropy of Hawking radiation,''}
\hri{2006.06872}{[hep-th]}.
%24 citations counted in INSPIRE as of 30 Sep 2020







 \bibitem{review}
  E.~Kiritsis,
  {\em ``D-branes in standard model building, gravity and cosmology,''}
  Phys.\ Rept.\  {\bf 421} (2005) 105
   Erratum: [Phys.\ Rept.\  {\bf 429} (2006) 121];
  \hre{hep-th}{0310001}.
  %%CITATION = doi:10.1016/j.physrep.2005.09.001;%%
  %208 citations counted in INSPIRE as of 25 Apr 2018


  \bibitem{Milgrom}
M.~Milgrom,
{\em ``A Modification of the Newtonian dynamics as a possible alternative to the hidden mass hypothesis,''}
\hrj{10.1086/161130}{Astrophys. J. \textbf{270} (1983), 365-370}.
%2045 citations counted in INSPIRE as of 13 Oct 2020

\bibitem{Skordis}
C.~Skordis and T.~Zlosnik,
{\em ``A new relativistic theory for Modified Newtonian Dynamics,''}
\hri{2007.00082}{[astro-ph.CO]}.
%4 citations counted in INSPIRE as of 13 Oct 2020

 \bibitem{quiros}
  A.~Delgado, J.~R.~Espinosa and M.~Quiros,
  {\em ``Unparticles Higgs Interplay,''}
  JHEP {\bf 0710} (2007) 094
  doi:10.1088/1126-6708/2007/10/094
  \hri{0707.4309}{[hep-ph]}.
  %%CITATION = doi:10.1088/1126-6708/2007/10/094;%%
  %106 citations counted in INSPIRE as of 20 Mar 2019


  %\cite{Kiritsis:2012ta}
\bibitem{lorentz}
  E.~Kiritsis,
  {\em ``Lorentz violation, Gravity, Dissipation and Holography,''}
  JHEP {\bf 1301} (2013) 030
  doi:10.1007/JHEP01(2013)030
\hri{1207.2325}{[hep-th]}.
  %%CITATION = doi:10.1007/JHEP01(2013)030;%%
  %38 citations counted in INSPIRE as of 18 Jun 2018



  \bibitem{wa}
  L.~Alvarez-Gaume and E.~Witten,
  {\em ``Gravitational Anomalies,''}
  \href{http://www.doi.org/10.1016/0550-3213(84)90066-X}{Nucl.\ Phys.\ B {\bf 234} (1984) 269}.

  %%CITATION = doi:10.1016/0550-3213(84)90066-X;%%
  %1524 citations counted in INSPIRE as of 19 Mar 2019


  %\cite{Warr:1986we}
\bibitem{warr}
  B.~J.~Warr,
  {\em ``Renormalization of Gauge Theories Using Effective Lagrangians. 1.,''}
  \href{http://www.doi.org/10.1016/0003-4916(88)90245-X}{Annals Phys.\  {\bf 183} (1988) 1};\\
  %%CITATION = doi:10.1016/0003-4916(88)90245-X;%%
  %87 citations counted in INSPIRE as of 19 Mar 2019
  {\em ``Renormalization of Gauge Theories Using Effective Lagrangians. 2.,''}
  \href{http://www.doi.org/10.1016/0003-4916(88)90246-1}{Annals Phys.\  {\bf 183} (1988) 59};\\
  %%CITATION = doi:10.1016/0003-4916(88)90246-1;%%
  %45 citations counted in INSPIRE as of 19 Mar 2019
  {\em ``The Regularization And Renormalization Of Gauge Theories,''}
  \href{https://search.proquest.com/docview/303552499}{PhD thesis: UMI-88-12034}.
  %%CITATION = UMI-88-12034;%%



  \bibitem{tom}
  J.~M.~Cornwall, R.~Jackiw and E.~Tomboulis,
  {\em ``Effective Action for Composite Operators,''}
  \href{http://www.doi.org/10.1103/PhysRevD.10.2428}{Phys.\ Rev.\ D {\bf 10} (1974) 2428}.
  %%CITATION = doi:10.1103/PhysRevD.10.2428;%%
  %1405 citations counted in INSPIRE as of 19 Mar 2019



  \bibitem{morris}
  T.~R.~Morris and A.~W.~H.~Preston,
  {\em ``Manifestly diffeomorphism invariant classical Exact Renormalization Group,''}
  JHEP {\bf 1606} (2016) 012
  doi:10.1007/JHEP06(2016)012
\hri{1602.08993}{[hep-th]}.;\\
  %%CITATION = doi:10.1007/JHEP06(2016)012;%%
  %20 citations counted in INSPIRE as of 06 Mar 2020
P.~Labus, T.~R.~Morris and Z.~H.~Slade,
  {\em ``Background independence in a background dependent renormalization group,''}
  Phys.\ Rev.\ D {\bf 94} (2016) no.2,  024007
  doi:10.1103/PhysRevD.94.024007
 \hri{1603.04772}{[hep-th]}.
  %%CITATION = doi:10.1103/PhysRevD.94.024007;%%


 \bibitem{a2}
  K.~A.~Intriligator,
  {\em ``Maximally supersymmetric RG flows and AdS duality,''}
  Nucl.\ Phys.\ B {\bf 580} (2000) 99
  doi:10.1016/S0550-3213(99)00803-2;
  \hre{hep-th}{9909082}.
  %%CITATION = doi:10.1016/S0550-3213(99)00803-2;%%
  %41 citations counted in INSPIRE as of 18 Jun 2018

  \bibitem{a3}
  M.~S.~Costa,
  {\em ``Absorption by double centered D3-branes and the Coulomb branch of N=4 SYM theory,''}
  JHEP {\bf 0005} (2000) 041
  doi:10.1088/1126-6708/2000/05/041;
  \hre{hep-th}{9912073};\\
  %%CITATION = doi:10.1088/1126-6708/2000/05/041;%%
  %15 citations counted in INSPIRE as of 18 Jun 2018
  {\em ``A Test of the AdS / CFT duality on the Coulomb branch,''}
  Phys.\ Lett.\ B {\bf 482} (2000) 287
   Erratum: [Phys.\ Lett.\ B {\bf 489} (2000) 439]
  doi:10.1016/S0370-2693(00)00484-6, 10.1016/S0370-2693(00)00939-4;
  \hre{hep-th}{0003289}.
  %%CITATION = doi:10.1016/S0370-2693(00)00484-6, 10.1016/S0370-2693(00)00939-4;%%
  %15 citations counted in INSPIRE as of 18 Jun 2018




  \bibitem{e2}
  E.~Kiritsis and V.~Niarchos,
  {\em ``Interacting String Multi-verses and Holographic Instabilities of Massive Gravity,''}
  Nucl.\ Phys.\ B {\bf 812} (2009) 488;   doi:10.1016/j.nuclphysb.2008.12.010;
  \hri{0808.3410}{[hep-th]};\\
  %%CITATION = doi:10.1016/j.nuclphysb.2008.12.010;%%
  %31 citations counted in INSPIRE as of 18 Jun 2018
{\em ``(Multi)Matrix Models and Interacting Clones of Liouville Gravity,''}
  JHEP {\bf 0808} (2008) 044
  doi:10.1088/1126-6708/2008/08/044;
  \hri{0805.4234}{[hep-th]}.
  %%CITATION = doi:10.1088/1126-6708/2008/08/044;%%
  %13 citations counted in INSPIRE as of 18 Jun 2018






\bibitem{Wet}
A.~Platania and C.~Wetterich,
{\em ``Non-perturbative unitarity and fictitious ghosts in quantum gravity,''}
\hri{2009.06637}{[hep-th]}.
%0 citations counted in INSPIRE as of 16 Sep 2020

%\cite{Ginsparg:1993is}
\bibitem{Ginsparg:1993is}
P.~H.~Ginsparg and G.~W.~Moore,
{\em ``Lectures on 2-D gravity and 2-D string theory,''}
\hri{hep-th/9304011}.
%526 citations counted in INSPIRE as of 30 Sep 2020

%\cite{Betzios:2020nry}
\bibitem{Betzios:2020nry}
P.~Betzios and O.~Papadoulaki,
{\em ``Liouville theory and Matrix models: A Wheeler DeWitt perspective,''}
\hrj{10.1007/JHEP09(2020)125}{JHEP \textbf{09} (2020), 125};
\hri{2004.00002}{[hep-th]}.
%9 citations counted in INSPIRE as of 30 Sep 2020


  \bibitem{DR}
  C.~Deffayet and J.~W.~Rombouts,
  {\em ``Ghosts, strong coupling and accidental symmetries in massive gravity,''}
  Phys.\ Rev.\ D {\bf 72} (2005) 044003
  doi:10.1103/PhysRevD.72.044003
  \hre{gr-qc}{0505134}.
  %%CITATION = doi:10.1103/PhysRevD.72.044003;%%
  %110 citations counted in INSPIRE as of 20 Mar 2019


  \bibitem{KR}
  A.~Karch and L.~Randall,
  {\em ``Locally localized gravity,''}
  JHEP {\bf 0105} (2001) 008
  doi:10.1088/1126-6708/2001/05/008
  \hre{hep-th}{0011156}.
  %%CITATION = doi:10.1088/1126-6708/2001/05/008;%%
  %365 citations counted in INSPIRE as of 20 Mar 2019


\bibitem{worm}
P.~Betzios, E.~Kiritsis and O.~Papadoulaki,
{\em ``Euclidean Wormholes and Holography,''}
\hrj{10.1007/JHEP06(2019)}{042JHEP \textbf{06} (2019), 042};
\hri{1903.05658}{[hep-th]}.
%10 citations counted in INSPIRE as of 10 Sep 2020


\bibitem{CM}
  G.~Compere and D.~Marolf,
  {\em ``Setting the boundary free in AdS/CFT,''}
  \hree{10.1088/0264-9381/25/19/195014}{Class.\ Quant.\ Grav.\  {\bf 25} (2008) 195014},
  \hri{0805.1902}{[hep-th]};\\
  %%CITATION = doi:10.1088/0264-9381/25/19/195014;%%
  %132 citations counted in INSPIRE as of 19 Aug 2019
T.~Andrade, D.~Marolf and C.~Deffayet,
{\em ``Can Hamiltonians be boundary observables in Parametrized Field Theories?,''}
\hrj{10.1088/0264-9381/28/10/105002}{Class. Quant. Grav. \textbf{28} (2011), 105002};
\hri{1010.2535}{[gr-qc]}.
%7 citations counted in INSPIRE as of 08 Oct 2020



\bibitem{solo}
  S.~de Haro, S.~N.~Solodukhin and K.~Skenderis,
  {\em ``Holographic reconstruction of space-time and renormalization in the AdS / CFT correspondence,''}
  \hree{10.1007/s002200100381}{Commun.\ Math.\ Phys.\  {\bf 217} (2001) 595},
  \hre{hep-th}{0002230}.
  %%CITATION = doi:10.1007/s002200100381;%%
  %1222 citations counted in INSPIRE as of 19 Aug 2019

\bibitem{Bianchi:2001kw}
  M.~Bianchi, D.~Z.~Freedman and K.~Skenderis,
  {\em ``Holographic renormalization,''}
  \hrj{10.1016/S0550-3213(02)00179-7}{Nucl.\ Phys.\ B {\bf 631} (2002) 159},
  \hre{hep-th}{0112119};\\
  %%CITATION = doi:10.1016/S0550-3213(02)00179-7;%%
  %479 citations counted in INSPIRE as of 19 Aug 2019
  {\em ``How to go with an RG flow,''}
  \hrj{10.1088/1126-6708/2001/08/041}{JHEP {\bf 0108} (2001) 041},
  \hre{hep-th}{0105276}.
  %%CITATION = doi:10.1088/1126-6708/2001/08/041;%%
  %298 citations counted in INSPIRE as of 19 Aug 2019

 \bibitem{large}
N.~Arkani-Hamed, S.~Dimopoulos and G.~R.~Dvali,
{\em ``The Hierarchy problem and new dimensions at a millimeter,''}
\hrj{10.1016/S0370-2693(98)00466-3}{Phys. Lett. B \textbf{429} (1998), 263-272};
\hre{hep-ph}{9803315};\\
I.~Antoniadis, N.~Arkani-Hamed, S.~Dimopoulos and G.~R.~Dvali,
{\em ``New dimensions at a millimeter to a Fermi and superstrings at a TeV,''}
\hrj{10.1016/S0370-2693(98)00860-0}{Phys. Lett. B \textbf{436} (1998), 257-263};
\hre{hep-ph}{9804398}.
%6782 citations counted in INSPIRE as of 28 Sep 2020


\bibitem{RS}
  L.~Randall and R.~Sundrum,
  {\em ``An Alternative to compactification,''}
  \hrj{10.1103/PhysRevLett.83.4690}{Phys.\ Rev.\ Lett.\  {\bf 83} (1999) 4690},
  \hre{hep-th}{9906064}.
  %%CITATION = doi:10.1103/PhysRevLett.83.4690;%%
  %6459 citations counted in INSPIRE as of 19 Aug 2019



 \bibitem{DN}
G.~R.~Dvali, G.~Gabadadze, M.~Kolanovic and F.~Nitti,
{\em ``Scales of gravity,''}
\hrj{10.1103/PhysRevD.65.024031}{Phys. Rev. D \textbf{65} (2002), 024031};
\hre{hep-th}{0106058}.
%161 citations counted in INSPIRE as of 17 Sep 2020

\bibitem{DGZ}
G.~Dvali, A.~Gruzinov and M.~Zaldarriaga,
{\em ``The Accelerated universe and the moon,''}
\hrj{10.1103/PhysRevD.68.024012}{Phys. Rev. D \textbf{68} (2003), 024012};
\hri{hep-ph}{0212069}.
%188 citations counted in INSPIRE as of 18 Sep 2020














\bibitem{vain}
A.~I.~Vainshtein,
{\em ``To the problem of nonvanishing gravitation mass,''}
\hrj{10.1016/0370-2693(72)90147-5}{Phys. Lett. B \textbf{39} (1972), 393-394}\\
%1268 citations counted in INSPIRE as of 29 Aug 2020
E.~Babichev and C.~Deffayet,
{\em ``An introduction to the Vainshtein mechanism,''}
\hrj{10.1088/0264-9381/30/18/184001}{Class. Quant. Grav. \textbf{30} (2013), 184001};
\hri{1304.7240}{ [gr-qc]}.


\bibitem{vai} E. Babichev, C. Markou, E. Kiritsis, F. Nitti, work in progress.


\bibitem{weak}
N.~Arkani-Hamed, L.~Motl, A.~Nicolis and C.~Vafa,
{\em ``The String landscape, black holes and gravity as the weakest force,''}
\hrj{10.1088/1126-6708/2007/06/060}{JHEP \textbf{06} (2007), 060};
\hre{hep-th}{0601001}.
%718 citations counted in INSPIRE as of 28 Sep 2020


\bibitem{rosen}
R.~A.~Rosen,
{\em ``Black Hole Mechanics for Massive Gravitons,''}
\hrj{10.1103/PhysRevD.98.104008}{Phys. Rev. D \textbf{98} (2018) no.10, 104008};
\hri{1805.12135}{ [hep-th]}.
%3 citations counted in INSPIR}{E as of 28 Sep 2020

























\end{thebibliography}
\end{document}